\documentclass[12pt,twoside]{report}

\usepackage{float}
\usepackage{makeidx}
\usepackage{eepic}
\usepackage{amsmath,amsfonts,amssymb}
\usepackage{umlaute}
\usepackage{fancyhdr}
\usepackage{theorem}
\usepackage{xspace}
\usepackage{psfig,epsfig}
\usepackage{rotating}
\usepackage{a4wide}
\usepackage{euscript}
\usepackage{fancyhdr}
\usepackage{myfancyheadings}
\usepackage{url}
\usepackage{theapa}

\floatstyle{ruled}
\restylefloat{figure}

\def\qed{{\unskip\nobreak\hfil\penalty50
   \hskip2em\hbox{}\nobreak\hfil
   \qedd
   \parfillskip=0pt \finalhyphendemerits=0
    \medskip\goodbreak\noindent}}
\def\qedd{\vrule height4pt width 4pt depth0pt}

\def\eod{{\unskip\nobreak\hfil\penalty50
   \hskip2em\hbox{}\nobreak\hfil
   \mbox{$\diamond$}
   \parfillskip=0pt \finalhyphendemerits=0
    \medskip\goodbreak\noindent}}

\theoremstyle{break}
\newtheorem{theorem}{Theorem}[chapter]
\newtheorem{definition}[theorem]{Definition}

\newtheorem{fact}[theorem]{Fact}
\newtheorem{lemma}[theorem]{Lemma}
\newtheorem{corollary}[theorem]{Corollary}

\newtheorem{algorithm}[theorem]{Algorithm}
\theorembodyfont{\rmfamily}
\theoremstyle{plain}
\newtheorem{claim}[theorem]{\normalfont \textsc{Claim}}

\newenvironment{proof}{\noindent \textbf{Proof.}}{}

\newcommand{\N}{\ensuremath{\mathbb{N}}\xspace}
\newcommand{\Z}{\ensuremath{\mathbb{Z}}\xspace}
\def\pos{\textit{pos}\xspace}
\def\xpos{\text{xpos}\xspace}
\def\ypos{\text{ypos}\xspace}

\newcommand{\pspace}{\textsc{PSpace}\xspace}
\newcommand{\npspace}{\textsc{NPSpace}\xspace}
\newcommand{\exptime}{\textsc{ExpTime}\xspace}
\newcommand{\nexptime}{\textsc{NExpTime}\xspace}

\newcommand{\alc}{\ensuremath{\mathcal{A}\!\mathcal{L}\!\mathcal{C}}\xspace}
\newcommand{\alcq}{\ensuremath{\mathcal{A}\!\mathcal{L}\!\mathcal{C}\!\mathcal{Q}}\xspace}
\newcommand{\alcqo}{\ensuremath{\mathcal{A}\!\mathcal{L}\!\mathcal{C}\!\mathcal{Q}\!\mathcal{O}}\xspace}

\newcommand{\alcnr}{\ensuremath{\mathcal{A}\!\mathcal{L}\!\mathcal{C}\!\mathcal{N}\!\mathcal{R}}\xspace}

\newcommand{\alcqi}{\ensuremath{\mathcal{A}\!\mathcal{L}\!\mathcal{C}\!\mathcal{Q}\!\mathcal{I}}\xspace}
\newcommand{\alci}{\ensuremath{\mathcal{A}\!\mathcal{L}\!\mathcal{C}\!\mathcal{I}}\xspace}
\newcommand{\alcqio}{\ensuremath{\mathcal{A}\!\mathcal{L}\!\mathcal{C}\!\mathcal{Q}\!\mathcal{I}\!\mathcal{O}}\xspace}

\newcommand{\alcqir}{\ensuremath{\mathcal{A}\!\mathcal{L}\!\mathcal{C}\!\mathcal{Q}\!\mathcal{I}\!b}\xspace}
\newcommand{\alcqiro}{\ensuremath{\mathcal{A}\!\mathcal{L}\!\mathcal{C}\!\mathcal{Q}\!\mathcal{I}\!b\!\mathcal{O}}\xspace}
\newcommand{\alcqib}{\ensuremath{\mathcal{A}\!\mathcal{L}\!\mathcal{C}\!\mathcal{Q}\!\mathcal{I}\!\mathcal{B}}\xspace}
\newcommand{\alcqibo}{\ensuremath{\mathcal{A}\!\mathcal{L}\!\mathcal{C}\!\mathcal{Q}\!\mathcal{I}\!\mathcal{B}\!\mathcal{O}}\xspace}
\newcommand{\alcqbb}{\ensuremath{\mathcal{A}\!\mathcal{L}\!\mathcal{C}\!\mathcal{Q}(\mathcal{I})\mathcal{B}}\xspace}
\newcommand{\alcqb}{\ensuremath{\mathcal{A}\!\mathcal{L}\!\mathcal{C}\!\mathcal{Q}\!\mathcal{B}}\xspace}
\newcommand{\alcb}{\ensuremath{\mathcal{A}\!\mathcal{L}\!\mathcal{C}\!\mathcal{B}}\xspace}

\newcommand{\shiq}{\ensuremath{\mathcal{S}\!\mathcal{H}\!\mathcal{I}\!\mathcal{Q}}\xspace}
\newcommand{\sh}{\ensuremath{\mathcal{S}\!\mathcal{H}\xspace}}
\newcommand{\shf}{\ensuremath{\mathcal{S}\!\mathcal{H}\!\mathcal{F}\xspace}}
\newcommand{\shn}{\ensuremath{\mathcal{S}\!\mathcal{H}\!\mathcal{N}\xspace}}
\newcommand{\shiqo}{\ensuremath{\mathcal{S}\!\mathcal{H}\!\mathcal{I}\!\mathcal{Q}\!\mathcal{O}}\xspace}
\newcommand{\shif}{\ensuremath{\mathcal{S}\!\mathcal{H}\!\mathcal{I}\!\mathcal{F}}\xspace}
\newcommand{\s}{\ensuremath{\mathcal{S}}\xspace}
\newcommand{\si}{\ensuremath{\mathcal{S}\!\mathcal{I}}\xspace}
\renewcommand{\sin}{\ensuremath{\mathcal{S}\!\mathcal{I}\!\mathcal{N}}\xspace}
\newcommand{\CPDL}{\ensuremath{\mathsf{CPDL}}\xspace}
 
\newcommand{\alcR}{\ensuremath{\mathcal{A}\!\mathcal{L}\!\mathcal{C}_{R^+}}\xspace} 
\newcommand{\ciq}{\ensuremath{\mathcal{C}\!\mathcal{I}\!\mathcal{Q}}\xspace}

\newcommand{\grkr}{\ensuremath{\mathbf{Gr}(\K_\mathcal{R})}\xspace}
\newcommand{\grkrri}{\ensuremath{\mathbf{Gr}(\K_{\mathcal{R}^{-1}_\cap})}\xspace}

\newcommand{\kr}{\ensuremath{\K_m}\xspace}

\newcommand{\Inv}{\ensuremath{\mathsf{Inv}}}
\newcommand{\Tr}{\ensuremath{\mathsf{Trans}}}

\newcommand{\alcRtrans}{
        \ensuremath{\alc_{R^{+}}}\xspace}

\newcommand{\K}{\textsf{K}\xspace}

\newcommand{\genericrule}[1]{\ensuremath{\mathbin{\rightarrow_#1}}\xspace}
\newcommand{\ruleand}{\genericrule{\sqcap}}
\newcommand{\ruleor}{\genericrule{\sqcup}}
\newcommand{\ruleleq}{\genericrule{{\leq}}}
\newcommand{\rulegeq}{\genericrule{{\geq}}}
\newcommand{\ruleex}{\genericrule{{\exists}}}
\newcommand{\rulefa}{\genericrule{{\forall}}}
\newcommand{\rulefatr}{\genericrule{{\forall_+}}}
\newcommand{\ruleleqz}{\genericrule{{\leq 0}}}
\newcommand{\rulechoose}{\genericrule{{\text{choose}}}}

\newcommand{\Names}{\ensuremath{\mathsf{NC}}\xspace}
\newcommand{\Roles}{\ensuremath{\mathsf{NR}}\xspace}
\newcommand{\RolesTrans}{\ensuremath{\mathsf{NR}^+}\xspace}
\newcommand{\Rplus}{\ensuremath{\mathsf{NR}^+}\xspace}
\newcommand{\Individuals}{\ensuremath{\mathsf{NI}}\xspace}

\newcommand{\all}[2]{\forall #1 . #2}
\newcommand{\some}[2]{\ensuremath{\exists #1 . #2}\xspace}
\newcommand{\qnrgeq}[3]{({\geq} #1 \; #2 \; #3)}
\newcommand{\qnrleq}[3]{({\leq} #1 \; #2 \; #3)}
\newcommand{\qnrgleq}[3]{({\bowtie} #1 \; #2 \; #3)}
\newcommand{\nrgeq}[2]{({\geq} #1 \; #2)}
\newcommand{\nrleq}[2]{({\leq} #1 \; #2)}
\newcommand{\feat}[1]{\nrleq 1 #1}
\newcommand{\crleq}[2]{({\leq} #1 \; #2)}
\newcommand{\crgeq}[2]{({\geq} #1 \; #2)}

\newcommand{\crall}[1]{(\forall \; #1)}

\newcommand{\I}{\mathcal{I}}
\newcommand{\domain}{\ensuremath{\Delta^\I}\xspace}
\newcommand{\ifunc}{^\I}
\newcommand{\Tree}{
        \ensuremath{\mathbf{T}}\xspace}
\newcommand{\Card}[1]{
        \ensuremath{\sharp#1}}

\newcommand{\mybigsqcap}{\mathop{\raisebox{1.3ex}{\begin{turn}{180}$\displaystyle\bigsqcup$\end{turn}}}}

\newcommand{\tuple}[2]{( #1, #2)}
\newcommand{\Implies}{\text{ implies }}
\renewcommand{\And}{\text{ and }}

\newcommand{\bnfor}{\; | \;}
\newcommand{\wrt}{w.r.t.\ }
\newcommand{\nneg}{\mathopen{\sim}}
\newcommand{\canI}{\ensuremath{{\I_\mcA}}}
\def\E{\exists}

\newcommand{\Edges}{\ensuremath{\EuScript{E}}\xspace}

\def\mfA{{\ensuremath{\mathfrak{A}}}}

\def\mfF{{\ensuremath{\mathfrak{F}}}}

\def\mfM{{\ensuremath{\mathfrak{M}}}}

\def\mcA{{\ensuremath{\mathcal{A}}}}
\def\mcC{{\ensuremath{\mathcal{C}}}}
\def\mcD{{\ensuremath{\mathcal{D}}}}
\def\mcR{{\ensuremath{\mathcal{R}}}}

\def\mcK{{\ensuremath{\mathcal{K}}}}
\def\mcS{{\ensuremath{\mathcal{S}}}}
\def\mcX{{\ensuremath{\mathcal{X}}}}

\def\mbP{{\ensuremath{\mathbf{P}}}}
\def\mcO{{\ensuremath{\mathcal{O}}}}
\def\mcT{{\ensuremath{\mathcal{T}}}}

\def\bfB{{\ensuremath{\mathbf{B}}}}
\def\bfE{{\ensuremath{\mathbf{E}}}}
\def\bfL{{\ensuremath{\mathbf{L}}}}
\def\bfS{{\ensuremath{\mathbf{S}}}}
\def\bfV{{\ensuremath{\mathbf{V}}}}

\def\bfC{\mathbf{C}}
\def\bfE{\mathbf{E}}
\def\bfT{\mathbf{T}}

\def\bfN{\mathbf{N}}
\def\bfx{\mathbf{x}}
\def\bfy{\mathbf{y}}
\def\bfz{\mathbf{z}}

\def\DLfont{\texttt}
\def\human{\DLfont{Human}\xspace}
\def\male{\DLfont{Male}\xspace}
\def\rich{\DLfont{Rich}\xspace}
\def\female{\DLfont{Female}\xspace}
\def\hasChild{\DLfont{has\_child}\xspace}
\def\hasParent{\DLfont{has\_parent}\xspace}
\def\hasSibling{\DLfont{has\_sibling}\xspace}
\def\hasOffspring{\DLfont{has\_offspring}\xspace}
\def\hasAncestor{\DLfont{has\_ancestor}\xspace}
\def\marriedTo{\DLfont{married\_to}\xspace}
\def\rich{\DLfont{Rich}\xspace}
\def\parent{\DLfont{Parent}\xspace}

\def\husband{\DLfont{Husband}\xspace}
\def\Mary{\DLfont{MARY}\xspace}
\def\Peter{\DLfont{PETER}\xspace}

\def\Bob{\DLfont{BOB}\xspace}

\newcommand{\ndoteq}{\mathbin{\not \doteq}}
\newcommand{\TabLab}{\mathcal{L}}
\newcommand{\Lab}{\bfL}
\newcommand{\BLab}{\bfB}
\newcommand{\sub}{\mathit{sub}}
\newcommand{\clos}{\mathit{clos}}
\newcommand{\klone}{\textsc{kl-one}\xspace}
\newcommand{\FaCT}{\text{FaCT}\xspace}
\newcommand{\iFaCT}{\text{iFaCT}\xspace}

\newcommand{\classic}{\textsc{classic}}

\newcommand{\ctwo}{\ensuremath{C^2}\xspace}
\newcommand{\ltwo}{\ensuremath{L^2}\xspace}

\newcommand{\tab}[1]{\textsf{(T#1)}}
\newcommand{\aut}[1]{\textsf{(A#1)}}
\newcommand{\run}[1]{\textsf{(R#1)}}

\newcommand{\bfif}{\textbf{if}}
\newcommand{\bfthen}{\textbf{then}}
\newcommand{\bfor}{\textbf{or}}

\def\sss{{\sqsubseteq^*}}

\newcommand{\dotminus}{\mathbin{\dot -}}

\def\nmax{n_\text{max}} 
\def\excon{\textsf{succ}} 
\def\lc{\textsf{lc}}
\def\limit{\textsf{limit}}
\def\NNF{\textit{NNF}}
\def\trans{{\textit{tr}}}
\def\Ru{{R^\uparrow}}
\def\Su{{S^\uparrow}}
\def\Tu{{T^\uparrow}}

\def\safe{{\textit{safe}}}

\makeatletter
\def\@dotted#1#2{\m@th\ooalign{$\hfil#1\mkern0.27mu\ast\hfil$\crcr$#1#2$}}
\newcommand*{\capast}{\mathbin{\m@th\mathpalette\@dotted\cap}}
\makeatother

\newcommand{\iemph}[1]{\emph{#1}\index{#1}}
\newcommand{\iiemph}[2]{\emph{#1}\index{#2}}

\def\EXPTIME{{\ensuremath{\mbox{\sc ExpTime}}}}

\def\FO{{\ensuremath{\mathrm{FO}}}}
\def\GF{{\ensuremath{\mathrm{GF}}}}

\def\LGF{{\ensuremath{\mathrm{LGF}}}}
\def\CGF{{\ensuremath{\mathrm{CGF}}}}

\def\NNF{{\ensuremath{\mathrm{NNF}}}}

\def\free{{\ensuremath{\mathrm{free}}}}
\def\width{{\ensuremath{\mathrm{width}}}}

\def\mfA{{\ensuremath{\mathfrak{A}}}}

\def\mfF{{\ensuremath{\mathfrak{F}}}}

\def\mcA{{\ensuremath{\mathcal{A}}}}
\def\mcR{{\ensuremath{\mathcal{R}}}}

\def\mcO{{\ensuremath{\mathcal{O}}}}
\def\mcT{{\ensuremath{\mathcal{T}}}}
\def\mcL{{\ensuremath{\mathcal{L}}}}

\def\mbR{{\ensuremath{\mathbf{R}}}}

\def\setN{\ensuremath{\mathbb{N}}}

\newcommand{\nc}{\newcommand}

\newcounter{menum}

\newcounter{aenum}

\nc{\look}{\begin{proof}}
\nc{\hx}{\end{proof}}

\nc{\eqa}{\begin{eqnarray}}
\nc{\eeqa}{\end{eqnarray}}
\nc{\nn}{\nonumber}
\nc{\neqa}{\nonumber\end{eqnarray}}

\nc{\ex}{\exists}
\nc{\fa}{\forall}
\nc{\mb}{\mbox}

\nc{\bfa}{\mathbf{a}}
\nc{\bfb}{\mathbf{b}}
\nc{\bfc}{\mathbf{c}}
\nc{\bfp}{\mathbf{p}}
\nc{\bfq}{\mathbf{q}}
\nc{\bsbfa}{{\tilde\bfa}}
\nc{\bsbfb}{{\tilde\bfb}}
\nc{\bsbfc}{{\tilde\bfc}}
\nc{\bsa}{{\tilde a}}
\nc{\bsb}{{\tilde b}}
\nc{\bsc}{{\tilde c}}
\nc{\bsd}{{\tilde d}}

\nc{\ra}{\rightarrow}
\nc{\Ra}{\Rightarrow}
\nc{\Lra}{\Longrightarrow}
\nc{\lra}{\leftrightarrow}
\nc{\la}{\leftarrow}
\nc{\La}{\Leftarrow}
\nc{\Lla}{\Longleftarrow}
\nc{\lmt}{\longmapsto}
\nc{\IFF}{\Longleftrightarrow}
\nc{\dla}{{\dashleftarrow}}
\nc{\dra}{{\dashrightarrow}}
\nc{\lrq}{{\leftrightsquigarrow}}
\nc{\rsquig}{{\rightsquigarrow}}

\newcommand{\Paths}{\ensuremath{\mathsf{Paths}}}
\newcommand{\Tail}{\mathsf{Tail}}
\newcommand{\apclass}[2]{[#1,#2]_\approx}
\newcommand{\pair}[2]{\frac{#1}{#2}}
\newcommand{\ubpair}[1]{\frac{#1}{#1}}
\newcommand{\bpair}[1]{\frac{#1}{#1'}}

\newcommand{\gtast}{\mathbin{\geq^*}}

\newcommand{\inast}{\mathbin{{{\in^*}}}}

\title{Complexity Results and Practical Algorithms for Logics in Knowledge 
Representation}

\author{Von der Fakult\"at f\"ur Mathematik, Informatik und
  Naturwissenschaften
  der\\
  Rheinisch-Westf\"alischen Technischen Hochschule Aachen\\
  zur Erlangung des akademischen Grades\\
 eines Doktors der Naturwissenschaften genehmigte Dissertation\\[20ex]
  von\\Diplom-Informatiker Stephan Tobies\\
  aus Essen\\[5ex]
  Berichter:\\
  Universit\"atsprofessor Dr. Franz Baader\\
  Universit\"atsprofessor Dr. Erich Gr\"adel\\[5ex]
  Tag der m\"undlichen Pr\"ufung: 31.05.2001\\[5ex]
  Diese Dissertation ist auf den Internetseiten der
  Hochschulbibliothek\\ online verf\"ugbar.  }

\date{}

\begin{document}

\vspace{3cm}

\noindent \textsf{\Huge
  \begin{tabular}{l}
    Complexity Results and \\ Practical Algorithms for \\ Logics in Knowledge 
Representation
\end{tabular}
}

\vfill

\noindent \hfill \textsf{\Large
   \begin{tabular}{r}
    Stephan Tobies\\
    \texttt{tobies@cs.rwth-aachen.de}\\
    LuFG Theoretical Computer Science\\
    RWTH Aachen, Germany
  \end{tabular}
  }

\cleardoublepage

\maketitle

\cleardoublepage

\thispagestyle{empty}

~ 

\vspace{4cm}

\hfill {\Large \textsl{For Antje, Johanna, and Annika.}}

\clearpage


\thispagestyle{empty}

~

\vfill

I would not have been able to write this thesis without the help of my family. I
would like to thank my wife Antje, who gave me love and support; my parents, on
whose assistance I could always count throughout the years; my sister, who
didn't stop nagging me about finishing this thesis; Marianne, who shared Antje's
burden of taking care of our twin daughters Johanna and Annika while I was
working overtime. And I would like to thank you, Annika and Johanna, for being
my favourite distraction. Without you, writing this thesis would have
been much less interesting and challenging.

I am also greatly indebted to my colleagues who have contributed to this thesis
in numerous ways. Many of the discussion we had have found there way into this
thesis.  I would like to thank Franz Baader, my thesis advisor, whose
encouraging ``haeh?''  forced me to write better papers; Ulrike Sattler and
Carsten Lutz, who never grew too annoyed with my sloppiness; and all my
colleagues, who happily endured the constant interruptions when I needed to
discuss some new ideas. Thanks for the good time I had while working with you.


\clearpage

\pagestyle{plain}
\pagenumbering{roman}
\setcounter{tocdepth}{2}
\tableofcontents

\cleardoublepage

\setcounter{page}{1}
\pagenumbering{arabic}
\pagestyle{fancy}

\renewcommand{\chaptermark}[1]%
  {\markboth{Chapter \thechapter. #1}{}}
\renewcommand{\sectionmark}[1]%
  {\markright{\thesection\ #1}}

\lhead[\fancyplain{}{\rm \thepage}]{\fancyplain{}{\sl\rightmark}}
\chead{}
\rhead[\fancyplain{}{\sl\leftmark}]{\fancyplain{}{\rm \thepage}}
\cfoot{}



\chapter{Introduction}

Description Logics (DLs) are used in knowledge-based systems to represent and
reason about terminological knowledge of the application domain in a
semantically well-defined manner. They allow the definition of complex concepts
(i.e., classes, unary predicates) and roles (relations, binary predicates) to be
built from atomic ones by the application of a given set of constructors. A DL
system allows concept descriptions to be interrelated and implicit knowledge can
be derived from the explicitly represented knowledge using inference services.

This thesis is concerned with issues of reasoning with DLs and Guarded Logics,
which generalise many of the good properties of DLs to a large fragment of
first-order predicate logic. We study inference algorithms for these logics, both from the
viewpoint of (worst-case) complexity of the algorithms and their usability in
system implementations. This chapter gives a brief introduction to DL systems
and reasoning in DLs. After that, we introduce the specific aspects of DLs we
will be dealing with and motivate their use in knowledge representation.  We
also introduce Guarded Logics and describe why they are interesting from
the viewpoint of DLs. We finish with an overview of the structure of this thesis
and the results we establish.

\section{Description Logic Systems}

\iemph{Description Logics} (DLs) are logical formalisms for representing and
reasoning about conceptual and terminological knowledge of a problem domain.
They have evolved from the knowledge representation formalism of Quillian's
\iemph{semantic networks} \citeyear{quillian67:_word_concep} and Minsky's
\iemph{frame systems} \citeyear{minsky81:_framew_repres_knowl}, as an answer to
the poorly defined semantics of these formalisms~\cite{woods75:_link}. Indeed,
one of the distinguishing features of DLs is the well-defined---usually
Tarski-style, extensional---semantics. DLs are based on the notions of concepts
(classes, unary predicates) and roles (binary relations) and are mainly
characterized by a set of operators that allow complex concepts and roles to be
built from atomic ones.  As an example consider the following concept that
describes fathers having a daughter whose children are all rich, using concept
conjunction ($\sqcap$), and universal ($\forall)$ and existential ($\exists$)
restriction over the role $\hasChild$:
\[
\male \sqcap \some{\hasChild}{(\female \sqcap \all \hasChild \rich)}
\]

\iemph{DL systems} \cite{woods92:_kl_one_famil} employ DLs to represent
knowledge of an application domain and offer inference services based on the
formal semantics of the DL to derive \emph{implicit} knowledge from the
\emph{explicitly} represented facts.


In many DL systems, one can find the following components:
\begin{itemize}
\item a \emph{terminological component} or \iemph{TBox}, which uses the DL to
  formalise the terminological domain knowledge. Usually, such a TBox at least
  allows to introduce abbreviations for complex concepts but also more general
  statements are available in some systems. As an example consider the following
  TBox that formalizes some knowledge about relationships of people, where
  $\bot$  denotes the concept with empty extension (the empty class):
  \begin{align*}
    \parent & = \human \sqcap \some{\hasChild}{\human} \sqcap \all \hasChild
    \human\\
    \husband & = \male \sqcap \some \marriedTo \human\\
    \human & = \male \sqcup \female\\
    \husband & \sqsubseteq \all \marriedTo \female\\
    \male \sqcap \female & = \bot
  \end{align*}
  The first three statements introduce \parent, \husband, and \male as
  abbreviations of more complex concepts. The fourth statement additionally
  requires that instances of \husband must satisfy $\all \marriedTo \female$,
  i.e., that a man, if married, must be married to a woman. Finally, the last statement
  expresses that the concepts \male and \female must be disjoint as their
  intersection is defined to be empty.
\item an \emph{assertional component} or \iemph{ABox}, which formalizes (parts of)
  a concrete situation involving certain individuals. A partial description of
  a concrete family, e.g., might look as this:
  \begin{align*}
    \Mary & : \female \sqcap \parent\\
    \Peter &  : \husband\\
    (\Mary,\Peter) & : \hasChild  
  \end{align*}
  Note, that it is allowed to refer to concepts mentioned in the TBox.
\item an \iemph{inference engine}, which allows implicit knowledge to be derived
  from what has been explicitly stated. One typical inference service is the
  calculation of the \iemph{subsumption hierarchy}, i.e., the arrangement of the
  concepts that occur in the TBox into a quasi-order according to their
  specialisation/generalisation relationship. In our example, this service could
  deduce that both $\male$ and $\female$ are a specialisation of (are subsumed
  by) $\human$. Another example of an inference service is \iemph{instance
    checking}, i.e., determining, whether an individual of the ABox is an
  instance of a certain concept. In our example, one can derive that $\Mary$ has
  a daughter in law (i.e., is an instance of $\some \hasChild {\some \marriedTo
    \female}$) and is an instance of $\lnot \husband$ because the TBox axiom
  $\male \sqcap \female = \bot$ require $\male$ and $\female$ to be disjoint. We
  do not make a \iemph{closed world assumption}, i.e., assertions not present in
  the ABox are not assumed to be false by default.  This makes it impossible to
  infer whether \Peter is an instance of \parent or not because the ABox does
  not contain information that supports or circumstantiates this.
\end{itemize}

\klone~\cite{brachman85:_overv_kl_one_knowl_repres_system}
\index{kl-one@\klone} is usually considered
to be the first DL system. Its representation formalism possesses a formal
semantics and the system allows for the specification of both terminological and
assertional knowledge. The inference services provided by \klone include
calculation of the subsumption hierarchy and instance checking.
Subsequently, a number of systems has been developed that
followed the general layout of \klone. 


\section{Reasoning in Description Logics}

To be useful for applications, a DL system must at least satisfy the following
three criteria: the implemented DL must be capable of capturing an interesting
proportion of the domain knowledge, the system must answer queries in a timely
manner, and the inferences computed by the systems should be accurate. At least,
the inferences should be \iiemph{sound}{sound inferences}, so that every drawn
conclusion is correct.  It is also desirable to have \iiemph{complete}{complete
  inferences} inference, so that every correct conclusion can be drawn.
Obviously, some of these requirements stand in conflict, as a greater
expressivity of a DL makes sound and complete inference more difficult or even
undecidable.  Consequently, theoretical research in DL has mainly focused on the
expressivity of DLs and decidability and complexity of their inference
algorithms.

When developing such inference algorithms, one is interested in their
computational complexity, their behaviour for ``real life'' problems, and, in
case of incomplete algorithms, their ``degree'' of completeness. From a
theoretical point of view, it is desirable to have algorithms that match the
known worst-case complexity of the problem. From the viewpoint of the
application, it is more important to have an easily implementable procedure that
is amenable to optimizations and hence has good run-time behaviour in realistic
applications.

\section{Expressive Description Logics}

Much research in Description Logic has been concerned with the expressivity and
computational properties of various DLs \cite<for an overview of current issues
in DL research, e.g., %
see>{baader01:_descr_logic_handb}.  These investigations were often triggered by
the provision of certain constructors in implemented systems
\cite{Nebel88,Borgida94}, or by the need for these operators in specific
knowledge representation tasks
\cite{BaaderHanschke-GWAI-92,franconi94:treatment_of_plurals,Sattler-Diss-1998}.

In the following we introduce the specific features of the DLs that are
considered in this thesis.

\subsection{Counting}

Since people tend to describe objects by the \emph{number} of other objects they
are related to (``Cars have four wheels, spiders have eight legs, humans have
one head, etc.'') it does not come as a surprise that most DL systems offer
means to capture these aspects.  \iiemph{Number restrictions}{number
  restrictions}, which allow to specify the number of objects related via
certain roles, can already be found in \textsc{kl-one} and have subsequently
been present in nearly all DL systems.  More recent systems, like \FaCT
\index{fact@\FaCT} \cite{Horrocks98c} or \iFaCT \index{ifact@\iFaCT}
\cite{horrocks99:_fact} also allow for \iemph{qualifying number restrictions}
\cite{Hollunder91b}, which, additionally, state requirements on the related
objects. Using number restrictions, it is possible, e.g., to define the concept
of parents having at least two children $(\human \sqcap \nrgeq 2 \hasChild)$, or
of people having exactly two sisters $(\human \sqcap \qnrleq 2 \hasSibling
\female \sqcap \qnrgeq 2 \hasSibling \female)$.

It is not hard to see that, at least for moderately expressive DLs,
reasoning with number restrictions is more involved than reasoning
 with universal or existential restrictions only, as number restrictions
 enforce interactions between role successors. The following concept
describes humans having two daughters and two rich children but at most three
children:
\[
\human \sqcap \qnrgeq 2 \hasChild \female \sqcap \qnrgeq 2 \hasChild \rich
\sqcap \nrleq 3 \hasChild ,
\]
which implies that at least one of the daughters must be rich. This form of
interaction between role successors cannot be created without number
restrictions and has to be dealt with by inference algorithms.

Number restrictions introduce a form of \emph{local} counting into DLs: for an
object it is possible to specify the number of other objects it is related to via
a given role. There are also approaches to augment DLs with a form of
\emph{global} counting. Baader, Buchheit, and Hollunder
\citeyear{BaaderBuchheit+-AIJ-1996} introduce \iiemph{cardinality restrictions
  on concepts}{cardinality restriction} as a terminological formalism that
allows to express constraints on the number of instances that a specific concept
may have in a domain. To stay with our family examples, using cardinality
restrictions it is possible to express that there are at most two earliest
ancestors:
\[
\crleq 2 {( \human \sqcap \nrleq 0 \hasParent)} .
\]

\subsection{Transitive Roles, Role Hierarchies, and Inverse Roles}

In many applications of knowledge representation, like configuration
\cite{wache96:_using_description_logic_for_configuration,sattler96:_know_repr_in_proc_engineering,mcguinness98:_concep},
ontologies \cite{rector:_exper_build_large_re_medic} or various applications in
the database area
\cite{CaLN98,CDLNR98,calvanese99:_data_integ,FrancBaadSattvass-DWQ-Buch},
\emph{aggregated objects} are of key importance. Sattler
\citeyear{Sattler-ECAI-2000} argues that \iemph{transitive roles} and \emph{role hierarchies}
provide elegant means to express various kinds of part-whole relations that can
be used to model aggregated objects. Again, to stay with our family example, it
would be natural to require the $\hasOffspring$ or $\hasAncestor$ roles to be
transitive as this corresponds to the intuitive understanding of these roles.
Without transitivity of the role $\hasOffspring$, the concept
\[
\all \hasOffspring \rich \sqcap \some \hasOffspring {\some \hasOffspring {\lnot
    \rich}}
\]
that describes someone who has only rich offsprings and who has an offspring
that has a poor offspring, would not be unsatisfiable, which is
counter-intuitive

\iiemph{Role hierarchies}{role hierarchies}
\cite{HorrocksGough-DL-1997} provide a mean to state
sub-role relationship between roles, e.g., to state that $\hasChild$ is a
sub-role of $\hasOffspring$, which makes it possible to infer that, e.g., a
grandchild of someone with only rich offsprings must be rich. Role hierarchies
also play an important role when modelling sub-relations of the general
part-whole relation \cite{Sattler-KI-1996}.

Role hierarchies only allow to express an approximation of the intuitive
understanding of the relationship between the roles $\hasChild$ and
$\hasOffspring$. Our intuitive understanding is that $\hasOffspring$ is the
transitive closure of $\hasChild$, whereas role hierarchies with transitive
roles are limited to state that $\hasOffspring$ is an \emph{arbitrary}
transitive super-role of $\hasChild$. Yet, this approximation is sufficient for
many knowledge representation tasks and there is empirical evidence that it
allows for faster implementations than inferences that support transitive
closure \cite{HorrSattTob-IGPL}.

Above we have used the roles $\hasOffspring$ and $\hasAncestor$ and the
intuitive understanding of these roles  requires them to be mutually
inverse. Without the expressive means of \iemph{inverse roles}, this cannot be captured
by a DL so that the concept
\[
\lnot \rich \sqcap \some \hasOffspring \top \sqcap \all \hasOffspring {\all
  \hasAncestor \rich}
\]
that describes somebody poor who has an offspring and whose offsprings only have
rich ancestors would not be unsatisfiable. This shortcoming of expressive power
is removed by the introduction of inverse roles into a DL, which would allow to
replace $\hasAncestor$ by $\hasOffspring^{-1}$, which denotes the inverse of
$\hasOffspring$.

\subsection{Nominals}

\iiemph{Nominals}{nominals}, i.e., atomic concepts referring to single
individuals of the domain, are studied both in the area of DLs
\cite{schaerf94:_reason_indiv_concep_languag,Borgida94,DeGiacomo96a} and modal
logics
\cite{gargov93:_modal_logic_with_names,blackburn95:_hybrid_languages,ArecesBlackburnMarx-JSL-2000}.
Nominals play an important r{\^o}le in knowledge representation because they
allow to capture the notion of uniqueness and identity. Coming back to the ABox
example from above, for a DL with nominals, the names \Mary or \Peter may not
only be used in ABox assertions but can also be used in place of atomic concept,
which, e.g., allows to describe \Mary's children by the concept $\some
{\hasChild^{-1}} \Mary$. Modeling named individuals by pairwise disjoint atomic
concepts, as it is done in the DL system \classic \index{classic@\classic}
\cite{Borgida94}, is not adequate and leads to incorrect inferences. For
example, if \Mary does not name a single individual, it would be impossible to
infer that every child of \Mary must be a sibling of \Peter (or \Peter himself),
and so the concept\[ \some {\hasChild^{-1}} \Mary \sqcap \all {\hasChild^{-1}}
{(\all \hasChild {\neg \Peter})}
\]
together with the example ABox would be incorrectly satisfiable. It is clear that
cardinality restrictions on concepts can be used to express nominals and and we
will see in this thesis that also the converse holds.

For decision procedures, nominals cause problems because they destroy the tree
model property of a logic, which has been proposed as an explanation for the
good algorithmic behaviour of modal and description
logics \cite{Vardi97,Graedel99c} and is often exploited by decision procedures.

\section{Guarded Logics}

The \iemph{guarded fragment} of first-order predicate logic, introduced by
Andr\'eka, van Benthem, and N\'emeti \citeyear{AndrekaBenNem98}, is a successful
attempt to transfer many good properties of modal, temporal, and Description
Logics to a larger fragment of predicate logic. Among these are decidability,
the finite model property, invariance under an
appropriate variant of bisimulation, and other nice model
theoretic properties~\cite{AndrekaBenNem98,Graedel99a}.

The Guarded Fragment (GF) is obtained from full first-order logic through
relativization of quantifiers by so-called \iemph{guard formulas}.  Every appearance of
a quantifier in \GF\ must be of the form
\[
\ex \bfy (\alpha(\bfx,\bfy) \wedge \phi(\bfx,\bfy)) \ \text{or} \ \fa
\bfy (\alpha(\bfx,\bfy) \rightarrow \phi(\bfx,\bfy)) ,
\]
where $\alpha$ is a positive atomic formula, the \emph{guard}, that contains all
free variables of $\phi$. This generalizes quantification in description, modal,
and temporal logic, where quantification is restricted to those elements
reachable via some accessibility relation. For example, in DLs, quantification
occurs in the form of existential and universal restrictions like $\all
\hasChild \rich$, which expresses that those individuals \emph{reachable via the
  role} (guarded by) \hasChild must be rich.

By allowing for more general formulas as guards while preserving the idea of
quantification only over elements that are \emph{close together} in the model,
one obtains generalisations of \GF\ which are still well-behaved in the sense of
GF.  Most importantly, one can obtain the \iiemph{loosely guarded
  fragment}{guarded fragment} (\LGF)~\cite{Benthem97} and the \iiemph{clique
  guarded fragment}{guarded fragment} (\CGF)~\cite{Graedel99b}, for which
decidability, invariance under clique guarded bisimulation, and some other
properties have been shown by Gr\"adel \citeyear{Graedel99b}. For other extension of \GF\ the
picture is irregular. While \GF\ remains decidable under the extension with
fixed point operators \cite{GraedelWal99}, adding counting constructs or
transitivity statements leads to undecidability
\cite{Graedel99a,GanzingerMeyerVeanes-99-lics}.

Guarded fragments are of interest for the DL community because many DLs are
readily embeddable into suitable guarded fragments. This allows the transfer of
results for guarded fragments to DLs.  For example, Goncalves and Gr{\"a}del
\citeyear{GoncalvesGra00} show decidability of the guarded fragment $\mu$AGFCI,
into which a number of expressive DLs is easily embeddable, yielding
decidability for these DLs.

\section{Outline and Structure of this Thesis}

This thesis deals with reasoning in expressive DLs and Guarded Logics. We supply
a number of novel complexity results and practical algorithms for inference
problems. Generally, we are more interested in the algorithmic properties of the
logics we study than their application in concrete knowledge representation
tasks. Consequently, the examples given in this thesis tend to be terse and
abstract and are biased towards computational characteristics. For more
information on how to use DLs for specific knowledge representation tasks, e.g.,
refer to \cite{BMPAB91,Borg95,CaLN98,Sattler-ECAI-2000}. 

This thesis is structured as follows:
\begin{itemize}
\item We start with a more formal introduction to DLs in Chapter \ref{chap:alc}.
  We introduce the standard DL \alc and define its syntax and semantics.  We
  specify the relevant inference problems and show how they are interrelated.
\item Chapter \ref{chap:reasoning} briefly surveys techniques employed for
  reasoning with DLs. We then describe a tableau algorithm that decides
  satisfiability of \alc-concepts with optimum worst-case complexity (\pspace) to
  introduce important notions and methods for dealing with tableau algorithms.
\item In Chapter \ref{chap:alcq} we consider the complexity of a number of DLs
  that allow for qualifying number restrictions. The DL \alcq is obtained from
  \alc by, additionally, allowing for qualifying number restrictions. We give a
  tableau algorithm that decides concept satisfiability for \alcq. We show how
  this algorithms can be modified to run in \pspace, which fixes the complexity
  of the problem as \pspace-complete. Previously, the exact complexity of the
  problem was only known for the (unnaturally) restricted case of unary coding
  of numbers~\cite{Hollunder91b} and the problem was conjectured to be
  \exptime-hard for the unrestricted case \cite{VanderHoekdeRijke-JLC-1995}.  We
  use the methods developed for \alcq to obtain a tableau algorithm that decides
  concept satisfiability for the DL \alcqir, which adds expressive role
  expressions to \alcq, in \pspace, which solves an open problem
  from~\cite{DLNN97}.
  
  We show that, for \alcqir, reasoning \wrt general TBoxes and knowledge bases
  is \exptime-complete. This extends the known result for \alcqi
  \cite{DeGiacomo95a} to a more expressive DL and, unlike the proof in
  \cite{DeGiacomo95a}, our proof is not restricted to the case of unary coding
  of numbers in the input.
  
\item The next chapter deals with the complexity of reasoning with cardinality
  restrictions on concepts.  We study the complexity of the combination of the
  DLs \alcq and \alcqi with cardinality restrictions. These combinations can
  naturally be embedded into \ctwo, the two variable fragment of predicate logic
  with counting quantifiers \cite{GraedelOttRos97}, which yields decidability in
  \nexptime~\cite{LICS::PacholskiST1997} (in the case of unary coding of
  numbers). We show that this is a (worst-case) optimal solution for \alcqi, as
  \alcqi with cardinality restrictions is already \nexptime-hard.  In contrast,
  we show that for \alcq with cardinality restrictions, all standard inferences
  can be solved in \exptime. This result is obtained by giving a mutual
  reduction from reasoning with cardinality restrictions and reasoning with
  nominals. Based on the same reduction, we show that already concept
  satisfiability for \alcqi extended with nominals is \nexptime-complete.  The
  results for \alcqi can easily be generalised to \alcqir.

\item In Chapter \ref{chap:shiq} we study DLs with transitive and inverse roles.
  For the DL \si---the extension of \alc with inverse and transitive roles---we
  describe a tableau algorithm that decides concept satisfiability in \pspace,
  which matches the known lower bound for the worst-case complexity of the
  problem and extends Spaan's results for the modal logic $\mathsf{K4}_t
  $\citeyear{spaan_e:1993a}. 
  
  \si is then extended to \shiq, a DL which, additionally, allows for role
  hierarchies and qualifying number restrictions. We determined the worst-case
  complexity of reasoning with \shiq as \exptime-complete. The \exptime upper
  bound has been an open problem so far. Moreover, we show that reasoning becomes
  \nexptime-complete if nominals are added to \shiq.
  
  The algorithm used to establish the \exptime-bound for \shiq employs a highly
  inefficient automata construction and cannot be used for efficient
  implementations.  Instead, we describe a tableau algorithm for \shiq that 
  promises to be amenable to optimizations and forms the basis of the
  highly-optimized DL system \iFaCT \cite{horrocks99:_fact}.  
\item In Chapter \ref{chap:gf} we develop a tableau algorithm for the clique
  guarded fragment of FOL, based on the same ideas usually found in algorithms
  for modal logics or DLs. Since tableau algorithms form the basis of some of
  the fastest implementations of DL systems, we believe that this algorithm is
  a viable starting point for an efficient implementation of a decision
  procedure for \CGF. Since many DLs are embeddable into \CGF, such an
  implementation would be of high interest.
\item In a final chapter, we conclude.
\end{itemize}

Some of the results in this thesis have previously been published. The
\pspace-algorithm for \alcq has been reported in \cite{Tobies-CADE-99} and is
extended to deal with inverse roles and conjunction of roles in
\cite{Tobies-JLC-2000}.  \nexptime-completeness of \alcqi with cardinality
restrictions is presented in \cite{Tobies-CSL-99,Tobies-JAIR-2000}, where the
latter publication establishes the connection of reasoning with nominals and
with cardinality restrictions. The \si-algorithm is presented in
\fullcite{HorrSattTob-IGPL}, a description of the tableau algorithm for \shiq can be
found in \cite{HorrocksSattlerTobies-LPAR-99}. Finally, the tableau algorithm
for \CGF\ has previously been published in \cite{HirschTobies-AIML-2000}.


\cleardoublepage


\chapter{Preliminaries}
\label{chap:alc}

In this chapter we give a more formal introduction to Description Logics
\index{Description Logics} and their inference problems. We define syntax and
semantics of the ``basic'' DL \alc and of the terminological and assertional
formalism used in this thesis. Based on these definitions, we introduce a
number of inference problems and show how they are interrelated.

\section{The Basic DL \alc}\label{sec:alc}

Schmidt-Schau{\ss}\ and Smolka \citeyear{Schmidt-Schauss91} introduce the DL
\alc, \index{alc@\alc}  which is
distinguished in that it is the ``smallest'' DL that is closed under all Boolean
connectives, and give a sound and complete subsumption algorithm. Unlike the
other DL inference algorithms developed at that time, they deviated from the
structural paradigm and used a new approach, which, due to its close resemblance
to first-order logic tableau algorithms, was later also called tableau algorithm.
Later, Schild's \citeyear{Schi91} discovery that \alc is a syntactic variant of
the basic modal logic \K made it apparent that Schmidt-Schau{\ss}\ and Smolka
had re-invented in DL notation the tableau-approach that had been successfully
applied to modal inference problems \cite<see, e.g.,>{Lad77,HalpernMoses92,Gore-Tableau-Handbook-1998}.

The DL \alc allows complex concepts to be built from concept and relation names
using the propositional constructors $\sqcap$ (\emph{and, class intersection}),
$\sqcup$ (\emph{or, class union}), and $\lnot$ (\emph{not, class
  complementation}). Moreover, concepts can be related using universal and
existential quantification along role names.

\begin{definition}[Syntax of \alc]\label{def:alc-syntax} 
  Let \Names be a set of  \emph{concept names} and \Roles be a set of
   \emph{role names}.  The set of \emph{\alc-concepts} is built
  inductively from these using the following grammar, where $A \in \Names$ and 
  $R \in \Roles$:
  \[
  C ::= A  \bnfor \lnot C \bnfor  C_1 \sqcap C_2 \bnfor  C_1 \sqcup C_2 \bnfor
  \all R C \bnfor \some R C  .
  \] \eod
\end{definition}

For now, we will use an informal definition of the \emph{size} $|C|$
\index{00size@$"| C "|$} of a concept $C$: we define $|C|$ to be the number of
symbols necessary to write down $C$ over the alphabet $\Names \cup \Roles \cup
\{ \sqcap, \sqcup, \neg, \forall, \exists, (, ) \}$. This will not be the
definitive definition of the size of the concept because it relies on an
unbounded alphabet ($\Names$ and $\Roles$ are arbitrary sets), which makes it
unsuitable for complexity considerations. We will clarify this issue in
Definition~\ref{def:concept-size}.

Starting with \cite{brachman84:_tract_subsum_frame_based_descr_languag},
semantics of DLs model concepts as sets and roles as binary relations. Starting
from an interpretation of the concept and role names, the semantics of arbitrary
concepts are defined by induction over their syntactic structure.
For \alc, this is done as follows.

\begin{definition}[Semantics of \alc]\label{def:alc-semantics}
  The semantics of \alc-concepts is defined relative to an
  \emph{interpretation} $\I = (\domain, \cdot^\I)$, where $\domain$ is a
  non-empty set, called the \emph{domain} of $\I$, and $\cdot^\I$ is a
  \emph{valuation} that defines the interpretation of concept and relation
  names by mapping every concept name to a subset of $\domain$ and every role
  name to a subset of $\domain \times \domain$. To obtain the semantics of a
  complex concept this valuation is inductively extended by setting:
  \begin{align*}
    (\lnot C)^\I & = \domain \setminus C^\I \qquad (C_1 \sqcap C_2)^\I = 
    C_1^\I \cap C_2^\I \qquad  (C_1 \sqcup C_2)^\I =  C_1^\I \cup C_2^\I\\
    (\all R C)^\I & =  \{ x \in \domain \mid \text{for all
      } y \in \domain, \tuple x y \in R^\I \Implies y \in C^\I \}\\
    (\some R C)^\I & = \{ x \in \domain \mid \text{there is a } y \in \domain
    \text{ with } \tuple x y \in R^\I \And y \in C^\I \} .
  \end{align*}
  
  A concept $C$ is \emph{satisfiable} \index{satisfiability!of a concept} iff
  there is an interpretation $\I$ such that $C^\I \neq \emptyset$. A concept $C$
  is \emph{subsumed} \index{subsumption of concepts} by a concept $D$ (written
  $C \sqsubseteq D$) \index{00csubsumedbyd@$C \sqsubseteq D$} iff, for every
  interpretation $\I$, $C^\I \subseteq D^\I$.  Two concepts $C,D$ are
  \emph{equivalent} \index{equivalence of concepts} (written $C \equiv D$)
  \index{00cequivalentd@$C \equiv D$} iff $C \sqsubseteq D \And D \sqsubseteq
  C$. \eod
\end{definition}

From this definition it is apparent, as noticed by Schild \citeyear{Schi91},
that \alc is a syntactic variant of the propositional (multi-) modal logic
$\K_m$. More precisely, for a set of concept names $\Names$ and role names
$\Roles$, the logic \alc corresponds to the modal logic $\K_m$ with
propositional atoms $\Names$ and modal operators $\{ \langle R \rangle, [R] \mid
R \in \Roles \}$ where the Boolean operators of \alc $(\sqcap, \sqcup, \neg)$
correspond to the Boolean operators of $\K_m$ \index{km@$\K_m$} $(\wedge, \vee,
\neg)$, existential restrictions over a role $R$ to the diamond modality
$\langle R \rangle$, and universal restrictions over a role $R$ to the box
modality $[R]$.  Applying this syntactic transformation in either direction
yields, for every \alc-concept $C$, an equivalent $\K_m$-formula $\phi_C$ and,
for every $\K_m$-formula $\phi$, an equivalent \alc-concept $C_\phi$. A similar
correspondence exists also for more expressive DLs.

We will often use $\top$ \index{00top@$\top$} as an abbreviation for an
arbitrary tautological concept, i.e., a concept with $\top^\I = \Delta^\I$ for
every interpretation $\I$.  E.g., $\top = A \sqcup \lnot A$ for an arbitrary
concept name $A \in \Names$.  Similarly, we use $\bot$ \index{00bottom@$\bot$}
as an abbreviation for an unsatisfiable concept ($\bot^\I = \emptyset$ for every
interpretation $\I$). E.g., $\bot = A \sqcap \lnot A$ for an arbitrary $A \in
\Names$. Also, we will use the standard logical abbreviations $C
\rightarrow D$ for $\neg C \sqcup D$ and $C \leftrightarrow D$ for $C
\rightarrow D \sqcap D \rightarrow C$.

\section{Terminological and Assertional Formalism}

Starting from the initial \klone system
\cite{brachman85:_overv_kl_one_knowl_repres_system}, DL systems allow to express
two categories of knowledge about the application domain:
\begin{itemize}
\item \emph{terminological} knowledge, which is stored in a so-called
  \emph{TBox} \index{TBox} and consists of general definition of concepts and
  knowledge about their interrelation, and
\item \emph{assertional} knowledge, which is stored in a so-called \emph{ABox}
  \index{ABox} and consist of a (partial) description of a specific situation
  consisting of elements of the application domains.
\end{itemize}

It should be noted that there are DL systems, e.g., \FaCT~\cite{Horrocks98c},
that do not support ABoxes but are limited to TBoxes only. In contrast to this,
all systems that support ABoxes also have some kind of support for TBoxes.

Different DL systems allow for different kinds of TBox formalism, which has an
impact on the difficulty of the various inference problems. Here, we define the
most general form of TBox formalism \index{general axiom} usually
studied---\emph{general axioms}---and describe other possibilities as a
restriction of this formalism.

\begin{definition}[General Axioms, TBox]
  A \emph{general axiom} \index{general axiom} is an expression of the
  form $C \sqsubseteq D$ or $C \doteq D$ where $C$ and $D$ are concepts. A
  \emph{TBox} \index{TBox} \mcT\ is a finite set of general axioms.

  An interpretation $\I$ satisfies a general axiom $C \sqsubseteq D$ ($C
  \doteq D$) iff $C^\I \subseteq D^\I$ ($C^\I = D^\I$). It satisfies $\mcT$ iff 
  it satisfies every axiom in $\mcT$. In this case, $\mcT$ is called
  \emph{satisfiable}, \index{satisfiability!of a TBox} $\I$ is called a
  \emph{model} of $\mcT$ and we write $\I  \models \mcT$. \index{00imodelst@$\I
    \models \mcT$} 
  
  Satisfiability, subsumption and equivalence of concepts can also be defined
  \wrt TBoxes: a concept $C$ is \emph{satisfiable \wrt $\mcT$} iff there is a
  model $\I$ of $\mcT$ with $C^\I \neq \emptyset$. $C$ is \emph{subsumed by
    $D$ \wrt $\mcT$} iff $C^\I \subseteq D^\I$ for every model $\I$ of $\mcT$.
  Equivalence \wrt $\mcT$ is defined analogously and denoted with $\sqsubseteq_\mcT$. \eod
  \index{satisfiability!of a concept!w.r.t.\ a TBox}
  \index{subsumption of concepts!w.r.t.\ a TBox}
  \index{equivalence of concepts!w.r.t.\ a TBox}
\end{definition}

Most DL systems, e.g., \textsc{kris}~\cite{BaaderHollunder-SIGART-91}, allow
only for a limited form of TBox that essentially contains only macro
definitions.  This is captured by the following definition.

\begin{definition}[Simple TBox]
  A TBox \mcT is called \emph{simple} iff
  \begin{itemize}
  \item the left-hand side of axioms consist only of concept names, that is,
    \mcT consists only of axioms of the form $A \sqsubseteq D$ and $A \doteq D$
    for $A \in \Names$,
  \item a concept name occurs at most once as the left-hand side of an axiom in
    \mcT, and
  \item \mcT\ is \emph{acyclic}. Acyclicity is defined as follows: $A \in
    \Names$ is said to ``\emph{directly use}'' $B \in \Names$ if $A \doteq D \in
    \mcT$ or $A \sqsubseteq D \in \mcT$ and $B$ occurs in $D$; ``\emph{uses}''
    is the transitive closure of ``directly uses''.  We say that $\mcT$ is
    \emph{acyclic} if there is no $A \in \Names$ that uses itself.
  \end{itemize} \eod
\end{definition}

Partial descriptions of the application domain can be given as an ABox.

\begin{definition}[ABox]\label{def:abox}
  \index{ABox}
  Let \Individuals be a set of individual names. For individual names $x,y \in
  \Individuals$, a concept $C$, and a role name $R$, the expressions $x : C$, $(x,y)
  : R$ and $x \ndoteq y$ are \iemph{assertional axioms}. An \emph{ABox} \mcA\ is a
  finite set of assertional axioms.

  To define the semantics of ABoxes we require interpretations, additionally,
  to map every individual name $x \in \Individuals$ to an element $x^\I$ of
  the domain $\Delta^\I$.
  
  An interpretation $\I$ satisfies an assertional axiom $x : C$ iff $x^\I \in
  C^\I$, it satisfies $(x,y) : R$ iff $(x^\I, y^\I) \in R^\I$, and it satisfies
  $x \ndoteq y$ iff $x^\I \neq y^\I$. $\I$ satisfies $\mcA$ iff it satisfies
  every assertional axiom in $\mcA$. If such an interpretation $\I$ exists, then
  $\mcA$ is called \emph{satisfiable}, \index{satisfiability!of an ABox} $\I$ is
  called a \emph{model} of $\mcA$, and we write $\I \models
  \mcA$. \index{00imodelsa@$\I \models \mcA$} \eod
\end{definition}

To decide whether $\I \models \mcA$ for an interpretation $\I$ and an ABox
$\mcA$, the interpretation of those individuals that do not occur in $\mcA$ is
irrelevant~\cite{nebel90:_reason_revis_hybrid_repres_system,Buchheit93}. Thus, to define a model of an ABox $\mcA$ it is sufficient to
specify the interpretation of those individuals occurring in $\mcA$.  Our
definition of ABoxes is slightly different from what can usually be found in the
literature, in that we do not impose the \iemph{unique name assumption}.  The
unique name assumption requires that every two distinct individuals must be
mapped to distinct elements of the domain. We do not have this requirement
but include explicit inequality assertions between two individuals as
assertional axioms. It is clear that our approach is more general than the
unique name assumption because inequality can be asserted selectively only for
some individual names. We use this approach due to its greater flexibility and
since it allows for a more uniform treatment of ABoxes in the context of
tableau algorithms, which we will encounter in Chapter~\ref{chap:reasoning}.

\begin{definition}[Knowledge Base]
  A \iemph{knowledge base} (KB) $\mcK = (\mcT, \mcA)$ consists of a TBox $\mcT$ 
  and an ABox $\mcA$. An interpretation $\I$ satisfies $\mcK$ iff $\I \models
  \mcT$ and $\I \models \mcA$. In this case, $\mcK$ is called \emph{satisfiable}, $\I$ 
  is called a \emph{model} of $\mcK$ and we write $\I \models \mcK$. \eod
  \index{satisfiability!of a knowledge base}
  \index{00imodelsk@$\I \models \mcK$}
\end{definition}

\section{Inference Problems}

From the previous definitions, one can immediately derive a number of (so called
\emph{standard}) inference problems \index{inference problems} for DL systems
that are commonly studied.  Here, we quickly summarize the most important of
them and show how they are interrelated.
\begin{itemize}
\item \textbf{Concept satisfiability}, \index{satisfiability!of a concept} i.e.,
  given a concept $C$, is $C$ satisfiable (maybe \wrt a TBox \mcT)? This
  inference allows to determine if concepts in the KB are contradictory
  (describe the empty class).
\item \textbf{Concept subsumption}, \index{subsumption of concepts} i.e., given two concepts $C,D$, is $C$
  subsumed by $D$ (maybe \wrt a TBox \mcT)? Using this inference, concepts
  defined in a TBox can be arranged in a subsumption quasi-order that reflects the
  specialisation/generalisation hierarchy of the concepts. Calculation of the
  subsumption hierarchy is one of the main inferences used by many applications of DL
  systems \cite<e.g.,>{rector:_exper_build_large_re_medic,schulz:_knowl_engin_large_scale_knowl_reuse,bechhofer:_drivin_user_inter_fact,FranconiNg-KRDB-2000}.

  For any DL that is closed under Boolean operations, subsumption and
  \mbox{(un-)}sat\-is\-fia\-bility are mutually reducible: a concept $C$ is unsatisfiable
  \wrt a TBox $\mcT$ iff $C \sqsubseteq_\mcT \bot$.  Conversely, $C
  \sqsubseteq_\mcT D$ iff $C \sqcap \lnot D$ is unsatisfiable \wrt $\mcT$.
  
  Concept satisfiability and subsumption are problems that are usually considered
  only \wrt TBoxes rather than KBs.  The reason for this is the fact
  \cite{nebel90:_reason_revis_hybrid_repres_system,Buchheit93} that the ABox
  does not interfere with these problems as long as the KB is satisfiable.
  W.r.t.\ unsatisfiable KBs, obviously every concept is unsatisfiable and every
  two concepts mutually subsume each other.
\item \textbf{Knowledge Base Satisfiability}, \index{satisfiability!of a
    knowledge base} i.e., given a KB $\mcK$, is $\mcK$ satisfiable? This
  inference allows to check whether the knowledge stored in the KB is free of
  contradictions, which is maybe the most fundamental requirement for knowledge
  in DL systems. For a KB that contains a contradiction, i.e., is not
  satisfiable, arbitrary conclusion can be drawn.
  
  Concept satisfiability (and hence concept subsumption) can be reduced to KB
  satisfiability: a concept $C$ is satisfiable \wrt a (possibly empty) TBox $\mcT$
  iff the KB $(\mcT, \{ x : C \})$ is satisfiable.
\item \textbf{Instance Checking}, \index{instance checking} i.e., given a KB $\mcK$, an individual name
  $x$, and a concept $C$, is $x^\I \in C^\I$ for every model $\I$ of
  $\mcK$? In this case, $x$ is called an \iemph{instance} of $C$ \wrt $\mcK$.
  Using this inference it is possible to deduce knowledge from a KB that is
  only implicitly present, e.g., it can be deduced that an individual $x$ is an
  instance of a concept $C$ in every model of the knowledge base even though $x
  : C$ is not asserted explicitly in the ABox---it follows from the other
  assertions in the KB.

  Instance checking can be reduced to KB \mbox{(un-)}satisfiability. For a KB $\mcK = (\mcT,
  \mcA)$, $x$ is an instance of $C$ \wrt $\mcK$ iff the KB $(\mcT, \mcA 
  \cup \{ x : \lnot C \})$ is unsatisfiable.
\end{itemize}

All the mentioned reductions are obviously computable in linear time. Hence, KB
satisfiability can be regarded as the most general of the mentioned inference
problems. As we will see in a later chapter, for some DLs, it is also possible
to polynomially reduce KB satisfiability to concept satisfiability.


\cleardoublepage

\chapter{Reasoning in Description Logics}
\label{chap:reasoning}

This chapter starts with an overview of methods that have been developed to
solve DL inference problems. We then describe a tableau algorithm that decides
concept satisfiability and subsumption for \alc and which can be implemented to
run in \pspace. Albeit this is a well-known result \cite{Schmidt-Schauss91}, it
is repeated here because it allows us to introduce important notions and methods
for dealing with tableau algorithms before these are applied to obtain results
for more expressive logics in the subsequent chapters.

\section{Reasoning Paradigms}

\label{sec:reasoning-methodologies}

Generally speaking, there are four major and some minor approaches to reasoning
with DLs that will be briefly described here. Refer to
\cite{BaaderSattler-StudiaLogica-2000} for a more history-oriented introduction
to reasoning with DLs.

\paragraph{Structural algorithms} 
The early DL systems like
\klone~\cite{brachman85:_overv_kl_one_knowl_repres_system} and its successor
systems \textsc{back}~\cite{TUB-FB13//TUB-FB13-KIT-78}, \index{back@\textsc{back}}
\textsc{k-rep}~\cite{mays-etal:1991a}, \index{k-rep@\textsc{k-rep}} or
  \textsc{loom}~\cite{macgregor-91a} \index{loom@\textsc{loom}}
used \iemph{structural algorithms} that rely on syntactic comparison of concepts
in a suitable normal form to decide subsumption. Nebel
\citeyear{nebel90:_reason_revis_hybrid_repres_system} gives a formal description
of an algorithm based on this approach. Usually, these algorithms had very good
(polynomial) run-time behaviour. Tractability was a major concern in the
development of DL systems and algorithms with super-polynomial runtime were
considered unusable in practical
applications~\cite{levesque87:_expres_tract_knowl_repres_reason}.  Yet, as it
turned out, even DLs with very limited expressive power prohibit tractable
inference
algorithms~\cite{brachman84:_tract_subsum_frame_based_descr_languag,Nebel90} and
for some, like \klone, subsumption is even
undecidable~\cite{schmidt-schausss89:_subsum_klone_undec}. Consequently,
complete structural algorithms are known only for DLs of very limited
expressivity.

This limitations were addressed by DL researchers in three general ways: some
system developers deliberately committed to incomplete algorithms to preserve the
good run-time behaviour of their systems. Others proceeded by carefully tailoring
the DL to maximise its expressivity while maintaining sound and complete
structural algorithms. Representatives of the former approach are the \textsc{back}
and the \textsc{loom} system while \classic
\cite{patel-schneider91:_classic_kr_system} \index{classic@\classic} follows the latter approach
with a ``nearly'' complete structural subsumption algorithm \cite{Borgida94}. The
third approach was to develop algorithms that are capable to deal with more
expressive DLs despite the higher complexity. This required departure from the
methods employed so far.

\paragraph{Tableau algorithms} 
The first such algorithm was developed by Schmidt-Schau{\ss} and Smolka
\citeyear{Schmidt-Schauss91} for the DL \alc, and the employed methodology
proved to be useful to decide subsumption and other inference problems like
concept satisfiability also for other
DLs~\cite{hollunder90:_subsum_algor_concep_descr_languag,Hollunder91b,Baader91c,hanschke92:_specif_role_inter_concep_languag}.
Due to their close resemblance to tableau algorithms for first-order predicate
logic (FOL) they were also called \iiemph{tableau algorithms}{tableau  algorithm}. For many DLs, it was possible to obtain algorithms based on the
tableau approach that match the known worst-case complexity of the problem
\cite<see, e.g.,>[for a systematic study]{DLNN91,DLNN91b,DHLM92}.  Although the
inference problems for these DLs are usually at least \textsc{NP}- or even
\pspace-hard, systems implementing the tableau approach, like
\textsc{kris}~\cite{BaaderHollunder-SIGART-91} or \textsc{crack}~\cite{BrFT95},
show reasonable runtime performance on application problems and more recent
systems that employ sophisticated optimization techniques, like Horrock's \FaCT
system~\citeyear{Horrocks98c} or Patel-Schneider's DLP
\citeyear{Patel-Schneider-Tableaux2000}, can deal with problems of considerable
size, even for \exptime-hard DLs. For theses logics, the employed tableau
algorithms exceed the known worst-case complexity of the problems, but are
rather biased towards optimizability for ``practical'' cases.  Indeed, for
\exptime-complete DLs, it turns out to be very involved to obtain tableau
algorithms with optimum worst-case complexity \cite{donini00:_exptim_alc}.

\paragraph{Translational approaches} \index{translation method}
Schild's discovery~\citeyear{Schi91} that DLs are syntactic variants of modal
logics made it possible to obtain inference procedures for DLs by simply
borrowing the methods from the corresponding modal logic.  This approach has
been refined for more expressive DLs and a number of (worst-case) optimal
decision procedures for very expressive---usually \exptime-complete---DLs were
obtained by sophisticated translation into
PDL~\cite{DeLe94,DeLe94c,DeGiacomo96a,DeGiacomo95a} or the modal $\mu$-calculus
\cite{Schi94,GiLe94}. While many interesting complexity results could be
obtained in this manner, there exists no implementation of a DL system that
utilizes this approach. Experiments indicate \cite{HorrSattTob-IGPL} that it
will be very hard to obtain efficient implementations based on this kind of
translations.  More recently, modal logicians like Areces and de Rijke
\citeyear{ArecesDeRijkeAIML00} have advocated hybrid modal logics \index{hybrid
  modal logic} \cite{ArecesBlackburnMarx-JSL-2000,areces00:_logic_engin} as a
suitable target for the translation of DLs and obtain novel theoretical results
and decision procedures.  It is unclear if these decision procedures can be
implemented efficiently.

A different approach utilizes translation into FOL. Already Brachman and
Levesque \citeyear{brachman84:_tract_subsum_frame_based_descr_languag} use FOL
to specify the semantics of their DL and inference problems for nearly all DLs
(and their corresponding modal logics) are easily expressible in terms of
satisfiability problems for (extensions of) FOL.  Since FOL is
undecidable, one does not immediately obtain a decision procedure in this
manner.  So, these approaches use more restricted and hence decidable fragments
of FOL as target of their translation. Borgida~\citeyear{Borgida96a} uses the
two-variable fragment of FOL to prove decidability of an expressive DL in
\nexptime while De Nivelle \citeyear{nivelle00:_trans_s4_k5_gf} gives a
translation of a number of modal logics into the guarded fragment to facilitate
the application of FOL theorem proving methods to these logics.
Schmidt~\citeyear{Schmidt98f} uses a non-standard translation into a fragment of
FOL for which decision procedures based on a FOL theorem prover exist. Areces
et. al \citeyear{arec:tree00} show that careful tuning of standard FOL theorem
proving methods also yields a decision procedure for the standard translation.
The latter approaches are specifically biased towards FOL theorem provers and
make it possible to utilize the massive effort spent on the implementation and
optimization of FOL theorem provers to reason with DLs.  It seems though, that
the translation approach leads to acceptable but inferior runtime when compared
with tableau
systems~\cite{MassacciDonini-Tableaux2000,HorrocksPatelSSebastiani-IGPL-2000}.

\paragraph{Automata based methods} \index{automata approach}
Many DL and modal logics possess the so-called \iemph{tree model property}, i.e.,
every satisfiable concept has---under a suitable abstraction---a tree-shaped
model. This makes it possible to reduced the satisfiability of a concept to the
existence of a tree with certain properties dependent on the formula. If it is
possible to capture these properties using a \emph{tree automaton}
\cite{Gecseg-Steinby:84,ThomasHTCS92}, satisfiability and hence subsumption of
the logic can be reduce to the emptiness problem of the corresponding class of
tree automata~\cite{VaWo86}. Especially for DLs with \exptime-complete inference
problems, where it is difficult to obtain tableau algorithms with optimum
complexity, exact complexity results can be obtained elegantly using the
automaton approach~\cite{CaDL99,LutzSattlerAIML00}. Yet, so far it seems
impossible to obtain efficient implementations from automata-based algorithms.
The approach usually involves an exponential step that occurs in \emph{every}
case independent of the ``difficulty'' of the input concept and cannot be
avoided by existing methods. This implies that such an algorithm will exhibit
exponential behaviour even for ``easy'' instances, which so far prohibits the
use of the approach in practice.

\paragraph{Other approaches}

In addition to these approaches, there also exist further, albeit less
influential, approaches.  Instead of dealing with DLs as fragments of a more
expressive formalism, the \iemph{SAT-based method} developed by Giunchiglia and
Sebastiani \citeyear{GiunchigliaSeb96b} uses the opposite approach and extend
reasoning procedures for the less expressive formalism of propositional logic to
DLs.  Since highly sophisticated SAT-solvers are available, this approach has
proven to be rather successful. Yet, it cannot compete with tableau based
algorithms \cite{Massacci-Tableaux-1999} and so far is not applicable to DLs
more expressive than \alc.

The \iemph{inverse method}~\cite{Voronkov-LICS-2000} takes a radically
different approach to satisfiability testing. It tries to prove unsatisfiability
of a formula in a bottom-up manner, by trying to derive the input formula
starting from ``atomic'' contradictions. There exists only a very early
implementation of the inverse method \cite{CADE99*383}, which shows promising
run-time performance but at the moment it cannot compete with tableau based
systems \cite{Massacci-Tableaux-1999}. 

Both the SAT-based approach and the inverse method have so far not been studied
with respect to their worst-case complexity.

\section{Tableau Reasoning for \alc-satisfiability}

Even though KB satisfiability is the most general standard inference problem, it
is also worthwhile to consider solutions for the less general problems. For many
applications of DL systems, the ABox does not play a role and reasoning is
performed solely on the terminological level
\cite{rector:_exper_build_large_re_medic,schulz:_knowl_engin_large_scale_knowl_reuse,bechhofer:_drivin_user_inter_fact,FranconiNg-KRDB-2000}.
For these applications, the additional overhead of dealing with ABoxes is
unnecessary.  Additionally, ABoxes do not have a resemblance in the modal world
(with the exception of \emph{hybrid modal logics}, see
\cite{ArecesDeRijkeAIML00}) and hence theoretical results obtained for KB inferences do not
transfer as easily as results for reasoning with TBoxes, which often directly
apply to modal logics.  From a pragmatic point of view, since full KB reasoning
is at least as hard as reasoning \wrt TBoxes, it is good to know how to deal
(efficiently) with the latter problem before trying to solve the former.
Finally, as we will see in Section~\ref{sec:other-infer-probl}, sometimes
concept satisfiability suffices to solve the more complicated inference
problems.

Schmidt-Schau{\ss} and Smolka \citeyear{Schmidt-Schauss91} were the first to
give a complete subsumption algorithm for \alc. The algorithm they used followed
a new paradigm for the development of inference algorithms for DLs that proved
to be applicable to a vast range of DL inference problems and, due to its
resemblance to tableau algorithms for FOL, was later called the tableau approach
\cite<see>[for an overview of tableau algorithms for
DLs]{BaaderSattler-StudiaLogica-2000}. After the correspondence of DLs and modal
logics had been pointed out by Schild \citeyear{Schi91}, it became apparent that the
tableau algorithms developed for DLs also closely resembled those used by modal
logicians. The tableau approach has turned out to be particularly amenable to
optimizations and some of the most efficient DL and modal reasoner currently
available are based on tableau algorithms \cite<see>[for a system
comparison]{MassacciDonini-Tableaux2000}.

Generally speaking, a tableau algorithm for a DL tries to prove satisfiability of a
concept (or a knowledge base) by trying to explicitly construct a model or some
kind of structure that induces the existence of a model (a \emph{pre-model}).
This is done by manipulating a \iemph{constraint system}---some kind of data
structure that contains a partial description of a model or pre-model---using a
set of \iemph{completion rules}.  Such constraint systems usually consist of a
number of individuals for which role relationships and membership in the extension
of concepts are asserted, much like this is done in an ABox. Indeed, for \alc and
the DLs considered in the next chapter, it is convenient to use the ABox
formalism to capture the constraints. For more expressive DLs, it will be more
viable to use a different data structure, e.g., to emphasise the graph structure
underlying the ABox.

Independent of the formalism used to express the constraints, completion of such
a constraint system is performed starting from an initial constraint system,
which depends on the input concept (or knowledge base), until either an obvious
contradiction (a clash) has been generated or no more rules can be applied. In
the latter case, the rules have been chosen in a way that a model of the concept
(or knowledge base) can be immediately derived from the constraint system.

\begin{definition}[Negation Normal Form]\label{def:nnf}
  In the following, we will  consider concepts in \iemph{negation normal
    form} (NNF), a form in which negations ($\lnot$) appear only in front of
  concept names. Every \alc-concept can be equivalently transformed into NNF by
  pushing negation inwards using the following equivalences:
\begin{equation*}
  \begin{array}{r@{\;}c@{\;}l@{\qquad}r@{\;}c@{\;}l}
    \lnot (C_1 \sqcap C_2) & \equiv & \lnot C_1 \sqcup \lnot C_2 & \lnot \all R C
    & \equiv & \some R {\lnot C}\\ 
    \lnot (C_1
    \sqcup C_2) & \equiv & \lnot C_1 \sqcap \lnot C_2 & \lnot \some R C & \equiv & \all 
    R {\lnot C}\\
    \lnot \lnot C & \equiv & C
  \end{array}
\end{equation*}

Note that every \alc-concept can be transformed into NNF in linear time.

For a concept $C$ in NNF, we denote the set of sub-concepts of $C$ by $\sub(C)$.
Obviously, the size of $\sub(C)$ is bounded by $|C|$. \eod \index{00subc@$\sub(C)$}
\end{definition}

\subsection{Deciding Concept Satisfiability for \alc}

We will now describe the \alc-algorithm that decides concept satisfiability (and
hence concept subsumption) for \alc. As mentioned before, we use ABoxes to
capture the constraint systems generated by the \alc-algorithm.

\begin{algorithm}[\alc-algorithm]\label{alg:alc}
  \index{tableau algorithm!for alc@for \alc}
  \index{alc@\alc!tableau algorithm}
  An ABox $\mcA$ contains a \emph{clash} iff, for an individual name $x \in
  \Individuals$ and a concept name $A \in \Names$, $\{ x : A, x : \lnot A \}
  \subseteq \mcA$. Otherwise, $\mcA$ is called \iemph{clash-free}.
  
  To test the satisfiability of an \alc-concept $C$ in NNF, the \alc-algorithm
  works as follows. Starting from the \emph{initial ABox} \index{ABox!initial} $\mcA_0 = \{ x_0 : C
  \}$ it applies the \iemph{completion rules} from Figure~\ref{fig:alc-rules}, which modify the
  ABox. It stops when a clash has been generated or when no rule is
  applicable. In the latter case, the ABox is \emph{complete}. The algorithm
  answers ``$C$ is satisfiable'' iff a complete and clash-free ABox has been
  generated.

  \begin{figure}[tbh]
    \begin{tabular}{lll}
      \ruleand: &  \bfif \hfill 1. & $x : C_1 \sqcap C_2 \in \mcA$ and\\
      & \hfill 2. &$\{ x:C_1, x:C_2 \} \not \subseteq \mcA$\\
      & \bfthen & $\mcA \ruleand \mcA \cup \{ x:C_1, x:C_2 \}$\\[1ex]
      \ruleor: & \bfif \hfill 1.& $x : C_1 \sqcup C_2 \in \mcA$ and \\
      & \hfill 2. & $\{ x:C_1,
      x:C_2 \} \cap \mcA = \emptyset$\\
      & \bfthen & $\mcA \ruleand \mcA \cup \{ x:D \}$ for some $D \in \{ C_1, C_2
      \}$\\[1ex]
      \ruleex: &  \bfif \hfill 1. & $x : \some R D \in \mcA$ and \\
      & \hfill 2. & there is no $y$ with $
      \{ (x,y) : R, y : D \} \subseteq \mcA$\\
      & \bfthen & $\mcA \ruleex \mcA \cup \{ (x,y) : R,\ y : D \}$ for a fresh
      individual $y$\\[1ex]
      \rulefa: &  \bfif \hfill 1. & $x : \all R D \in \mcA$ and\\
      & \hfill 2. & there is a $y$ with
      $(x,y) : R \in \mcA$ and $y : D \not \in \mcA$\\
      & \bfthen & $\mcA \rulefa \mcA \cup \{ y : D \}$
    \end{tabular}
    \caption{The completion rules for \alc}
    \index{completion rules!for alc@for \alc}
    \index{alc@\alc!completion rules}
    \index{00ruleand@\ruleand}
    \index{00ruleor@\ruleor}
    \index{00ruleex@\ruleex}
    \index{00rulefa@\rulefa}
    \label{fig:alc-rules}
  \end{figure}
  
  From the rules in Figure~\ref{fig:alc-rules}, the \ruleor-rule is called
  \emph{non-deterministic} while the other rules are called \emph{deterministic}.
  The \ruleex-rule is called \emph{generating}, while the other rules are
  called \emph{non-generating}. \eod
\end{algorithm}

The \alc-algorithm is a non-deterministic algorithm due to the \ruleor-rule,
which non-de\-ter\-min\-is\-tic\-al\-ly chooses which disjunct to add for a disjunctive
concept. Also, we have not specified a precedence that determines which rule to
apply if there is more than one possibility. To prove that such a
non-deterministic algorithm is indeed a decision procedure for satisfiability of
\alc-concepts, we have to establish three things:
\begin{enumerate}
\item \textbf{Termination}, i.e., every sequence of rule-applications
  terminates after a finite number of steps. 
\item \textbf{Soundness}, i.e., if the algorithm has generated a complete and
  clash-free ABox for $C$, then $C$ is satisfiable.
\item \textbf{Completeness}, i.e., for a satisfiable concept $C$ there is a
  sequence of rule applications that leads to a complete and clash-free ABox
  for $C$. 
\end{enumerate}

When dealing with non-deterministic algorithms, one can distinguish two
different kinds of non-determinism, namely \emph{don't-know} and
\emph{don't-care} non-determinism. \index{non-determinism!don't-know}
\index{non-determinism!don't-care} Choices of an algorithm that may affect the
result are called don't-know non-deterministic. For the \alc-algorithm, the
choice of which disjunct to add by the \ruleor-rule is don't-know
non-deterministic. When dealing with the initial ABox
\[
\mcA = \{x_0 : A \sqcup (B \sqcap \lnot B) \},
\]
the algorithm will only find a clash-free completion of $\mcA$ if the
\ruleor-rule chooses to add the assertion $x_0 : A$. In this sense, adding $x_0
: A$ is a ``good'' choice while adding $x_0 : B \sqcap \lnot B$ would be a
``bad'' choice because it prevents the discovery of a clash-free completion of
$\mcA$ even though there is one. For a (necessarily deterministic)
implementation of the \alc-algorithm, this implies that exhaustive search over
all possibilities of don't-know non-deterministic choices is required to obtain a
complete algorithm.

Non-deterministic choices that don't effect the outcome of the algorithm in the
sense that any choice is a ``good'' choice are called don't-care
non-deterministic. Don't-care non-determinism is also (implicitly) present in
the \alc-algorithm. Even though in an ABox several rules might be applicable at
the same time, the algorithm does not specify which rule to apply to which
constraint in which order. On the contrary, it will turn out that, whenever a
rule is applicable, it can be applied in a way that leads to the discovery of a
complete and clash-free ABox for a satisfiable concept
(Lemma~\ref{lem:alc-local-correctness}). This implies that in case of a
(deterministic) implementation of the \alc-algorithm one is free to choose an
arbitrary strategy which rule to apply where and when without sacrificing the
completeness of the algorithm, although the efficiency of the implementation
might depend on the choice of the employed strategy.

\subsubsection{Termination}

\index{termination!of the tableau algorithm}
\index{tableau algorithm!termination of}

The general idea behind the termination proofs of the tableau algorithms we will
encounter in this thesis is the following:
\begin{itemize}
\item The concepts and roles appearing in a constraint are taken from a finite
  set.
\item Paths in the constraint system are of bounded length and every individual
  has a bounded number of successors.
\item The application of a rule either adds a constraint for an individual
  already present in the constraint system, or it adds new individuals. No
  constraints or individuals are ever deleted, or, if deletion takes place, the
  number of deletions is bounded.
\end{itemize}

Together, this implies termination of the tableau algorithm since an infinite
sequence of rule applications would either lead to an unbounded number of
constraints for a single individual or to infinitely many individuals in the
constraint system. Both stand in contradiction to the mentioned properties.

To prove the termination of the \alc-algorithm, it is convenient to ``extract''
the underlying graph-structure from an ABox and to view it as an edge and node
labelled graph.

\begin{definition}
  Let $\mcA$ be an ABox. The graph $G_\mcA$ \index{00ga@$G_\mcA$} \emph{induced
    by} $\mcA$ is an edge and node labelled graph $G_\mcA = (V,E,\Lab)$ defined
  by
  \begin{align*}
    V & = \{ x \in \Individuals \mid \text{$x$ occurs in \mcA} \},\\
    E & = \{ (x,y) \mid (x,y) : R \in \mcA \},\\
    \Lab(x) & = \{ D \mid x : D \in \mcA \},\\
    \Lab(x,y) & = \{ R \mid (x,y) : R \in \mcA \} .
  \end{align*}     \eod
  \index{00lx@$\Lab(x)$}
  \index{00lxy@$\Lab(x,y)$}
\end{definition}

It is easy to see that, for any ABox $\mcA$ generated by a sequence of
applications of the completion rules for \alc from an initial ABox $\{ x_0 : C
\}$, the induced graph $G_\mcA$ satisfies the following properties:
\begin{itemize}
\item $G_\mcA$ is a tree rooted at $x_0$.
\item For any node $x \in V$, $\Lab(x) \subseteq \sub(C)$.
\item For any edge $(x,y) \in E$, $\Lab(x,y)$ is a singleton $\{R\}$ for a role
  $R$ that occurs in $C$.
\end{itemize}
A proof of this properties can easily be given by induction on the number of
rule applications and is left to the reader. Moreover, it is easy to show the
following lemma that states that the graph generated by the \alc-algorithm is
bounded in the size of the input concept.

\begin{lemma}\label{lem:alc-tree-bounds}
  Let $C$ be an \alc-concept in NNF and $\mcA$ an ABox generated by the
  \alc-algorithm by a sequence of rule applications from the initial
  ABox $\{ x_0 : C \}$. Then the following holds:
  \begin{enumerate}
  \item For every node $x$, the size of $\Lab(x)$ is bounded by $|C|$.
  \item The length of a directed path in $G_\mcA$ is bounded by $|C|$.
  \item The out-degree of $G_\mcA$ is bounded by $|C|$.
  \end{enumerate}
\end{lemma}

\begin{proof}
  \begin{enumerate}
  \item For every node $x$, $\Lab(x) \subseteq \sub(C)$. Hence, $|\Lab(x)|
    \leq |\sub(C)| \leq |C|$.
  \item For every node $x$ we define $\ell(x)$ as the maximum nesting of
    existential or universal restrictions in a concept in $\Lab(x)$. Obviously,
    $\ell(x_0) \leq |C|$. Also, if $(x,y) \in E$, then $\ell(x) > \ell(y)$.
    Hence, any path $x_1, \dots x_k$ in $G_\mcA$ induces a sequence $\ell(x_1) >
    \dots > \ell(x_k)$ of non-negative integers. Since $G_\mcA$ is a tree rooted
    at $x_0$, the longest path starts with $x_0$ and is bounded by $|C|$.
  \item Successors of a node $x$ are only generated by an application of the
    \ruleex-rule, which generates at most one successor for each concept of the
    form $\some R D$ in $\Lab(x)$. Together with (1), this implies that the
    out-degree is bounded by $|C|$. \qed
  \end{enumerate}
\end{proof}

From this lemma, termination of the \alc-algorithm is a simple corollary:

\begin{corollary}[Termination]\label{cor:alc-termination}
  \index{termination!of the alc-algorithm@of the \alc-algorithm}
  \index{alc@\alc!tableau algorithm!termination}
  Any sequence of rule-applications of the \alc-algorithm terminates after a
  finite number of steps.
\end{corollary}

\begin{proof}
  A sequence of rule-applications induces a sequence of trees whose depth and
  out-degree is bounded by the size of the input concept by
  Lemma~\ref{lem:alc-tree-bounds}. Moreover, every rule application adds a
  concept to the label of a node or adds a node to the tree. No nodes are ever
  deleted from the tree and no concepts are ever deleted from the label of a
  node.
  
  Hence, an unbounded sequence of rule-applications would either lead to an
  unbounded number of nodes or to an unbounded label of one of the nodes. Both 
  cases contradict  Lemma~\ref{lem:alc-tree-bounds}. \qed
\end{proof}

\subsubsection{Soundness and Completeness}

Soundness and completeness of a tableau algorithm is usually proved by
establishing the following properties of the algorithm based on an appropriate 
notion of satisfiability of constraint systems, which is tailored for the needs
of every specific DL and tableau algorithm.

\begin{enumerate}
\item A constraint system that contains a clash is necessarily unsatisfiable.
\item The initial constraint system is satisfiable iff the input concept (or
  knowledge base) is satisfiable. 
\item A complete and clash-free constraint systems is satisfiable.
\item For every applicable deterministic rule, its application preserves
  satisfiability of the constraint systems. For every applicable
  non-deterministic rule, there is a way of applying the rule that preserves
  satisfiability.
\item For every rule, no satisfiable constraint system can be
  generated from an unsatisfiable one, or, alternatively,
\item [5'.] a complete and clash-free constraint system implies satisfiability of
  the initial constraint system.
\end{enumerate}

Property 4 and 5 together are often referred to as \emph{local correctness} of
the rules.

\begin{theorem}[Generic Correctness of Tableau Algorithms]\label{theo:generic-correctness}
  A terminating tableau algorithm that satisfies the properties mentioned
  above is correct.
  \index{tableau algorithm!correctness of}
\end{theorem}

\begin{proof}
  Termination is required as a precondition of the theorem.  The tableau
  algorithm is sound because a complete and clash-free constraint system is
  satisfiable (Property 3) which implies satisfiability of the initial
  constraint system (either by Property 5 and induction over the number of rule
  applications of directly by Property 5') and hence (by Property 2) the
  satisfiability of the input concept (or knowledge base).
  
  It is complete because, given a satisfiable input concept (or knowledge base),
  the initial constraint system is satisfiable (Property 2). Each rule can be
  applied in a way that maintains the satisfiability of the constraint system
  (Property 4) and, since the algorithm terminates, any sequence of
  rule-applications is finite.  Hence, after finitely many steps a satisfiable
  and complete constraint system can be derived from the initial one. This
  constraint system must be clash-free because (by Property 1) a clash would
  imply unsatisfiability. \qed
\end{proof}

Specifically, for \alc we use the usual notion of satisfiability of
ABoxes. Clearly, for a satisfiable concept $C$, the initial ABox $\{ x_0 : C
\}$ is satisfiable and a clash in an ABox implies unsatisfiability.

It remains to prove that a complete and clash-free ABox is satisfiable and that
the rules preserve satisfiability in the required manner. The following definition
extracts a model from a complete and clash-free ABox.

\begin{definition}[Canonical Interpretation]\label{def:canonical-interpretation}
  For an ABox $\mcA$, the \iemph{canonical interpretation} $\I_\mcA=(\Delta^{\I_\mcA},
  \cdot^{\I_\mcA})$ is defined  by
  \begin{align*}
    \Delta^{\I_\mcA} & = \{ x \in \Individuals \mid \text{$x$ occurs in $\mcA$} \},\\
    A^{\I_\mcA} & = \{ x \mid x : A \in \mcA \} \quad \text{for every $A \in
      \Names$},\\
    R^{\I_\mcA} & = \{ (x,y) \mid (x,y) : R \in \mcA \} \quad \text{for every
      $R \in \Roles$},\\
    x^{\I_\mcA} & = x \quad \text{for every individual $x$ that occurs in
      $\mcA$} .
  \end{align*} \eod
\end{definition}

\begin{lemma}\label{lem:alc-abox-yields-model}
  Let $\mcA$ be a complete and clash-free ABox. Then $\mcA$ has a model.
\end{lemma}

\begin{proof}
  It is obvious that, for an arbitrary ABox $\mcA$, the canonical interpretation
  satisfies all assertion of the form $(x,y) : R \in \mcA$. $\mcA$ does not
  contain any assertions of the form $x \ndoteq y$.

  By induction on the structure of concepts occurring in $\mcA$, we show that
  the canonical interpretation $\I_\mcA$ satisfies any assertion of the form
  $x : D \in \mcA$ and hence is a model of $\mcA$.
  \begin{itemize}
  \item For the base case $x : A$ with $A \in \Names$, this holds by definition
    of ${\I_\mcA}$.
  \item For the case $x : \lnot A$, since $\mcA$ is clash free, $x  : A \not
    \in \mcA$ and hence $x \not \in A^{\I_\mcA}$.
  \item If $x : C_1 \sqcap C_2 \in \mcA$, then, since $\mcA$ is complete, also
    $\{x : C_1, x:C_2\} \subseteq \mcA$. By induction this implies $x \in
    C_1^{\I_\mcA}$ and $x \in C_2^{\I_\mcA}$ and hence $x \in (C_1 \sqcap C_2)^{\I_\mcA}$.
  \item If $x : C_1 \sqcup C_2 \in \mcA$, then, again due the completeness of
    $\mcA$, either $x:C_1 \in \mcA$ or $x:C_2 \in \mcA$. By induction this
    yields $x \in C_1^{\I_\mcA}$ or $x \in C_2^{\I_\mcA}$ and hence $x \in (C_1 \sqcup
    C_2)^{\I_\mcA}$.
  \item If $x : \some R D \in \mcA$, then completeness yields $\{ (x,y) : R, y 
    : D \} \subseteq \mcA$ for some $y$. By construction of ${\I_\mcA}$, $(x,y) \in
    R^{\I_\mcA}$ holds and by induction we have $y \in D^{\I_\mcA}$. Together this implies
    $x \in (\some R D)^{\I_\mcA}$.
  \item If $x : \all R D \in \mcA$, then, for any $y$ with $(x,y) \in
    R^{\I_\mcA}$, $(x,y) : R \in \mcA$ must hold due to the construction of
    ${\I_\mcA}$. Then, due to completeness, $y : D \in \mcA$ must hold and
    induction yields $y \in D^{\I_\mcA}$. Since this holds for any such $y$,
    $x \in (\all R D)^{\I_\mcA}$. \qed
  \end{itemize}
\end{proof}

\begin{lemma}[Local Correctness]\label{lem:alc-local-correctness}
  \begin{enumerate}
  \item If $\mcA$ is an ABox and $\mcA'$ is obtained from $\mcA$ by an
    application of a completion rule, then satisfiability of $\mcA'$ implies
    satisfiability of $\mcA$.
  \item If $\mcA$ is satisfiable and $\mcA'$ is obtained from $\mcA$ by an
    application of a deterministic rule, then $\mcA'$ is satisfiable.
  \item If $\mcA$ is satisfiable and the \ruleor-rule is applicable, then there
    is a way of applying the \ruleor-rule such that the obtained ABox $\mcA'$ is
    satisfiable.
  \end{enumerate}
\end{lemma}

\begin{proof}
  \begin{enumerate}
  \item Since $\mcA$ is a subset of $\mcA'$, satisfiability of $\mcA'$
    immediately implies satisfiability of $\mcA$.
  \item Let $\I$ be a model of $\mcA$. We distinguish the different rules:
    \begin{itemize}
    \item The application of the \ruleand-rule is triggered by an assertion
      $x : C_1 \sqcap C_2 \in \mcA$. Since $x^\I \in (C_1 \sqcap
      C_2)^\I$, also $x^\I \in C_1^\I \cap C_2^\I$. Hence, $\I$ is also a
      model for $\mcA' = \mcA \cup \{ x: C_1, x: C_2\}$.
    \item The \ruleex-rule is applied due to an assertion $x : \some R D \in
      \mcA$. Since $\I$ is a model of $\mcA$, there exists an $a \in \Delta^\I$
      with $(x^\I,a) \in R^\I$ and $a \in D^\I$. Hence, the interpretation $\I[y
      \mapsto a]$, which maps $y$ to $a$ and behaves like $\I$ on all other
      names, is a model of $\mcA' = \mcA \cup \{ (x,y) : R, y : D\}$. Note, that
      this requires $y$ to be fresh.
    \item The \rulefa-rule is applied due to an  assertions $\{ x : \all R D,
      (x,y) : R \} \subseteq \mcA$. Since $\I \models \mcA$, $y^\I \in
      D^\I$ must hold. Hence, $\I$ is also a model of $\mcA' = \mcA \cup \{ y
      : D \}$.
    \end{itemize}
  \item Again, let $\I$ be a model of $\mcA$. If an assertion $x : C_1 \sqcup
    C_2$ triggers the application of the \ruleor-rule, then $x^\I \in (C_1
    \sqcup C_2)^\I$ must hold. Hence, at least for one of the possible choices
    for $D \in \{ C_1, C_2\}$, $x^\I \in D^\I$ holds. For this choice, adding $x :
    D$ to $\mcA$ leads to an ABox that is satisfied by $\I$. \qed
  \end{enumerate} 
\end{proof}

\begin{theorem}[Correctness of the \alc-algorithm]
  The \alc-algorithm is a non-deterministic decision procedure for
  satisfiability of \alc-concepts.
  \index{alc@\alc!tableau algorithm!correctness}
\end{theorem}

\begin{proof}
  Termination was shown in Corollary~\ref{cor:alc-termination}. In
  Lemma~\ref{lem:alc-abox-yields-model} and
  Lemma~\ref{lem:alc-local-correctness}, we have established the conditions
  required to apply Theorem~\ref{theo:generic-correctness}, which yields
  correctness of the \alc-algorithm. \qed
\end{proof}

\subsection{Complexity}\label{sec:alc-complexity}

Now that we know that the \alc-algorithm is a non-deterministic decision
procedure for satisfiability of \alc-concepts, we want to analyse the
computational complexity of the algorithm to make sure that it matches the
known worst-case complexity of the problem. 

\subsubsection{Some Basics from Complexity Theory}

First, we briefly introduce the notions from complexity theory that we will
encounter in this thesis. For a thorough introduction to complexity theory we
refer to \cite{papadimitriou94:_compuat_compl}.

Let $M$ be a Turing Machine (TM) with input alphabet $\Sigma$. For a function
$f: \N \rightarrow \N$, we say that $M$ \emph{operates within time} $f(n)$ if,
for any input string $x \in \Sigma^*$, $M$ terminates on input $x$ after at most
$f(|x|)$ steps, where $|x|$ denotes the length of $x$. $M$ \emph{operates within
  space} $f(n)$ if, for any input $x \in \Sigma^*$, $M$ requires space at most
$f(|x|)$. For an arbitrary function $f(n)$ we define the following classes of
languages:
{\footnotesize
  \begin{align*}
  \textsc{Time}(f(n)) & = \{ L \subseteq \Sigma^* \mid \text{$L$ is decided by
    a deterministic TM  that  operates\ within time $f(n)$} \},\\
  \textsc{NTime}(f(n)) & = \{ L \subseteq \Sigma^* \mid \text{$L$ is decided by
    a non-deterministic TM  that operates\ within time $f(n)$} \},\\ 
  \textsc{Space}(f(n)) & = \{ L \subseteq \Sigma^* \mid \text{$L$ is
    decided by a deterministic TM  that  operates\ within space $f(n)$},\\
  \textsc{NSpace}(f(n)) & = \{ L \subseteq \Sigma^* \mid \text{$L$ is decided
    by a non-deterministic TM  that operates\ within space $f(n)$} \}.
\end{align*}
}
Since every deterministic TM is a non-deterministic TM, $\textsc{Time}(f(n))
\subseteq \textsc{NTime}(f(n))$ and $\textsc{Space}(f(n)) \subseteq
\textsc{NSpace}(f(n))$ hold trivially for an arbitrary function $f$. Also,
$\textsc{Time}(f(n)) \subseteq \textsc{Space}(f(n))$ and $\textsc{NTime}(f(n))
\subseteq \textsc{NSpace}(f(n))$  hold trivially for an arbitrary $f$
because within time $f(n)$ a TM can consume at most $f(n)$ units of space. 

In this thesis, we will encounter complexity classes shown in
Figure~\ref{fig:complexity-classes}.

\begin{figure}[tbh]
  \begin{align*}
    \pspace & = \bigcup_{k \in \N} \textsc{Space}(n^k)\\
    \npspace & = \bigcup_{k \in \N} \textsc{NSpace}(n^k)\\
    \exptime & = \bigcup_{k \in \N} \textsc{Time}(2^{n^k})\\
    \nexptime & = \bigcup_{k \in \N} \textsc{NSpace}(2^{n^k})\\
    2\mbox{-}\exptime & = \bigcup_{k \in \N} \textsc{Time}(2^{2^{n^k}})\\
    2\mbox{-}\nexptime & = \bigcup_{k \in \N} \textsc{NTime}(2^{2^{n^k}})
  \end{align*}
  \caption{Some complexity classes}
  \label{fig:complexity-classes}
  \index{pspace@\pspace}
  \index{npspace@\npspace}
  \index{exptime@\exptime}
  \index{nextime@\nexptime}
  \index{2-exptime@2-\exptime}
  \index{2-nexptime@2-\nexptime}
\end{figure}

It is known that the following relationships hold for these classes:
\[
\pspace = \npspace \subseteq \exptime \subseteq \nexptime \subseteq 2\mbox{-}\exptime
\subseteq 2\mbox{-}\nexptime,
\]
where the fact that $\pspace = \npspace$ is a corollary of Savitch's theorem
\cite{Savitch}.

We employ the usual notion of polynomial many-to-one reductions and completeness:
let $L_1, L_2 \subseteq \Sigma^*$ be two languages. A function $r : \Sigma^*
\rightarrow \Sigma^*$ is a \emph{polynomial reduction from $L_1$ to $L_2$} iff
there exists a $k \in \N$ such that $r(x)$ can be compute within time
$\mcO(|x|^k)$ and $x \in L_1$ iff $r(x) \in L_2$. A language $L$ is \emph{hard}
for a complexity class $\mcC$ if, for any $L' \in \mcC$, there exists a
polynomial reduction from $L'$ to $L$. The language $L$ is \emph{complete} for
$\mcC$ if it is $\mcC$-hard and $L \in \mcC$.

All these definitions are dependent on the arbitrary but fixed finite input
alphabet $\Sigma$. The choice of this alphabet is inessential as long as it
contains at least two symbols. This allows for succinct encoding of arbitrary
problems and a larger input alphabet can reduce the size of the encoding of a
problem only by a polynomial amount. All defined complexity classes are
insensitive to these changes. From now on, we assume an arbitrary but fixed
finite input alphabet $\Sigma$ with at least two symbols.

Note that this implies that there is not necessarily a distinct symbol for every
concept, role, or individual name in $\Sigma$. Instead, we assume that the
names appearing in concepts are suitably numbered. The results we are going to
present are insensitive to this (logarithmic) overhead and so we ignore this
issue from now on.

\begin{definition}\label{def:concept-size}
  For an arbitrary syntactic entity $X$, like a concept, TBox assertion,
  knowledge base, etc., we denote the length of a suitable encoding of $X$ in the
  alphabet $\Sigma$ with $|X|$. \eod
\end{definition}

\subsubsection{The Complexity of \alc-Satisfiability}

\begin{fact}[\npcite{Schmidt-Schauss91}, Theorem 6.3]
  Satisfiability of \alc-concepts is \pspace-complete.
\end{fact}

Since we are aiming for a \pspace-algorithm, we do not have to deal explicitly
with the non-determinism because $\pspace = \npspace$. Yet, if naively executed,
the algorithm behaves worse because it generates a model for a satisfiable
concept and there are \alc-concepts that are only satisfiable in exponentially
large interpretations, i.e., it is possible to give a concept $C_n$ of size
polynomially in $n$ such that any model of $C_n$ essentially contains a full
binary tree of depth $n$ and hence at least $2^n - 1$ nodes
\cite{HalpernMoses92}. Since the tableau generates a full description of a model, a
naive implementation would require exponential space.

To obtain an algorithm with optimal worst case complexity, the \alc-algorithm
has to be implemented in a certain fashion using the so-called
\iemph{trace technique}. The key idea behind this technique is that instead of
keeping the full ABox $\mcA$ in memory simultaneously, it is sufficient to
consider only a single path in $G_\mcA$ at one time.  In
Lemma~\ref{lem:alc-tree-bounds} we have seen that the length of such a path is
linearly bounded in the size of the input concept and there are only linearly
many constraints for every node on such a path.  Hence, if it is possible to
explore $G_\mcA$ one path at a time, then polynomial storage suffices. This can
be achieved by a depth-first expansion of the ABox that selects the rule to
apply in a given situation according to a specific strategy (immediately
stopping with the output ``unsatisfiable'' if a clash is generated).

\begin{figure}[tbh]
  \begin{center}
    \leavevmode
    \begin{tabbing}
      \hspace{2em}\=\hspace{2em}\=\hspace{2em}\= \kill
      $\alc$-$\textsc{Sat}(C) := \texttt{sat}(x_0,\{x_0 : C\})$\\
      $\texttt{sat}(x,\mcA)$: \\
      \> \texttt{while} (the \ruleand- or the \ruleor-rule can be
      applied) \texttt{and}  ($\mcA$ is clash-free) \texttt{do}\\
      \>\> apply the \ruleand- or the \ruleor-rule to $\mcA$.\\
      \> \texttt{od}\\
      \> \texttt{if} $\mcA$ contains a clash \texttt{then}
      \texttt{return} ``not satisfiable''.\\
      \> $\bfE := \{ x : \some R D \mid x : \some R D \in \mcA\}$\\
      \> \texttt{while} $\bfE \neq \emptyset$  \texttt{do}\\
      \>\> pick an arbitrary $x : \some R D \in \bfE$\\
      \>\> $\mcA_{\textit{new}} := \{ (x,y) : R, y : D\}$ \texttt{where} $y$ is
      a fresh individual\\ 
      \>\> \texttt{while} (the \rulefa-rule can be applied to $\mcA \cup
      \mcA_{\textit{new}})$ \texttt{do}\\
      \>\>\> apply the \rulefa-rule and add the new constraints to
      $\mcA_{\textit{new}}$\\ 
      \>\> \texttt{od}\\
      \>\> \texttt{if} $\mcA \cup \mcA_{\textit{new}}$ contains a clash \texttt{then}
      \texttt{return} ``not satisfiable''.\\
      \>\> \texttt{if} $\texttt{sat}(y,\mcA \cup \mcA_{\textit{new}}) =
      \text{``not satisfiable''}$ \texttt{then return} ``not satisfiable''\\
      \>\> $\bfE := \bfE \setminus \{ x : \some R D \mid y : D \in \mcA_{\textit{new}}\}$\\
      \>\> discard $\mcA_{\textit{new}}$ from memory\\
      \> \texttt{od}\\
      \> \texttt{return} ``satisfiable''
    \end{tabbing}
    \caption{A non-deterministic \pspace decision procedure for \alc.}
    \label{fig:alc-decision-proc}
  \end{center}
\end{figure}

\begin{lemma}\label{lem:alc-in-pspace}
  The \alc-algorithm can be implemented in \pspace.
  \index{alc@\alc!tableau algorithm!complexity}
\end{lemma}

\begin{proof}
  Let $C$ be the \alc-concept to be tested for satisfiability. We can
  assume $C$ to be in NNF because transformation into NNF can be
  performed in linear time.  Figure~\ref{fig:alc-decision-proc}
  sketches an implementation of the \alc-algorithm that uses the
  trace-technique to preserve memory and runs in polynomial space.
  
  The algorithm generates the constraint system in a depth-first
  manner: before generating any successors for an individual $x$, the
  \ruleand- and \ruleor-rule  are applied exhaustively. Then successors are
  considered for every existential restriction in $\mcA$ one after
  another re-using space. This has the consequence that a clash
  involving an individual $x$ must be present in $\mcA$ by the time
  generation of successors for $x$ is initiated or will never occur.
  This also implies that it is safe to delete parts of the constraint
  system for a successor $y$ as soon as the existence of a complete
  and clash-free ``sub'' constraint system has been determined. Of
  course, it then has to be ensured that we do not consider the same
  existential restriction $x : \some R D$ more than once because this
  might lead to non-termination.  Here, we do this using the set $\bfE$
  that records which constraints still have to be considered.  Hence,
  the algorithm is indeed an implementation of the \alc-algorithm.
  
  Space analysis of the algorithm is simple: since
  $\mcA_{\textit{new}}$ is reset for every successor that is
  generated, this algorithm stores only a single path at any given
  time, which, by Lemma~\ref{lem:alc-tree-bounds}, can be done using
  polynomial space only. \qed
\end{proof}

As a corollary, we get an exact classification of the complexity of
satisfiability of \alc-concepts.

\begin{theorem}\label{theo:alc-pspace-complete}
  Satisfiability of \alc-concepts is \pspace-complete.
\end{theorem}

\begin{proof}
  Satisfiability of \alc-concepts is known to be
  \pspace-hard~\cite{Schmidt-Schauss91}, which is shown by reduction from the
  well-known \pspace-complete problem QBF \cite{Stockmeyer-Meyer-73}.
  Lemma~\ref{lem:alc-in-pspace} together with the fact that $\pspace = \npspace$
  (Savitch's theorem \citeyear{Savitch}) yields the corresponding upper
  complexity bound. \qed
\end{proof}

It is possible to give an even tighter bound for the complexity of \alc-concept
satisfiability and to show that the problem is solvable in deterministic linear
space. This was already claimed in \cite{Schmidt-Schauss91}, but a closer
inspection of that algorithm by Hemaspaandra reveals that it consumes memory in
the order of $\mathcal{O}(n \log n)$ for a concept with length $|C| = n$.
Hemaspaandra \citeyear{hemaspaandra00:_modal_satis_is_deter_linear_space} gives
an algorithm that decides satisfiability for the modal logic $\K$ in
deterministic linear space and which is easily applicable to \alc.

\subsection{Other Inference Problems for \alc}
\label{sec:other-infer-probl}

Concept satisfiability is only one inference that is of interest for DL
systems. In the remainder of this chapter we give a brief overview of solutions
for the other standard inferences for \alc.

\subsubsection{Reasoning with ABoxes}

To decide ABox satisfiability of an \alc-ABox $\mcA$ (\wrt an empty TBox), one can
simply apply the \alc-algorithm starting with $\mcA$ as the initial ABox. One
can easily see that the proofs of soundness and completeness uniformly apply
also to this case. Yet, since the generated constraint system is no longer of
tree-shape, termination and complexity have to be reconsidered. Hollunder
\citeyear{hollunder96:_consis_check_reduc_satis_concep_termin_system} describes
\iemph{pre-completion}---a technique that allows reduction of ABox satisfiability
directly to \alc-concept satisfiability. The general idea is as follows: all
non-generating rules are applied to the input ABox $\mcA$ exhaustively yielding
a pre-completion $\mcA'$ of $\mcA$. After that, the \alc-algorithm is
called for every individual $x$ of $\mcA'$ to decide satisfiability of the
concept 
\[
C_x := \mybigsqcap_{x:D \in \mcA'} D .
\]
It can be shown that $\mcA'$ is satisfiable iff $C_x$ is satisfiable for every
individual $x$ in $\mcA'$ and that $\mcA$ is satisfiable iff the non-generating
rules can be applied in a way that yields a satisfiable pre-completion.  Since
ABox satisfiability is at least as hard as concept satisfiability, we get:

\begin{corollary}
  Consistency of \alc-ABoxes \wrt an empty TBox is \pspace-complete.
\end{corollary}

\subsubsection{Reasoning with Simple TBoxes}

For a simple TBox $\mcT$, concept satisfiability \wrt $\mcT$ can be reduced to
concept satisfiability by a process called \iemph{unfolding}:

Let $C$ be an \alc-concept and $\mcT$ a simple TBox. The \emph{unfolding
  $C_\mcT$ of $C$ \wrt $\mcT$} is obtained by successively replacing every
defined name in $C$ by its definition from $\mcT$ until only primitive (i.e.,
undefined) names occur. It can easily be shown that $C$ is satisfiable \wrt
$\mcT$ iff $C_\mcT$ is satisfiable. Unfortunately, this does not yield a
\pspace-algorithm, as the size of $C_\mcT$ may be exponential in the size of $C$
and $\mcT$. Lutz \citeyear{Lutz-99d} describes a technique called \iemph{lazy  unfolding} 
that performs the unfolding of $C$ \wrt $\mcT$ \emph{on demand},
which yields:

\begin{fact}[\npcite{Lutz-99d}, Theorem 1]
  Satisfiability of \alc-concepts \wrt to a simple TBox is \pspace-complete.
\end{fact}

Finally, the techniques of pre-completion and lazy-unfolding can be combined,
which yields:

\begin{corollary}
  Consistency of \alc knowledge bases with a simple TBox is \pspace-complete.
\end{corollary}

\subsubsection{Reasoning with General TBoxes}

If general TBoxes are considered instead of simple ones, the complexity of the
inference problems rises.

\begin{theorem}\label{theo:alc-general-tboxes-exptime-complete}
  Satisfiability of \alc-concepts (and hence of ABoxes) \wrt general TBoxes is
  \exptime-hard.
\end{theorem}

\begin{proof}
  As mentioned before, \alc is a syntactic variant of the propositional modal
  logic $\mathsf{K}_m$~\cite{Schi91}. As a simple consequence of the proof of
  \exptime-completeness of $\mathsf{K}$ with a universal modality
  \cite{Spaan93a}] (i.e., in DL terms, a role linking every two individuals), one
  obtains that the \emph{global satisfaction problem} for $\mathsf{K}$ is an
  \exptime-complete problem. The global satisfaction problem is defined as
  follows:
  \begin{quote}
    Given a $\mathsf{K}$-formula $\phi$, is there a Kripke model $\mfM$ such
    that $\phi$ holds at every world in $\mfM$?
  \end{quote}
  
  \index{global satisfiability}
  \index{satisfiability!global}

  Using the correspondence between $\alc$ and $\mathsf{K}_m$, this can be
  re-stated as an \exptime-complete problem for \alc:
  \begin{quote}
    Given an \alc-concept $C$, is there an interpretation $\I$ such that $C^\I 
    = \Delta^\I$?
  \end{quote}
  
  Obviously, this holds iff the tautological concept $\top$ is satisfiable \wrt
  the (non-simple) TBox $\mcT = \{ \top \doteq C \}$, which implies that
  satisfiability of \alc-concepts (and hence of ABoxes) \wrt general TBoxes is
  \exptime-hard. \qed
\end{proof}

A matching upper bound for \alc is given by De Giacomo and Lenzerini
\citeyear{DeGiacomo96a} by a reduction to PDL, which yields:

\begin{corollary}\label{cor:tbox-exptime-complete}
  Satisfiability and subsumption \wrt general TBoxes, knowledge base
  satisfiability and instance checking for \alc are \exptime-complete problems.
\end{corollary}



\cleardoublepage


\chapter{Qualifying Number Restrictions}
\label{chap:alcq}

In this chapter we study the complexity of reasoning with \alcq, the extension
of \alc with qualifying number restrictions. While for \alc, or, more precisely,
for its syntactic variant $\K$, \pspace-completeness has already been
established quite some time ago by Ladner \citeyear{Lad77}, the situation is
entirely different for \alcq or its corresponding (multi-)modal logic
$\grkr$. \index{grkr@\grkr} 
For \alcq, decidability of concept satisfiability has been shown only rather
recently by Baader and Hollunder~\citeyear{Hollunder91b} and the known \pspace
upper complexity bound for \alcq is only valid if we assume unary coding of
numbers in the input, which is an unnatural restriction.  For binary coding no
upper bound was known and the problem had been conjectured to be \exptime-hard by
van der Hoek and de Rijke \citeyear{VanderHoekdeRijke-JLC-1995}. This coincides
with the observation that a straightforward adaptation of the translation
technique leads to an exponential blow-up in the size of the first-order
formula. This is because it is possible to store the number $n$ in $\log_k
n$ bits if numbers are represented in $k$-ary coding.

We show that reasoning for \alcq is not harder than
reasoning for \alc (\wrt worst-case complexity) by presenting an
algorithm that decides satisfiability in \pspace, even if the numbers
in the input are binary coded.  It is based on the tableau algorithm
for \alc and tries to prove the satisfiability of a given concept by
explicitly constructing a model for it.  When trying to generalise the
tableau algorithms for \alc to deal with \alcq, there are some
difficulties: (1) the straightforward approach leads to an incorrect
algorithm; (2) even if this pitfall is avoided, special care has to be
taken in order to obtain a space-efficient solution. As an example for
(1), we will show that the algorithm presented in
\cite{VanderHoekdeRijke-JLC-1995} to decide satisfiability of \grkr, a
syntactic variant of \alcq, is incorrect.  Nevertheless, this
algorithm will be the basis of our further considerations.  Problem
(2) is due to the fact that tableau algorithms try to prove the
satisfiability of a concept by explicitly building a model for it. If
the tested formula requires the existence of $n$ accessible role successors, a
tableau algorithm will include them in the constructed model, which
leads to exponential space consumption, at least if the numbers in the
input are not unarily coded or memory is not re-used.  An example for
a correct algorithm which suffers from this problem can be found
in~\cite{Hollunder91b} and is briefly presented in this thesis. As we
will see, the trace technique alone is not sufficient to obtain an
algorithm that runs in polynomial space. Our algorithm overcomes this
additional problem by organising the search for a model in a way that allows for
the re-use of space \emph{for each successor}, thus being capable of
deciding satisfiability of \alcq in \pspace.

Using an extension of these techniques we obtain a \pspace algorithm
for the logic \alcqir, which extends \alcq by expressive role
expressions. This solves an open problem from~\cite{DLNN97}.

Finally, we study the complexity of reasoning \wrt general knowledge bases for
\alcqir and establish \exptime-completeness. This extends the
\exptime-completeness result for the more ``standard'' DL \alcqi
\cite{DeGiacomo95a}. Moreover, the proof in \cite{DeGiacomo95a} is only valid in
case of unary coding of numbers in the input whereas our proof also applies in
the case of binary coding.

\section{Syntax and Semantics of \alcq}

The DL \alcq is obtained from \alc by adding so-called \emph{qualifying number
  restrictions},
\index{qualifying number restrictions|textbf} i.e., concepts restricting the
number of individuals 
that are related via a given role instead of allowing  for existential or
universal restrictions only like in \alc. \alcq is a syntactic variant of the graded 
propositional modal logic \grkr.

\begin{definition}[Syntax of \alcq]
  \index{alcq@\alcq}
  Let \Names be a set of atomic \emph{concept names} and \Roles be a
  set of atomic \emph{role names}.  The set of \emph{\alcq-concepts}
  is built inductively from these according to the following grammar,
  where $A \in \Names$, $R \in \Roles$, and $n \in \N$:
  \[
  C ::= A  \bnfor \neg C \bnfor  C_1 \sqcap C_2 \bnfor  C_1 \sqcup C_2 \bnfor
  \all R C \bnfor \some R C  \bnfor \qnrleq n R C \bnfor  \qnrgeq n R C .
  \] \eod
\end{definition}

Thus, the set of \alcq-concepts is defined similar to the set of
\alc-concepts, with the additional rule that, if $R \in \Roles$, $C$ is an
\alcq-concept, and $n \in \N$, then also $\qnrleq n R C$ and $\qnrgeq n R C$
are \alcq-concepts. To define the semantics of \alcq-concepts, we extend
Definition~\ref{def:alc-semantics} to deal with these additional concept
constructors:

\begin{definition}[Semantics of \alcq]\label{def:alcq-semantics}
  For an interpretation $\I = (\domain, \cdot^\I)$, the semantics of
  \alcq-concepts is defined inductively as for \alc-concepts with the
  additional rules:
  \begin{align*}
    \qnrleq n R C ^\I & = \{ x \in \domain \mid \sharp R^\I(x,C) \leq n \} \text{ and}\\
    \qnrgeq n R C ^\I & = \{ x \in \domain \mid \sharp R^\I(x,C) \geq n \},
  \end{align*}
  where $\sharp R^\I(x,C) = \{ y \mid (x,y) \in R^\I \text{ and } y \in C^\I \}$
  and $\sharp$ denotes set cardinality. \eod
  \index{00rixc@$R^\I(x,C)$}
  \index{00sharp@$\sharp$}
\end{definition}

From these semantics, it is immediately clear that we can dispose of
existential and universal restriction in the syntax without changing the
expressiveness of \alcq, since the following equivalences allow the
elimination of universal and existential restrictions in linear time:
\[
\some R C \equiv \qnrgeq 1 R C \qquad \all R C \equiv \qnrleq 0 R {\neg C}
\]
In the following, we assume that \alcq-concepts are built without 
existential or universal restrictions. To obtain the NNF of an \alcq-concept,
one can ``apply'' the following equivalences (in addition to de Morgan's
laws): \index{negation normal form}
\begin{align*}
  \neg \qnrleq n R C & \equiv  \qnrgeq {(n+1)} R C\\
  \neg \qnrgeq n R C & \equiv  \begin{cases}
    \bot & \text{if $n=0$},\\
    \qnrgeq {(n+1)} R C & \text{otherwise} .
  \end{cases}
\end{align*}
Like for \alc, one can obtain the NNF of an \alcq-concept in linear time. For an
\alcq-concept $C$, we denote the NNF of $\lnot C$ by $\nneg
C$. \index{00nnegc@$\nneg C$}


\section{Counting Pitfalls}

Before we present our algorithm for deciding satisfiability of \alcq, for
historic and didactic reasons, we present two other solutions: an incorrect one
\cite{VanderHoekdeRijke-JLC-1995}, and a solution that is less
efficient~\cite{Hollunder91b}.


\subsection{An Incorrect Solution}
\label{sec:wrong-solution}

Van der Hoek and de Rijke \citeyear{VanderHoekdeRijke-JLC-1995} give
an algorithm for deciding satisfiability of the graded modal logic
\grkr. Since \grkr is a notational variant of \alcq, such an
algorithm could also be used to decide concept satisfiability for
\alcq. Unfortunately, the given algorithm is incorrect.  Nevertheless,
it will be the basis for our further considerations and thus it is
presented here. It will be referred to as the \emph{incorrect}
algorithm. It is based on a tableau algorithm given in~\cite{DLNN97}
to decide the satisfiability of the DL $\mathcal{ALCN}$, but overlooks
an important pitfall that distinguishes reasoning for qualifying
number restrictions from reasoning with number restrictions. This
mistake leads to the incorrectness of the algorithm. To fit our
presentation, we use DL syntax in the presentation of the algorithm.
Refer to~\cite{Tobies-CADE-99} for a presentation in modal syntax.

Similar to the \alc-algorithm presented in Section~\ref{sec:alc}, the flawed
solution is a tableau algorithm that tries to build a model for a concept $C$ by
manipulating sets of constraints with certain completion rules. Again, ABoxes
are used to capture constraint systems.

\begin{algorithm}[Incorrect Algorithm for \alcq, \npcite{VanderHoekdeRijke-JLC-1995}]\label{def:incorrect-abox}
  \index{alcq@\alcq!incorrect tableau algorithm}
  For an ABox $\mcA$, a role name $R$, an individual $x$, and a concept $D$,
  let $\sharp R^\mcA(x,D)$ be the number of individuals $y$ for which $\{(x,y)
  : R, y : D \} \subseteq \mcA$.  The ABox $[z/y]\mcA$ is obtained from $\mcA$
  by replacing every occurrence of $y$ by $z$; this replacement is said to be
  \emph{safe} iff, for every individual $x$, concept $C$, and role name $R$
  with $\{x : \qnrgeq n R D , (x,y):R, (x,z):R\} \subseteq S$ we have $ \sharp
  R^{[z/y]\mcA}(x,D) > n$.
  
  \index{00raxc@$R^\mcA(x,C)$}
  \index{safe replacement}

  The definition of a clash is slightly extended from the
  \alc-case to deal with obviously contradictory number restrictions:
  An ABox $\mcA$ is said to contain a \emph{clash}, iff 
  \[
  \{ x : A, x : \neg A \} \subseteq \mcA \ \text{ or } \ \{ x : \qnrleq m R D, x
  : \qnrgeq n R D \} \subseteq \mcA.
  \]
  for a concept name $A$, a concept $D$, and two integers $m < n$. Otherwise,
  $\mcA$ is called \emph{clash-free}.  An ABox $\mcA$ is called \emph{complete}
  iff none of the rules given in Fig.~\ref{fig:expansion-rules-derijke} is
  applicable to $\mcA$.
  
  To test the satisfiability of a concept $C$, the incorrect algorithm
  works as follows: it starts with the initial ABox $ \{ x_0 : C\}$
  and successively applies the rules given in
  Fig.~\ref{fig:expansion-rules-derijke}, stopping when a clash
  occurs. Both the rule to apply and the concept to add (in the
  \ruleor-rule) or the individuals to identify (in the \ruleleq-rule)
  are selected non-deterministically. The algorithm answers ``$C$ is
  satisfiable'' iff the rules can be applied in a way that yields a
  complete and clash-free ABox. \eod
  \index{ABox!initial}
\end{algorithm}

The notion of \emph{safe} replacement of variables is needed to ensure the
termination of the rule application \cite<see>{Hollunder91b}. The same purpose
could be achieved by explicitly asserting all successors generated to satisfy an
at-least restriction to be unequal and preventing the identification of unequal
elements. Yet, since this notion of safe replacement recurs in the algorithm of
Baader and Hollunder \citeyear{Hollunder91b}, which we are going to describe
later on, and since we want to outline an error in the incorrect algorithm, we
stay as close to the original description as possible.

Since we are interested in \pspace algorithms, as for \alc, non-determinism
poses no problem due to Savitch's Theorem, which implies that deterministic and
non-deterministic polynomial space coincide \cite{Savitch}.

\begin{figure}[tbh]
  \begin{minipage}[t]{\textwidth}
    \begin{tabular}{lll}
      \ruleand: &  \bfif \hfill 1. & $x : C_1 \sqcap C_2 \in \mcA$ and\\
      & \hfill 2. &$\{ x:C_1, x:C_2 \} \not \subseteq \mcA$\\
      & \bfthen & $\mcA \ruleand \mcA \cup \{ x:C_1, x:C_2 \}$\\[1ex]
      \ruleor: & \bfif \hfill 1.& $x : C_1 \sqcup C_2 \in \mcA$ and \\
      & \hfill 2. & $\{ x:C_1,
      x:C_2 \} \cap \mcA = \emptyset$\\
      & \bfthen & $\mcA \ruleor \mcA \cup \{ x:D \}$ for some $D \in \{ C_1, C_2
      \}$\\[1ex]
      \rulegeq: & \bfif \hfill 1. & $x : \qnrgeq n R D$ and\\
      & \hfill 2. & $\sharp R^\mcA(x,D) < n$ \\
      & \bfthen & $\mcA  \rulegeq \mcA  \cup \{ (x,y) : R, y : D \}$ where $y$ is a fresh
      variable.\\[1ex]
      \ruleleqz: & \bfif \hfill 1. & $\{ x : \qnrleq 0 R D, (x,y) : R\} \subseteq \mcA$
      and\\
      & \hfill 2. & $y : \nneg D\not\in \mcA$\\
      & \bfthen & $\mcA \ruleleqz \mcA \cup \{ y : \nneg D \}$\\[1ex]
      \ruleleq: & \bfif \hfill 1. & $x : \qnrleq n R D \in \mcA$, $R^\mcA(x,D) > n > 0$ and \\
      & \hfill 2. & $\{ (x,y) : R, (x,z) : R \} \subseteq  \mcA$ for some $y \neq
      z$ and\footnote{The rules in \cite{VanderHoekdeRijke-JLC-1995} do not
        require $\{ y : D, z : D \} \in \mcA$, as one might expect.}\\
      & \hfill 3.& replacing $y$ by $z$ is safe in $\mcA$\\
      & \bfthen & $\mcA \ruleleq [z/y]\mcA$
    \end{tabular}
  \end{minipage}
  \caption{The incorrect completion rules for \alcq}
  \label{fig:expansion-rules-derijke}
  \index{00ruleand@\ruleand}
  \index{00ruleor@\ruleor}
  \index{00rulegeq@\rulegeq}
  \index{00ruleleqz@\ruleleqz}
  \index{00ruleleq@\ruleleq}
  \index{alcq@\alcq!incorrect completion rules}
  \index{completion rules!for alcq@for \alcq!incorrect}

\end{figure}

As described in Section~\ref{sec:alc}, to prove the correctness of
such a tableau algorithm, we need to show three properties of the
completion:
\begin{enumerate}
\item Termination: Any sequence of rule applications is finite.
\item Soundness: If the algorithm terminates with a complete and clash-free ABox
  $\mcA$, then the tested concept is satisfiable.
\item Completeness: If the concept is satisfiable, then there is a sequence of
  rule applications that yields a complete and clash-free ABox.
\end{enumerate}

The error of the incorrect algorithm is, that is does not satisfy Property 2,
even though the opposite is claimed:

\begin{quote}
  \noindent \textsc{Claim}~\cite{VanderHoekdeRijke-JLC-1995}: (\emph{Restated in DL
    terminology}) Let $C$ be an \alcq-concept in NNF.  $C$ is satisfiable iff $\{
  x_0 : C \}$ can be transformed into a clash-free complete ABox using the
  rules from Figure~\ref{fig:expansion-rules-derijke}.
\end{quote}

Unfortunately, the \emph{if}-direction of this claim is not true. The
problem lies in the fact that, while a clash causes unsatisfiability, a
complete and clash-free ABox is not necessarily satisfiable. The
following counterexample exhibits this problem. Consider the concept
\[
C = \qnrgeq 3 R A \sqcap \qnrleq 1 R B \sqcap \qnrleq 1 R {\neg B} .
\]
On the one hand, $C$ is clearly not satisfiable. Assume an interpretation $\I$
with $x \in C^\I$.  This implies the existence of at least three
$R$-successors $y_1,y_2,y_3$ of $x$. For each of the $y_i$ either $y_i \in
B^\I$ or $y_i \in (\neg B)^\I$ holds by the definition of $\cdot^\I$.  Without
loss of generality, there are two elements $y_{i_1}, y_{i_2}$ such that $\{
y_{i_1}, y_{i_2} \} \subseteq B^\I$, which implies $x \not\in  \qnrleq
1 R B^\I$ and hence $x \not \in C^\I$.

On the other hand, the ABox $\mcA = \{ x_0 : C \}$ can be turned into a
complete and clash-free ABox using the rules from
Fig.~\ref{fig:expansion-rules-derijke}, as is shown in
Fig.~\ref{fig:wrong-run}.  Clearly this invalidates the claim and thus its
proof.

\begin{figure}[tbh]
  \begin{center}
    \begin{align*}
      \{ x : C \} \ruleand \cdots \ruleand & \underbrace{\{ x : C, \ x :
        \qnrgeq 3 R A, \ x : \qnrleq 1 R B, \ x : \qnrleq 1 R {\neg B} \}}_{=\mcA_1}\\
      \rulegeq \cdots \rulegeq & \underbrace{\mcA_1 \cup \{ (x,y_i) : R , \ y_i
        : A \mid i = 1, 2, 3 \}}_{= \mcA_2}
    \end{align*}
    $\mcA_2$ is clash-free and complete, because $\sharp R^{\mcA_2}(x,A) = 3$ and
    $\sharp R^{\mcA_2}(x,B) = 0$.
    \caption{A run of the incorrect algorithm.}
    \label{fig:wrong-run}
  \end{center}
\end{figure}

To understand the mistake of the incorrect algorithm, it is useful to recall
how soundness is usually established for tableau algorithms. The central
idea is that a complete and clash-free ABox $\mcA$ is ``obviously''
satisfiable, in the sense that a model of $\mcA$ can directly be constructed
from $\mcA$. For a complete and clash-free \alcq-ABox $\mcA$ we define the
canonical interpretation $\I_\mcA$ as in Definition~\ref{def:canonical-interpretation}.

The mistake of the incorrect algorithm is due to the fact that it did not take
into account that, in the canonical interpretation induced by a complete and
clash-free ABox, there are concepts satisfied by the individuals even though
these concepts do not appear as constraints in the ABox.  In our example, all of
the $y_i$, for which $B$ is not explicitly asserted, satisfy $\neg B$ in the
canonical interpretation but this is not reflected in the generated ABox.

\subsection{A Correct but Inefficient Solution}

This problem has already been noticed in~\cite{Hollunder91b}, where an
algorithm very similar to the incorrect one is presented that
correctly decides the satisfiability of $\alcq$-concepts.

The algorithm essentially uses the same definitions and rules. The
only substantial difference is the introduction of the
\rulechoose-rule, which makes sure that all ``relevant'' concepts that
are implicitly satisfied by an individual are made explicit in the
ABox. Here, relevant concepts for an individual $y$ are those occurring
in qualifying number restrictions in constraints for variables $x$
such that $(x,y) : R$ appears in the ABox.

\begin{algorithm}[The Standard Algorithm for \alcq,  \npcite{Hollunder91b}]\label{alg:alcq-standard}
  \index{alcq@\alcq!standard tableau algorithm} The rules of the \emph{standard
    algorithm} are given in Figure~\ref{fig:mod-rules}. The definition of
  \emph{clash} is modified as follows: an ABox $\mcA$ contains a clash iff
  \begin{itemize}
  \item $\{ x : A, x : \neg A \} \subseteq \mcA$ for some individual $x$ and
    a concept name  $A$, or
  \item $x : \qnrleq n R D \in \mcA$ and $\sharp R^\mcA(x,D) > n$ for some
    variable $x$, relation $R$, concept $D$, and $n \in \N$.
  \end{itemize}

  \begin{figure}[tbh]
    \begin{center}
      \begin{tabular}{@{ }l@{ }l@{ }l@{}}
        \ruleand, \ruleor: & \multicolumn{2}{@{}l}{see Fig.~\ref{fig:expansion-rules-derijke}}\\[1ex]      
        \rulechoose: & if \hfill 1. & $\{x : \qnrgleq n R D, (x,y) : R \}  \subseteq \mcA$ and\\
        & \hfill 2. & $\{ y : D, y : \nneg D \} \cap \mcA =
        \emptyset$\\
        & then & $\mcA \rulechoose \mcA \cup \{ y : E \}$ where $E \in \{
        D, \nneg D\}$\\[1ex]
        \rulegeq: & if \hfill 1. & $x : \qnrgeq n R D \in \mcA$ and
        \\
        & \hfill 2. & $\sharp R^\mcA(x,D) < n$\\
        & then & $\mcA \rulegeq \mcA \cup \{ (x,y) : R, y : D \}$ where $y$ is a new
        variable.  \\[1ex]
        \ruleleq: & if \hfill 1. & $x : \qnrleq n R D \in \mcA$, $\sharp
        R^\mcA(x,D) > n$, and \\
        & \hfill 2. &  $\{ (x,y) : R, (x,z) : R, y : D , z : D\}
        \subseteq \mcA$, for some $y \neq z$      and \\
        & \hfill 3. & for every $u$ with $(x,u) : R \in \mcA$, $\{ u : D, u :
        \nneg D \} \cap \mcA \neq \emptyset$, and\\
        & \hfill 4. & the replacement of $y$ by $z$
        is safe in $\mcA$\\
        & then &$\mcA \ruleleq [y/z]\mcA$
      \end{tabular}
      \caption{The standard completion rules for \alcq}
      \label{fig:mod-rules}
    \end{center}
    \index{00ruleand@\ruleand}
    \index{00ruleor@\ruleor}
    \index{00rulegeq@\rulegeq}
    \index{00rulechoose@\rulechoose}
    \index{alcq@\alcq!standard completion rules}
    \index{completion rules!for alcq@for \alcq!standard}
  \end{figure}
    
  The algorithm works like the incorrect algorithm with the following
  differences: (1) it uses the completion rules from
  Fig.~\ref{fig:mod-rules} (where $\bowtie$ \index{00bowtie@$\bowtie$} is used as a placeholder
  for either $\leq$ or $\geq$); (2) it uses the definition of clash
  from above; and (3) it does not immediately stop when a clash has
  been generated but always generates a complete ABox.  \eod
\end{algorithm}

The standard algorithm is a decision procedure for \alcq-concept
satisfiability:

\begin{theorem}[\npcite{Hollunder91b}]\label{theo:modified-sound-and-correct}
  Let $C$ be an \alcq-concept in NNF. $C$ is satisfiable iff $\{ x_0 : C \}$
  can be transformed into a clash-free complete ABox using the rules in
  Figure~\ref{fig:mod-rules}. Moreover, each sequence of these
  rule-applications is finite.
\end{theorem}

While no complexity result is explicitly given in
\cite{Hollunder91b}, it is easy to see that a \pspace result could be
derived from the algorithm using the trace technique from
Section~\ref{sec:alc}.

Unfortunately this is only true if we assume the numbers in the input to be
unary coded. The reason for this lies in the \rulegeq-rule, which generates
$n$ successors for a concept of the form $\qnrgeq n R D$. If $n$ is unary
coded, these successors consume at least polynomial space in the size of the
input concept. If we assume binary (or $k$-ary with $k>1$) encoding, the space
consumption is exponential in the size of the input because a number $n$ can
be represented in $\log_k n$ bits in $k$-ary coding. This blow-up cannot be
avoided because the completeness of the standard algorithm relies on the
generation \emph{and identification} of these successors, which makes it
necessary to keep them in memory \emph{at one time}.



\section{An Optimal Solution}

In the following, we will now present the algorithm with optimal worst case
complexity, which will be used to prove the exact complexity result for \alcq:

\begin{theorem}\label{theo:completeness-binary-coding}
  Satisfiability of \alcq-concepts is \pspace-complete, even if numbers in the
  input are represented using \textbf{binary} coding.
\end{theorem}

When aiming for a \pspace algorithm, it is impossible to generate all
successors of an individual in the ABox simultaneously at a given stage as this may
consume space that is exponential in the size of the input concept. We will
give an optimal rule set for \alcq-satisfiability that does not rely on the
identification of successors.  Instead we will make stronger use of
non-determinism to guess the assignment of the relevant concepts to the
successors by the time of their generation.  This will make it possible to
generate the completion tree in a depth-first manner, which facilitates re-use
of space.

\begin{algorithm}[The Optimal Algorithm for \alcq]\label{alg:alcq-optimal}
  \index{alcq@\alcq!optimal tableau algorithm}
  The definition of \emph{clash} is taken from
  Algorithm~\ref{alg:alcq-standard}.
  
  To test the satisfiability of a concept $C$, the optimal algorithm starts with
  the initial ABox $ \{ x_0 : C\}$ and successively applies the rules given in
  Fig.~\ref{fig:alcq-opt-rules}, stopping when a clash occurs. The algorithm
  answers ``$C$ is satisfiable'' iff the rules can be applied in a way that
  yields a complete and clash-free ABox. \eod 
  \index{ABox!initial}
\end{algorithm}

\begin{figure}[tbh]
    \begin{center}
      \begin{tabular}{@{ }l@{ }l@{ }l@{}}
        \multicolumn{3}{@{ }l}{\ruleand-, \ruleor: see
          Fig.~\ref{fig:expansion-rules-derijke}}\\[1ex]       
        \rulegeq: & if \hfill 1. & $x : \qnrgeq n R D \in \mcA$, and
        \\
        & \hfill 2. & $\sharp R^\mcA(x,D) < n$, and \\
        & \hfill 3. & neither the $\ruleand$- nor the $\ruleor$-rule apply to a constraint
        for $x$ \\
        & then & $ \mcA \rulegeq \mcA \cup \{ (x,y) : R, y : D , y : D_1, \dots, y
        : D_k \}$ 
        where \\
        & & $\{ E_1, \dots, E_k \} = \{ E \mid x : \qnrgleq m R E \in \mcA \}$,
        $D_i \in \{ E_i, \nneg E_i \}$, and\\ 
        & & $y$ is a fresh individual.
      \end{tabular}
    \end{center}
    \caption{The optimal completion rules for \alcq.}
    \label{fig:alcq-opt-rules}
    \index{00rulegeq@\rulegeq}
    \index{00ruleand@\ruleand}
    \index{00ruleor@\ruleor}
    \index{alcq@\alcq!optimal completion rules}
    \index{completion rules!for alcq@for \alcq!optimal}
  \end{figure}  
  
  For the different kinds on non-determinism present in this algorithm, compare
  the discussion below Algorithm~\ref{alg:alc}. In the proof of
  Lemma~~\ref{lem:alcq-local-correctness}, it is shown that the choice
  of which rule to apply when is don't-care non-deterministic. Any strategy that
  decides which rule to apply if more than one is applicable will yield a
  complete algorithm.

At first glance, the $\rulegeq$-rule may appear to be complicated and therefore it
is explained in more detail: like the standard $\rulegeq$-rule, it is applicable
to an ABox that contains the constraint $x : \qnrgeq n R D$ if there are less
than $n$ $R$-successors $y$ of $x$ with $y : D \in \mcA$.  The rule then adds a
new successor $y$ of $x$ to $\mcA$.  Unlike the standard algorithm, the
optimal algorithm also adds additional constraints of the form $y : (\nneg)E$
to $\mcA$ for each concept $E$ appearing in a constraint of the form $x :
\qnrgleq m R E$. Since  application of the \rulegeq-rule is suspended
until no other rule applies to $x$, by this time $\mcA$ contains all constraints
of the form $x : \qnrgleq m R E$ it will ever contain.  This combines the
effects of both the $\rulechoose$- and the $\ruleleq$-rule of the standard
algorithm.



\subsection{Correctness of the Optimized Algorithm}

To establish the correctness of the optimal algorithm, we will show
its termination, soundness, and completeness. Again, it is convenient
to view $\mcA$ as the graph $G_\mcA=(V,E,\Lab)$ as defined in
Section~\ref{sec:alc}.  Since the \rulegeq-rule not only adds
sub-concepts of $C$ but in some cases also the NNF of sub-concepts,
the label $\Lab(x)$ of a node $x$ is no longer a subset of $\sub(C)$
but rather of the larger set $\clos(C)$ defined below.

\begin{definition}
  For an \alcq-concept $C$ we define the closure $\clos(C)$ as the smallest
  set of \alcq-concepts that
  \begin{itemize}
  \item contains $C$, 
  \item is closed under sub-concepts, and
  \item is closed under the application of $\nneg$.
  \end{itemize} \eod
  \index{00closC@$\clos(C)$}
\end{definition}

It is easy to see that the size of $\clos(C)$ is linearly bounded in $|C|$:

\begin{lemma}\label{lem:alcq-closure-linear}
  For an \alcq-concept $C$ in NNF,
  \[
  \sharp \clos(C) \leq 2 \times |C|
  \]
\end{lemma}

\begin{proof}
  This is an immediate consequence of the fact that
  \[
  \clos(C) \subseteq \sub(C) \cup \{ \nneg D \mid D \in \sub(C) \}
  \]
  which can be shown as follows. Obviously, the set $\sub(C) \cup \{ \nneg D
  \mid D \in \sub(C) \}$ contains $C$ and is closed under the application of
  $\nneg$ (Note that, for a sub-concept $D$ of a concept in NNF, $\nneg \nneg
  D = D$). Closure under sub-concepts for the concepts in $\sub(C)$ is also
  immediate, and can be established for $ \{ \nneg D \mid D \in \sub(C) \}$ by
  considering the various possibilities for \alcq-concepts. \qed
\end{proof}

Similar to \alc, it is easy to show that the graph $G_\mcA$ for an
ABox $\mcA$ generated by the optimal algorithm from an initial ABox
$\{ x_0 : C \}$ is a tree with root $x_0$, and for each edge $(x,y)
\in E$, the label $\Lab(x,y)$ is a singleton.  Moreover, for each node
$x$ it holds that $\Lab(x) \subseteq \clos(C)$.

\subsubsection{Termination}

First, we will show that the optimal algorithm always terminates, i.e., each
sequence of rule applications starting from the ABox $\{ x_0 : C \}$ is
finite. The next lemma will also be helpful when we  consider the
complexity of the algorithm.

\begin{lemma}\label{lem:path-is-poly-long}
  Let $C$ be a concept in NNF and $\mcA$ an ABox that is generated by the
  optimal algorithm starting from $\{ x_0 : C \}$.
  \begin{itemize}
  \item The length of a path in $G_\mcA$ is limited by $|C|$.
  \item The out-degree of $G_\mcA$ is bounded by $|C| \times 2^{|C|}$.
  \end{itemize}
\end{lemma}

\begin{proof}
  The linear bound on the length of a path in $G_\mcA$ is established as for
  the \alc-algorithm using the fact that the nesting of qualifying number
  restrictions strictly decreases along a path in $G_\mcA$.

  
  Successors in $G(\mcA)$ are only generated by the $\rulegeq$-rule. For an
  individual $x$ this rule will generate at most $n$ successors for each
  $\qnrgeq n R D \in \Lab(x)$. There are at most $|C|$ such
  concepts in $\Lab(x)$. Hence the out-degree of $x$ is bounded by
  $|C|\times 2^{|C|}$, where $2^{|C|}$ is a limit for the
  biggest number that may appear in $C$ if binary coding is used.  \qed
\end{proof}

\begin{corollary}[Termination]
\label{lem:optimised-alg-terminates}
Any sequence of rule applications starting from an ABox $\mcA = \{ x_0 : C \}$
of the optimal algorithm is finite. \index{alcq@\alcq!optimal tableau
  algorithm!termination}
\index{termination!of the alcq-algorithm@of the \alcq-algorithm}
\end{corollary}

\begin{proof}
  The sequence of rules induces a sequence of trees. The depth and the
  out-degree of these trees is bounded by some function in $|C|$ by
  Lemma~\ref{lem:path-is-poly-long}. For each individual $x$ the label
  $\Lab(x)$ is a subset of the finite set $\clos(C)$. Each application of
  a rule either
  \begin{itemize}
  \item adds a new constraint of the form $x : D$ and hence adds an element
    to $\Lab(x)$, or
  \item adds fresh individuals to $\mcA$ and hence adds additional nodes to the
    tree $G_\mcA$.
  \end{itemize}
  Since constraints are never deleted and individuals are never deleted or identified, an
  infinite sequence of rule application must either lead to an infinite
  number of nodes in the trees which contradicts their boundedness, or it
  leads to an infinite label of one of the nodes $x$ which contradicts
  $\Lab(x) \subseteq \clos(C)$.  \qed
\end{proof}

\subsubsection{Soundness and Completeness}

We establish soundness and completeness of the optimal algorithm
along the lines of Theorem~\ref{theo:generic-correctness}. We use a slightly
modified notion of ABox satisfiability, which is already implicitly present in
the definition of clash. If we want to apply
Theorem~\ref{theo:generic-correctness} to prove the correctness of the
algorithm, then we need that a clash in an ABox causes unsatisfiability
of that ABox. For an arbitrary ABox and the definition of clash used
by the optimal algorithm, this is not the case. For example the ABox
\[
\mcA = \{ x : \qnrleq 1 R A, (x,y) : R, (x,z) : R, y : A, z: A \}
\]
contains a clash but is satisfiable. Yet, if we require, that for all individuals
$x,y,z$, if $(x,y) : R, (x,z) : R \in \mcA$ and $y \neq z$, then $y$ and $z$ must
be interpreted with different elements of the domain, then a clash obviously
implies unsatisfiability. This is captured by the definition of the function $\widehat
\cdot$ \index{00widehata@$\widehat \mcA$} that maps an \alcq-ABox to its \iemph{differentiation} 
\index{ABox!differentiation} $\widehat \mcA$ defined by
\[
\widehat \mcA = \mcA \cup \{ y \ndoteq z \mid \{ (x,y) : R, (x,z) : R \} \subseteq \mcA, y
\neq z \} .
\]

For the proof of soundness and completeness of Algorithm~\ref{alg:alcq-optimal},
an ABox $\mcA$ is called satisfiable iff $\widehat \mcA$ is satisfiable in the
(standard) sense of Definition~\ref{def:abox}. Since $\widehat \cdot$ is idempotent,
the standard and the modified notion of satisfiability coincide for a
differentiated ABox $\widehat \mcA$ and we can unambiguously speak of the
satisfiability of $\widehat \mcA$ without specifying if we refer to the modified or
the standard notion.


Consider the properties required by Theorem~\ref{theo:generic-correctness}.  As
discussed before, with this definition of satisfiability of ABoxes, it is
obvious that, for an ABox $\mcA$ generated by the optimal algorithm that contains
a clash, $\widehat \mcA$ (and hence, by our definition, \mcA) must be
unsatisfiable (Property 1) and that $\{ x_0 : C \}$ is satisfiable iff $C$ is
satisfiable (Property 2).  It remains to establish Property 3 (a clash-free and
complete ABox is satisfiable, Lemma~\ref{lem:soundness-optimised-algorithm}) and
the local correctness (Properties 4,5) of the rules
(Lemma~\ref{lem:alcq-local-correctness}).

For \alc, to prove satisfiability of a complete and clash-free ABox
$\mcA$, we used induction over the structure of concepts appearing in
constraints in $\mcA$. This was possible because the \alc-rules, when
triggered by an assertion $x : D$, only add constraints to $\mcA$ that
involve sub-concepts of $D$. For \alcq, and specifically for the
\rulegeq-rule, this is no longer true  and hence a proof by
induction on the structure of concepts is not feasible. Instead, we
will use induction on following \iemph{norm} of concepts.

\begin{definition}\label{def:alcq-concept-norm}
  For an \alcq-concept $D$ in NNF, then norm $\| D \|$ is inductively defined
  by:
  \[
  \begin{array}{lclcl}
    \|A\| & := & \|\lnot A\| & := & 0 \quad \text{for $A\in\Names$}\\ 
    \| C_1 \sqcap C_2 \| & := &  \| C_1 \sqcup C_2 \| & := & 
    1+\|C_1\|+\|C_2\|\\
    \|\qnrgleq n R D\| & & & := & 1+\|D\|
  \end{array} 
  \] \eod
  \index{00normc@$"\"|C"\"|$}
\end{definition}

The reader may verify that this norm satisfies $\| D\| = \| \nneg D
\|$ for every concept $D$.

\begin{lemma}
  \label{lem:soundness-optimised-algorithm}
  Let $\mcA$ be a complete and clash-free ABox generated by the optimal
  algorithm. Then $\widehat \mcA$ is satisfiable.
\end{lemma}

\begin{proof}
  Let $\mcA$ be a complete and clash-free ABox generated by applications of the
  optimal rules and $\widehat \mcA$ its differentiation. We show that the canonical
  interpretation $\I_\mcA$, as defined in
  Definition~\ref{def:canonical-interpretation}, is a model of $\widehat \mcA$.
  
  By definition of $\I_\mcA$, all constraints of the form $(x,y) : R$ are
  trivially satisfied.  Also, $y \neq z$ implies $y^\canI \neq z^\canI$ by
  construction of $\canI$. Thus, all remaining assertions in $\widehat \mcA$ are of
  the form $x : D$ and are also present in $\mcA$. Thus, it is sufficient to
  show that $x : D \in \mcA$ implies $x^{\I_\mcA} \in D^\canI$, which we will
  do by induction on the norm $\|\cdot\|$ of a concept $D$. Note that, by the
  definition of $\canI$, $x^\canI = x$ for every individual $x$ that occurs in
  $\mcA$.
  \begin{itemize}
  \item The first base case is $D = A$ for $A \in \Names$. $x : A \in \mcA$
    immediately implies $x \in A^\canI$ by the definition of $\canI$.  The
    second base case is $x : \lnot A \in \mcA$. Since $\mcA$ is clash-free,
    this implies $x : A \not\in \mcA$ and hence $x \not\in A^\canI$.  This
    implies $x \in (\lnot A)^\canI$
  \item For the conjunction and disjunction of concepts this follows exactly
    as in the proof of Lemma~\ref{lem:alc-abox-yields-model}.
  \item $x : \qnrgeq n R D\in \mcA$ implies $\sharp R^\mcA(x,D) \geq n$
    because otherwise the \rulegeq-rule would be applicable and $\mcA$ would
    not be complete. By induction, we have $y \in D^\canI$ for each $y$ with
    $y : D \in \mcA$. Hence $\sharp R^\canI(x,D) \geq n$ and thus $x \in
    \qnrgeq n R D ^\canI$.
  \item $x : \qnrleq n R D \in \mcA$ implies $\sharp R^\mcA(x,D) \leq n$ because
    $\mcA$ is clash-free.  Hence it is sufficient to show that $\sharp
    R^\canI(x,D) \leq \sharp R^\mcA(x,D)$ holds. On the contrary, assume $\sharp
    R^\canI(x,D) > \sharp R^\mcA(x,D)$ holds. Then there is an individual $y$
    such that $(x,y) : R \in \mcA$ and $y \in D^\canI$ but $y : D \not\in \mcA$.
    The application of the $\rulegeq$-rule is suspended until the propositional
    rules are no longer applicable to $x$ and hence, by the time $y$ is
    generated by an application of the \rulegeq-rule, $\mcA$ contains the
    assertion $x : \qnrleq n R D$. Hence, the \rulegeq-rule ensures $y : D \in
    \mcA$ or $y : \nneg D \in \mcA$.  Since we have assumed that $y : D \not \in
    \mcA$, this implies $y : \nneg D \in \mcA$ and, by the induction hypothesis,
    $y \in (\nneg D)^\canI$ holds, which is a contradiction. \qed
  \end{itemize} 
\end{proof}

\begin{lemma}[Local Correctness]
  \label{lem:alcq-local-correctness}
  Let $\mcA,\mcA'$ be ABoxes generated by the optimal algorithm from an ABox
  of the form $\{ x_0 : C \}$.
  \begin{enumerate}
  \item If $\mcA'$ is obtained from $\mcA$ by application of the (deterministic)
    $\ruleand$-rule, then $\widehat \mcA$ is satisfiable iff $\widehat \mcA'$ is satisfiable.
  \item If $\mcA'$ is obtained from $\mcA$ by application of the
    (non-deterministic) $\ruleor$- or $\rulegeq$-rule, then $\widehat \mcA$ is
    satisfiable if $\widehat \mcA'$ is satisfiable. Moreover, if $\widehat \mcA$ is
    satisfiable, then the rule can always be applied in such a way
    that it yields a satisfiable $\widehat \mcA''$.
  \end{enumerate}
\end{lemma}

\index{alcq@\alcq!optimal tableau algorithm!local correctness}

\begin{proof}
  $\mcA \to \mcA'$ for any rule $\to$ implies $\mcA \subseteq \mcA'$ and, by the
  definition of $\widehat \cdot$, $\widehat \mcA \subseteq \widehat \mcA'$,
  hence, if $\widehat \mcA'$ is satisfiable then so is $\widehat \mcA$. For the
  other direction, the \ruleand- and \ruleor-rule can be handled  as
  in the proof for \alc in Lemma~\ref{lem:alc-local-correctness}.
    
  It remains to consider the \rulegeq-rule. Let $\I$ be a model of $\widehat
  \mcA$ and let $x : \qnrgeq n R D$ be the constraint that triggers the
  application of the \rulegeq-rule. Since the \rulegeq-rule is applicable, we
  have $\sharp R^\mcA(x,D) < n$. We claim that there is an $a \in \Delta^\I$ with
  \begin{equation}
    \tag{$*$}
    (x^\I,a) \in R^\I, a \in D^\I \ \text{and} \ a \not\in \{ z^\I \mid
    (x,z) : R \in \mcA \}.
  \end{equation}
  Before we prove this claim, we show how it can be used to finish the proof.
  The element $a$ is used to ``select'' a choice of the \rulegeq-rule that
  preserves satisfiability: let $\{ E_1, \dots, E_k \}$ be an
  enumeration of the set $\{ E \mid x : \qnrgleq m R E \in \mcA \}$. We set
  \begin{align*}
    \mcA'' = \mcA & \cup \{ (x,y) : R , y : D \} \cup \{ y : E_i \mid a \in E_i^\I \}
    \cup \{ y : \nneg E_i \mid a \not\in E_i^\I \}
  \end{align*}
  Obviously, $\I[y \mapsto a]$, the interpretation that maps $y$ to $a$ and
  agrees with $\I$ on all other names, is a model for $\widehat \mcA''$, since
  $y$ is a fresh individual and $a$ satisfies $(*)$. The ABox $\mcA''$ is a
  possible result of the application of the \rulegeq-rule to $\mcA$, which
  proves that the \rulegeq-rule can indeed be applied in a way that maintains
  satisfiability of the ABox.
  
  We will now come back to the claim. It is obvious that there is an $a$ with
  $(x^\I,a) \in R^\I$ and $a \in D^\I$ that is not contained in $\{ z^\I \mid
  (x,z) : R, z : D \in \mcA \}$, because $\sharp R^\I(x,D) \geq n > \sharp
  R^\mcA(x,D)$.  Yet $a$ might appear as the image of an individual $z$ such
  that $(x,z) : R \in \mcA$ but $z : D \not\in \mcA$.
  
  Now, $(x,z) : R \in \mcA$ and $z : D \not\in \mcA$ implies $z : \nneg D\in
  \mcA$.  This is due to the fact that the constraint $(x,z) : R$ must have
  been generated by an application of the \rulegeq-rule because it has not
  been an element of the initial ABox. The application of this rule was
  suspended until neither the \ruleand- nor the \ruleor-rule were applicable to
  $x$.  Hence, if $x : \qnrgeq n R D$ is an element of $\mcA$ now, then
  it has already been in $\mcA$ when the \rulegeq-rule that generated $z$ was
  applied. The \rulegeq-rule guarantees that either $z : D$ or $z : \nneg D$
  is added to $\mcA$, hence $z : \nneg D \in \mcA$.  This is a contradiction
  to $z^\I = a$ because under the assumption that $\I$ is a model of $\mcA$
  this would imply $a \in (\nneg D)^\I$ while we initially assumed $a \in
  D^\I$.  \qed
\end{proof}

As an immediate consequence of the Lemmas~\ref{lem:optimised-alg-terminates},
\ref{lem:soundness-optimised-algorithm}, and \ref{lem:alcq-local-correctness}
together with Theorem~\ref{theo:generic-correctness} we get:

\begin{corollary}\label{theo:correctness-optimised-algorithm}
  The optimal algorithm is a non-deterministic decision procedure for
   satisfiability of \alcq-concepts.
\end{corollary}

\subsection{Complexity of the Optimal Algorithm}

The optimal algorithm will enable us to prove
Theorem~\ref{theo:completeness-binary-coding}. We will give a proof by sketching
an implementation of this algorithm that runs in polynomial space.

\begin{lemma}\label{lem:optimised-algorithm-in-pspace}
  The optimal algorithm can be implemented in \pspace
\end{lemma}

\begin{proof}
  Let $C$ be an \alcq-concept to be tested for satisfiability. We can assume $C$
  to be in NNF because the transformation of a concept to NNF can be performed
  in linear time.
  
  The key idea for the \pspace implementation is the \iemph{trace technique}
  \cite{Schmidt-Schauss91} we have already used for the \alc-algorithm in
  Section~\ref{sec:alc-complexity}, and which is based on the fact that it is
  sufficient to keep only a single path (a trace) of $G_\mcA$ in memory at a
  given stage if $\mcA$ is generated in a depth-first manner. This idea has been
  the key to a \pspace upper bound for \kr and \alc
  in~\cite{Lad77,Schmidt-Schauss91,HalpernMoses92}. To do this we need to store
  the values for $\sharp R^\mcA(x,D)$ for each individual $x$ in the path, each
  $R$ that appears in $\clos(C)$, and each $D \in \clos(C)$.  By storing these
  values in binary form, we are able to keep information \emph{about}
  exponentially many successors in memory while storing only a single path at a
  given stage.
  
  Consider the algorithm in Fig.~\ref{fig:decision-proc}, where $\Roles_C$
  \index{00nrc@$\Roles_C$} denotes the set of role names that appear in
  $\clos(C)$.  It re-uses the space needed to check the satisfiability of a
  successor $y$ of $x$ once the existence of a complete and clash-free
  ``subtree'' for the constraints on $y$ has been established.  This is
  admissible since, as was the case for \alc, the optimal rules will never
  modify this subtree once it is completed.  Constraints in this subtree also
  have no influence on the completeness or the existence of a clash in the rest
  of the tree, with the exception that constraints of the form $y : D$ for
  $R$-successors $y$ of $x$ contribute to the value of $\sharp R^\mcA(x,D)$.
  These numbers play a role both in the definition of a clash and for the
  applicability of the \rulegeq-rule. Hence, in order to re-use the space
  occupied by the subtree for $y$, it is necessary and sufficient to store these
  numbers.

  \begin{figure}[tbh]
    \begin{tabbing}
      \hspace{2em}\=\hspace{2em}\=\hspace{2em}\= \kill
      $\alcq\mbox{-}\textsc{Sat}(C) := \texttt{sat}(x_0,\{x_0 : C\})$\\
      $\texttt{sat}(x,\mcA)$: \\
      \> allocate counters  $\sharp R^\mcA(x,D) :=0$ for all $R \in
      \Roles_C$ and $D \in \clos(C)$. \\
      \> \texttt{while} (the \ruleand- or the \ruleor-rule can be
      applied) \texttt{and}  ($\mcA$ is clash-free) \texttt{do}\\
      \>\> apply the \ruleand- or the \ruleor-rule to $\mcA$.\\
      \> \texttt{od}\\
      \> \texttt{if} $\mcA$ contains a clash \texttt{then}
      \texttt{return} ``not satisfiable''.\\
      \> \texttt{while} (the \rulegeq-rule applies to a constraint $x :
      \qnrgeq n R D \in \mcA$) \texttt{do} \\
      \>\> $\mcA_{\textit{new}} := \{(x,y):R, y : D, y : D_1, \dots, y : D_k \}$ \\
      \>\> \texttt{where}\\
      \>\>\> $y$ is a fresh individual,\\
      \>\>\> $\{ E_1, \dots, E_k \} = \{ E \mid x  : \qnrgleq m R E \in \mcA \}$, and \\
      \>\>\> $D_i$ is chosen non-deterministically from $\{ E_i, \nneg E_i \}$ \\
      \>\> \texttt{for each} $y : E \in \mcA_{\textit{new}}$ \texttt{do}
      increment $\sharp R^\mcA(x,E)$\\
      \>\> \texttt{if} $x : \qnrleq m R E \in \mcA$ and $\sharp
      R^\mcA(x,E) > m$ \texttt{then}
      \texttt{return} ``not satisfiable''.\\
      \>\> \texttt{if} $\texttt{sat}(y,\mcA_{\textit{new}}) = 
      \text{``not satisfiable''}$ \texttt{then} \texttt{return} ``not
      satisfiable'' \\[0.25ex]
      \>\> discard $\mcA_{\textit{new}}$ from memory\\
      \> \texttt{od}\\
      \> discard the counters for $x$ from memory.\\
      \> \texttt{return} ``satisfiable''
    \end{tabbing}
    \caption{A non-deterministic \pspace decision procedure for \alcq.}
    \label{fig:decision-proc}
  \end{figure}
  
  Let us examine the space usage of this algorithm. Let $n = |C|$. The
  algorithm is designed to keep only a single path of $G_\mcA$ in memory at a
  given stage.  For each individual $x$ on a path, constraints of the form $x
  : D$ have to be stored for concepts $D \in \clos(C)$. The size of $\clos(C)$
  is bounded by $2n$ and hence the constraints for a single individual can be
  stored in $\mathcal{O}(n)$ bits.  For each individual, there are at most
  $|\Roles_C| \times |\clos(C)| = \mathcal{O}(n^2)$ counters to be stored.
  The numbers to be stored in these counters do not exceed the out-degree of
  $x$, which, by Lemma~\ref{lem:path-is-poly-long}, is bounded by
  $|\clos(C)|\times 2^{|C|}$. Hence each counter can be stored using
  $\mathcal{O}(n^2)$ bits when binary coding is used to represent the
  counters, and all counters for a single individual require $\mathcal{O}(n^4)$
  bits.  Due to Lemma~\ref{lem:path-is-poly-long}, the length of a path is
  limited by $n$, which yields an overall memory consumption of
  $\mathcal{O}(n^5 + n^2) = \mcO(n^5)$. \qed
\end{proof}

Theorem~\ref{theo:completeness-binary-coding} now is a simple Corollary from the
\pspace-hardness of \alc, Lemma~\ref{lem:optimised-algorithm-in-pspace}, and
Savitch's Theorem \cite{Savitch}.



\section{Extensions of \alcq}

It is possible to augment the DL \alcq without loosing the \pspace property of
the concept satisfiability problem. In this section we extend the techniques to
obtain a \pspace algorithm for the logic \alcqir, which extends \alcq with
inverse roles and \emph{safe} Boolean combinations of roles. This extends the
results from \cite{Tobies-JLC-2000} for the modal logic \grkrri, which
corresponds to \alcq extended with inverse roles and intersection of roles.

\begin{definition}[Syntax of \alcqir]
  \label{def:alcqir-syntax+semantics}
  \index{alcqib@\alcqir}
  Let \Names be a set of atomic \emph{concept names} and \Roles be a set of
  atomic \emph{role names}.  With $\overline \Roles := \Roles \cup \{ R^{-1}
  \mid R \in \Roles \}$ we denote the set of \emph{\alcqir-roles}. 
  \index{00nr@$\overline{\Roles}$}
  \index{alcqib-role@\alcqir-role}
  
  A role $S$ of the form $S = R^{-1}$ with $R \in \Roles$ is called
  \iemph{inverse role}.
  
  An \emph{\alcqir-role expression} \index{alcqib-role expression@\alcqir-role
    expression} $\omega$ is built from \alcqir-roles using the operators
  $\sqcap$ (role intersection), $\sqcup$ (role union), and $\lnot$ (role
  complement), with the restriction that, when transformed into disjunctive
  normal form, every disjunct contains at least one non-negated conjunct. A role
  expression that satisfies this constraint is called \emph{safe}.
  
  \index{role!intersection}
  \index{role!union}
  \index{role!complement}
  \index{role expression}
  \index{role expression!safe}

  The set of \emph{\alcqir-concepts} is built inductively from these using the
  following grammar, where $A \in \Names$, $\omega$ is an \alcqir-role
  expression, and $n \in \N$:
  \[
  C ::= A \bnfor \lnot C \bnfor C_1 \sqcap C_2 \bnfor C_1 \sqcup C_2 \bnfor
  \qnrleq n \omega C \bnfor \qnrgeq n \omega C .
  \] 
  \alcqi \index{alcqi@\alcqi} is the fragment of \alcqir, where every role
  expression in a number restriction consists of a single (possibly inverse)
  \alcqir-role. \eod
\end{definition}

The role-expressions $\lnot(\lnot R_1 \sqcup (R_2^{-1} \sqcap \lnot R_3)) \sqcup
(\neg R_2 \sqcap R_2^{-1})$ is safe (its DNF is $(R_1 \sqcap \neg R_2^{-1})
\sqcup (R_1 \sqcap R_3) \sqcup (\neg R_2 \sqcap R_2^{-1})$) while $R \sqcup
\lnot R$ is not an \alcqir role expression since it is already in DNF and $\lnot
R$ occurs as single element in one of the disjuncts. The latter example also
shows that some kind of restrictions on role expressions is indeed necessary if
we want to obtain a \pspace algorithm: the concept $\qnrleq 0 {R \sqcup \neg R}
{\lnot C}$ is satisfiable iff $C$ is globally satisfiable, which is an
\exptime-complete problem (see the proof of
Theorem~\ref{theo:alc-general-tboxes-exptime-complete}.  Indeed, for
unrestricted role expressions, the problem in the presence of qualifying number
restrictions is of even higher complexity. It is \nexptime-complete (see
Corollary~\ref{cor:alcq-alcqib-nexptime-complete}).

The syntactic restriction we have chosen enforces that, for a pair $(x,y)$ to
appear in the extension of a role expression $\omega$, they must occur at least
in the extension of one of the roles that occur in $\omega$. Hence, if no role
relation holds between $x$ and $y$, concepts asserted for $x$ do not impose any
restrictions on $y$.

A similar restriction can be found in the database world in conjunction with the
notion of \emph{safe-range queries} \cite[Chapter 5]{abiteboul95:_found_datab}.
To decide whether a role expression $\omega$ is safe, it is not necessary to
calculate its DNF (which might require exponential time). One can rather use the
following algorithm: first, compute the NNF $\omega'$ of $\omega$ by pushing
negation inwards using de Morgan's law. Second, test whether $\safe(\omega')$
holds, where the function $\safe$ is defined inductively on the structure of
role expressions as follows \cite<compare>[Algorithm 5.4.3]{abiteboul95:_found_datab}:
\begin{align*}
  \safe(R) & = \texttt{true} \text{ for } R \in \overline{\Roles}  \\
  \safe(\neg R) & = \texttt{false} \text{ for } R \in \overline{\Roles} \\
  \safe(\omega_1 \sqcap \omega_2) & = \safe(\omega_1) \vee \safe(\omega_2)\\
  \safe(\omega_1 \sqcup \omega_2) & = \safe(\omega_1) \wedge \safe(\omega_2)
\end{align*}
It is easy to see that a role expression is safe iff this algorithm yields
\texttt{true}. Hence, a role expression can be tested for safety in polynomial
time.

The semantics of \alcq-concepts can be extended to \alcqir-concepts by
fixing the interpretations of the role expressions. This is done in
the obvious way.

\begin{definition}[Semantics of \alcqir]\label{ref:alcirq-semantics}
  For an interpretation $\I = (\domain, \cdot^\I)$, the semantics of
  \alcqir-concepts is defined inductively as for \alcq-concepts with the
  additional rules:
  \begin{align*}
    \qnrleq n \omega C ^\I & = \{ x \in \domain \mid \sharp \{ y \mid (
    x,y ) \in \omega^\I \text{ and } y \in C^\I \} \leq n \},\\
    \qnrgeq n \omega C ^\I & = \{ x \in \domain \mid \sharp \{ y \mid (
    x,y ) \in \omega^\I \text{ and } y \in C^\I \} \geq n \},
  \end{align*}
  where the interpretation of a role expression $\omega$ is obtained
  by extending the valuation $\I$ inductively to role expressions by setting:
  \begin{align*}
    R^{-1} & = \{ (y,x) \mid (x,y) \in R^\I \},\\ 
    (\lnot \omega)^\I & = (\Delta^\I \times \Delta^\I) \setminus \omega^\I,\\
    (\omega_1 \sqcap \omega_2)^\I & = \omega_1^\I \cap \omega_2^\I,\\
    (\omega_1 \sqcup \omega_2)^\I & = \omega_1^\I \cup \omega_2^\I .
  \end{align*}
 \eod
\end{definition}

Obviously every \alcq concept is also a \alcqir concept.  We will use the
letters $\omega,\sigma$ to range over \alcqir-role-expressions. To avoid dealing
with roles of the form $(R^{-1})^{-1}$ we use the convention that $(R^{-1})^{-1}
= R$ for any $R \in \overline \Roles$. This is justified by the semantics.  The
definition of NNF and $\clos(\cdot)$ can be extended from \alcq to \alcqir in a
straightforward manner. Moreover, we use the following notation:

\begin{definition}
  Let $\mbR$ a set of (possibly inverse) roles and $\omega$ a role
  expression. We view then roles in $\overline \Roles$ as
  propositional variables and $\mbR$ as the propositional interpretation
  that maps exactly the elements of $\mbR$ to true and all other roles
  to false. We write $\mbR \models \omega$ iff $\omega$,
  viewed as a propositional formula, evaluates to true under $\mbR$. \eod
  \index{00rmodelsomega@$\mbR \models \omega$}
\end{definition}

The intended use of this definition is captured by the following simple lemma: 

\begin{lemma}\label{lem:alcqir-role-expression}
  Let $\I$ be an interpretation, $x,y \in \Delta^\I$ and $\omega$ a role expression.
  \[
  (x,y) \in \omega^\I \text{ iff } \{ R \in \overline \Roles \mid
  (x,y) \in R^\I \} \models \omega .
  \]
  For two individuals $x,y$ in an ABox $\mcA$ and a role expression $\omega$,
  \[
  \{ R \mid (x,y) : R \in \mcA\} \models \omega 
  \]
  implies $(x,y) \in \omega^\canI$ for the canonical interpretation $\canI$.
\end{lemma}

\subsection{Reasoning for \alcqir}

We will use similar techniques as in the previous section to obtain a
\pspace-algorithm for \alcqir. We still use ABoxes to capture the constraints
generate by completion rules, with the only change that we allow inverse roles
$R^{-1}$ to appear in role assertions and require that, for any $R \in
\overline{\Roles}$, an ABox contains the constraint $(x,y) : R$ iff it contains the
constraint $(y,x) : R^{-1}$.  For an ABox $\mcA$, a role-expression $\omega$,
and a concept $D$, let $\sharp \omega^\mcA(x,D)$ be the number of individuals
$y$ such that $\{ R \mid (x,y) : R \in \mcA\} \models \omega$ and $ y : D \in
\mcA$. Due to the syntactic restriction on role expressions, an individual $y$
may only contribute to $\sharp \omega^\mcA(x,D)$ if $(x,y) : R \in \mcA$ for
some (possibly inverse) role $R$ that occurs in $\omega$.

\begin{algorithm}[The \alcqir-algorithm]\label{alg:alcqir}
  \index{alcqib@\alcqir!tableau algorithm} \index{tableau algorithm!for
    alcqib@for \alcqir} We modify the definition of \emph{clash} to deal with
  safe role expressions as follows. An ABox $\mcA$ contains a clash iff
  \begin{itemize}
  \item $\{x : A, x : \lnot A \} \subseteq \mcA$ for some individual $x$ and $A
    \in \Names$, or
  \item $x : \qnrleq n \omega D \in \mcA$ and $\sharp \omega^\mcA(x,D) > n$
    for some individual $x$, role expression $\omega$, concept $D$,
    and $n \in \N$.
  \end{itemize}
  
  The set of rules dealing with \alcqir is shown in
  Figure~\ref{fig:expansion-rules-alcqir}. The algorithm maintains a binary
  relation $\prec_\mcA$ \index{00preca@$\prec_\mcA$} between the individuals in
  an ABox $\mcA$ with $x \prec_\mcA y$ iff $y$ was inserted by the \rulegeq-rule
  to satisfy a constraint for $x$. When considering the graph $G_\mcA$, the
  relation $\prec_\mcA$ corresponds to the successor relation between nodes.
  Hence, when $x \prec_\mcA y$ holds we will call $y$ a successor of $x$ and $x$
  a predecessor of $y$. We denote the transitive closure of $\prec_\mcA$ by
  $\prec^+_\mcA$. \index{00precatr@$\prec^+_\mcA$}
  
  For a set of individuals $\mathcal{X}$ and an ABox $\mcA$, we denote the
  subset of $\mcA$ in which no individual from $\mathcal{X}$ occurs in a
  constraint by $\mcA - \mathcal{X}$. \index{00aminusx@$\mcA - \mathcal{X}$} The
  \ruleand-, \ruleor- and \rulechoose-rule are called \emph{non-generating
    rules} while the \rulegeq-rule is called a \emph{generating rule}. 
  
  Let $C$ be an \alcqir-concept in NNF and $\overline \Roles_C$ the set of roles
  that occur in $C$ together with their inverses.
  \index{00nrc@$\overline{\Roles}_C$} To test the satisfiability of $C$,
  the \alcqir-algorithm starts with the initial ABox \index{ABox!initial} $\{
  x_0 : C \}$ and successively applies the rules from
  Figure~\ref{fig:expansion-rules-alcqir}. stopping when a clash occurs or the
  \rulegeq-rule fails. The algorithm answers ``$C$ is satisfiable'' iff the rule
  can be applied in a way that yields a complete ABox. \eod
\end{algorithm}


\begin{figure}[tbh]
  \begin{center}
    \begin{tabular}{@{ }l@{ }l@{ }l@{}}
      \multicolumn{3}{@{ }l}{\ruleand, \ruleor:  see 
      Fig.~\ref{fig:expansion-rules-derijke}}\\[1ex]      
      \rulechoose: & if \hfill 1. & $x : \qnrgleq n  \omega D \in \mcA$ and\\
      & \hfill 2. & for some $R$ that occurs in $\omega$ there is a $y$ with  $(x,y) : R
      \in \mcA$, and \\ 
      && $\{ y : D, y : \nneg D \} \cap \mcA = \emptyset$\\
      & then & $\mcA \rulechoose \mcA' \cup \{ y  :  E \}$ where $E \in \{ D,
      \nneg D \}$\\
      && and $\mcA' = \mcA - \{ z \mid y \prec^+_\mcA z \}$\\
      \rulegeq: & if \hfill 1. & $x : \qnrgeq n \omega D \in \mcA$, and\\
      & \hfill 2. & $\sharp \omega^\mcA(x,D) < n$, and \\
      & \hfill 3. & no non-generating rule can be applied to a constraint
      for $x$ \\
      & then & guess a set $\mbR = \{ R_1, \dots, R_m \} \subseteq \overline \Roles_C$\\
      && if $\mbR \not\models \omega$ then \textit{fail}\\
      && else $ \mcA \rulegeq \mcA \cup \{ y : D \} \cup \mcA' \cup \mcA''$
      and set $x \prec_\mcA y$ where\\ 
      & & 
      \begin{tabular}{l}
        $\mcA' = \{ y : D_1, \dots, y : D_k \}$, $D_i \in \{ E_i, \nneg
        E_i \}$, and\\
        \quad $\{ E_1, \dots, E_k \} = \{ E \mid x : \qnrgleq m \sigma E \in S \}$\\
        $\mcA'' = \{ (x,y) : R_1, (y,x) : R^{-1}_1, \dots, (x,y) : R_m, (y,x) :  
        R^{-1}_m \}$ \\ 
        $y$ is a fresh individual
      \end{tabular}
    \end{tabular}
  \end{center}
  \caption{The completion rules for \alcqir.}
  \label{fig:expansion-rules-alcqir}
  \index{00ruleand@\ruleand}
  \index{00ruleor@\ruleor}
  \index{00rulechoose@\rulechoose}
  \index{00rulegeq@\rulegeq}
  \index{completion rules!for alcqib@for \alcqir}
  \index{alcqib@\alcqir!completion rules}
\end{figure}

For the different kinds on non-determinism present in this algorithm, compare
the discussion below Algorithm~\ref{alg:alc}. Similar to the case for \alcq, it
is shown in the proof of Lemma~\ref{lem:alcqir-completeness} that the choice of
which rule to apply when is don't-care non-deterministic. This implies that one
is free to choose an arbitrary strategy that decides which rule to apply if more
than one is applicable.

For the different kinds of non-determinism present in the \alcqir-algorithm,
refer to the discussion below

The \rulegeq-rule, while looking complicated, is a straightforward
extension of the \rulegeq-rule for \alcq, which takes into account
that we need to guess a set of roles between the old individual $x$
and the freshly introduced individual $y$ such that these roles
satisfy the role expression $\omega$ currently under consideration.
The \rulechoose-rule requires more explanation.

For \alcq, the optimal algorithm generates an ABox $\mcA$ in a way that,
whenever $x : \qnrgleq n R D \in \mcA$, then, for any $y$ with $(x,y) : R \in
\mcA$, either $y : D$ or $y : \nneg D \in \mcA$.  This was achieved by
suspending the generation of any successors $y$ of $x$ until $\mcA$ contained
all constraints of the from $x : D$ it would ever contain.  In the presence of
inverse relations, this is no longer possible because $y$ might have been
generated as a predecessor of $x$ and hence before it was possible to know
which concepts $D$ might be relevant.  There are at least two possible ways to
overcome this problem. One is, to guess, for every $x$ and \emph{every} $D \in
\clos(C)$, whether $x : D$ or $x : \nneg D$. In this case, since the
termination of the optimal algorithm as proved in
Lemma~\ref{lem:optimised-alg-terminates} relies on the fact that the nesting
of qualifying number restrictions strictly decreases along a path in the
induced graph $G_\mcA$, termination would no longer be guaranteed. It would
have to be enforced by different means.

Here, we use another approach. We can distinguish two different situations where
$\{x : \qnrgleq n \omega D, (x,y) : R \} \subseteq \mcA$ for some $R$ that
occurs in $\omega$, and $ \{ y : D, y : \nneg D \} \cap \mcA = \emptyset$: $y$
is a predecessor of $x$ ($y \prec_\mcA x$) or a successor of $x$ ($x \prec_\mcA
y$).  The second situation will never occur. This is due to the interplay of the
\rulegeq-rule and the \rulechoose-rule. The former is suspended until all known
relevant information has been added for $x$, the latter deletes certain parts of
the ABox whenever new constraints are added for predecessor individuals.

The first situation is resolved by non-deterministically adding either $y : D$
or $y : \nneg D$ to $\mcA$. The subsequent deletion of all constraints involving
individuals from $\{ z \mid y \prec^+_\mcA z \}$, which correspond to the
deletion of all subtrees of $G_\mcA$ rooted below $y$, is necessary to make this
rule ``compatible'' with the trace technique \index{trace technique} we want to
employ in order to obtain a \pspace-algorithm. The correctness of the trace
approach relies on the property that, once we have established the existence of
a complete and clash-free ``subtree'' for a node $x$, we can remove this tree
from memory because it will not be modified by the algorithm.  In the presence
of inverse roles this can be no longer taken for granted as can be illustrated
by the concept
\[
C = \qnrleq 0 {R_1} B \; \sqcap \; \qnrgeq 1 {R_1} {A \sqcup B} \; \sqcap \;
\qnrgeq 1 {R_2} {\qnrleq 0 {R_2^{-1}} {\qnrgeq 1 {R_1} A}} .
\]
Figure~\ref{fig:reset-restart} shows the beginning of a run of the
\alcqir-algorithm. After a number of steps, a successor $y$ of $x$ has been
generated and the expansion of constraints has produced a complete and
clash-free subtree for $y$. Nevertheless, the concept $C$ is not satisfiable.
The expansion of $\qnrgeq 1 {R_2} {\qnrleq 0 {R_2^{-1}} {\qnrgeq 1 {R_1} A}}$
will eventually lead to the generation of the constraint $x : \nneg \qnrgeq 1
{R_1} A = \qnrleq 0 {R_1} A$ in $\mcA_5$, which disallows $R_1$-successors that
satisfy $A$. This conflicts with the constraints $x : \qnrleq 0 {R_1} B$ and $x
: \qnrgeq 1 {R_1} {A \sqcup B}$ , which require a successor of $x$ that
satisfies $A$. Consider an implementation of the algorithm that employs tracing:
the ABox $\mcA_3$ contains a complete and clash-free subtree for $y$, which is
deleted from memory and it is recorded that the constraint $x : \qnrgeq 1 {R_1}
{A \sqcup B}$ has been satisfied and this constraint is never reconsidered---the
conflict goes undetected. To make tracing possible, the \rulechoose-rule deletes
all information about $y$ when stepping from $\mcA_4$ to $\mcA_5$, which, while
duplicating some work, makes it possible to detect this conflict even when
tracing through the ABox. An implementation that uses tracing can safely discard
the information about $y$ from memory once the existence of a complete and
clash-free subtree has been established in $\mcA_3$ because, whenever the effect
of an application of the \rulechoose-rule might conflict with assertions for a
successor $y$, all required successors of $x$ have to be re-generated anyway.

A similar technique will be used in a subsequent chapter to obtain a
\pspace-result for another DL with inverse roles.

\begin{figure}[tbh]
  \begin{center}
    \begin{align*}
      \{ x : C \}& \ruleand \dots \\
      & \ruleand \underbrace{\{ x : C, x : \qnrleq 0 {R_1} B, x : \qnrgeq 1
        {R_1} {A \sqcup B}, x :
        \qnrgeq 1 {R_2} {\qnrleq 0 {R_2^{-1}}{\qnrgeq 1 {R_1} A}} \}}_{\mcA_1}\\
      & \rulegeq \underbrace{\mcA_1 \cup \{(x,y) : R_1, (y,x) : R^{-1}_1, y : A
        \sqcup B, y : \lnot B \}}_{\mcA_2} \ruleor \underbrace{\mcA_2 \cup \{y :
        A
        \}}_{\mcA_3} \\
      & \ruleand \underbrace{\mcA_3 \cup \{ (x,z) : R_2, (z,x) : R_2^{-1}, z :
        \qnrleq 0
        {R_2^{-1}}{\qnrgeq 1 {R_1} A} \}}_{\mcA_4} \\
      & \rulechoose \underbrace{\mcA_1 \cup \{ \qnrleq 0 {R_1} A \}}_{\mcA_5}
    \end{align*}
  \end{center}
  \caption{Inverse roles make tracing difficult.}
  \label{fig:reset-restart}
\end{figure}

\subsection{Correctness of the Algorithm}

Like for \alcq, we show correctness of the \alcqir-algorithm along the lines of Theorem~\ref{theo:generic-correctness}. 

\subsubsection{Termination}

Obviously, the deletion of constraints in $\mcA$ makes a new proof of
termination necessary, since the proof of
Lemma~\ref{lem:optimised-alg-terminates} relied on the fact that constraints
were never removed from the ABox. Note, however, that the
Lemma~\ref{lem:path-is-poly-long} still holds for \alcqir.

\begin{lemma}[Termination]
  \label{lem:alcqri-alg-terminates}
  Any sequence of rule applications starting from an ABox $\mcA = \{ x_0 : C \}$
  of the \alcqir algorithm is finite.
  \index{alcqib@\alcqir!tableau algorithm!termination}
\end{lemma}

\begin{proof}
  The sequence of rule applications induces a sequence of trees. As before,
  the depth and out-degree of this tree is bounded in $|C|$ by
  Lemma~\ref{lem:path-is-poly-long}. For each individual $x$, $\Lab(x)$ is a
  subset of the finite set $\clos(C)$. Each application of a rule either
  \begin{itemize}
  \item adds a constraint of the form $x : D$ and hence adds an element
    to $\Lab(x)$, or
  \item adds fresh individuals to $\mcA$ and hence adds additional nodes to
    the tree $G_\mcA$, or
  \item adds a constraint to a node $y$ and deletes all subtrees
    rooted below $y$.
  \end{itemize}
  
  Assume that algorithm does not terminate. Due to the mentioned facts
  this can only be because of an infinite number of deletions of
  subtrees.  Each node can of course only be deleted once, but the
  successors of a single node may be deleted several times. The root
  of the constraint system cannot be deleted because it has no
  predecessor.  Hence there are nodes that are never deleted.  Choose
  one of these nodes $y$ with maximum distance from the root, i.e.,
  which has a maximum number of ancestors in $\prec_\mcA$.  Suppose
  that $y$'s successors are deleted only finitely many times. This can
  not be the case because, after the last deletion of $y$'s
  successors, the ``new'' successors were never deleted and thus $y$
  would not have maximum distance from the root. Hence $y$ triggers
  the deletion of its successors infinitely many times. However, the
  \rulechoose-rule is the only rule that leads to a deletion, and it
  simultaneously leads to an increase of $\Lab(y)$, namely by the
  missing concept which caused the deletion of $y$'s successors.  This
  implies the existence of an infinitely increasing chain of subsets
  of $\clos(C)$, which is clearly impossible. \qed
\end{proof}

\subsubsection{Soundness and Completeness}

We start by proving an important property of the interplay of the \rulegeq-rule and the
\rulechoose-rule.

\begin{lemma}\label{lem:deletion-works}
  Let $\mcA_1,\mcA_2, \mcA_3$ be ABoxes generated by the \alcqir-algorithm, such
  that $\mcA_2$ is derived from $\mcA_1$ by application of the \rulegeq-rule to
  an individual $x$ in a way that creates the new successor $y$ of $x$, and
  $\mcA_3$ is derived from $\mcA_2$ by zero or more rule applications. If both
  $x,y$ occur in $\mcA_3$, then $\{ D \mid x : D \in \mcA_1 \} = \{ D \mid x : D
  \in \mcA_3 \}$ and the \rulechoose-rule is not applicable to $x$ in $\mcA_3$
  in a way that adds a concept assertion for $y$.
\end{lemma}

\begin{proof}
  Assume that $x,y$ occur in $\mcA_3$.  Then they also occur in all intermediate
  ABoxes because, once an individual is deleted from the constraint system, it
  is never re-introduced.  The proof is by induction on the number of rule
  applications necessary to derive $\mcA_3$ from $\mcA_2$. If no rule must be
  applied, then $\mcA_2 = \mcA_3$ holds, and since application of the
  \rulegeq-rule to $x$ does not alter the concepts asserted for $x$, we are done. Now
  assume that the lemma holds for every ABox $\mcA'$ derivable from $\mcA_2$ by
  $n$ rule applications.
  
  Let $\mcA_3$ be derivable from $\mcA_2$ in $n+1$ steps and let $\mcA'$ be an
  ABox such that $\mcA_2 \rightarrow^n \mcA' \rightarrow \mcA_3$. Since $\{ D
  \mid x : D \in \mcA_1 \} = \{ D \mid x : D \in \mcA' \}$ holds by induction,
  also $\{ D \mid x : D \in \mcA_1 \} = \{ D \mid x : D \in \mcA_3 \}$ holds as
  long as the rule application that derives $\mcA_3$ from $\mcA'$ does not alter
  the concepts asserted for $x$. 
  
  The \rulegeq-rule does not alter the constraints for any individual that is
  already present in the ABox because it introduces a fresh individual.  
  
  The \ruleand- or \ruleor-rule cannot be applicable to $x$ because, if the rule
  is applicable in $\mcA'$, then, since $\{ D \mid x : D \in \mcA_1 \} = \{ D
  \mid x : D \in \mcA' \}$, it is also applicable in $\mcA_1$ and the
  \rulegeq-rule that creates $y$ is not applicable. Assume that an application
  of the \rulechoose-rule asserts an additional concept for $x$. Any
  application of the \rulechoose-rule that adds a constraint for $x$ removes the
  individuals $\{ z \mid x \prec_{\mcA'} z \}$ from $\mcA'$.  This includes $y$
  and hence $y$ would not occur in $\mcA_3$, in contradiction to the assumption
  that $x,y$ occur in $\mcA_3$. 
  
  Since the concept assertions for $x$ have not changed since the generation of
  $y$, it holds that $x : \qnrgleq n \omega D \in \mcA_3$ iff $x : \qnrgleq n
  \omega D \in \mcA_1$ and so $\{ y : D, y : \nneg D \} \cap \mcA_1$ is ensured
  by the \rulegeq-rule that creates $y$. The individual $y$ still occurs in
  $\mcA_3$ and hence $\{ y : D, y : \nneg D \} \cap \mcA_3$ holds, which implies that
  the \rulechoose-rule cannot be applied for the constraint $x : \qnrgleq n \omega D \in
  \mcA_3$ in a way that adds $y : D$ or $y : \nneg D$ to $\mcA_3$. \qed
\end{proof}

The correctness of the \alcqir-algorithm is again proved along the lines of
Theorem~\ref{theo:generic-correctness}, but in a slightly different manner than
it was proved for \alcq. Instead of proving local correctness of the rules,
which is difficult to establish due to the deletion of constraints by the
\rulechoose-rule, we use Property 5'.  Additionally, we require a stronger
notion of satisfiability than standard ABox satisfiability. Similar as for
\alcq, we define the differentiation \index{differentiation} $\widehat \mcA$
\index{00widehata@$\widehat \mcA$} of an ABox $\mcA$ by setting
\[
\widehat \mcA = \mcA \cup \{ y \ndoteq z \mid \{ (x,y) : R, (x,z) : S \}
\subseteq \mcA, y \neq z \} .
\]
Note the slight difference to the definition of \alcq, where only those
individuals reachable from $x$ via the same role $R$ were asserted to be
distinct. Here, all individuals reachable from $x$ via an arbitrary role are
asserted to be distinct. We say that an ABox $\mcA$ is satisfiable iff there
exists a model $\I$ of its differentiation $\widehat \mcA$ that, in addition to
what is required by the standard notion of ABox satisfiability from
Definition~\ref{def:abox}, satisfies:
\begin{equation}
  \tag{$\S$}
  \text{if } (x,y) : R \in \widehat \mcA \text{ then } \{ R \mid (x,y) : R \in
  \widehat \mcA \} = \{ R \mid (x^\I,y^\I)) \in R^\I \} \cap \overline \Roles_C .
\end{equation}
Note that this additional property is trivially satisfied by a
canonical interpretation.

Obviously, Properties 1 and 2 of Theorem~\ref{theo:generic-correctness} hold for
every ABox generated by the \alcqir-algorithm.

\begin{lemma}[Soundness]
  \label{lem:soundness-alcqir}
  Let $\mcA$ be a complete and clash-free ABox generated by the
  \alcqir-algorithm. Then $\mcA$ is satisfiable, i.e., there exists a model $\I$
  of $\widehat \mcA$ that additionally satisfies $(\S)$.
  \index{alcqib@\alcqir!tableau algorithm!soundness}
\end{lemma}

\begin{proof}
  Let $\mcA$ be a complete and clash-free ABox obtained by a sequence of rule
  applications starting from $\{ x_0 : C \}$. We show that the canonical
  interpretation $\canI$ (as defined in
  Definition~\ref{def:canonical-interpretation}) is indeed a model of $\widehat
  \mcA$ that satisfies $(\S)$. Please note that we need the condition ``$(x,y)
  : R \in \mcA$ iff $(y,x) : R^{-1} \in \mcA$'', which is maintained by the
  algorithm, to make sure that all information from the ABox is reflected in the
  canonical interpretation.
  
  Every canonical interpretation trivially satisfies $(\S)$ and also every two
  different individuals are interpreted differently, which takes care of the
  additional assertions in $\widehat \mcA$. So, it remains to show that $x : D
  \in \mcA$ implies $x \in D^\canI$ for all individuals $x$ in $\mcA$ and all
  concepts $D \in \clos(C)$. This is done by induction over the norm of concepts
  $\| \cdot \|$. The only interesting cases that are different from the
  \alcq-case are the qualifying number restrictions.
  \begin{itemize}
  \item $x : \qnrgeq n \omega D \in \mcA$ implies $\omega^\mcA(x,D) \geq n$
    because $\mcA$ is complete. Hence, there are $n$ distinct individuals $y_1,
    \dots, y_n$ with $y_i : D \in \mcA$ and $\{ R \mid (x,y_i) : R \in \mcA \}
    \models \omega$ for each $1 \leq i \leq n$. By induction and
    Lemma~\ref{lem:alcqir-role-expression}, we have $y_i \in D^\canI$ and
    $(x,y_i) \in \omega^\canI$ and hence $x \in \qnrgeq n \omega D^\canI$.
  \item $x : \qnrleq n \omega D \in \mcA$ implies, for any $R$ that occurs in
    $\omega$ and any $y$ with $(x,y) : R \in \mcA$, $y : D \in \mcA$ or $y :
    \nneg D \in \mcA$. For any predecessor of $x$, this is guaranteed by the
    \rulechoose-rule. For any successor, this follows from
    Lemma~\ref{lem:deletion-works}. Hence, $x : \qnrleq n \omega D$ is present
    in $\mcA$ by the time $y$ is generated and the \rulegeq-rule ensures $y : D
    \in \mcA$ or $y : \nneg D \in \mcA$.
  
    We show that $\sharp \omega^\canI(x,D) \leq \sharp \omega^\mcA(x,D)$: assume
    $\sharp \omega^\canI(x,D) > \sharp \omega^\mcA(x,D)$.  This implies the
    existence of some $y$ with $(x,y) \in \omega^\canI$ and $y \in D^\canI$ but
    $y : D \not\in \mcA$.  Due to the syntactic restriction on role expressions,
    $(x,y) \in \omega^\canI$ implies $(x,y) \in R^\canI$ for some $R$ that
    occurs in $\omega$ and and hence $(x,y) : R \in \mcA$ must hold by
    construction of $\canI$. The \rulechoose-rule and the \rulegeq-rule then
    guarantee that $y : D \not \in \mcA$ implies $y : \nneg D \in \mcA$. By induction
    this yields $y \in (\nneg D)^\canI$ in contradiction to $y \in D^\canI$.
    \qed
  \end{itemize}
\end{proof}


\begin{lemma}[Local Completeness]
  \label{lem:alcqir-completeness}
  If $\mcA$ is a satisfiable ABox generated by the \alcqir-algorithm and a rule
  is applicable to $\mcA$, then it can be applied in a way that yields a
  satisfiable $\mcA'$.
\end{lemma}

\begin{proof}
  Let $\I$ be a model of $\widehat \mcA$ that satisfies $(\S)$, as required by
  our notion of satisfiability. We distinguish the different rules. For most rules
  $\I$ can remain unchanged, in all other cases we explicitly state how ${\I}$
  must be modified in order to witness the satisfiability of the modified ABox.

  \begin{itemize}
  \item The \ruleand-rule: if $x : C_1 \sqcap C_2 \in \mcA$, then $x^\I \in (C_1 \sqcap
    C_2)^\I$.  This implies $x^\I \in C_i^\I$ for $i=1,2$, and hence satisfiability
    is preserved.
  \item The \ruleor-rule: if $x : C_1 \sqcup C_2 \in \mcA$, then $x^\I \in (C_1 \sqcup
    C_2)^\I$. This implies $x^\I \in C_1^\I$ or $x^\I \in C_2^\I$. Hence the
    \ruleor-rule can add a constraint $x : D$ with $D \in \{ C_1, C_2 \}$ and
    maintains satisfiability.
  \item The \rulechoose-rule: obviously, either $y^\I \in D^\I$ or $y^\I \not\in
    D^\I$ for any individual $y$ in $\mcA$.  Hence, the rule can always be
    applied in a way that maintains satisfiability.  Deletion of constraints as
    performed by the \rulechoose-rule cannot cause unsatisfiability.
  \item The \rulegeq-rule: if $x : \qnrgeq n \omega D \in \mcA$, then
    $x^\I \in \qnrgeq n \omega D^\I$. This implies $\sharp
    \omega^\I(x^\I,D) \geq n$.  We claim that there is an element $a
    \in \Delta^\I$ such that
    \begin{equation}
      \tag{$*$}
      \begin{array}{l}
        (x^\I,a) \in  \omega^\I,\; a \in
        D^\I, \text{ and }\\  
        a \not \in \{ z^\I \mid (x,z) : S  \in \mcA \text{ for some $S \in
          \overline \Roles_C$} \} .
      \end{array}
      \quad 
    \end{equation}
    We will prove this claim later. Let $E_1, \dots, E_k$ be an enumeration of
    the set $\{E \mid x : \qnrgleq m \sigma E \in \mcA \}$. The \rulegeq-rule can
    add the constraints
    \begin{align*}
      \mcA' & = \{ y : E_i \mid a \in E_i^\I \} \cup \{ y : \nneg E_i
      \mid a \not\in E_i^\I \}\\
      \mcA'' & = \{ (x,y) :R \mid R \in \overline \Roles_C, (x^\I,a)
      \in R^\I \} \cup \{ (y,x) : R \mid R \in \overline \Roles_C,
      (a,x^\I) \in R^\I \}
    \end{align*}
    as well as $\{ y : D \}$ to $\mcA$. If we set $\I' := \I[y \mapsto a]$, then
    $\I'$ is a model of the differentiation of the ABox obtained this way that
    satisfies $(\S)$.
  
    Why does there exists an element $a$ that satisfies $(*)$? Let $b \in
    \Delta^\I$ be an individual with $(x^\I,b) \in \omega^\I$ and $b \in
    D^\I$ that appears as an image of an arbitrary element $z$ with $(x,z) : S
    \in \mcA$ for some $S \in \overline \Roles_C$.  The requirement $(\S)$
    implies that $\{ R \mid (x,z) : R \in \mcA\} \models \omega$ and also $z : D \in
    \mcA$ must hold. This can be shown  as follows:
  
    Assume $z : D \not \in \mcA$.  This implies $z : \nneg D \in \mcA$: either
    $z \prec_\mcA x$, then in order for the \rulegeq-rule to be applicable, no
    non-generating rules and especially the \rulechoose-rule is not applicable
    to $x$ and its ancestor, which implies $\{ z : D, z : \nneg D \} \cap \mcA
    \neq \emptyset$. If not $z \prec_\mcA x$, then $z$ must have been generated
    by an application of the \rulegeq-rule to $x$.
    Lemma~\ref{lem:deletion-works} implies that at the time of the generation of
    $z$ already $x : \qnrgeq n \omega D \in \mcA$ held and hence the
    \rulegeq-rule ensures $\{z : D ,z : \nneg D\} \cap \mcA \neq \emptyset$.
    
    In any case $z : \nneg D \in \mcA$ holds, which implies $b \not \in D^\I$, in
    contradiction to $b \in D^\I$.
    
    Together this implies that, whenever an element $b$ with $(x^\I,b) \in \omega^\I$
    and $b \in D^\I$ is assigned to an individual $z$ with $(x,z) : S \in \mcA$,
    then it must be assigned to an individual that contributes to $\sharp
    \omega^\mcA(x,D)$. Since the \rulegeq-rule is applicable, there are less than $n$
    such individuals and hence there must be an unassigned element $a$ as
    required by $(*)$. \qed
  \end{itemize}
\end{proof}

The \rulechoose-rule deletes only assertions for successors of a node and hence
never deletes any assertions for the root $x_0$. Hence, for any ABox $\mcA$
generated by application of the completion rules from an initial ABox $\{ x_0 :
C\}$, $\{ x_0 : C\} \subseteq \mcA$ holds and hence we get the following.

\begin{lemma}\label{lem:alcqir-property-five}
  If a complete and clash-free ABox $\mcA$ can be generated from an initial ABox
  $\mcA_0$, then $\mcA_0$ is satisfiable.
\end{lemma}

\begin{proof}
  From Lemma~\ref{lem:alcqir-completeness}, it follows that $\mcA$ is
  satisfiable and every model of $\mcA$ is also a model of $\mcA_0 =\{ x_0 :
  C\}$ because $\mcA_0 \subseteq \mcA$ and $\mcA_0$ contains no role assertions,
  which implies $\widehat \mcA_0 = \mcA_0$ and every interpretation trivially
  satisfies $(\S)$ for $\widehat \mcA_0$. \qed
\end{proof}

Hence, we can apply Theorem~\ref{theo:generic-correctness}  and get:

\begin{corollary}
  The \alcqir-algorithm is a non-deterministic decision procedure for
  satisfiability of \alcqir-concepts.
\end{corollary}

\begin{proof}
  Termination has been shown in Lemma~\ref{lem:alcqri-alg-terminates}. As
  mentioned before, Property 1 and 2 of Theorem~\ref{theo:generic-correctness}
  are trivially satisfied due to the chosen notion of ABox satisfiability.
  Property 3 has been shown in Lemma~\ref{lem:soundness-alcqir}, Property 4 in
  Lemma~\ref{lem:alcqir-completeness} and Property 5' in Lemma~\ref{lem:alcqir-property-five}. \qed
\end{proof}

\subsection{Complexity of the Algorithm}

Like for the optimal algorithm for \alcq, we have to show that the
\alcqir-algorithm can be implemented in a way that consumes only polynomial
space. This is done similarly to the \alcq-case, but we have to deal with two
additional problems: we have to find a way to implement the ``reset-restart''
caused by the \rulechoose-rule, and we have to store the values of the relevant
counters $\omega^\mcA(x,D)$. It is impossible to store the values for every
possible role expression $\omega$ because there are exponentially many
inequivalent of these. Fortunately, storing only the values for those $\omega$
that actually appear in $C$ is sufficient.

\begin{lemma}
  \label{lem:alcqir-algo-in-pspace}
  The \alcqir-algorithm can be implemented in \pspace.
\end{lemma}

\begin{proof}
  Consider the algorithm in Figure~\ref{fig:alcqir-decision-proc}, where
  $\Omega_C$ \index{00omegac@$\Omega_C$} denotes all role expressions that occur
  in the input concept $C$.  Like the algorithm for \alcq, the \alcqir-algorithm
  re-uses the space used to check for the existence of a complete and clash-free
  ``subtree'' for each successor $y$ of an individual $x$ and keeps only a
  single path in memory at one time.  Counter variables are used to keep track
  of the values $\sharp \omega^\mcA(x,D)$ for all $\omega \in \Omega_C$ and $D
  \in \clos(C)$.
  
  Resetting a node and restarting the generation of its successors is achieved
  by jumping to the label $\textsf{restart}$ in the algorithm, which
  re-initializes all successor counters for a node $x$.  Note, how the
  predecessor of a node is taken into account when initializing the counter
  individuals. Since $G_\mcA$ is a tree, every newly generated node has a
  uniquely determined predecessor and since only safe role expressions occur in
  $\Omega_C$, it is sufficient to take only this predecessor node into account
  when initializing the counter.
  
  Let $n = |C|$. For every node $x$ of a path in $G_\mcA$, $\mcO(n)$ bits
  suffice to store the constraints of the form $x : D$ and $\mcO(n^4)$ suffice
  to store the counters (in binary representation) because  $\sharp \Omega_C
  = \mcO(n)$, $\sharp \clos(C) = \mcO(n)$, and the out-degree of $G_\mcA$ is
  bounded by $\mcO(n) \times 2^n$ (by Lemma~\ref{lem:path-is-poly-long}, which
  also holds for \alcqir). Also by Lemma~\ref{lem:path-is-poly-long}, the length
  of a path in $G_\mcA$ is bounded by $\mcO(n)$, which yields an overall memory
  requirement of $\mcO(n^5)$ for a path. \qed
\end{proof}

\begin{figure}
  \begin{center}
    \leavevmode
    \begin{tabbing}
      \small
      \hspace{2em}\=\hspace{2em}\=\hspace{2em}\= \kill
      $\alcqir\mbox{-}\textsc{Sat}(C) := \texttt{sat}(x_0,\{x_0 : C \})$\\
      $\texttt{sat}(x,S)$: \\
      \> allocate counters  $\sharp \omega^\mcA(x,D)$ for all $\omega \in
      \Omega_C$ and $D \in \clos(C)$. \\
      \hspace{5pt} \textsf{restart:}\\
      \> \texttt{for each} counter $\sharp \omega^\mcA(x,D)$:\\
      \>\> \texttt{if} ($x$ has a predecessor  $y \prec_\mcA x$ with $\{ R \mid (x,y) : R \in \mcA \} \models \omega$ and $y : D \in \mcA$)\\
      \>\>\> \texttt{then} $\sharp \omega^\mcA(x,D) := 1$ \texttt{else}
      $\sharp \omega^\mcA(x,D):=0$\\ 
      \> \texttt{while} (the \ruleand- or the \ruleor-rule can be
      applied at $x$) \texttt{and} ($\mcA$ is clash-free) \texttt{do}\\
      \>\> apply the \ruleand- or the \ruleor-rule to $\mcA$.\\
      \> \texttt{od}\\
      \> \texttt{if} $\mcA$ contains a clash \texttt{then}
      \texttt{return} ``not satisfiable''.\\
      \> \texttt{if} the \rulechoose-rule is applicable to the constraint
      $x : \qnrgleq n \omega D \in \mcA$\\
      \>\> \texttt{then return} ``restart with $D$''\\
      \> \texttt{while} (the \rulegeq-rule applies to a constraint $x :
      \qnrgeq n \omega D \in \mcA$) \texttt{do} \\
      \>\> non-deterministically choose $\mbR \subseteq \overline \Roles_C$\\
      \>\> \texttt{if} $\mbR \not \models \omega$ \texttt{then return} ``not satisfiable''\\
      \>\> $\mcA_{\text{new}} := \{ y : D \} \cup \mcA' \cup \mcA''$\\
      \>\> \texttt{where}\\
      \>\>\> $y$ is a fresh individual\\
      \>\>\> $\{ E_1, \dots, E_k \} = \{  E \mid x : \qnrgleq m \sigma E \in \mcA \}$\\
      \>\>\> $\mcA' = \{y : D_1, \dots, y : D_k \}$, and\\
      \>\>\> $D_i$ is chosen non-deterministically from $\{ E_i, \nneg
      E_i \}$ \\
      \>\>\> $\mcA'' = \{ (x,y) : R, (y,x) : R^{-1} \mid R \in \mbR \}$\\
      \>\> \texttt{for each} $E$ with $y : E \in \mcA'$
      \texttt{and} $\sigma \in \Omega_C$ with $\mbR \models \sigma$ \texttt{do}\\
      \>\>\> increment $\sharp \sigma^\mcA(x,E)$\\
      \>\> \texttt{if} $x : \qnrleq m \sigma E \in \mcA$ and $\sharp
      \sigma^\mcA(x,E) > m$ \\
      \>\>\> \texttt{then return} ``not satisfiable''.\\
      \>\> $result := \texttt{sat}(y,\mcA  \cup \mcA_{\text{new}})$\\
      \>\> \texttt{if} $result =$ ``not satisfiable'' \texttt{then return}
      ``not satisfiable''\\
      \>\> \texttt{if} $result =$ ``restart with $D$'' \texttt{then} \\
      \>\>\> $\mcA := \mcA \cup \{ x : E \}$\\
      \>\>\> \texttt{where} $E$ is chosen non-deterministically from $\{D,
      \nneg D \}$ \\
      \>\>\> \texttt{goto} \textsf{restart}\\
      \> \texttt{od} \\
      \> discard the counters for $x$ from memory.\\
      \> \texttt{return} ``satisfiable''
    \end{tabbing}
    \caption{A  non-deterministic \pspace  decision procedure for \alcqir-satisfiability.}
    \label{fig:alcqir-decision-proc}
  \end{center}
\end{figure}

Obviously, satisfiability of \alcqir-concepts is \pspace-hard, hence
Lemma~\ref{lem:alcqir-algo-in-pspace} and Savitch's
Theorem~\cite{Savitch} yield:

\begin{theorem}\label{theo:alcqir-pspace-complete}
  Satisfiability of \alcqir-concepts is \pspace-complete if the numbers in the
  input are represented using binary coding.
\end{theorem}

As a simple corollary, we get the solution of an open problem
in~\cite{DLNN97}:

\begin{corollary}
  Satisfiability of $\alcnr$-concepts is \pspace-complete if the
  numbers in the input are represented using binary coding.
\end{corollary}

\begin{proof}
  The DL $\alcnr$ is a syntactic restriction of the DL $\alcqir$, where we do not
  allow for inverse roles and in number restrictions $\qnrgleq n \omega D$,
  $\omega$ must be a conjunction of positive roles and $D$ the tautological
  concept $\top$.  Hence, the \alcqir-algorithm can immediately be applied to
  $\alcnr$-concepts, which yields decidability in \pspace. \qed
\end{proof}



\section{Reasoning with \alcqir-Knowledge Bases}

So far, we have only dealt with the problem of concept satisfiability rather
than satisfiability of knowledge bases. In this section, we will examine the
complexity of reasoning with knowledge bases for the DL \alcqir.  For the more
``standard'' DL \alcqi, this problem has been shown to be \exptime-complete by
De Giacomo~\citeyear{DeGiacomo95a}, but this result does not easily transfer to
\alcqir because of the role expressions and the proof in \cite{DeGiacomo95a} is
only valid in case of unary coding of numbers in the input. Here, we are aiming
for a proof that is valid also in the case of binary coding of numbers.

In a first step, we deal with concept satisfiability \wrt general TBoxes and prove
that this problem can be solved in \exptime using an automata approach. ABoxes
are then handled by a pre-completion algorithm similar to the one presented by
Hollunder \citeyear{hollunder96:_consis_check_reduc_satis_concep_termin_system}
(see also Section~\ref{sec:other-infer-probl}). It should be mentioned that the
algorithms developed in this section are by no means intended for implementation.
They are used only to obtain tight worst-case complexity results. We are also very
generous in size or time estimates.

The lower complexity bound for \alcqir with general TBoxes is an immediate
consequence of Theorem~\ref{theo:alc-general-tboxes-exptime-complete} because
\alc is strictly contained in \alcqir.

\begin{lemma}\label{lem:alcqir-tbox-exptime-hard}
  Satisfiability of \alcqir-concepts (and hence of ABoxes) \wrt general TBoxes
  is \exptime-hard.
\end{lemma}

To establish a matching upper complexity bound, we employ an automata approach,
where \mbox{(un-)}satisfiability of concepts is reduced to emptiness of suitable
finite automata, usually B{\"u}chi word or tree automata \cite{ThomasHTCS92}.
This approach is a valuable tool to establish exact complexity results for DLs
and modal logics \cite{VaWo86,LutzSattlerAIML00}, particularly for
\exptime-complete logics, where tableau approaches---due to their
non-deterministic nature---either fail entirely or require very sophisticated
techniques \cite{donini00:_exptim_alc} to prove decidability of the decision
problem in \exptime.

In general, the automata approach works as follows. To test satisfiability of a
concept $C$ \wrt a TBox $\mcT$, an automaton $\mfA_{C,\mcT}$ is constructed that
accepts exactly (abstractions of) models of $C$ and $\mcT$, so that
$\mfA_{C,\mcT}$ accepts a non-empty language iff $C$ is satisfiable \wrt $\mcT$.
For \alcqir, we do not require the full complexity of B{\"u}chi tree
automata---the simpler formalism of \emph{looping tree automata}
\cite{vardi94:_reason_infin_comput} suffices.

\begin{definition}[Looping Tree Automata]
  \index{looping tree automaton}
  For a natural number $n$, let $[n]$ denote the set $\{1, \dots, n\}$. An
  \emph{$n$-ary infinite tree over the alphabet $\Sigma$} is a mapping $t :
  [n]^* \rightarrow \Sigma$, where $[n]^*$ denotes the set of finite strings
  over $[n]$.
  
  \index{00n@$[n]$}
  \index{n-ary infinite tree@$n$-ary infinite tree}
  
  An \emph{$n$-ary looping tree automaton} is a tuple $\mfA = (Q,\Sigma,I,
  \delta)$, where $Q$ is a finite set of states, $\Sigma$ is a finite alphabet,
  $I \subseteq Q$ is the set of initial states, and $\delta \subseteq Q \times
  \Sigma \times Q^n$ is the transition relation.  Sometimes, we will view
  $\delta$ as a function from $Q \times \Sigma$ to $2^{Q^n}$ and write
  $\delta(q,\sigma)$ for the set of tuples $\{ \bfq \mid (q,\sigma, \bfq) \in
  \delta \}$.
  
  A \emph{run} of $\mfA$ on an $n$-ary infinite tree $t$ over $\Sigma$ is an
  $n$-ary infinite tree $r$ over $Q$ such that, for every $p \in [n]^*$,
  \[
  (r(p),t(p),(r(p1),\dots,r(pn))) \in \delta .
  \]
  An automaton $\mfA$ \emph{accepts} $t$ iff there is a run $r$ of $\mfA$ on $t$
  with $r(\epsilon) \in I$.  With $L(\mfA)$ we denote the \emph{language
    accepted} by $\mfA$ defined by $L(\mfA) : = \{ t \mid \text{$\mfA$ accepts
    $t$} \}$. \eod
\end{definition}

For a looping automaton $\mfA$, emptiness of $L(\mfA)$ can be decided
efficiently.

\begin{fact}\label{fact:looping-automaton-polynomial-emptiness}
  Let $\mfA = (Q,\Sigma,I,\delta)$ be an $n$-ary looping tree automaton. Emptiness
  of $L(\mfA)$ can be decided in time $\mcO(\sharp Q + \sharp \delta)$.
\end{fact}

A polynomial bound directly follows from the quadratic time algorithm for
B{\"u}chi tree automata \cite{VaWo86} of which looping tree automata are special
cases. A closer inspection of this algorithm shows that one can even obtain a
linear algorithm using the techniques from
\cite{dowling84:_linear_time_testin_horn}. For our purposes also the mentioned
quadratic and really every \emph{polynomial} algorithm suffices.

Before we formally define $\mfA_{C,\mcT}$ we give an informal description of the
employed construction of the automaton and the abstraction from an
interpretation $\I$ to a tree $T$ we use.  Generally speaking, nodes of $T$
correspond to elements of an unraveling of $\I$. In the label of the node, we
record the relevant (sub-)concepts from $C$ and $\mcT$ that are satisfied by
this element, and also which roles connect the element to its unique predecessor
in $\I$. This information has to be recorded at the node since edges of a tree
accepted by a looping automaton are unlabelled.  Hence, the label of a node is a
\emph{locally consistent} set of ``relevant'' concepts (as defined below) paired
with a set of ``relevant'' roles.

For now, we fix an \alcqir-concept $C$ in NNF and an \alcqir-TBox $\mcT$. Let
$\overline{\Roles}_{C,\mcT}$ \index{00nrct@$\overline{\Roles}_{C,\mcT}$} be the
set of role names that occur in $C$ and $\mcT$ together with their inverse and
$\Omega_{C,\mcT}$ \index{00omegact@$\Omega_{C,\mcT}$} the set of
role-expressions that occur in $C$ and $\mcT$. The closure $\clos(C,\mcT)$
\index{00closct@$\clos(C,\mcT)$} of ``relevant'' concepts is defined as the
smallest set $X$ of concepts such that
\begin{itemize}
\item $C \in X$ and $\NNF(\neg C_1 \sqcup C_2) \in X$ for every $C_1 \sqsubseteq
  C_2 \in \mcT$
\item $X$ is closed under sub-concepts and the application of $\nneg$, the
  operator that maps ever concept $C$ to $\NNF(\neg C)$. \index{00nnegc@$\nneg C$}
\end{itemize}
Obviously, $\sharp \clos(C,\mcT) = \mcO(|C| + |\mcT|)$ (compare
Lemma~\ref{lem:alcq-closure-linear}).

A subset $\Phi \subseteq \clos({C,\mcT})$ is \emph{locally consistent}
\index{locally consistent set} iff
\begin{itemize}
\item for every $D \in \clos({C,\mcT})$, $\Phi \cap \{ D, \nneg D \} \neq \emptyset$
  and $\{ D, \nneg D \} \not \subseteq \Phi$,
\item for every $C_1 \sqsubseteq C_2 \in \mcT$, $\NNF(\neg C_1 \sqcup C_2) \in \Phi$,
\item if $C_1 \sqcap C_2 \in \Phi$ then $\{ C_1 , C_2 \} \subseteq \Phi$, and
\item if $C_1 \sqcup C_2 \in \Phi$ then $\Phi \cap \{ C_1, C_2 \} \neq \emptyset$.
\end{itemize}

The set of locally consistent subsets of $\clos({C,\mcT})$ is defined by
$\lc({C,\mcT}) = \{ \Phi \subseteq \clos({C,\mcT}) \mid \Phi \text{ is l.c.}
\}$. \index{00lcct@$\lc({C,\mcT})$} Obviously, for every element $x$ in a model
of $\mcT$, there exists a set $\Phi \subseteq \lc(C,\mcT)$ such that all concepts
from $\Phi$ are satisfied by $x$.

It remains to describe how the role relationships in $\I$ are mapped to $T$.
Unfortunately, it is not possible to simply map successors in $\I$ to
successors in $T$ due to the presence of binary coding of numbers in number
restrictions. A number restriction of the form $\qnrgeq n \omega D$ requires the
existence of $n$ successors, where $n$ may be exponential in the size of $C$ if
numbers are coded binarily. In this case, the transition table of the
corresponding automaton requires double-exponential space in the size of $C$ and
the automata approach would not yield the \exptime-result we desire.

We overcome this problem as follows. Instead of using a $k$-ary tree, where $k$
somehow depends on the input $C$ and $\mcT$, we use a binary tree.  Required
successors $t_i$ of an element $s$ in $\I$ are not mapped to direct successors
of the node corresponding to $s$ but rather to nodes that are reachable by zero
or more steps to the left and a single step to the right. The dummy label
$\langle *,* \rangle$ is used for the auxiliary states that are reachable by
left-steps only because these do not correspond to any elements of $\I$. If $n$
successors must be mapped, the subtree rooted $n$ left-steps from the current
node is not needed to map any more successors and hence is  labelled entirely
with $\langle *,* \rangle$.  Figure~\ref{fig:model-to-tree} illustrates this
construction, where $\Phi_x$ denotes the concepts from $\clos(C,\mcT)$ that are
satisfied by $x$ and $\mbR_x$ the set of roles connecting $x$ with its
predecessor.

\begin{figure}[tbh]
  \begin{center}
    \vspace{2ex}
    \input{tree-construction.pstex_t}
  \end{center}
  \caption{Transforming a model for $C$  into a tree accepted by $\mfA_{C,\mcT}$}
  \label{fig:model-to-tree}
\end{figure}

In order to accept exactly the abstractions of models generated by this
transformation, it is necessary to perform additional book-keeping in the
states.  Since successors of the element $s$ are spread through the tree, we
must equip the states of $\mfA_{C,\mcT}$ ``responsible'' for the auxiliary nodes
with enough information to ensure that the number restrictions are ``obeyed''.
For this purpose, we use counters to record the minimal and maximal number of
$\omega$-successors satisfying $D$ that a node may have. This information is
initialized whenever stepping to a right successor and updated when moving to a
left successor in the tree.  The counters are modelled as functions as follows.

The maximum number $\nmax(C,\mcT)$ \index{00nmaxct@$\nmax(C,\mcT)$} occurring in
a qualifying number restriction in $\clos(C,\mcT)$ is defined by $\nmax(C,\mcT)
= \max \{ n \in \N \mid \qnrgleq n \omega D \in \clos(C,\mcT) \}$ with
$\max(\emptyset) := 0$.

The set of concepts that occur in number restrictions and hence must be
considered at successor and predecessor nodes is defined by
\[
\excon({C,\mcT}) = \{ D \mid \qnrgleq n \omega D \in \clos(C,\mcT) \} .
\]
\index{00succct@$\excon({C,\mcT})$}

In the automaton, we keep track of the numbers of witnesses for every occurring
role expression $\omega \in \Omega_{C,\mcT}$ and concept from
$\excon({C,\mcT})$.  This is done using a set of \emph{limiting functions}
\index{limiting function}
$\limit({C,\mcT})$ defined by
\[
\limit({C,\mcT}) := \{ f \mid f : \Omega_{C,\mcT} \times \excon({C,\mcT})
\rightarrow \{0, \dots, \nmax({C,\mcT}), \infty \} \} .
\]
\index{00limitct@$\limit({C,\mcT})$}
  
The maximum/minimum number of allowed/required $\omega$-successors satisfying
a certain concept $D$ imposed by number restrictions in a set of concepts is
captured by the functions 
\[
\min, \max : \lc({C,\mcT}) \times \overline{\Roles}_{C,\mcT} \times
\excon({C,\mcT}) \rightarrow \{ 0, \dots, \nmax({C,\mcT}), \infty \}
\]
defined by
\begin{align*}
  \max (\Phi,\omega,D) & = \min \{ n \mid \qnrleq n \omega D \in \Phi \}\\
  \min (\Phi,\omega,D) & = \max \{ n \mid \qnrgeq n \omega D \in \Phi \}
\end{align*}
with $\min(\emptyset) := \infty$.

In the automaton $\mfA_{C,\mcT}$, each state consists of
a locally consistent set, a set of roles, and two limiting functions for the
upper and lower bounds. There are three kinds of states. 
\begin{itemize}
\item states that label nodes of $T$ corresponding to elements of the
  interpretation.  These states record the locally consistent set $\Phi$
  labelling that node, the set of roles that connect the corresponding element to
  its unique predecessor in $\I$ and the appropriate initial values of the
  counters for this node---taking into account the concepts satisfied by the
  predecessor. This is necessary due to the presence of inverse roles.
\item states labelling nodes that are reachable from a node $s$ corresponding to
  an element of $\I$ by one or more steps to the left. These states are marked
  by an empty set of roles and record the locally consistent set labelling $s$ to
  allow for the correct initialization of the counters for nodes corresponding
  to successors of $s$. Moreover, their limiting functions record the upper and
  lower bound of $\omega$-successors of $s$ still allowed/required. According to
  these functions, their right successor state ``expects'' a node corresponding
  to a successor of $s$ and their left successor state a further auxiliary node.
  The limiting functions of this auxiliary state are adjusted according to the
  right successor.  Once sufficiently many successors have been generated, the
  automaton switches to the following dummy state.
\item a dummy state $\langle *,*,*,* \rangle$, which reproduces itself and
  accepts a tree entirely labelled with $\langle *,* \rangle$.
\end{itemize}

For a role $R \in \overline{\Roles}_{C,\mcT}$, we define $\Inv(R)$ by setting
\[
\Inv(R) = \begin{cases}
  R^{-1} & \text{if $R \in \Roles$},\\
  S & \text{if $R = S^{-1}$ for some $S \in \Roles$},
\end{cases}
\]
\index{00invr@$\Inv(R)$}

and for a set of roles $\mbR$ we define $\Inv(\mbR) = \{ \Inv(R) \mid R \in \mbR
\}$.  We are now ready to define the automaton $\mfA_{C,\mcT}$.

\begin{definition}\label{def:alcqir-atomaton}
  Let $C$ be an \alcqir-concept in NNF and $\mcT$ an \alcqir-TBox.  The binary
  looping tree automaton $\mfA_{C,\mcT} = (Q,\Sigma,I,\delta)$ for $C$ and
  $\mcT$ is defined by
  \begin{align*}
    Q & = \left ( \lc({C,\mcT}) \times 2^{\overline{\Roles}_{C,\mcT}} \times
      \limit({C,\mcT}) \times \limit({C,\mcT}) \right ) \cup
    \{ \langle *,*,*,* \rangle \} \\
    \Sigma & = \left (\lc({C,\mcT}) \times 2^{\overline{\Roles}_{C,\mcT}} \right
    )\cup \{
    \langle *,* \rangle \} \\
    I & = \{ \langle \Phi, \overline{\Roles}_{C,\mcT}, \ell, h \rangle \in Q
    \mid C \in \Phi, \ell = \lambda \omega D . \min(\Phi,\omega,D), h = \lambda
    \omega D .
    \max(\Phi,\omega,D) \}\\
    \delta & \subseteq Q \times \Sigma \times Q^2 , 
  \end{align*}
  such that $\delta$ is the maximal transition relation with $(\langle
  *,\!*,\!*,\!* \rangle, \langle *,\!* \rangle, \langle *,\!*,\!*,\!* \rangle,
  \langle *,\!*,\!*,\!* \rangle ) \in \delta$ and if $(q_0,\sigma,q_1,q_2) \in
  \delta$ with $q_0 \neq \langle *,*,*,*\rangle$ and $q_i = \langle \Phi_i,
  \mbR_i, \ell_i, h_i \rangle$ then
  \begin{itemize}
  \item [\aut 1] if $\mbR_0 \neq \emptyset$ then $\sigma = \langle \Phi_0,
    \mbR_0 \rangle$ else $\sigma = \langle *,* \rangle$
  \item [\aut 2] if, for all $\omega \in \Omega_{C,\mcT}$ and $D \in \excon({C,\mcT})$,
    $\ell_0(\omega,D) = 0$, then $q_1 = q_2 = \langle *,*,*,* \rangle$
  \item [\aut 3] otherwise, $\Phi_2 \in \lc(C,\mcT)$ and $\mbR_2 \subseteq
    \overline{\Roles}_{C,\mcT}$ such that there is a $\omega \in
    \Omega_{C,\mcT}$ and a $D \in \Phi_2$ with $\mbR_2 \models \omega$ and
    $\ell_0(\omega,D) > 0$.  As an auxiliary function, we define
    \[
    e(\Phi,\mbR,\omega,D) = \begin{cases},
      1 & \text{if $\mbR \models \omega$ and $D \in \Phi$}\\
      0 & \text{otherwise},
    \end{cases}
    \]
    \index{00ephi@$e(\Phi,\mbR,\omega,D)$}
    and require, for all $\omega \in \Omega_{C,\mcT}$ and $D \in \clos(C,\mcT)$,
    \begin{equation}
      \tag{$*$}
      \begin{array}{cccc}
        \text{if} & \max(\Phi_2,\omega,D) = 0 & \text{then} &
        e(\Phi_0,\Inv(\mbR_2),\omega,D) = 0 \\ 
        \text{if} & h_0(\omega,D)=0 & \text{then} &  e(\Phi_2,\mbR_2,\omega,D) =0 . 
      \end{array}
    \end{equation}
    Finally, $\Phi_1 = \Phi_0, \mbR_1 = \emptyset$ and
    \begin{align*}
      \ell_1 & = \lambda \omega D . \ell_0(\omega,D) \dotminus
      e(\Phi_2,\mbR_2,\omega,D),\\
      h_1 & = \lambda \omega D . h_0(\omega,D) - e(\Phi_2,\mbR_2,\omega,D),\\
      \ell_2 & = \lambda \omega D . \min(\Phi_2,\omega,D) \dotminus
      e(\Phi_0,\Inv(\mbR_2),\omega,D), \text{ and}\\
      h_2 & = \lambda \omega D . \max(\Phi_2,\omega,D) -
      e(\Phi_0,\Inv(\mbR_2),\omega,D)
    \end{align*}
    must hold,
    where $\dotminus$ denotes subtraction in $\N$, i.e., $x \dotminus
    y = \max(0, x - y)$. \index{00dotminus@$\dotminus$} \eod
  \end{itemize}
  \index{00act@$\mfA_{C,\mcT}$}
\end{definition} 

The choice of $\overline{\Roles}_{C,\mcT}$ as the role component of the initial
states in $I$ is arbitrary and indeed every non-empty set of could be used instead
of $\overline{\Roles}_{C,\mcT}$.  Note that the subtraction in the requirements
for $h_1$ and $h_2$ never yields a negative value because of $(*)$.  Moreover,
$\mfA_{C,\mcT}$ is small enough (i.e., exponential in the input) to be of use in
our further considerations:

\begin{lemma}\label{lem:shiq-automaton-exponential-size}
  Let $C$ be a \alcqir-concept in NNF, $\mcT$ an \alcqir-TBox, $m = |C| + | \mcT
  |$, and $\mfA_{C,\mcT} = (Q,\Sigma, I, \delta)$ the looping tree automaton for
  $C$ and $\mcT$. Then
  \[
  \sharp Q + \sharp \delta = \mcO(2^{m^5}) .
  \]
\end{lemma}
  
\begin{proof}
  The cardinality of $\lc({C,\mcT})$ is bounded by $2^{\sharp \clos({C,\mcT})} =
  \mcO(2^m)$.  The cardinality of $\overline{\Roles}_{C,\mcT}$ is bounded by
  $2m$ and hence $\sharp 2^{\overline{\Roles}_{C,\mcT}} = \mcO(2^m)$.
  Finally, the cardinality of $\limit({C,\mcT}) \{ f \mid f : \Omega_{C,\mcT}
  \times \excon({C,\mcT}) \rightarrow \{0, \dots, \nmax(C,\mcT), \infty \} \}$ is
  bounded by $(\nmax({C,\mcT})+2)^{(\sharp \Omega_{C,\mcT} \times \sharp
    \excon({C,\mcT}))} = \mcO((2^m)^{m^2}) = \mcO(2^{m^3})$, where $2^m$ is an
  upper bound for $\nmax({C,\mcT})$ if numbers are coded binarily in the input.
  Summing up, we get $ \mcO(2^m \times 2^m \times 2^{m^3}) = \mcO(2^{m^4})$ as a
  bound for $\sharp Q$ and $\mcO(2^{m^5})$ as a bound for $\sharp \delta$,
  which dominates $\sharp Q$. \qed
\end{proof}

We now show that emptiness of $L(\mfA_{C,\mcT})$ is indeed equivalent to
unsatisfiability of $C$ \wrt $\mcT$.

\begin{lemma}\label{lem:shiq-automaton-emptiness}
  For an \alcqir-concept $C$ in NNF and a \alcqir-TBox $\mcT$, $L(\mfA_{C,\mcT})
  \neq \emptyset$ iff $C$ is satisfiable \wrt $\mcT$.
\end{lemma}

\begin{proof}
  Assume $L(\mfA_{C,\mcT}) \neq \emptyset$, $T$ is a tree accepted by
  $\mfA_{C,\mcT}$, and $r$ is an arbitrary run of $\mfA_{C,\mcT}$ on $T$ with
  $r(\epsilon) \in I$. From $T$, we will construct a model $\I = (\Delta^\I,
  \cdot^\I)$ for $C$ and $\mcT$, which proves satisfiability of $C$ \wrt $\mcT$.
  For every path $p \in \{1,2\}^*$ with $r(p) = \langle \Phi, \mbR, \ell, h
  \rangle$, we define $\Phi_p := \Phi, \mbR_p := \mbR, \ell_p := \ell$, and $h_p :=
  h$.
  
  The domain $\Delta^\I$ of $\I$ is defined by $\Delta^\I = \{ p \in \{1,2\}^*2
  \mid T(p) \neq \langle *,* \rangle \} \cup \{ \epsilon \}$. Hence, $\Delta^\I$
  contains only ``right successors'' and the root.  For concept names $A$, we
  define
  \[
  A^\I = \{ p \in \Delta^\I \mid A \in \Phi_p \} .
  \]
  For the interpretation of roles, we define
  \begin{align*}
    R^\I = & \{ (p,p') \in \Delta^\I \times \Delta^\I \mid p' \in p 1^* 2, R
    \in \mbR_{p'} \} \cup {} \\
    & \{ (p',p) \in \Delta^\I \times \Delta^\I \mid p' \in p 1^* 2, R \in
    \Inv(\mbR_{p'}) \} .
  \end{align*}
  
  Before we prove that $\I$ is indeed a model for $C$ and $\mcT$, we state
  some general properties of the automaton and  this construction.

  \begin{itemize}
  \item [\run 1] Due to the construction of $\Delta^\I$, for every $p \in \Delta^\I$,
    $\mbR_p \neq \emptyset$ and hence $r(p) \neq \langle *,*,*,* \rangle$.
  \item [\run 2] ``Once $\langle *,*,*,* \rangle$, always $\langle *,*,*,*
    \rangle$.'' For a path $p \in \{ 1,2 \}^*$, if $r(p) = \langle *,*,*,*
    \rangle$, then, for all $p'$ with $p' \in p \{1,2\}^*$, $T(p') = \langle *,*
    \rangle$ and $r(p') = \langle *,*,*,* \rangle$.
  \item [\run 3] ``A left successor is either $\langle *,*,*,* \rangle$ or an
    auxiliary state, in which case it is labelled with the same set from
    $\lc(C,\mcT)$.'' For a path $p \in \{ 1,2 \}^*$, if $r(p) = \langle \Phi,
    \mbR, \ell, h\rangle$, then, for all $p' \in p 1^*$, if $r(p') \neq \langle
    *,*,*,* \rangle$ then $r(p')$ is of the form
    $r(p') = \langle \Phi, \emptyset, \ell', h' \rangle$.
  \item [\run 4] ``$h$ and $\ell$ are a lower an upper bounds on the number of
    successors of a node.''  For a path $p \in \{ 1,2 \}^*$ with $T(p) \neq
    \langle *,* \rangle$, $\omega \in \Omega_{C,\mcT}$, and $D \in
    \excon(C,\mcT)$,
    \[
    \ell_p(\omega,D) \leq \sharp \{ p' \in p 1^* 2 \mid  \mbR_{p'} \models \omega, D \in
    \Phi_{p'} \} \leq h_p(\omega,D) .
    \]
    
    This property is less obvious than the others and we give a proof by
    induction on
    \[
    \|p\| =\sum_{\omega \in \Omega_{C,\mcT}, D \in \excon(C,\mcT)}
    \ell_{p}(\omega,D) .
    \]
    If $\|p\| = 0$, then $\ell_p(\omega,D) = 0$ for all $\omega \in
    \Omega_{C,\mcT}$ and $D \in \excon(C,\mcT)$ and hence, by \aut 2 and \run 2,
    for all ancestors $p' \in p 1^* 2$ of $p$, $r(p') = \langle *,*,*,* \rangle$
    and $T(p') = \langle *,* \rangle$.  Thus
    \[
    0 = \ell_p(\omega,D) = \sharp \{ p' \in p 1^* 2 \mid \mbR_{p'} \models \omega, D
    \in \Phi_{p'} \} \leq h_p(\omega,D) 
    \]
    holds for all $\omega \in \Omega_{C,\mcT}$ and $D \in \excon(C,\mcT)$.
    
    If $\|p\| > 0$ then there is an $\omega \in \Omega_{C,\mcT}$ and a $D \in
    \excon(C,\mcT)$ with $\ell_p(\omega,D) > 0$, $\mbR_{p2} \models \omega$, and
    $D \in \Phi_{p2}$. Hence, $\ell_{p1}(\omega,D) = \ell_p(\omega,D) - 1$ and
    $\|p1\| < \|p\|$ by \aut 3 and we can use the induction hypothesis for $p1$.

    For all $\omega \in \Omega_{C,\mcT}$ and $D \in \excon(C,\mcT)$, 
    \begin{align*}
      \ell_p(\omega,D) & \leq \ell_{p1}(\omega,D) + e(\Phi_{p2},\mbR_{p2},\omega,D) \\
      & \leq^{(*)} \sharp \{ p' \in p11^*2 \mid \mbR_{p'} \models \omega, D \in
      \Phi_{p'} \} +
      \sharp \{ p' \in p2 \mid \mbR_{p'} \models \omega, D \in \Phi_{p'} \} \\
      & = \sharp \{ p' \in p1^*2 \mid \mbR_{p'}\models \omega, D \in \Phi_{p'} \}\\
      & = \sharp \{ p' \in p11^*2 \mid \mbR_{p'}\models \omega, D \in \Phi_{p'}
      \} +
      \sharp \{ p' \in p2 \mid \mbR_{p'}\models \omega, D \in \Phi_{p'} \}\\
      & \leq^{(*)} h_{p1}(\omega,D) + e(\Phi_{p2},\mbR_{p2},\omega,D) \\
      & = h_p(\omega,D),
    \end{align*}
    where the steps marked with $(*)$ use the induction hypothesis. This is what
    we needed to show.
  \item [\run 5] For two paths $p,q \in \Delta^\I$ and a role expression $\omega
    \in \Omega_{C,\mcT}$, if $(p,q) \in \omega^\I$ then $q \in p1^*2$ or $p \in
    q1^*2$. 
    
    Because of the syntactic restriction to safe role expressions in \alcqir,
    for $(p,q) \in \omega^\I$ to hold there must be a role $R \in
    \overline{\Roles}_{C,\mcT}$ such that $(p,q) \in R^\I$. By construction of
    $R^\I$, this can only be the case if $q \in p1^*2$ or $p \in q1^*2$. 
    
  \item [\run 6] For two paths $p,q \in \Delta^\I$ with $q \in p1^*2$ and a role
    $\omega$, $(p,q) \in \omega^\I$ iff $\mbR_q \models \omega$ and $(q,p) \in
    \omega^\I$ iff $\Inv(\mbR_q) \models \omega$.
    
    For every $R \in \overline{\Roles}_{C, \mcT}$, $(p,q) \in R^\I$ iff $R \in
    \mbR_q$ holds as follows. For a (non-inverse) role $R \in
    \overline{\Roles}_{C, \mcT} \cap \Roles$, immediately by the construction of
    $R^\I$, $(p,q) \in R^\I$ iff $R \in \mbR_q$. For an inverse role $R =
    S^{-1}$ with $S \in \overline{\Roles}_{C, \mcT} \cap \Roles$, $(p,q) \in
    R^\I$ iff $(q,p) \in S^\I$ iff $\Inv(S) = R \in \mbR_q$. Hence, $(p,q) \in
    \omega^\I$ iff $\mbR_q \models \omega$. Similarly, for every $R \in
    \overline{\Roles}_{C, \mcT}$, $(q,p) \in R^\I$ iff $\Inv(R )\in \mbR_q$, and
    hence $(q,p) \in \omega^\I$ iff $\Inv(\mbR_q) \models \omega$.
  \end{itemize}

  Using these properties we can now show:

  \begin{claim}
    For all $p \in \Delta^\I$ and $D \in \Phi_p$, $p \in  D^\I$.
  \end{claim}
  
  The proof is by induction on the norm $\|\cdot \|$ of the concepts (as defined
  Definition~\ref{def:alcq-concept-norm}). The base
  cases are $D = A$ or $D = \neg A$ for a concept name $A \in \Names$. For $D =
  A$ this is immediate by the definition of $A^\I$. For the case $D = \neg A$,
  since $\Phi_p \in \lc(C,\mcT)$, $\neg A \in \Phi_p$ implies $A \not \in
  \Phi_p$ and hence $p \in (\neg A)^\I$.  For the induction step, we distinguish
  the different concept operators of \alcqir.
  \begin{itemize}
  \item If $D = C_1 \sqcap C_2 \in \Phi_p$ then, since $\Phi_p \in \lc(C,\mcT)$,
    also $\{ C_1, C_2 \} \subseteq \Phi_p$. Hence, by induction, $p \in C_1^\I$,
    $p \in C_2^\I$ and thus $p \in D^\I$.
  \item The case $D = C_1 \sqcup C_2$ is similar to the previous one.
  \item Now assume $D = \qnrgleq n \omega E$. For every $q \in \Delta^\I$,
    $\Phi_q \in \lc(C,\mcT)$ and hence $E \in \Phi_q$ iff $\nneg E \not \in
    \Phi_q$. Since $\| E \| = \| \nneg E \| < \| D \|$, by induction, $q \in
    E^\I$ iff $E \in \Phi_q$ holds for every $q \in \Delta^\I$.
                                                                        
    If $p = \epsilon$ is the root of $T$ then, by \run 5 and \run 6,
    \[ 
    \sharp \{ q \mid (p,q) \in \omega^\I, q \in E^\I \} = \sharp \{ q \in p 1^*
    2 \mid \mbR_q \models \omega, E \in \Phi_q \}
    \]
    and hence, by \run 4, 
    \[
    \min(\Phi_p,\omega,E) \leq \sharp \{ q \mid (p,q) \in \omega^\I, q \in E^\I
    \} \leq \max(\Phi_p,\omega,E) .
    \]
    If $p \neq \epsilon$, then $p \in \{1,2\}^*2$ is a ``right successor''. Let
    $q_0$ be the unique path in $\{1,2\}^*2 \cup \{ \epsilon \}$ with $p = q_0
    1^k 2$, i.e., $p$'s ``predecessor'' in $\I$.
    \begin{align*}
      & \sharp \{ q \mid (p,q) \in \omega^\I, q \in E^\I \} \\
      = & \sharp \{ q \in p 1^* 2 \mid \mbR_q \models \omega, E \in \Phi_q \} +
      e(\Phi_{q_0}, \Inv(\mbR_p),\omega,E) \\
      = & \sharp \{ q \in p 1^* 2 \mid \mbR_q \models \omega, E \in \Phi_q \} +
      e(\Phi_{q_0 1^k}, \Inv(\mbR_p), \omega,E) .
    \end{align*}
    If $E \not \in \Phi_{q_0}$ or $\Inv(\mbR_p) \not \models \omega$, then
    $e(\Phi_{q_0 1^k}, \Inv(\mbR_p), \omega,E) = 0$ and
    \begin{align*}
      \min(\Phi_p,\omega,E) & = \ell_p(\omega,E)\\
      & \leq \sharp \{ q \mid q \in p 1^* 2, \mbR_q \models \omega, E \in \Phi_q \} \\
      & = \sharp \{ q \mid (p,q) \in \omega^\I, q \in E^\I \}\\
      & \leq h_p(\omega,E) = \max(\Phi_p,\omega,E) 
    \end{align*}
    holds because of induction, \run 4, \run 5, and \run 6.
    If $E \in \Phi_{q_0}$ and $\Inv(\mbR_p) \models \omega$, then
    $e(\Phi_{q_0 1^k}, \Inv(\mbR_p), \omega,E) = 1$ and
    \begin{align*}
      \min(\Phi_p,\omega,E) & \leq \ell_p(\omega,E) + 1\\
      & \leq \sharp \{  q \in p 1^* 2 \mid  \mbR_q \models \omega, E \in \Phi_q \} + 1 \\
      & = \sharp \{ q \mid (p,q) \in \omega^\I, q \in E^\I \}\\
      & \leq h_p(\omega,E) + 1 = \max(\Phi_p,\omega,E)
    \end{align*}
    again holds by \run 4, \run 5, and \run 6.
    
    If $D = \qnrgeq n \omega E$ then $n \leq \min(\Phi_p,\omega,E) \leq
    \sharp \{ q \mid (p,q) \in \omega^\I, q \in E^\I \}$ and hence $p \in D^\I$.
    If $D = \qnrleq n \omega E$ then $n \geq \max(\Phi_p,\omega,E) \geq \sharp
    \{ q \mid (p,q) \in \omega^\I, q \in E^\I \}$ and hence $p \in D^\I$.
  \end{itemize}
  
  This finishes the proof of the claim, which yields the \emph{only-if}
  direction of the lemma: if $L(\mfA_{C,\mcT}) \neq \emptyset$ then there exists
  a tree $T \in L(\mfA_{C,\mcT})$ and a corresponding interpretation $\I$ that
  satisfies the claim. Since $C \in \Phi_\epsilon$, $\epsilon \in C^\I$ and hence $C^\I
  \neq \emptyset$. Also, for every $p \in \Delta^\I$ and every $C_1 \sqsubseteq
  C_2 \in \mcT$, $\NNF(\neg C_1 \sqcup C_2) \in \Phi_p$. Hence $(\neg C_1 \sqcup
  C_2)^\I = \Delta^\I$ and $\I \models \mcT$.
  
  For the \emph{if}-direction, let $C$ be satisfiable \wrt $\mcT$ and $\I =
  (\domain,\cdot^\I)$ a model of $\mcT$ with $C^\I \neq \emptyset$. We construct
  a tree $T$ from $\I$ that is accepted by $\mfA_{C,\mcT}$.  To this purpose, we
  define a function $\pi : \{1,2\}^* \rightarrow \Delta^\I \mathbin{\dot \cup}
  \{ * \}$ and maintain an agenda of paths $p \in \{1,2\}$ whose successors
  still need consideration.
  
  Let $s \in \Delta^\I$ be an arbitrary element such that $s \in C^\I$. Set
  $\pi(\epsilon) = s$ and $T(\epsilon) = \langle \Phi_s,
  \overline{\Roles}_{C,\mcT} \rangle $ with $\Phi_s = \{ D \in \clos({C,\mcT})
  \mid s \in D^\I \}$. Initialize the agenda with $\epsilon$.
  
  Pick the first element $p \in \{ 1,2 \}^*$ off the agenda. For $s = \pi(p)$,
  let $\Phi_s = \{ D \in \clos({C,\mcT}) \mid s \in D^\I \}$ and let $X
  \subseteq \Delta^\I$ be a set such that
  \begin{itemize}
  \item $X \subseteq \{ t \in \Delta^\I \mid (s,t) \in R^\I \text{ for
      some $R \in \overline{\Roles}_{C,\mcT}$ } \}$ .
  \item For every $\qnrgeq n \omega D \in \Phi_s$ there are $t_1, \dots t_n \in
    X$ with $(s,t_i) \in \omega^\I$,  $t_i \in D^\I$ for $1 \leq i \leq n$ and $t_i
    \neq t_j$ for $1\leq i < j \leq n$ .
  \item $X$ is minimal w.r.t.\ set cardinality with these properties.
  \end{itemize}
  Such a set $X$ exists, is finite, possibly empty, and not necessarily uniquely
  defined.  Let $\{ t_1, \dots, t_n \}$ be an enumeration of $X$.
  \begin{itemize}
  \item For every $1 < i \leq n$, we set $\pi(p1^{i-1}) = *$ and
    $T(p1^{i-1}) = \langle *,* \rangle$.
  \item For every $1 \leq i \leq n$, we set $\pi(p1^{i-1}2) = t_i$ and
    \begin{align*}
      T(p1^{i-1}2) & = \langle \Phi_{t_i}, \mbR_{t_i} \rangle
      \intertext{where}
      \Phi_{t_i} & = \{ D \in \clos({C,\mcT}) \mid t_i \in D^\I \}\\
      \mbR_{t_i} & = \{ R \in \overline{\Roles}_{C,\mcT}  \mid (s,t_i) \in R^\I
      \} .
    \end{align*}
    Put $p1^{i-1}2$ at the end of the agenda.
  \item Finally, for all $p' \in p 1^n \{1,2\}^*$ we define $\pi(p') = *$ and
    $T(p') = \langle *,*\rangle$.
  \end{itemize}
  
  Figure~\ref{fig:model-to-tree} illustrates this construction.

  Continuing this process until the agenda runs empty (or indefinitely if it
  never does) eventually defines $T(p)$ for every $p \in \{ 1,2\}^*$ (since the
  agenda is organised as a queue, every element will eventually be taken off the
  agenda). The proof that $T \in L(\mfA_{C,\mcT})$ (and hence $L(\mfA_{C,\mcT})
  \neq \emptyset$) is relatively simple and omitted here.  \qed
\end{proof}

\begin{theorem}\label{theo:alcqir-tbox--exptime-complete}
  Satisfiability of \alcqir-concepts \wrt general TBoxes is \exptime-complete,
  even if numbers in the input are represented in binary coding.
\end{theorem}

\begin{proof}
  \exptime-hardness was established in Lemma~\ref{lem:alcqir-tbox-exptime-hard}.
  By Lemma~\ref{lem:shiq-automaton-emptiness}, generating $\mfA_{C,\mcT}$ and
  testing $L(\mfA_{C,\mcT})$ for emptiness decides satisfiability of $C$ \wrt
  $\mcT$. Due to Lemma~\ref{lem:shiq-automaton-exponential-size} and
  Fact~\ref{fact:looping-automaton-polynomial-emptiness} this can be done in
  time exponential in $|C| + |\mcT|$. \qed
\end{proof}

Now that we know how to deal with satisfiability of \alcqir-concept \wrt TBoxes,
we show how satisfiability of full knowledge bases can reduced to that problem
using a pre-completion technique similar to the one in
\cite{hollunder96:_consis_check_reduc_satis_concep_termin_system} for
\alcq-knowledge bases (see also Section~\ref{sec:other-infer-probl}).

The definition of $\clos(\cdot)$ is extended to \alcqir-knowledge bases as
follows. For a \alcqir-knowledge base $\mcK = (\mcT, \mcA)$, we define
$\clos(\mcK)$ \index{00closk@$\clos(\mcK)$} as the smallest set $X$ that
satisfies the following properties:
\begin{itemize}
\item for every $x : D \in \mcA$, $\NNF(D) \in X$
\item for every $C_1 \sqsubseteq C_2 \in \mcT$, $\NNF(\neg C_1 \sqcup C_2) \in
  X$
\item $X$ is closed under sub-concepts and the application of $\nneg$.
\end{itemize}
Again, $\sharp\clos(\mcK) = \mcO(|\mcK|)$ holds (compare
Lemma~\ref{lem:alcq-closure-linear}).

\begin{definition}\label{def:alcqir-precompletion}
  Let $\mcK = (\mcT, \mcA)$ be an \alcqir-knowledge base. A knowledge base $\mcK'
  = (\mcT, \mcA')$ is a \iemph{pre-completion of} $\mcK$, if
  \begin{enumerate}
  \item there is a surjective function $$f : \{ x \in \Individuals \mid x \text{
      occurs in $\mcA$} \} \rightarrow \{ x \in \Individuals \mid x \text{ occurs
      in $\mcA'$} \}$$ such that
    \begin{itemize}
    \item if $x : C \in \mcA$ then $f(x) : C \in \mcA'$
    \item if $(x,y) : R \in \mcA$ then $(f(x),f(y)) : R \in \mcA'$
    \end{itemize}
  \item for every $x$ that occurs in $\mcA'$ and every $D \in \clos(\mcK)$, $x :
    D \in \mcA'$ or $x : \nneg D \in \mcA'$
  \item for every $x$ that occurs in $\mcA'$, if $x: C_1 \sqcap C_2 \in \mcA'$
    then $x : C_1 \in \mcA'$ and $x : C_2 \in \mcA'$
  \item for every $x$ that occurs in $\mcA'$, if $x : C_1 \sqcup C_2 \in \mcA'$
    then $x : C_1 \in \mcA'$ or $x : C_2 \in \mcA'$
  \item for every two distinct $x,y$ that occur in $\mcA'$, $x \ndoteq y \in
    \mcA'$
  \end{enumerate}
  A knowledge base $\mcK'$ that satisfies 2--5 is called \emph{pre-completed}. \eod
\end{definition}

It is easy to see that a knowledge base is satisfiable iff it has as
pre-completion that has a model that \emph{exactly} satisfies the role
assertions:

\begin{lemma}\label{lem:alcqir-kb-consistent-iff-precompletion}
  Let $\mcK = (\mcT, \mcA)$ be an \alcqir-knowledge base and
  $\overline{\Roles}_\mcK$ \index{00nrk@$\overline{\Roles}_\mcK$} the set of
  roles that occur in $\mcK$ together with their inverse. $\mcK$ is satisfiable
  iff there exists pre-completion $\mcK' = (\mcT, \mcA')$ of $\mcK$ and a model
  $\I$ of $\mcK'$ such that, for every $x,y \in \Individuals$ that occur in
  $\mcA'$,
  \[
  \{ R \in \overline{\Roles}_\mcK \mid (x,y) : R \in \mcA' \} = \{ R \in
  \overline{\Roles}_\mcK \mid (x,y) \in R^\I \} .
  \]
\end{lemma}

For a pre-completion $\mcK' = (\mcT,\mcA')$, the existence of such a model can
be reduced to concept satisfiability \wrt $\mcT$. For an individual $x$ that
occurs in $\mcA'$, we define $C_x$ by
\begin{align*}
  C_x = & \mybigsqcap \{ A \mid A \in \Names, x : A \in \mcA' \} \; \sqcap \\
  & \mybigsqcap \{ \neg A \mid A \in \Names, x : \neg A \in \mcA' \} \; \sqcap \\
  & \mybigsqcap \{ \qnrgeq {(n - m)} \omega D \mid x : \qnrgeq n \omega D \in \mcA',
  m = \sharp \omega^{\mcA'}(x,D) \} \; \sqcap\\
  & \mybigsqcap \{ \qnrleq {(n - m)} \omega D \mid x : \qnrleq n \omega D \in \mcA',
  m = \sharp \omega^{\mcA'}(x,D) \} .
\end{align*}

\begin{lemma}\label{lem:alcqir-precompletion-to-concept-satisfiability}
  Let $\mcK' = (\mcT, \mcA')$ be a pre-completed \alcqir-knowledge base. $\mcK'$ has
  a model that satisfies,
  \[
  \{ R \in \overline{\Roles}_\mcK \mid (x,y) : R \in \mcA' \} = \{ R \in
  \overline{\Roles}_\mcK \mid (x,y) \in R^\I \} ,
  \]
  for every $x,y \in \Individuals$ that occur in $\mcA'$ iff, for every $x$ that
  occurs in $\mcA'$, the concept $C_x$ is satisfiable \wrt $\mcT$.
\end{lemma}

The proof of this lemma is straightforward and omitted here.

Putting together Lemma~\ref{lem:alcqir-kb-consistent-iff-precompletion} and
Lemma~\ref{lem:alcqir-precompletion-to-concept-satisfiability}, we have the
steps of a reduction from knowledge-base satisfiability to concept
satisfiability \wrt general TBoxes---a problem that we know how to solve in
\exptime (Theorem~\ref{theo:alcqir-tbox--exptime-complete}). But how do we obtain
an \exptime-algorithm from these lemmas?
Lemma~\ref{lem:alcqir-kb-consistent-iff-precompletion} involves a
non-deterministic step since it talks about the \emph{existence} of a
completion. Since it is generally assumed that $\exptime \neq \nexptime$ we have
to show how to search for such a completion in exponential time.

\begin{theorem}\label{theo:alcqir-kb-exptime-complete}
  Knowledge base satisfiability and instance checking for \alcqir are
  \exptime-complete, even if numbers in the input are represented using binary
  coding.
\end{theorem}

\begin{proof}
  \exptime-hardness is immediate from
  Theorem~\ref{theo:alcqir-tbox--exptime-complete}. It remains to show that
  these problems can be decided in exponential time.
  
  Let $\mcK = (\mcT, \mcA)$ be an \alcqir-knowledge base,
  $\overline{\Roles}_\mcK$ the set of roles that occur in $\mcK$ together with
  their inverse, and $\clos(\mcK)$ defined as above. Let $m = | \mcK |$.  Only
  ABoxes $\mcA'$ with no more individuals than $\mcA$ are candidates for
  pre-completions because the mapping $f$ must be surjective. The number of
  individuals in $\mcA$ is bounded by $m$.
  
  For an ABox $\mcA'$ with $i\leq m$ individuals, concept assertions ranging
  over $\clos(\mcK)$, and role assertions ranging over $\overline{\Roles}_\mcK$,
  there are at most $2^{i \times m} \times 2^{i^2 \times 2m} = \mcO(2^{m^5})$
  different possibilities, and each such ABox contains at most $i \times m + i^2
  \times 2m + i^2 = \mcO(m^3)$ assertions.  For an ABox $\mcA'$ with $i$
  individuals there are at most $i^m = \mcO(2^{m^2})$ different possibilities of
  mapping the individuals from $\mcA$ (of which there are at most $m$ many) into
  the $i$ individuals of $\mcA'$. Given a fixed $\mcA'$ and a fixed mapping $f$,
  testing whether the requirement of Definition~\ref{def:alcqir-precompletion}
  are satisfied can be done in polynomial time in $m$ and hence certainly in
  time $\mcO(2^m)$.

  Summing up, it is possible to enumerate all potential pre-completions of
  $\mcK$, generate all possible mappings $f$, and test whether all requirements
  from Definition~\ref{def:alcqir-precompletion} are satisfied in time bounded
  in 
  \[
  \sum_{i=1}^m \left ( \mcO(2^{m^5}) \times \mcO(2^{m^2}) \times \mcO(2^m)
  \right ) =    \mcO(2^{m^6}) .
  \]
  Due to Lemma~\ref{lem:alcqir-kb-consistent-iff-precompletion} and
  Lemma~\ref{lem:alcqir-precompletion-to-concept-satisfiability}, $\mcA$ is
  satisfiable iff this enumeration yields a pre-completion $\mcA'$ such that
  $C_x$ is satisfiable \wrt $\mcT$ for every $x$ that occurs in $\mcA'$.  Since
  all candidate pre-completions $\mcA'$ from the enumeration contain at most
  $\mcO(m^3)$ assertions, this can be checked for in time exponential in $m$ for
  every candidate pre-completion $\mcA'$. This yields an overall decision
  procedure that runs in time exponentially bounded in $m$.
  
  Instance checking is at least as hard as concept satisfiability \wrt general
  TBoxes and not harder than knowledge base satisfiability, hence
  \exptime-completeness of instance checking for \alcqir is immediate from what
  we have just proved. \qed
\end{proof}



\cleardoublepage


\chapter{Cardinality Restrictions and Nominals}

\label{chap:cadres}

In this chapter, we study the complexity of the combination of the DLs \alcq and
\alcqi with a terminological formalism based on cardinality restrictions on
concepts.  Cardinality restrictions were first introduced by Baader, Buchheit,
and Hollunder \citeyear{BaaderBuchheit+-AIJ-1996} as a terminological formalism
that is particularly useful for configuration applications. They allow to
restrict the number of instances of a (possibly complex) concept $C$
\emph{globally} using expressions of the form $(\geq n \; C)$ or $(\leq n \;
C)$. In a configuration application, the cardinality restriction $(\geq 100 \;
\texttt{Parts})$ can be used to limit the overall number of $\texttt{Parts}$ by
$100$, the cardinality restrictions $(\geq 1 \; \texttt{PowerSource})$ and
$(\leq 1 \; \texttt{PowerSource})$ together state that there must be exactly one
\texttt{PowerSource}, etc.

As it turns out, cardinality restrictions are closely connected to nominals,
i.e., atomic concepts referring to single individuals of the domain. Nominals
are studied both in the context of DLs \cite{Borgida94,DeGiacomo96a} and of modal
logics
\cite{gargov93:_modal_logic_with_names,blackburn95:_hybrid_languages,ArecesBlackburnMarx-JSL-2000}.

After introducing cardinality restrictions and nominals, we show that, in the
presence of nominals, reasoning \wrt cardinality restrictions can be
polynomially reduced to reasoning \wrt TBoxes. In general the latter is a
simpler problem. This allows to determine the complexity of \alcq with
cardinality restrictions as \exptime-complete as a corollary of a result in
\cite{DeGiacomo95a}, if unary coding of numbers in the input is assumed. For
binary coding, we will show that the problem becomes \nexptime-hard. Of all
logics studied in this thesis, \alcq with number restrictions is the only logic
for which it has been shown that the coding of numbers effects the complexity of
the inference problems.

For \alcqi with cardinality restrictions, we show that reasoning becomes
\nexptime-hard and is \nexptime-complete if unary coding of numbers is
assumed. By the connection  to reasoning with nominals, this implies that
reasoning \wrt general TBoxes for \alcqi with nominals has the same complexity
and we sharpen this result to pure concept satisfiability.

Finally, we generalise the results for reasoning with \alcqi to \alcqir, with
little effort, and show that, for \alcqib, i.e., \alcqir without the restriction
to safe role expressions, concept satisfiability is \nexptime-complete (this is
also a simple corollary of the \nexptime-completeness of Boolean Modal Logic
\cite{LutzSattlerAIML00}).

\section{Syntax and Semantics}

Cardinality restrictions can be defined independently of a particular DL as long
as it has standard extensional semantics. In this thesis, we will mainly study
cardinality restrictions in combination with the DLs \alcq and \alcqi. To make
our considerations here easier, we assume that concepts are built using only the
restricted set of concept constructors $\neg, \sqcap, \qnrgeq n R C$. Using de
Morgan's laws and the duality of the at-least and at-most-restriction (see below
Definition~\ref{def:alcq-semantics}) the other constructors can be defined as
abbreviations.

\begin{definition}[Cardinality Restrictions]
  A \iemph{cardinality restriction} is an expression of the form
  $\crleq n C$ or $\crgeq n C$ where $n \in \N$ and $C$ is a concept.
  
  A \iemph{CBox} is a finite set of cardinality restrictions.
  
  An interpretation $\I$ \emph{satisfies} a cardinality restriction $\crleq n C$
  iff $\sharp (C^{\I}) \leq n$, and it satisfies $\crgeq n C$ iff $\sharp
  (C^{\I}) \geq n$. It satisfies a CBox $\mcC$ iff it satisfies all cardinality
  restrictions in $\mcC$; in this case, $\I$ is called a \emph{model} of $\mcC$
  and we will denote this fact by $\I \models \mcC$. \index{00imodelsc@$\I
    \models \mcC$} \index{satisfiability!of a CBox} A CBox that has a model is
  called \emph{satisfiable}.
  
  Since $\I \models \crleq 0 {\neg C}$ iff $C$ is satisfied by all
  elements of $\I$, we will use $\crall C$ as an abbreviation for the
  cardinality restriction $\crleq 0 {\neg C}$. \eod
\end{definition}

It is obvious that, for DLs that are closed under Boolean combinations of
concepts, reasoning with cardinality restrictions is at least as hard as
reasoning with TBoxes, as $\I \models C \sqsubseteq D$ iff $\I \models \crleq 0
{(C \sqcap \neg D)}$. As we will see, CBoxes can also be used to express ABoxes
and even the stronger formalism of \emph{nominals}.  In this thesis, we have
already encountered nominals in a restricted form, namely, as individuals that
may occur in ABox assertions. DLs that allow for nominals allow those
individuals to appear in arbitrary concept expressions, which, e.g., makes it
possible to define the concept of parents of \Bob by $\some \hasChild \Bob$ or
the concept of \Bob's siblings by $\neg \Bob \sqcap \some {\hasChild^{-1}}
{\some \hasChild \Bob}$.

\begin{definition}[Nominals]
  Let \Individuals be a set of \emph{individual names} or
  \emph{nominals}. \index{nominal} For an arbitrary DL $\mcL$, its extension with
  nominals (usually denoted by $\mcL\mcO$) is obtained by,
  additionally, defining that every $i \in \Individuals$ is a concept.
  
  For the semantics, we require an interpretation $\I$ to map every $i
  \in \Individuals$ to a singleton set $i^\I$ and extend the semantics of
  $\mcL$ to $\mcL\mcO$ canonically. \eod
\end{definition}

Nominals in a DL makes ABoxes superfluous, since these can be
captured using nominals. Indeed, in the presence of nominals, it
suffices to consider satisfiability of TBoxes as the ``strongest''
inference required.

\begin{lemma}\label{lem:nominals-vs-abox}
  For an arbitrary DL $\mcL$, KB-satisfiability can be polynomially
  reduced to satisfiability of $\mcL\mcO$-TBoxes.
\end{lemma}

\begin{proof}
  Let $\mcK= (\mcT,\mcA)$ be an $\mcL$-knowledge base, where the 
  individuals in the ABox coincide with the individuals of $\mcL\mcO$. The ABox
  $\mcA$ is transformed into a TBox as follows. We define 
  \[
  \mcT_\mcA = \{ i \sqsubseteq C \mid i : C \in \mcA \} \cup
  \{ i \sqsubseteq \some R j \mid (i,j) : R \in \mcA \} \cup
  \{ i \sqsubseteq \neg j \mid i \ndoteq j \in \mcA \} .
  \]

  \begin{claim}
    $\mcK$ is satisfiable iff $\mcT \cup \mcT_\mcA$ is satisfiable. 
  \end{claim}
  
  If $\mcK$ is satisfiable with $\I \models \mcK$, it is easy to verify that
  $\I'$, which is obtained from $\I$ by setting $i^{\I'} = \{ i^\I \}$ and
  preserving the interpretation of the concept and role names, is a model for
  $\mcT \cup \mcT_\mcA$.
  
  Conversely, any model $\I$ of $\mcT \cup \mcT_\mcA$ can be turned
  into a model $\I'$ of $\mcK$ by setting, for every individual $i \in
  \Individuals$, $i^{\I'} = x$ for the unique $x \in i^\I$ and
  preserving the interpretation of concept and role names.  \qed
\end{proof}

Now that we have seen how to get rid of ABoxes in the presence of
nominals, we show how cardinality restrictions and nominals can
emulate each other.

\begin{lemma}\label{lem:cardres-vs-nominals}
  For an arbitrary DL $\mcL$, satisfiability of $\mcL$-CBoxes and
  $\mcL\mcO$-TBoxes are mutually reducible. The reduction from $\mcL\mcO$ to
  $\mcL$ is polynomial.  The reduction from $\mcL$ to $\mcL\mcO$ is polynomial
  if unary coding of numbers in the cardinality restrictions is assumed.
\end{lemma}

\begin{proof}
  It is obvious that the cardinality restrictions $\crleq 1 C$ and $\crgeq
  1 C$ enforce the interpretation of a concept name $C$ to be a
  singleton, which can now serve as a substitute for a nominal. Also,
  an interpretation satisfies a general axiom $C \sqsubseteq D$ iff it
  satisfies $\crleq 0 {(C \sqcap \neg D)}$. In this manner, every
  nominal can be replaced by a concept and every general axiom by a
  cardinality restriction, which yields the reduction from reasoning
  with nominals and TBoxes to reasoning with cardinality restrictions. For the
  converse direction, the reduction works as follows.
  
  Let $\mcC = \{ (\bowtie_1 n_1 \; C_1), \dots (\bowtie_k n_k \; C_k)
  \}$ be an $\mcL$-CBox. W.l.o.g., we assume that $\mcC$ contains no
  cardinality restriction of the form $(\geq 0 \; C)$ because these
  are trivially satisfied by any interpretation. The translation of
  $\mcC$, denoted by $\Phi(\mcC)$, is the $\mcL\mcO$-TBox defined by:
  \[
  \Phi(\mcC) = \bigcup \{ \Phi(\bowtie_i n_i \; C_i) \mid 1 \leq i \leq k \} , 
  \]
  where $\Phi(\bowtie_i n_i \; C_i)$ is defined depending on whether
  ${\bowtie_i} = {\leq}$ or ${\bowtie_i} = {\geq}$. 
  \[
  \Phi(\bowtie_i n_i \; C_i) = 
  \begin{cases}
    \{ C_i \sqsubseteq o_i^1 \sqcup \dots \sqcup o_i^{n_i} \} & \text{if }
    {\bowtie_i} = {\leq} \\
    \{ o_i^j \sqsubseteq C_i \mid 1 \leq j \leq n_i \} \cup \{ o_i^j
    \sqsubseteq \neg o_i^\ell \mid 1 \leq j < \ell \leq n_i \} & \text{if }
    {\bowtie_i} = {\geq} 
  \end{cases} ,
  \]
  where $o_i^1, \dots, o_i^{n_i}$ are fresh and distinct nominals and
  we use the convention that the empty disjunction is interpreted as
  $\bot$ to deal with the case $n_i = 0$.
  
  Assuming unary coding of numbers, the translation of a CBox $\mcC$ is
  obviously computable in polynomial time.

  \begin{claim}
    $\mcC$ is satisfiable iff $\Phi(\mcC)$ is satisfiable.
  \end{claim}
  
  If $\mcC$ is satisfiable then there is a model $\I$ of $\mcC$ and $\I
  \models (\bowtie_i n_i \; C_i)$ for each $1 \leq i \leq k$.  We show
  how to construct a model $\I'$ of $\Phi(\mcC)$ from $\I$.  $\I'$
  will be identical to $\I$ in every respect except for the
  interpretation of the nominals $o_i^j$ (which do not appear in
  $\mcC$).
  
  If $\bowtie_i = \leq$, then $\I \models \mcC$ implies $\sharp
  C_i^\I \leq n_i$. If $n_i = 0$, then we have not introduced new
  nominals, and $\Phi(\mcC)$ contains $C_i \sqsubseteq \bot$.
  Otherwise, we define $(o_i^j)^{\I'}$ such that $C_i^\I \subseteq \{
  (o_i^j)^{\I'} \mid 1 \leq j \leq n_i \}$. This implies $C_i^{\I'}
  \subseteq (o_i^1)^{\I'} \cup \dots \cup (o_i^{n_i})^{\I'}$ and
  hence, in either case, $\I' \models \Phi(\leq n_i \; C_i)$.
  
  If $\bowtie_i = \geq$, then $n_i > 0$ must hold, and $\I
  \models \mcC$ implies $\sharp C_i^\I \geq n_i$.  Let $x_1, \dots
  x_{n_i}$ be $n_i$ distinct elements from $\Delta^\I$ with $\{ x_1,
  \dots, x_{n_i} \} \subseteq C_i^\I$.  We set $(o_i^j)^{\I'} = \{x_j
  \}$. Since we have chosen distinct individuals to interpret
  different nominals, we have $\I' \models o_i^j \sqsubseteq \neg
  o_i^\ell$ for every $1 \leq i < \ell \leq n_i$. Moreover, $x_j \in
  C_i^\I$ implies $\I' \models o_i^j \sqsubseteq C_i$ and hence $\I'
  \models \Phi(\geq n_i \; C_i)$.
  
  We have chosen distinct nominals for every cardinality restrictions,
  hence the previous construction is well-defined and, since $\I'$
  satisfies $\Phi(\bowtie_i n_i \; C_i)$ for every $i$, $\I' \models
  \Phi(\mcC)$.
  
  For the converse direction, let $\I$ be a model of $\Phi(\mcC)$.
  The fact that $\I \models \mcC$ (and hence the satisfiability of
  $\mcC$) can be shown as follows: let $(\bowtie_i n_i \; C_i)$ be an
  arbitrary cardinality restriction in $\mcC$.  If $\bowtie_i =
  \leq$ and $n_i = 0$, then we have $\Phi(\leq 0 \; C_i) =
  \{ C_i \sqsubseteq \; \bot \}$ and, since $\I \models
  \Phi(\mcC)$, we have $C_i^\I = \emptyset$ and $\I \models
  (\leq 0 \; C_i)$. If $\bowtie_i = \leq$ and $n_i > 0$, we
  have $\{ C_i \sqsubseteq o_i^1 \sqcup \dots \sqcup o_i^{n_i} \}
  \subseteq \Phi(\mcC)$. From $\I \models \Phi(\mcC)$ follows $\sharp
  C_i^\I \leq \sharp (o_i^1 \sqcup \dots \sqcup o_i^{n_i})^\I \leq
  n_i$.  If $\bowtie_i = \geq$, then we have $ \{ o_i^j
  \sqsubseteq C_i \mid 1 \leq j \leq n_i \} \cup \{ o_i^j \sqsubseteq
  \neg o_i^\ell \mid 1 \leq j < \ell \leq n_i \} \subseteq
  \Phi(\mcC)$. From the first set of axioms we get $\{ (o_i^j)^\I \mid
  1 \leq j \leq n_i \} \subseteq C_i^\I$.  From the second set of
  axioms we get that, for every $1 \leq j < \ell \leq n_i$,
  $(o_i^j)^\I \neq (o_i^\ell)^\I$.  This implies that $n_i = \sharp
  \bigcup \{ (o_i^j)^\I \mid 1 \leq j \leq n_i \} \leq \sharp C_i^\I$.
  \qed
\end{proof}

\section{The Complexity of Cardinality Restrictions and Nominals} 

We will now study the complexity of reasoning with cardinality restrictions both
for \alcq and \alcqi. Baader, Buchheit and Hollunder
\citeyear{BaaderBuchheit+-AIJ-1996} give an algorithm that decides satisfiability
of CBoxes for \alcq but they do not give complexity results. Yet, it is easy to
see that their algorithm runs in non-deterministic exponential time, which gives
us a first upper bound for the complexity of the problem. For the lower bound,
it is obvious that the problem is at least \exptime-hard, due to
Lemma~\ref{lem:cardres-vs-nominals} and
Theorem~\ref{theo:alc-general-tboxes-exptime-complete} Lemma~\ref{lem:cardres-vs-nominals}
also yields \exptime as an upper bound for the complexity of this problem using the
following result established by De Giacomo~\citeyear{DeGiacomo95a}.

\begin{fact}[\npcite{DeGiacomo95a}, Section 7.3]
  Satisfiability and logical im\-pli\-ca\-tion for $\mathcal{C\!N\!O}$
  knowledge bases (TBox and ABox) are \exptime-complete.
\end{fact}

The DL $\mathcal{C\!N\!O}$ studied by the author is a strict extension of
\alcqo. Unary coding of numbers is assumed throughout his thesis.  Although the
author imposes a unique name assumption, it is not inherent to the utilized
reduction and must be explicitly enforced.  It is thus possible to eliminate the
formulas that require a unique interpretation of individuals from the
reduction. Hence, according to Lemma~\ref{lem:cardres-vs-nominals}, reasoning
with cardinality restrictions for \alcq can be reduced to $\mathcal{C\!N\!O}$,
which yields:

\begin{corollary}\label{cor:alcq-exptime-complete}
  Consistency of \alcqo-CBoxes is \exptime-complete if unary coding of
  number is assumed.
\end{corollary}

For binary coding of numbers, the reduction used in the proof of
Lemma~\ref{lem:cardres-vs-nominals} is no longer polynomial and, indeed,
reasoning for \alcq-CBoxes becomes at least \nexptime-hard if binary coding is
assumed (Corollary~\ref{theo:alcq-sat-is-nexptime-hard}).

\subsection{Cardinality Restrictions and \alcqi}
\label{sec:cardres-and-alcqi}

The algorithm developed by Baader et. al.  \citeyear{BaaderBuchheit+-AIJ-1996}
for \alcq with number restrictions cannot easily be extended to \alcqi with
cardinality restrictions. One indication for this is that the algorithm from
\cite{BaaderBuchheit+-AIJ-1996}  is a tableau algorithm that always
constructs a finite model for a satisfiable CBox; yet, \alcqi with cardinality
restriction no longer has the finite model property. The CBox
\[
\crgeq 1 {\neg A}, \ (\forall \;  (\some R \top \sqcap \nrleq 1 {R^{-1}} \sqcap \all R A))
\]
is satisfiable, but does not have a finite model. The first cardinality
restriction requires the existence of an instance $x$ of $\neg A$ in the model.
The second cardinality restriction requires every element of the model to have an
$R$-successor, so from $x$ there starts an infinite path of $R$-successors. This
path must either run into a cycle or there must be infinitely many elements in
the model. It cannot cycle back to $x$ because this would conflict with the
requirement that every element satisfies $\all R A$. It cannot cycle back
to another element of the path because in that case, this element would have two
incoming $R$-edges, which conflicts with $\nrleq 1 {R^{-1}}$.

There exists no dedicated decision procedure for \alcqi with number
restrictions, but it is easy to see that the problem can be solved by a
reduction to $\ctwo$, \index{00ctwo@\ctwo} the two-variable fragment of FOL
extended with counting quantifiers. \index{two-variable fragment of FOL}
\index{counting quantifiers} Let $\ltwo$ \index{00ltwo@\ltwo} denote the
fragment of FOL that only has the variable symbols $x$ and $y$. Then $\ctwo$ is
the extension of $\ltwo$ that admits all counting quantifiers $\exists^{\geq m}$
and $\exists^{\leq m}$ for $m \geq 1$, rather than only $\exists$. Gr{\"a}del,
Otto, and Rosen \citeyear{GraedelOttRos97} show that \ctwo is decidable. Based
on their decision procedure \cite{LICS::PacholskiST1997} determine the
complexity of \ctwo:

\begin{fact}[\npcite{LICS::PacholskiST1997}]
  \label{fact:ctwo-nexptime}
  Satisfiability of \ctwo is decidable in 2-\nexptime for binary
  coding of number and is \nexptime-complete for unary coding of numbers.
\end{fact}

Figure~\ref{fig:translation} shows how the standard translation of
\alcqi into \ctwo due to Borgida~\citeyear{Borgida96a} can be extended
to cardinality restrictions. It is obviously a satisfiability preserving
translation, which yields:

\begin{figure}[t]
  \begin{center}
      $\begin{array}{l@{\;}c@{\;}l}
        \Psi_x(A) & := & Ax \hfill \text{for $A \in N_C$}\\
        \Psi_x(\neg C) & := & \neg \Psi_x(C)\\
        \Psi_x(C_1 \sqcap C_2) & := & \Psi_x(C_1) \wedge \Psi_x(C_2)\\
        \Psi_x \qnrgeq n R C & := & \E^{\geq n} y.(Rxy \wedge \Psi_y(C))\\
        \Psi_x \qnrgeq n {R^{-1}} C & := & \E^{\geq n} y.(Ryx \wedge
        \Psi_y(C))\\[1ex]
        \Psi_y(A) & := & Ay \hfill \text{for $A \in N_C$}\\
        \Psi_y(\neg C) & := & \neg \Psi_y(C)\\
        \Psi_y(C_1 \sqcap C_2) & := & \Psi_y(C_1) \wedge \Psi_y(C_2)\\
        \Psi_y \qnrgeq n R C & := & \E^{\geq n} y.(Ryx \wedge \Psi_x(C))\\
        \Psi_y \qnrgeq n {R^{-1}} C & := & \E^{\geq n} y.(Rxy \wedge
        \Psi_x(C))\\[1ex]
        \Psi(\bowtie n \ C) & := & \E^{\bowtie n} x.\Psi_x(C) \hfill
        \text{for $\bowtie \;\in \{ \geq,\leq\}$}\\[1ex]
        \Psi(\mcC) & := & \bigwedge \{ \Psi(\bowtie \; n \; C) \mid (\bowtie \
        n \ C) \in \mcC \}
      \end{array}$
  \end{center}
  \caption{The translation from \alcqi into \ctwo}
  \label{fig:translation}
\end{figure}

\begin{lemma}\label{lem:cardres-into-ctwo}
  An-\alcqi CBox is satisfiable iff $\Psi(\mcC)$ is satisfiable.
\end{lemma}

The translation from Figure~\ref{fig:translation} is obviously
polynomial, and so we obtain, from Lemma~\ref{lem:cardres-into-ctwo} and Fact~\ref{fact:ctwo-nexptime}:

\begin{lemma}\label{lem:cadres-alcqi-in-nexptime}
  Satisfiability of \alcqi-CBoxes can be decided in \nexptime, if unary coding
  of numbers in the input is assumed.
\end{lemma}

We will see that, from the viewpoint of worst-case complexity, this is an
optimal result, as the problem is also \nexptime hard. To prove this, we use a
bounded version of the domino problem.  Domino problems
\cite{wang63:_domin_aea_decis_probl,berger66:_undecidability_dominoe_problem}
have successfully been employed to establish undecidability and complexity
results for various description and modal logics
\cite{Spaan93a,BaaderSattler-JLC-99}.


\subsubsection{Domino Systems}

\begin{definition}\label{def:domino-system}
  For $n \in \N$, let $\Z_n$ denote the set $\{0,\dots,n-1\}$ and $\oplus_n$
  denote the addition modulo $n$.  A \iemph{domino system} is a triple
  $\mathcal{D} = (D,H,V)$, where $D$ is a finite set (of tiles) and $H,V
  \subseteq D \times D$ are relations expressing horizontal and vertical
  compatibility constraints between the tiles.  For $s,t \in \N$, let $U(s,t)$ be
  the torus $\mathbb{Z}_s \times \mathbb{Z}_t$, and let $w = w_0 \dots w_{n-1}$ be a
  word over $D$ of length $n$  (with $n \leq s$). We say that $\mathcal{D}$ \emph{tiles}
  $U(s,t)$ \emph{with initial condition} $w$ iff there exists a mapping $\tau:
  U(s,t) \to D$ such that, for all $(x,y) \in U(s,t)$,
  \begin{itemize}
  \item if $\tau(x,y) = d$ and $\tau(x
    \oplus_s 1,y) = d'$, then $(d,d') \in H$ (horizontal constraint);
  \item if $\tau(x,y) = d$ and $\tau(x, y \oplus_t 1) = d'$, then $(d,d')
    \in V$ (vertical constraint);
  \item $\tau(i,0) = w_i$ for $0\leq i < n$ (initial condition). \eod
  \end{itemize}
  \index{00zn@$\Z_n$}
  \index{00oplusn@$\oplus_n$}
\end{definition}

Bounded domino systems are capable of expressing the computational behaviour of
restricted, so-called \emph{simple}, Turing Machines (TM). This restriction is
non-essential in the following sense: Every language accepted in time $T(n)$ and
space $S(n)$ by some one-tape TM is accepted within the same time and space
bounds by a simple TM, as long as $S(n),T(n) \geq 2n$~\cite{BoGrGu97}.

\begin{theorem}[\npcite{BoGrGu97}, Theorem 6.1.2]\label{theo:simple-turing-machine-domino}
  Let $M$ be a simple TM with input alphabet $\Sigma$.  Then there exists a
  domino system $\mathcal{D} = (D,H,V)$ and a linear time reduction which takes
  any input $x \in \Sigma^*$ to a word $w \in D^*$ with $|x| = |w|$ such that
  \begin{itemize}
  \item If $M$ accepts $x$ in time $t_0$ with space $s_0$, then $\mathcal{D}$
    tiles $U(s,t)$ with initial condition $w$ for all $s\geq s_0+2, t\geq t_0
    +2$;
  \item if $M$ does not accept $x$, then $\mathcal{D}$ does not tile $U(s,t)$
    with initial condition $w$ for any $s,t \geq 2$.
  \end{itemize}
\end{theorem}


\begin{corollary}\label{cor:tiling-is-nexptime-hard}
  There is a domino system $\mathcal{D}$ such that the following is a
  \nexptime-hard problem:

  \begin{quote}
    Given an initial condition $w = w_0 \dots w_{n-1}$ of length $n$. Does
    $\mathcal{D}$ tile the torus $U(2^{n+1},2^{n+1})$ with initial
    condition $w$? 
  \end{quote}
\end{corollary}

\proof Let $M$ be a (w.l.o.g. simple) non-deterministic TM with time- (and
hence space-) bound $2^n$ deciding an arbitrary \nexptime-complete language
$\mathcal{L}(M)$ over the alphabet $\Sigma$. Let $\mathcal{D}$ be the
according domino system and $\textit{trans}$ the reduction from
Theorem~\ref{theo:simple-turing-machine-domino}.
  
  The function $\textit{trans}$ is a linear reduction from $\mathcal{L}(M)$ to
  the problem above: For $v \in \Sigma^*$ with $|v| = n$, it holds that $v \in
  \mathcal{L}(M)$ iff $M$ accepts $v$ in time and space $2^{|v|}$ iff
  $\mathcal{D}$ tiles $U(2^{n+1},2^{n+1})$ with initial condition
  $\textit{trans}(v)$. \qed

\subsubsection{Defining a Torus of Exponential Size}

Similar to proving undecidability by reduction of unbounded domino problems,
where defining infinite grids is the key problem, defining a torus of
exponential size is the key to obtain a \nexptime-completeness proof by
reduction of bounded domino problems.



To be able to apply Corollary~\ref{cor:tiling-is-nexptime-hard} to CBox
satisfiability for \alcqi, we must characterize the torus $\mathbb{Z}_{2^n}
\times \mathbb{Z}_{2^n}$ with a CBox of polynomial size. To characterize
this torus, we use $2n$ concepts $X_0,\dots,X_{n-1}$ and $Y_0,\dots,Y_{n-1}$,
where $X_i$ ($Y_i$) codes the $i$th bit of the binary representation of
the X-coordinate (Y-coordinate) of an element $a$.

For an interpretation $\I$ and an element $a \in \Delta^\I$, we define $\pos(a)$
by
\begin{gather*}
  \pos(a) := (\xpos(a), \ypos(a)) := \biggl( \sum_{i = 0}^{n-1} x_i \cdot 2^i, \quad \sum_{i =
    0}^{n-1} y_i \cdot 2^i \biggr) \; , \; \text{where}\\
  \begin{array}{c@{\qquad}c}
    x_i = 
    \begin{cases}
      0, & \text{if $a \not\in X_i^\I$} \\
      1, & \text{otherwise}
    \end{cases} &
    y_i = 
    \begin{cases}
      0, & \text{if $a \not\in Y_i^\I$} \\
      1, & \text{otherwise} \ .
    \end{cases}
  \end{array} 
\end{gather*}

We use a well-known characterization of binary addition \cite<see, e.g.,>{BoGrGu97}
to interrelate the positions of the elements in the torus:
\begin{lemma}\label{lem:binary-addition}
  Let $x,x'$ be natural numbers with binary representations
  \[
  x = \sum_{i=0}^{n-1} x_i \cdot 2^i \quad \text{and} \quad x' =
  \sum_{i=0}^{n-1} x'_i \cdot 2^i .
  \]  
  Then
  \begin{multline*}
    \begin{split}
      x' \equiv x+1 \pmod{2^n} \quad \text{iff} \quad & \bigwedge_{k=0}^{n-1}
      (\bigwedge_{j=0}^{k-1} x_j = 1) \to (x_k = 1 \leftrightarrow x'_k = 0) \\
      \land & \bigwedge_{k=0}^{n-1} (\bigvee_{j=0}^{k-1} x_j = 0) \to
      (x_k = x'_k) \; ,
    \end{split}
  \end{multline*}
  where the empty conjunction and disjunction are interpreted as true
  and false, respectively.
\end{lemma}

\newcommand{\north}{{\textit{north}}\xspace}
\newcommand{\east}{{\textit{east}}\xspace}

\begin{figure}[tbh]
  \begin{center}
      $
      \begin{array}{r@{\;}c@{\;}ll}
        \mcC_n & = \bigl \{  & (\forall \ \exists {\east} . \top), & (\forall \
        \exists {\north} . \top), \\ 
        && (\forall \ (= 1 \; \east^{-1} \; \top)), &   (\forall \ (= 1 \;
        \north^{-1} \; \top)),\\
        && (\geqslant 1 \; C_{(0,0)}), &  (\geqslant 1 \; C_{(2^n-1,2^n-1)}),\\
        && (\leqslant 1 \; C_{(2^n-1,2^n-1)}), & (\forall \ D_\east \sqcap
        D_\north) \ \bigr \}\\[1ex]
        C_{(0,0)}  & = & \underset{k=0}{\overset{n-1}{\mybigsqcap}} \lnot X_k \sqcap
        \underset{k=0}{\overset{n-1}{\mybigsqcap}} \lnot Y_k\\[1ex]
        C_{(2^n-1,2^n-1)} & = & \underset{k=0}{\overset{n-1}{\mybigsqcap}} X_k \sqcap
        \underset{k=0}{\overset{n-1}{\mybigsqcap}} Y_k\\[1ex]
        D_\east & = & \multicolumn{2}{@{}l}{\underset{k=0}{\overset{n-1}{\mybigsqcap}} (
          \underset{j=0}{\overset{k-1}{\mybigsqcap}} X_j) \to ((X_k \to
          \forall \east . \lnot X_k) \sqcap (\lnot X_k \to \forall
          \east.X_k)) \; \sqcap}\\
        && \multicolumn{2}{@{}l}{\underset{k=0}{\overset{n-1}{\mybigsqcap}} (
          \underset{j=0}{\overset{k-1}{\displaystyle \bigsqcup}} \lnot X_j) \to ((X_k
          \to \forall \east.X_k) \sqcap (\lnot X_k \to \forall \east.
          \lnot X_k)) \; \sqcap}\\
        && \multicolumn{2}{@{}l}{\underset{k=0}{\overset{n-1}{\mybigsqcap}}((Y_k \to \forall 
          \east . Y_k) \sqcap (\lnot Y_k \to \forall \east.\lnot Y_k))}\\
        D_\north & = & \multicolumn{2}{@{}l}{\underset{k=0}{\overset{n-1}{\mybigsqcap}} (
          \underset{j=0}{\overset{k-1}{\mybigsqcap}} Y_j) \to ((Y_k \to
          \forall \north . \lnot Y_k) \sqcap (\lnot Y_k \to \forall
          \north.Y_k)) \; \sqcap}\\
        && \multicolumn{2}{@{}l}{\underset{k=0}{\overset{n-1}{\mybigsqcap}} (
          \underset{j=0}{\overset{k-1}{\displaystyle \bigsqcup}} \lnot Y_j) \to ((Y_k
          \to \forall \north.Y_k) \sqcap (\lnot Y_k \to \forall \north.
          \lnot Y_k)) \; \sqcap}\\
        && \multicolumn{2}{@{}l}{\underset{k=0}{\overset{n-1}{\mybigsqcap}}((X_k \to \forall 
          \north . X_k) \sqcap (\lnot X_k \to \forall \north.\lnot X_k))}\\
      \end{array}
      $
  \end{center}
  \caption{A CBox defining a torus of exponential size}
  \label{fig:tbox-torus}
\end{figure}

\def\Tn{\mcC_n}

The CBox $\Tn$ is defined in Figure~\ref{fig:tbox-torus}.  The concept
$C_{(0,0)}$ is satisfied by all elements $a$ of the domain for which $\pos(a)
= (0,0)$ holds. $C_{(2^n-1,2^n-1)}$ is a similar concept, whose instances $a$
satisfy $\pos(a) = (2^n-1, 2^n-1)$.

The concept $D_\north$ is similar to $D_\east$ where the role $\north$ has
been substituted for $\east$ and variables $X_i$ and $Y_i$ have been swapped.
The concept $D_\east$ ($D_\north$) enforces that, along the role $\east$
($\north$), the value of $\xpos$ ($\ypos$) increases by one while
the value of $\ypos$ ($\xpos$) is unchanged. They are analogous to the
formula in Lemma~\ref{lem:binary-addition}.

The following lemma is a  consequence of the definition of $\pos$ and
Lemma~\ref{lem:binary-addition}.

\begin{lemma}\label{lem:addition-along-successors}
  Let $\I = (\Delta^\I, \cdot^\I)$ be an interpretation, $D_\east, D_\north$
  defined as in Figure~\ref{fig:tbox-torus}, and $a,b \in \Delta^\I$.
  
  \begin{gather*}
    \begin{array}{lll}
      \text{$(a,b) \in \east^\I$ and $a \in D_\east^\I$ implies:} \quad &
      \xpos(b) & \equiv \xpos(a)+1 \pmod{2^n} \\
      & \ypos(b) & = \ypos(a)
    \end{array}\\
    \begin{array}{lll}
      \text{$(a,b) \in \north^\I$ and $a \in D_\north^\I$ implies:} \quad &
      \xpos(b) & =  \xpos(a) \\
      & \ypos(b) & \equiv \ypos(a) +1 \pmod{2^n}
    \end{array}
  \end{gather*}
\end{lemma}

The CBox $\Tn$ defines a torus of exponential size in the following
sense:

\begin{lemma}\label{lem:tbox-defines-torus}
  Let $\Tn$ be the CBox as defined in Figure~\ref{fig:tbox-torus}. Let $\I =
  (\Delta^\I,\cdot^\I)$ be a model of $\Tn$. Then
  \[
  (\Delta^\I,\east^\I,\north^\I) \cong (U(2^n,2^n),S_1,S_2) \; ,
  \]
  where $U(2^n,2^n)$ is the torus $\Z_{2^n} \times \Z_{2^n}$ and $S_1,S_2$ are
  the horizontal and vertical successor relations on this torus.
\end{lemma}

\proof
  We show that the function \pos is an isomorphism from $\Delta^\I$ to
  $U(2^n,2^n)$. Injectivity of $\pos$ is shown by induction on the
  ``Manhattan distance'' $d(a)$ of the $\pos$-value of an element $a$ to the
  $\pos$-value of the upper right corner.

  For an element $a \in \Delta^\I$ we define $d(a)$ by
  \[
  d(a) = (2^n-1 - \xpos(a)) + (2^n-1 - \ypos(a)) .
  \]
  Note that $\pos(a) = \pos(b)$ implies $d(a) = d(b)$. Since $\I \models
  (\leqslant 1 \; C_{(2^n-1,2^n-1)})$, there is at most one element $a \in
  \Delta^\I$ such that $d(a)=0$. Hence, there is at most one element $a$ such
  that $\pos(a) = (2^n-1,2^n-1)$.  Now assume there are elements $a,b \in
  \Delta^\I$ such that $\pos(a)=\pos(b)$ and $d(a) = d(b) > 0$. Then 
  $\xpos(a) < 2^n-1$ or $\ypos(a) < 2^n-1$.  W.l.o.g., we assume $\xpos(a) <
  2^n-1$. From $\I \models \Tn$, it follows that $a,b \in (\exists
  \east.\top)^\I$.  Let $a_1,b_1$ be elements such that $(a,a_1) \in \east^\I$
  and $(b,b_1) \in \east^\I$.  From
  Lemma~\ref{lem:addition-along-successors}, it follows that
  \begin{align*}
    \xpos(a_1) & \equiv \xpos(b_1) \equiv \xpos(a) + 1 \pmod{2^n} \\
    \ypos(a_1) & = \ypos(b_1) = \ypos(a). 
  \end{align*}
  This implies $\pos(a_1) = \pos(b_1)$ and, since $\xpos(a) < 2^n-1$, it holds
  that $\xpos(a_1) = \xpos(b_1) = \xpos(a)+1 > \xpos(a)$. Hence, $d(a_1) =
  d(b_1) < d(a)$ and the induction hypothesis is applicable, which yields $a_1 =
  b_1$.  This also implies $a=b$ because $a_1 \in (=1 \ \east^{-1}.\top)^\I$ and
  $\{(a,a_1), (b,a_1) \} \subseteq \east^\I$. Hence $\pos$ is injective.
  
  To prove that $\pos$ is also \emph{surjective} we use a similar technique.
  This time, we use an induction on the distance from the lower left corner.
  For each element $(x,y) \in U(2^n,2^n)$, we define:
  \[
  d'(x,y) = x+y .
  \]
  
  We show by induction that, for each $(x,y) \in U(2^n,2^n)$, there is an
  element $a \in \Delta^\I$ such that $\pos(a) = (x,y)$. If $d'(x,y) = 0$, then
  $x=y=0$. Since $\I \models (\geqslant 1 \; C_{(0,0)})$, there is an element $a \in
  \Delta^\I$ such that $\pos(a) = (0,0)$. Now consider $(x,y) \in U(2^n,2^n)$
  with $d'(x,y) >0$.  Without loss of generality we assume $x > 0$ (if $x = 0$
  then $y > 0$ must hold). Hence $(x-1,y) \in U(2^n,2^n)$ and $d'(x-1,y) <
  d'(x,y)$.  From the induction hypothesis, it follows that there is an element
  $a \in \Delta^\I$ such that $\pos(a) = (x-1,y)$. Then there must be an
  element $a_1$ such that $(a,a_1) \in \east^\I$ and 
  Lemma~\ref{lem:addition-along-successors} implies that $\pos(a_1) =
  (x,y)$.  Hence $\pos$ is also surjective.
  
  Finally, $\pos$ is indeed a homomorphism as an immediate consequence of
  Lemma~\ref{lem:addition-along-successors}.  \qed
  
  It is interesting to note that we need inverse roles only to guarantee that
  the function \pos is injective. The same can be achieved by adding the
  cardinality restriction $(\leqslant (2^n \cdot 2^n) \ \top)$ to $\Tn$, from
  which the injectivity of \pos follows from its surjectivity and simple
  cardinality considerations. Of course, the size of this cardinality restriction
  is polynomial in $n$ only if we assume binary coding of numbers. This has
  consequences for the complexity of \alcq-CBoxes if binary coding of numbers in
  the input is assumed (see Corollary~\ref{theo:alcq-sat-is-nexptime-hard}).

  Also note that we have made explicit use of the special expressive power of
  cardinality restrictions by stating that, in any model of $\Tn$, the extension
  of $C_{(2^n-1,2^n-1)}$ must have \emph{at most} one element. This cannot be
  expressed with a \alcqi-TBox consisting of terminological axioms.

\subsubsection{Reducing Domino Problems to CBox Satisfiability}
\label{sec:reduction}

Once Lemma~\ref{lem:tbox-defines-torus} has been proved, it is easy to reduce
the bounded domino problem to CBox satisfiability. We use the standard reduction
that has been applied in the DL context, e.g., by Baader and
Sattler~\citeyear{BaaderSattler-JLC-99}.

\def\TnDw{\mcC(n,\mcD,w)}

\begin{lemma}\label{lem:dominoe-reduces-to-alcqi-sat}
  Let $\mathcal{D} = (D,V,H)$ be a domino system. Let $w = w_0 \dots w_{n-1} \in
  D^*$. There is a CBox $\TnDw$ such that:
  \begin{itemize}
  \item $\TnDw$ is satisfiable iff $\mathcal{D}$ tiles $U(2^n,2^n)$
    with initial condition $w$, and
  \item $\TnDw$ can be computed in time polynomial in $n$.
  \end{itemize}
\end{lemma}

\proof We define $\TnDw := \Tn \cup \mcC_\mathcal{D} \cup \mcC_w $, where $\Tn$ is
defined in Figure~\ref{fig:tbox-torus}, $\mcC_\mathcal{D}$ captures the vertical
and horizontal compatibility constraints of the domino system $\mathcal{D}$,
and $\mcC_w$ enforces the initial condition. We use an atomic concept $C_d$ for
each tile $d \in D$.  $\mcC_\mathcal{D}$ consists of the following cardinality
restrictions:
  \begin{gather*}
    (\forall \ \underset{d \in D}{\bigsqcup} C_d), \quad
    (\forall \ \underset{d \in D}{\mybigsqcap} \ \ \underset{d' \in
      D\setminus\{d\}}{\mybigsqcap} \lnot (C_d \sqcap C_{d'})),\\
    (\forall \ \underset{d \in D}{\mybigsqcap} (C_d \to (\forall
    \east. \underset{(d,d') \in H}{\bigsqcup} C_{d'}))), \quad
    (\forall \ \underset{d \in D}{\mybigsqcap} (C_d \to (\forall \north.
    \underset{(d,d') \in V}{\bigsqcup} C_{d'}))).
  \end{gather*}
  $\mcC_w$ consists of the cardinality restrictions
  \[
  (\forall \ (C_{(0,0)}  \to C_{w_0})), \dots, 
  (\forall \ (C_{(n-1,0)} \to C_{w_{n-1}}) ,
  \]  
  where, for each $x,y$, $C_{(x,y)}$ is a concept that is satisfied by an
  element $a$ iff $\pos(a) = (x,y)$, defined similarly to $C_{(0,0)}$ and $C_{(2^n-1,2^n-1)}$.
  
  From the definition of $\TnDw$ and Lemma~\ref{lem:tbox-defines-torus}, it
  follows that each model of $\TnDw$ immediately induces a tiling of
  $U(2^n,2^n)$ and vice versa.  Also, for a fixed domino system $\mathcal{D}$,
  $\TnDw$ is obviously polynomially computable. \qed

  The main result of this section is now an immediate consequence of
  Lem\-ma~\ref{lem:cadres-alcqi-in-nexptime},
  Lem\-ma~\ref{lem:dominoe-reduces-to-alcqi-sat}, and
  Corollary~\ref{cor:tiling-is-nexptime-hard}:

\begin{theorem}\label{theo:alcqi-sat-is-nexptime-hard}
  Satisfiability of \alcqi-CBoxes is \nexptime-hard. It is \nexptime-complete if
  unary coding of numbers is used in the input.
\end{theorem}

Recalling the note below the proof of Lemma~\ref{lem:tbox-defines-torus},
we see that the same reduction also applies to $\alcq$ if we allow
binary coding of numbers.

\begin{corollary}\label{theo:alcq-sat-is-nexptime-hard}
  Satisfiability of $\alcq$-CBoxes is \nexptime-hard if binary coding is used to
  represent numbers in cardinality restrictions.
\end{corollary}

It should be noted that it is open whether the problem can be decided in \nexptime,
if binary coding of numbers is used. In fact, the reduction to \ctwo only yields
decidability in 2-\nexptime if binary coding is assumed.

We have already seen that, for unary coding of numbers, deciding satisfiability of
\alcq-CBoxes can be done in \exptime
(Corollary~\ref{cor:alcq-exptime-complete}). This shows that the coding of
numbers indeed has an influence on the complexity of the reasoning problem. For
the problem of concept satisfiability in \alcq this is not the case; in
Chapter~\ref{chap:alcq} we have shown that the complexity of the problem does
not rise when going from unary to binary coding.

For unary coding, we needed both inverse roles and cardinality restrictions
for the reduction. This is consistent with the fact that satisfiability for
\alcqi concepts with respect to TBoxes that consist of terminological axioms is
still in \exptime. This can be shown by a reduction to the \exptime-complete
logics $\mathcal{CIN}$~\cite{DeGiacomo95a} or \textsf{CPDL}~\cite{Pratt1979}.
This shows that cardinality restrictions on concepts are an additional source of
complexity.

Using Lemma~\ref{lem:cardres-vs-nominals} it is now also possible to determine
the complexity of reasoning with \alcqio TBoxes:

\begin{corollary}\label{cor:alcqio-nexptime}
  Satisfiability of \alcqio-TBoxes is \nexptime-hard. It is \nexptime-complete if
  unary coding of numbers in the input is assumed.
\end{corollary}

\begin{proof}
  Lemma~\ref{lem:cardres-vs-nominals} states that satisfiability of
  \alcqio-TBoxes and satisfiability of \alcqi-CBoxes are mutually
  polynomially reducible problems.  Hence, both the lower and the
  upper complexity bound follow from
  Theorem~\ref{theo:alcqi-sat-is-nexptime-hard}. \qed
\end{proof}

This result explains a gap in \cite{DeGiacomo95a}. There the author establishes
the complexity of satisfiability of knowledge bases consisting of TBoxes and
ABoxes both for $\mathcal{C\!N\!O}$, which allows for qualifying number
restrictions, and for $\mathcal{C\!I\!O}$, which allows for inverse roles, by
reduction to the \exptime-complete logic PDL. No results are given for the combination
$\mathcal{C\!I\!N\!O}$, which is a strict extension of \alcqio.
Corollary~\ref{cor:alcqio-nexptime} shows that, assuming $\exptime \neq
\nexptime$, there cannot be a polynomial reduction from the satisfiability
problem of $\mathcal{C\!I\!N\!O}$ knowledge bases to PDL.
A possible explanation for this leap in complexity is the loss of the tree model
property, which has been proposed by Vardi \citeyear{Vardi97} and Gr{\"a}del
\citeyear{Graedel99c} as an explanation for good algorithmic properties of a
logic. While, for $\mathcal{C\!I\!O}$ and $\mathcal{C\!N\!O}$, satisfiability is
decided by searching for tree-like pseudo-models even in the presence of
nominals, this seems no longer to be possible in the case of knowledge bases for
$\mathcal{C\!I\!N\!O}$.

\subsubsection{Unique Name Assumption}

It should be noted that our definition of nominals is non-standard for
DLs in the sense that we do not impose the unique name
assumption that is widely made, i.e., for any two individual names
$o_1,o_2 \in \Individuals$, $o_1^\I \neq o_2^\I$ is required. Even without a
unique name assumption, it is possible to enforce distinct
interpretation of nominals by adding axioms of the form $o_1
\sqsubseteq \lnot o_2$, which we have already used in the proof of
Lemma~\ref{lem:nominals-vs-abox}.  Moreover, imposing a unique name
assumption in the presence of inverse roles and number restriction
leads to peculiar effects. Consider the following TBox:
\begin{equation*}
  \mcT = \{ o \sqsubseteq \qnrleq k R \top, \ \top \sqsubseteq \exists R^{-1} . o \}
\end{equation*}
Under the unique name assumption, $\mcT$ is satisfiable iff $\Individuals$ contains at most
$k$ individual names, because each individual name must be interpreted by a
unique element of the domain, every element of the domain must be reachable
from $o^\I$ via the role $R$, and $o^\I$ may have at most $k$ $R$-successors.
We believe that this dependency of the satisfiability of a TBox on constraints
that are not explicit in the TBox is counter-intuitive and hence have not
imposed the unique name assumption.

Nevertheless, it is possible to obtain a tight complexity bound for
satisfiability of \alcqio-TBoxes with  unique name assumption
without using Lemma~\ref{lem:cardres-vs-nominals}, but by an immediate
adaptation of the proof of
Theorem~\ref{theo:alcqi-sat-is-nexptime-hard}.

\begin{corollary}
  Satisfiability of \alcqio-TBoxes \emph{with the unique name assumption} is
  \nexptime-hard. It is \nexptime-complete if unary coding of numbers in the
  input is assumed.
\end{corollary}

\def\create{\textit{create}}

\begin{proof}
  A simple inspection of the reduction used to prove
  Theorem~\ref{theo:alcqi-sat-is-nexptime-hard}, and especially of the
  proof of Lemma~\ref{lem:tbox-defines-torus} shows that only a single
  nominal, which marks the upper right corner of the torus, is
  sufficient to perform the reduction. If $o$ is an individual name
  and $\create$ is a role name, then the following TBox defines a
  torus of exponential size:
  \[
  \begin{array}{r@{\;}c@{\;}ll}
    \mcT_n & = \bigl \{  & \top \sqsubseteq \exists {\east} . \top, & \top \sqsubseteq
    \exists {\north} . \top, \\ 
    && \top \sqsubseteq  (= 1 \; \east^{-1} \; \top), &   \top \sqsubseteq  (= 1 \;
    \north^{-1} \; \top),\\
    && \top \sqsubseteq  \exists {\create} . {C_{(0,0)}}, &  \top \sqsubseteq
    D_\east \sqcap  D_\north,\\  
    && C_{(2^n-1,2^n-1)} \sqsubseteq o, & o \;\sqsubseteq
    C_{(2^n-1,2^n-1)}  \ \bigr \}
  \end{array}
  \]
  Since this reduction uses only a single individual name, the unique name
  assumption is irrelevant in this case. \qed
\end{proof}

\subsubsection{Internalization of Axioms}

In the presence of inverse roles and nominals, it is possible to
\emph{internalise} \index{internalisation of general axioms} general inclusion
axioms into concepts \cite{Baader91c,Schi91,BBNNS93} using the \emph{spy-point}
technique \index{spy-point technique} used, e.g., by Blackburn and Seligman
\citeyear{blackburn95:_hybrid_languages} and Areces, Blackburn, and Marx
\citeyear{areces:_road_map_compl_hybrid_logic}. The main idea of this technique
is to enforce that all elements in the model of a concept are reachable in a
single step from a distinguished point (the spy-point) marked by an individual
name.

\newcommand{\spy}{\textit{spy}}

\begin{definition}
  Let $\mcT$ be an \alcqio-TBox. W.l.o.g., we assume that $\mcT$ contains only a
  single axioms $\top \sqsubseteq D$.  Let $\spy$ denote a fresh role name and
  $i$ a fresh individual name.  We define the function $\cdot^{\spy}$
  inductively on the structure of concepts by setting $A^{\spy} = A $ for all $A
  \in \Names$, $o^{\spy} = o$ for all $o \in \Individuals$, $(\lnot C)^{\spy} =
  \lnot C^{\spy}$, $(C_1 \sqcap C_2)^{\spy} = C_1^{\spy} \sqcap C_2^{\spy}$, and
  ${\qnrgeq n R C}^{\spy} = \qnrgeq n R {(\exists {\spy^{-1}}.i) \sqcap C^{\spy}}$.
  
  The \emph{internalization} $C_\mcT$ of $\mcT$ is defined as follows:
  \begin{equation*}
    C_\mcT = i \sqcap  D^{\spy}
    \sqcap \forall \spy . D^{\spy}
  \end{equation*}
\end{definition}

\begin{lemma}\label{lem:spypoint}
  Let $\mcT$ be an \alcqio-TBox. $\mcT$ is satisfiable iff $C_\mcT$ is
  satisfiable.
\end{lemma}

\begin{proof}
  For the \emph{if}-direction let $\I$ be a model of $C_\mcT$ with $a \in
  (C_\mcT)^\I$.  This implies $i^\I = \{ a\}$. Let $\I'$ be defined by
  \[
  \Delta^{\I'} = \{ a\} \cup \{ x \in \Delta^\I \mid (a,x) \in \spy^\I
  \}
  \]
  and $\cdot^{\I'} = \cdot^\I |_{\Delta^{\I'}}$.

  \begin{claim}\label{claim:nexptime-1}
    For every $x \in \Delta^{\I'}$ and every
    \alcqio-concept $C$, we have $x \in (C^{\spy})^\I$ iff $x \in C^{\I'}$.
  \end{claim}
  
  We proof this claim by induction on the structure of $C$.  The only
  interesting case is $C = \qnrgeq n R D$. In this case $C^{\spy} =
  \qnrgeq n R {(\exists {\spy^{-1}}.i) \sqcap D^{\spy}}$. We have
  \begin{align*}
    & x \in \qnrgeq n R
    {(\exists {\spy^{-1}}.i) \sqcap D^{\spy}}^\I \\
    \text{iff } \; & \sharp \{ y \in \Delta^\I \mid (x,y) \in R^\I \text{ and } y
    \in (\exists \spy^{-1}.i)^\I \cap (D^{\spy})^\I \} \geqslant n\\
    (*) \text{ iff } \; & \sharp \{ y \in \Delta^{\I'} \mid (x,y) \in R^{\I'} \text{
      and } y \in D^{\I'} \} \geqslant n\\
    \text{iff } \; & x \in \qnrgeq n R D^{\I'} ,
  \end{align*}
  where the equivalence $(*)$ holds because, if $y \in (\some {\spy^{-1}} i)^\I
  \cap (D^\spy)^\I$ then $y \in \Delta^{\I'}$ and $y \in D^{\I'}$ by induction.
  Also, if $y \in \Delta^{\I'}$, then $(x,y) \in R^\I$ iff $(x,y) \in R^{\I'}$
  and hence the sets $\{ y \in \Delta^\I \mid (x,y) \in R^\I \text{ and } y \in
  (\exists \spy^{-1}.i)^\I \cap (D^{\spy})^\I \}$ and $\{ y \in \Delta^{\I'} \mid
  (x,y) \in R^{\I'} \text{ and } y \in D^{\I'} \}$ are equal.
  
  By construction, for every $x \in \Delta^{\I'}$, $x \in (D^{\spy})^\I$. Due to
  Claim~\ref{claim:nexptime-1}, this implies $x \in D^{\I'}$ and hence $\I'
  \models \top \sqsubseteq D$.
  
  For the \emph{only-if}-direction, let $\I$ be an interpretation with $\I
  \models \mcT$. We pick an arbitrary element $a \in \Delta^\I$ and define an
  extension $\I'$ of $\I$ by setting $i^{\I'} = \{ a \}$ and $\spy^{\I'} = \{
  (a,x) \mid x \in \Delta^\I \}$. Since $i$ and $\spy$ do not occur in $\mcT$, we
  still have that $\I' \models \mcT$.

  \begin{claim}\label{claim:nexptime-2}
    For every $x \in \Delta^{\I'}$ and every \alcqio-concept $C$ that does not
    contain $i$ or $\spy$, $x \in C^{\I'}$ iff $x \in (C^{\spy})^{\I'}$.
  \end{claim}
  
  Again, this claim is proved by induction on the structure of concepts and the
  only interesting case is $C = \qnrgeq n R E$. 
  \begin{align*}
    & x \in \qnrgeq n R E^{\I'}\\
    \text{iff } \; & \sharp \{ y \in \Delta^{\I'} \mid (x,y) \in R^{\I'}
    \text{ and } y \in E^{\I'} \} \geqslant n\\
    (*) \text{ iff } \; & \sharp \{ y \in \Delta^{\I'} \mid (x,y) \in R^{\I'},
    (a,y) \in \spy^{\I'}, \text{ and } y \in (E^{\spy})^{\I'} \} \geqslant n\\
    \text{iff } \; & x \in \qnrgeq n R {(\exists {\spy^{-1}}.i) \sqcap
      E^{\spy}}^{\I'} .
  \end{align*}
  The equivalence  $(*)$ holds because, by construction of $\I'$, $(a,y) \in
  \spy^{\I'}$ holds for every element $y$ of the domain and $y \in E^{\I'}$ iff
  $y \in (E^\spy)^{\I'}$ holds by induction.
  
  Since, $\I' \models \top \sqsubseteq D$, Claim~\ref{claim:nexptime-2} yields
  that $(D^{\spy} )^{\I'} = \Delta^{\I'}$ and hence $a \in (C_\mcT)^{\I'}$ \qed
\end{proof}

As a consequence, we obtain the sharper result that already pure concept
satisfiability for \alcqio is a \nexptime-complete problem.

\begin{corollary}\label{cor:alcqio-concept-nexptime}
  Concept satisfiability for \alcqio is \nexptime-hard. It is \nexptime-complete
  if unary coding of numbers in the input is assumed.
\end{corollary}

\begin{proof}
  From Lemma~\ref{lem:spypoint}, we get that the function mapping a \alcqio-TBox
  $\mcT$ to $C_\mcT$ is a reduction from satisfiability of \alcqio-TBoxes to
  satisfiability of \alcqio concepts. From Corollary~\ref{cor:alcqio-nexptime}
  we know that the former problem is \nexptime-complete. Obviously, $C_\mcT$ can
  be computed from $\mcT$ in polynomial time. Hence, the lower complexity bound
  transfers. The \nexptime upper bound is a consequence of
  Corollary~\ref{cor:alcqio-nexptime} and the fact that an \alcqio concept $C$
  is satisfiable iff, for an individual $j$ that does not occur in $C$, the TBox
  $\{j \sqsubseteq C\}$ is satisfiable. \qed
\end{proof}

\subsection{Boolean Role Expressions}

In Chapter~\ref{chap:alcq}, we have studied the DL \alcqir, which allowed for a
restricted---so called safe---form of Boolean combination of roles, and for
which concept satisfiability is decidable in polynomial space.  It is easy to
see that the results established for \alcqi in this chapter all transfer to
\alcqir and we state them here as (indeed, trivial) corollaries.

We have already argued that the restriction to safe role expressions is
necessary to obtain a DL for which satisfiability is still decidable in
polynomial space: the concept $\qnrleq 0 {(R \sqcup \lnot R)} {\lnot C}$ is
satisfiable iff $C$ is globally satisfiable, which is an \exptime-complete
problem (see, Theorem~\ref{theo:alc-general-tboxes-exptime-complete}). Indeed,
as a corollary of Theorem~\ref{theo:alcqi-sat-is-nexptime-hard}, it can be shown
that concept satisfiability becomes a \nexptime-hard in the presence of
arbitrary Boolean operations on roles. 

\begin{definition}
  The DL \alcqib is defined as \alcqir with the exception that arbitrary role
  expressions are allowed.  The DL \alcqb is the restriction of \alcqib that
  disallows inverse roles. The semantics of \alcqib and \alcqb are define as
  for \alcqir. \eod
  \index{alcqib@\alcqib}
  \index{alcqb@\alcqb}
\end{definition}

Decidability of concept and CBox satisfiability for \alcqir, \alcqb, and \alcqib
in \nexptime can easily be shown by extending the embedding $\Psi_x$ into \ctwo
from Figure~\ref{fig:translation} to deal with Boolean combination of roles.

\begin{lemma}\label{lem:boolean-roles-to-ctwo}
  Satisfiability of \alcqib-concepts and \alcqib-CBoxes is polynomially
  reducible to \ctwo-satisfiability.
\end{lemma}

\begin{proof}
  For a role expression $\omega$, we define $\Psi_{xy}(\omega)$ inductively by
  \begin{align*}
    \Psi_{xy}(R) & = Rxy\\
    \Psi_{xy}(R^{-1}) & = Ryx\\
    \Psi_{xy}(\lnot \omega) & = \lnot \Psi_{xy}(\omega)\\
    \Psi_{xy}(\omega_1 \sqcap \omega_2) & = \Psi_{xy}(\omega_1) \land \Psi_{xy}(\omega_2)\\
    \Psi_{xy}(\omega_1 \sqcup \omega_2) & = \Psi_{xy}(\omega_1) \lor
    \Psi_{xy}(\omega_2)
  \end{align*}
  and set $\Psi_x(\bowtie n \; \omega \; C) = \exists^{\bowtie n} x .
  \Psi_{xy}(\omega) \land \Psi_y(C)$. 
  
  This translation is obviously polynomial and satisfies, for every
  interpretation $\I$ and concept $C$, 
  \[
  C^\I = \{ a \in \Delta^\I \mid \I \models \Psi_x(C)(a) \} .
  \]
  Hence, a concept $C$ is satisfiable iff $\exists^{\geq 1} x . \Psi_x(C)$ is
  satisfiable. CBoxes can be reduced to \ctwo as shown in
  Figure~\ref{fig:translation}. This yields the desired reductions. \qed
\end{proof}

Since \alcqi is a subset of \alcqir, which, in turn, is a subset of \alcqib, the
following is a simple corollary of
Theorem~\ref{theo:alcqi-sat-is-nexptime-hard} and Theorem~\ref{lem:boolean-roles-to-ctwo}:

\begin{corollary}
  Satisfiability of \alcqir- or \alcqib-CBoxes is \nexptime-hard. The problems
  are \textsc{NExp\-Time}-complete if unary coding of numbers in the input is assumed.
\end{corollary}

\begin{proof}
  The lower bound is immediate from
  Corollary~\ref{theo:alcqi-sat-is-nexptime-hard} because the set of
  \alcqi-concepts is strictly included in the set of \alcqir- and
  \alcqib-concepts. In the case of unary coding of numbers in the input, the
  upper bound follows from Lemma~\ref{lem:boolean-roles-to-ctwo} and
  Fact~\ref{fact:ctwo-nexptime}. \qed
\end{proof}

Similarly, the results for reasoning with nominals also transfer from
Corollary~\ref{cor:alcqio-nexptime}.

\begin{corollary}
  Satisfiability of \alcqiro- and \alcqibo-concepts is \nexptime-hard. The problems
  are \nexptime-complete if unary coding of numbers in the input is assumed.
  \index{alcqibo@\alcqiro}
  \index{alcqibo@\alcqibo}
\end{corollary}

So, in the presence of cardinality restrictions or nominals, reasoning with
\alcqib is not harder than reasoning with \alcqir.  Without cardinality
restrictions or nominals, though, reasoning with \alcqir is less complex
(\exptime-complete, Theorem~\ref{theo:alcqir-kb-exptime-complete}) than
reasoning for \alcqib. The reason for this is that \alcqib can easily mimic
cardinality restrictions (and hence nominals) using a fresh role:

\begin{lemma}\label{lem:cadres-to-boolean-roles}
  CBox satisfiability for \alcq and \alcqi is polynomially reducible to concept
  satisfiability of \alcqb and \alcqib respectively.
\end{lemma}

\begin{proof}
  Let $\mcC$ be a $\alcqbb$-CBox and $R$ a role that does not occur in $\mcC$. We
  transform $\mcC$ into a \alcqbb concept $C_\mcC$ by setting
  \[
  C_\mcC = \qnrleq 0 {\lnot R} \top \sqcap \mybigsqcap_{i=1}^k (\bowtie_i n_i \;
  R \; C_i) .
  \]
  
  \begin{claim}
    $C_\mcC$ is satisfiable iff $\mcC$ is satisfiable.
  \end{claim}
  
  Let $\I$ be a model for $\mcC$ . We define a model $\I'$ of $C_\mcC$ by
  setting $R^{\I'} := \Delta^\I \times \Delta^\I$ and preserving the
  interpretation of all other names. Since $R$ does not occur in $\mcC$, $C_i^\I
  = C_i^{\I'}$ holds for every $i$. Since $R$ is interpreted by the universal
  relation, $\qnrleq 0 {\lnot R} \top ^{\I'} = \Delta^{\I'}$ holds. Also, again
  since $R^{\I'}$ is the universal relation, for every $x \in \Delta^{\I'}$, $\{
  y \mid (x,y) \in R^{\I'} \text{ and } y \in C^{\I'} \} = C^{\I'}$. Thus, if
  $\I \models (\bowtie_i {n_i} \; {C_i})$, then $(\bowtie {n_i} \; R \;
  {C_i})^{\I'} = \Delta^{\I'}$.  Hence, from $\I \models \mcC$ is follows that
  $C_\mcC^{\I'} = \Delta^{\I'}$, which proves its satisfiability.
  
  For the converse direction, if $C_\mcC$ is satisfiable with $x \in C_\mcC^\I$
  for an interpretation $\I$, then, since $x \in \qnrleq 0 {\lnot R_i} \top ^\I$,
  $\{ y \mid (x,y) \in R^\I \} = \Delta^\I$ must hold and hence $\{ y \mid (x,y)
  \in R^\I \text{ and } y \in C^\I \} = C^\I$. It immediately follows that $\I
  \models \mcC$.

  Obviously, the size of $C_\mcC$ is linear in the size of $\mcC$, which proves
  this lemma. \qed
\end{proof}



  

\begin{corollary}\label{cor:alcq-alcqib-nexptime-complete}
  Concept satisfiability for \alcqb and \alcqib is \nexptime-hard. The problems
  are \nexptime-complete if unary coding of numbers in the input is assumed.
\end{corollary}

\begin{proof}
  Concept satisfiability for \alcqib is \nexptime-hard by
  Lemma~\ref{lem:cadres-to-boolean-roles} and
  Theorem~\ref{theo:alcqi-sat-is-nexptime-hard}, it can be decided in \nexptime
  by Lemma~\ref{lem:boolean-roles-to-ctwo} and Fact~\ref{fact:ctwo-nexptime}.
  
  For \alcqb the situation is slightly more complicated because
  Lemma~\ref{lem:cadres-to-boolean-roles} yields \nexptime-hardness only for
  binary coding of numbers. Yet, Lutz and Sattler \citeyear{LutzSattlerAIML00}
  show that concept satisfiability even for \alcb, i.e., \alc extended with Boolean
  role expressions, is \nexptime-hard, which yields the lower bound also for the
  case of unary coding of numbers. The matching upper bound (in the case of
  unary coding) again follows from Lemma~\ref{lem:boolean-roles-to-ctwo} and
  Fact~\ref{fact:ctwo-nexptime}.  \qed
\end{proof}

Of course, \cite{LutzSattlerAIML00} yields the lower bound also for \alcqib.
Since the connection between reasoning with cardinality restrictions and full
Boolean role expression established in Lemma~\ref{lem:cadres-to-boolean-roles}
is interesting in itself and yields, as a simple corollary, the result for
\alcqib, we include this alternative proof of this fact in this thesis.




\cleardoublepage


\chapter{Transitive Roles and Role Hierarchies}

\label{chap:shiq}

This chapter explores reasoning with Description Logics that allow for
transitive roles. \index{transitive role} \index{role!transitive} Transitive
roles play an important r\^ole in knowledge representation because, as argued by
Sattler \citeyear{Sattler-ECAI-2000}, transitive roles in combination with role
hierarchies are adequate to represent aggregated objects, which occur in many
application areas of knowledge representation, like configuration
\cite{wache96:_using_description_logic_for_configuration,sattler96:_know_repr_in_proc_engineering,mcguinness98:_concep},
ontologies \cite{rector:_exper_build_large_re_medic}, or data modelling
\fullcite{CaLN98,CDLNR98,calvanese99:_data_integ,FrancBaadSattvass-DWQ-Buch}.

Baader \citeyear{Baader91c} and Schild \citeyear{Schi91} were the first to study
the \emph{transitive closure} \index{transitive closure of roles}
\index{roles!transitive closure} of roles in DLs that extend \alc, and they both
developed DLs that are notational variants of PDL
\cite{fischer79:_propos_dynam_logic_regul_progr}. Due to the expressive power of
the transitive closure, these logics allow for the internalisation
\index{internalization of general axioms} of
terminological axioms \cite{Baader91c,Schi91,BBNNS93} and hence reasoning for
these logics is at least \exptime-hard. Sattler \citeyear{Sattler-KI-1996}
studies a number of DLs with transitive constructs and identifies the DL
\s,\footnote{Previously, this logic has been called \alcR. Here, we use \s
  instead because of a vague correspondence of \alcR with the modal logic
  $\mathsf{S4}$.} i.e., \alc extended with \emph{transitive roles}, as an
extension of \alc that still permits a \pspace reasoning procedure.

Horrocks and Sattler \citeyear{HorrocksSattler-DL-1998} study \si, the extension of
\s with inverse roles, and develop a tableau based reasoning procedure. While
they conjecture that concept satisfiability  and subsumption can be decided
for \si in \pspace, their algorithm only yields an \nexptime upper bound.  We
verify their conjecture by refining their tableau algorithm so that it decides
concept satisfiability (and hence subsumption) in \pspace. A comparable
approach is used by Spaan \citeyear{spaan_e:1993a} to show that satisfiability of
the modal logic $\mathsf{K4}_t$---corresponding to \si with only a single,
transitive role---can be decided in \pspace.

Subsequently \si is extended with role hierarchies \cite{HorrocksGough-DL-1997}
and qualifying number restrictions, which yields the DL \shiq.  The expressive
power of \shiq is particularly well suited to capture many properties of
aggregated objects \cite{Sattler-ECAI-2000} and has applications in the area of
conceptual data models \cite{CalvLenzNardi-KR-1994,FranconiNg-KRDB-2000} and
query optimization \cite{HorrocksSattler+-LPAR-2000}.  Furthermore, there exists
the OIL approach \cite{fensel00:_oil_nutsh} to add \shiq-based inference
capabilities to the semantic web~\cite{berners-lee99:_weavin_web}.  These applications have only become feasible due to the availability
of the highly optimized reasoner $\iFaCT$ \cite{horrocks99:_fact} for \shiq.

We determine the worst-case complexity of reasoning with \shiq as
\exptime-com\-plete, even if binary coding of numbers in the number restrictions
is used. This result relies on a reduction of \shiq to \alcqir with TBoxes, a
problem we already know how to solve in \exptime
(Theorem~\ref{theo:alcqir-tbox--exptime-complete}). Using the same reduction we
show that reasoning for \shiqo, i.e., the extension of \shiq with nominals,
is \nexptime-complete (in the case of unary coding of numbers).

As the upper \exptime-bound for \shiq relies on a highly inefficient automata
construction, Section \ref{sec:shiq-practical} extends the tableau algorithm for
\shif \cite{HorrSat-JLC-99} to deal with full qualifying number restrictions.
While this algorithm does not meet the worst-case complexity of the problem (a
naive implementation of the tableau algorithm would run in 2-\nexptime), it is
amenable to optimizations and forms the basis of the highly optimised DL system
\iFaCT~\cite{horrocks99:_fact}. See Section~\ref{sec:reasoning-methodologies}
for a discussion of the different reasoning paradigms and issues of
practicability of algorithms.


\newcommand{\satalci}{\textsc{SAT}(\si)\xspace}
\newcommand{\satalcni}{\textsc{SAT}(\sin)\xspace}

\section{Transitive and Inverse Roles: \si}

In this section we study the complexity of reasoning with the DL \si, an
extension of the DL \alc with \iemph{transitive roles} and \iemph{inverse  roles}:

\index{roles!transitive}
\index{roles!inverse}

\begin{definition}[Syntax and Semantics of \si]\label{def:si-syntax}
  Let \Names be a set of atomic \emph{concept names}, \Roles a set of atomic
  \emph{role names}, and $\RolesTrans \subseteq \Roles$ a set of
  \emph{transitive role names}.  With $\overline \Roles := \Roles \cup \{ R^{-1}
  \mid R \in \Roles \}$ we denote the set of \emph{\si-roles}.  The set of
  \emph{\si-concepts} is built inductively from $\Names$ and $\overline{\Roles}$
  using the following grammar, where $A \in \Names$ and $R \in \overline
  \Roles$:
  \[
  C ::= A  \bnfor \lnot C \bnfor  C_1 \sqcap C_2 \bnfor  C_1 \sqcup C_2 \bnfor
  \all R C \bnfor \some R C  .
  \]
  The semantics of \si-concepts is defined similarly to the semantics of
  \alc-concepts \wrt an interpretation $\I$, where, for an inverse role $R^{-1}
  \in \overline \Roles$, we set $(R^{-1})^\I = \{ (y,x) \mid (x,y) \in R^\I \}$.
  Moreover, we only consider those interpretations that interpret transitive
  roles $R \in \RolesTrans$ by transitive relations. An \si-concept $C$ is
  \emph{satisfiable} iff there exists an interpretation $\I$ such that, for
  every $R \in \RolesTrans$, $R^\I$ is transitive, and $C^\I \neq \emptyset$.
  Subsumption is defined as usual, again with the restriction to interpretations
  that interpret transitive roles with transitive relations.
  
  \index{si@\si}
  \index{s@\s}
  \index{00nr@$\overline{\Roles}$}
  \index{00nrplus@$\RolesTrans$}

  With $\s$ we denote the fragment of $\si$ that does not contain any inverse
  roles.  \eod
\end{definition}

In order to make the following considerations easier, we introduce two functions
on roles:

\begin{enumerate}
\item The inverse relation on roles is symmetric, and to avoid considering roles
  such as $R^{-1 -1}$, we define a function $\Inv$ which returns the inverse of a
  role. More precisely, $\Inv(R) = R^{-1}$ if $R$ is a role name, and $\Inv(R) = S$
  if $R=S^{-1}$. \index{00invr@$\Inv(R)$}
  
\item Obviously, the interpretation $R^\I$ of a role $R$ is transitive if and
  only if the interpretation of $\Inv(R)$ is transitive. However, this may be
  required by either $R$ or $\Inv(R)$ being in $\Rplus$. We therefore define a
  function $\Tr$, \index{00transr@$\Tr(R)$} which is $\mathrm{true}$ iff $R$ must
  be interpreted with a transitive relation---regardless of whether it is a role
  name or the inverse of a role name.  More precisely, $\Tr(R) = \texttt{true}$
  iff $R\in \Rplus$ or $\Inv(R)\in\Rplus$.
\end{enumerate}

\subsection{The \si-algorithm}

We will now describe a tableau algorithm that decides satisfiability of
\si-concepts in \pspace, thus proving \pspace-completeness of \si-satisfiability.
Like other tableau algorithms, the \si-al\-go\-rithm tries to prove the
satisfiability of a concept $C$ by constructing a model for $C$. The model is
represented by a so-called \iemph{completion tree}, a tree some of whose nodes
correspond to individuals in the model, each node being labelled with two sets
of \si-concepts.  When testing the satisfiability of an \si-concept $C$, these
sets are restricted to subsets of $\sub(C)$, where $\sub(C)$ is
the set of subconcepts of $C$, which is defined in the obvious
way. \index{00subc@$\sub(C)$}
Before we formally present the algorithm, we first discuss some problems that
need to be overcome when trying to develop an \si-algorithm that can be
implemented to run in \pspace.

Dealing with transitive roles in tableau algorithms requires extra
considerations because transitivity of a role is, generally speaking, a
\emph{global} constraint whereas the expansion rules and clash conditions of the
tableau algorithms that we have studied so far are of a more \emph{local}
nature. They only take into account a single node of the constraint system or at
most a node and its direct neighbours. Also, many of our previous considerations
relied on the fact that satisfiable concepts have tree models, which, in the
presence of transitive roles is no longer the case. To circumvent these
problems, we use the solution that has already been used, e.g., by Halpern and
Moses \citeyear{HalpernMoses92} to deal with the modal logic $\mathsf{S4}$,
which possesses a reflexive and transitive accessibility relation. Instead of
directly dealing with models and transitive relations, we use abstractions of
models---so called \emph{tableaux}---that disregard transitivity of roles and
have the form of a tree. This is done in a way that allows to recover a model of
the input concept by transitively closing those role relations that are
explicitly asserted in the tableau. To prove satisfiability of the input
concept, the \si-algorithm now tries to build a tableau instead of trying to
construct a model. Apart from this difference, the \si-algorithm is very similar
to the tableau algorithms we have encountered so far: starting from an initial
constraint system it employs completion rules until the constraint system is
complete, in which case the existence of a tableau is evident, or until an
obvious contradiction indicates an unsuccessful run of the (non-deterministic)
algorithm.

While it would be possible to maintain the use of ABoxes to capture the
constraint system that we will encounter during our discussion of DLs with
transitive roles, it is much more convenient to emphasise the view of constraint
systems as node- and edge-labelled trees, so this view will prevail in the
remainder of this chapter.


\subsection{Blocking}

Sattler \citeyear{Sattler-KI-1996} shows that concept satisfiability for \s can
be determined in polynomial space using an adaptation of the techniques employed
by Halpern and Moses \citeyear{HalpernMoses92} to decide satisfiability for the
modal logic $\mathsf{S4}$.  To understand why these techniques cannot be
extended easily to deal with inverse roles, as we have done in
Chapter~\ref{chap:alcq} when generalizing from \alcq to \alcqir, we have to
discuss the role of \iemph{blocking}.

The key difference between the algorithms from the previous chapters and the
\si-algorithm lies in the way universal restrictions are propagated through the
constraint system: whenever $\all R C$ with $\Tr(R)$ appears in the label of a
node $x$ and $x$ has an $R$-neighbour $y$, then not only $C$ is asserted for
$y$, but also $\all R C$. This makes sure that $C$ is successively asserted for
every node reachable from $x$ via a \emph{chain} of $R$-edges. These are exactly
the nodes that are reachable from $x$ with a single $R$-step once $R$ has been
transitively closed; exactly these nodes must satisfy $C$ in order for
$\all R C$ to hold at $x$.

Previously the termination of the tableau algorithms relied on the fact that the
nesting of universal and existential restrictions strictly decreases along a
path in the tableau. When dealing with transitive roles in the described manner,
this is no longer guaranteed. For example, consider a node $x$ labelled $\{C,
\some{R}{C}, \all{R}{(\some{R}{C})}\}$, where $R$ is a transitive role. The
described approach would cause the new node $y$ that is created due for the
\some{R}{C} constraint to receive a label identical to the label of $x$. Thus,
the expansion process could be repeated indefinitely.

The way we deal with this problem is by \emph{blocking}: halting the expansion
process when a cycle is
detected~\cite{Baader91c,Buchheit93,HalpernMoses92,Sattler-KI-1996,BaaderBuchheit+-AIJ-1996,HorrSat-JLC-99}.
For logics without inverse roles, the general procedure is to check the
constraints asserted for each new node $y$, and if they are a \emph{subset} of
the constraints for an existing node $x$, then no further expansion of $y$ is
performed: $x$ is said to block $y$. The resulting constraint system
corresponds to a cyclic model in which $y$ is identified with
$x$.\footnote{For logics with a transitive closure operator it is necessary to
  check the validity of the cyclic model created by blocking~\cite{Baader91c},
  but for logics that only support transitive roles the cyclic model is always
  valid~\cite{Sattler-KI-1996}.} The validity of the cyclic model is an easy
consequence of the fact that each $\some{R}{D}$ constraint for $y$ must also be
satisfied by $x$ because the constraints for $x$ are a superset of the
constraints for $y$. Termination is now guaranteed by the fact that all
constraints for individuals in the constraint system are ultimately derived from
the decomposition of the input concept $C$, so every set of constraints for an
individual must be a subset of the subconcepts of $C$, and a blocking situation must
therefore occur within a finite number of expansion steps.

\subsubsection{Dynamic Blocking}
\label{sec:dynblk}

Blocking is more problematic when inverse roles are added to the logic, and a
key feature of the algorithms presented in~\cite{HorrSat-JLC-99} was the
introduction of a \iemph{dynamic blocking} strategy. It uses label equality
instead of the subset condition, and it allows blocks to be established,
broken, and re-established.  \index{blocking!dynamic blocking}

Label inclusion as a blocking criterion is no longer correct in the presence of
inverse roles because roles are now bi-directional, and hence universal
restrictions at the blocking node can conflict with the constraints for the
predecessor of the blocked node. 

\begin{figure}[tbh]
  \begin{center}
    \vspace{2ex}
    \parbox{5.75in}{\input{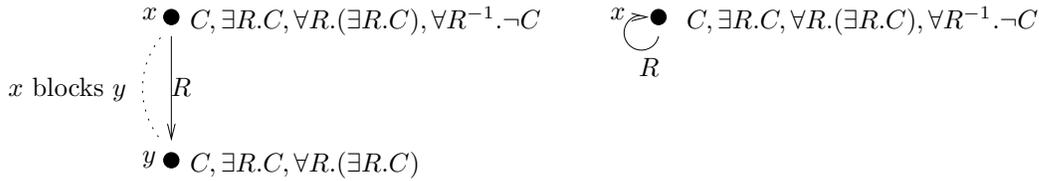}}
    \caption{An invalid cyclic model}
    \label{fig:block-cyclic}
  \end{center}
\end{figure}

Taking the above example of a node labelled $\{C, \some{R}{C},
\all{R}{(\some{R}{C})}\}$, if the successor of this node were blocked by a node
whose label additionally included $\all{R^{-1}}{\neg C}$, then the cyclic model
would clearly be invalid. This is shown in Figure~\ref{fig:block-cyclic}, where
$x$ blocks its $R$-successor $y$ (if subset-blocking is assumed) and hence in
the induced model (shown on the right), there exists an $R$-cycle from $x$ to $x$. Hence, $C$ and
$\all {R^{-1}} {\neg C}$, which have both been asserted for $x$, now stand in a
conflict.

In~\cite{HorrSat-JLC-99}, this problem was overcome by allowing a node $x$ to be
blocked by one of its ancestors if and only if they were labelled with the
same set of concepts.

\begin{figure}[tbh]
  \begin{center}
    \vspace{2ex}
    \parbox{3.75in}{\input{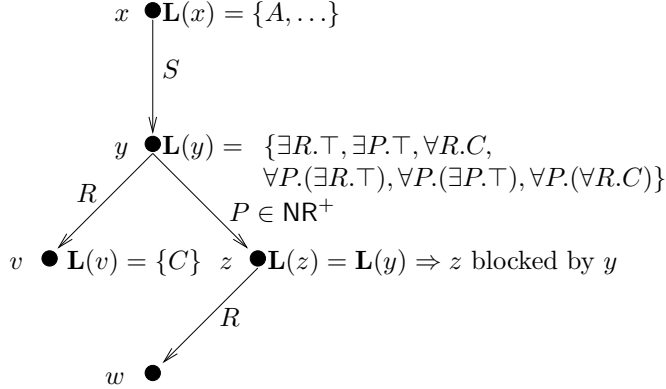}}
    \caption{A tableau where dynamic blocking is crucial}
    \label{fig:block-exa}
  \end{center}
\end{figure}

Another difficulty introduced by inverse roles is the fact that it is no longer
possible to establish a block on a once-and-for-all basis when a new node is
added to the tree. This is because further expansion in other parts of the tree
could lead to the labels of the blocking and/or blocked node being extended and
the block being invalidated. Consider the example sketched in
Figure~\ref{fig:block-exa}, which shows parts of a tableau that was generated
for the concept
$$A\sqcap \exists S. ( \exists R.\top \sqcap \exists P. \top \sqcap \forall R.C
\sqcap \forall P.(\exists R.\top) \sqcap \forall P.(\forall R.C) \sqcap \forall
P.(\exists P.\top)),
$$
where $C$ represents the concept
$$\forall R^{-1} . (\forall P^{-1} . (\forall S^{-1} . \neg A)).$$
This concept is clearly not satisfiable: $w$ has to be an instance of $C$, which
implies that $x$ is an instance of $\neg A$. This is inconsistent with $x$
being an instance of $A$.

Since $P$ is a transitive role, all universal value restrictions over $P$ are
propagated from $y$ to $z$, hence $y$ and $z$ are labelled with the same
constraints and hence $z$ is blocked by $y$.  If the blocking of $z$ was not
subsequently broken when $\forall P^{-1}.(\forall S^{-1}.\neg A)$ is added to
$\Lab(y)$ from $C\in \Lab(v)$, then $\neg A$ would never be added to $\Lab(x)$
and the unsatisfiability would not be detected.

As well as allowing blocks to be broken, it is also necessary to continue with
some expansion of blocked nodes, because $\all{R}{C}$ concepts in their labels
could affect other parts of the tree. Again, let us consider the example in
Figure~\ref{fig:block-exa}. After the blocking of $z$ is broken and $\forall
P^{-1}.(\forall S^{-1}.\neg A)$ is added to $\Lab(z)$ from $C \in \Lab(w)$, $z$ is
again blocked by $y$.  However, the universal value restriction $\forall
P^{-1}.(\forall S^{-1}.\neg A) \in \Lab (z)$ has to be expanded in order to detect the
unsatisfiability.

These problems are overcome by using \emph{dynamic} blocking: using label
equality as blocking criterion and allowing blocks to be dynamically established
and broken as the expansion progresses, and continuing to expand $\all{R}{C}$
concepts in the labels of blocked nodes.

\subsubsection{Refined blocking}

\index{refined blocking}
\index{blocking!refined blocking}

As mentioned before, in~\cite{HorrSat-JLC-99}, blocking of nodes is based on
label equality. This leads to major problems when trying to establish a
polynomial bound on the length of paths in the completion tree. If a node can only be
blocked by an ancestor when the labels coincide, then there could potentially be
exponentially many ancestors in a path before blocking actually occurs. Due to
the non-deterministic nature of the expansion rules, these subsets might
actually be generated; the algorithm would then need to store the node labels of
a path of exponential length, thus consuming exponential space.

This problem is already present when one tries to implement a tableau algorithm
for the logic \alcRtrans~\cite{Sattler-KI-1996}, where the non-deterministic nature
of the expansion rules for disjunction might lead to the generation of a chain
of exponential size before blocking occurs. Consider, for example, the concept
\begin{align*}
  C & = \some{R}{D} \sqcap \all{R}{(\some{R}{D})} \\ D & = (A_1 \sqcup B_1)
  \sqcap (A_2 \sqcup B_2) \sqcap \dots \sqcap (A_n \sqcup B_n)
\end{align*}
where $R$ is a transitive role. The concept $C$ causes the generation of a chain
of $R$-successors for all of which $D$ is asserted. There are $2^n$ possible
ways of expanding $D$ because for every disjunctive concept $A_i \sqcup B_i$ the
\ruleor-rule can choose to add $A_i$ or $B_i$. The completion tree for $D$ is
only complete once one node of this path is blocked and all unblocked nodes
(including the blocking node) are fully expanded. For \alcRtrans, a polynomial
bound on the length of paths  is obtained by applying a simple strategy: a new
successor is only generated when no other rule can be applied, and propositional
expansion of concepts only takes place if universal restrictions have been
exhaustively been dealt with. Once a node is blocked, it is not necessary to
perform its propositional expansion because it has already been ensured at the
blocked node that such an expansion is possible without causing a clash.

However, in the presence of inverse roles, this strategy is no longer possible.
Indeed, the expansion rules for \si as they have been presented
in~\cite{HorrocksSattler-DL-1998} based on set equality might lead to a tableau
with paths of exponential length for $C$---even though $C$ does not contain any
inverse roles. This is due to the fact that blocking is established on the basis
of label equality. Since the label of the blocked and blocking node must be
equal, this implies that, since the label of the blocking node must be fully
expanded, this also must hold for the label of the blocked. Since there are
$2^n$ possibilities for such an expansion, it might indeed take a path of
$2^n+1$ nodes before such a situation necessarily occurs and the completion
tree is complete.


In order to obtain a tableau algorithm that circumvents this problem and
guarantees blocking after a polynomial number of steps, we will keep the
information that is relevant for blocking separated from the ``irrelevant''
information (due to propositional expansion) in a way which allows for a simple
and comprehensible tableau algorithm.  In the following, we will explain this
``separation'' idea in more detail.

\begin{figure}[h]
  \begin{center}
    \vspace{2ex}
    \parbox{3.25in}{\input{block-bsp3.pstex_t}}
    \caption{Refined blocking}
    \label{fig:block-sit}
  \end{center}
\end{figure}

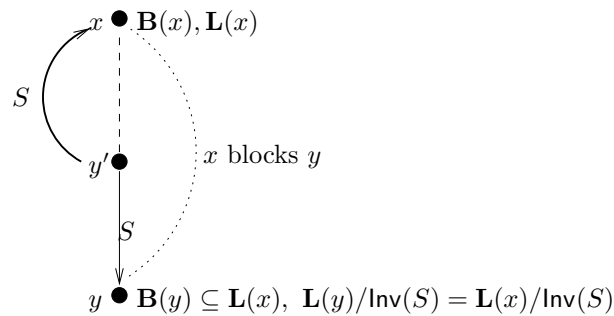
Figure~\ref{fig:block-sit} shows a blocking situation. Assume node $y$ to be
blocked by node $x$. When generating a model from this tree, the blocked node
$y$ will be omitted and $y'$ will get $x$ as an $S$-successor, which is
indicated by the backward arrow. On the one hand, this construction yields a
new $S$-successor $x$ of $y'$, a situation which is taken care of by the subset
blocking used in the normal \alcRtrans tableau algorithms. On the other hand,
$x$ receives a new $S^{-1}$-successor $y'$. Now blocking has to make sure that, if
$x$ must satisfy a concept of the form $\all{S^{-1}}{D}$, then $D$ (and
$\all{S^{-1}}{D}$ if $S$ is a transitive role) is satisfied by $y'$.

This was dealt with by equality blocking in \cite{HorrSat-JLC-99}. In the
following algorithm it will be dealt with using two labels per node and a
modified blocking condition that takes these two labels into account. In addition
to the label $\Lab$, each node now has a second label $\BLab$, where the latter
is always a subset of the former. The label $\Lab$ contains complete
information, whereas $\BLab$ contains only information relevant to blocking.
Propositional consequences of concepts in $\Lab$ and concepts being propagated
``upwards'' in the tree are stored in $\Lab$ only, as they are not important for
blocking as long as they are not universal restrictions that state requirements
on the predecessor in the completion tree.  The modified blocking condition now
looks as follows. For a node $y$ to be blocked by a node $x$ we require that

\begin{itemize}
\item the label $\BLab(y)$ of the blocked node $y$ is contained in the label
  $\Lab(x)$ of the blocking node $x$. Expansions of disjunctions are only stored
  in $\Lab$ and thus cannot prevent a node from being blocked.
\item if $y$ is reachable from its predecessor in the completion tree via the
  role $S$, then the universal restrictions along $\Inv(S)$ asserted
  for $y$ are the same as those asserted for $x$. This takes care of the fact
  that the predecessor $y'$ of the blocked node $y$ becomes a new
  $\Inv(S) $-successor of the blocking node $x$.
\end{itemize}



 
Summing up, we build a completion tree in a way that, for all nodes $x$,
 
\begin{itemize}
\item we have $\BLab(x) \subseteq \Lab(x)$,
\item $\BLab(x)$ contains only concepts which move \emph{down} the tree,
\item $\Lab(x)$ contains, additionally, all concepts which move \emph{up} the
tree, and
\item expansion of disjunctions and conjunctions only affects $\Lab(x)$.
\end{itemize}

\subsection{A Tableau Algorithm for \si}   

We now present a tableau algorithm derived from the one presented
in~\cite{HorrSat-JLC-99}. We shape the rules in a way that allows for the
separation of the concepts which are relevant for the two parts of the blocking
condition. For ease of construction, we assume all concepts to be in
\emph{negation normal form} (NNF), \index{negation normal form} that is, negation occurs only in front of
concept names. Any \si-concept can easily be transformed into an equivalent one in
NNF in the same way as this can be done for \alc-concepts
(Definition~\ref{def:nnf}).

The soundness and completeness of the algorithm will be proved by showing that
it creates a \emph{tableau} for $C$.  In contrast to the approach we have taken
in the previous chapters, where a constraint system stood in direct
correspondence to a model, here we introduce tableaux as intermediate
structures that encapsule the transition from the \emph{syntactic object} of a
completion tree to the \emph{semantical} object of a model and takes care of the
transitive roles on that way. This makes it possible for the algorithm to
operate on trees even though \si does not have a genuine tree model property due
to its transitive roles.

\begin{definition}[A Tableau for \si] \label{def:si-tableau}
  If $C$ is an \si-concept in NNF and $\overline{\Roles}_C$
  \index{00nrc@$\overline{\Roles}_C$} is the set of roles occurring in $C$
  together with their inverses, a \emph{tableau} \index{tableau!for si@for \si}
  $\mcT$ for $C$ is a triple $(\mcS,\TabLab,\Edges)$ such that $\mcS$ is a
  non-empty set, $\TabLab:\mcS \rightarrow 2^{\textit{sub}(C)}$ maps each
  element of $\mcS$ to a subset of $\textit{sub}(C)$, and
  $\Edges:\overline{\Roles}_C \rightarrow 2^{\mcS \times \mcS}$ maps each role
  in $\overline{\Roles}_C$ to a set of pairs of individuals. In addition, the
  following conditions must be satisfied:
  \begin{itemize}
  \item [\tab 1] There is an $s \in \mcS$ with $C \in \TabLab(s)$, and
  \end{itemize}
  for all $s,t \in \mcS$, $A,C_1,C_2,D\in \mathit{sub}(C)$, and $R\in \overline{\Roles}_C$,
  \begin{enumerate}
  \item [\tab 2] if $A \in \TabLab(s)$, then $\neg A \notin
    \TabLab(s)$, for $A \in \Names$,
  \item [\tab 3] if $C_1 \sqcap C_2 \in \TabLab(s)$, then $C_1\in\TabLab(s)$ and
    ${C_2\in\TabLab(s)}$, 
  \item [\tab 4] if $C_1 \sqcup C_2 \in \TabLab(s)$, then $C_1 \in \TabLab(s)$ or $C_2 \in \TabLab(s)$,
  \item [\tab 5] if $\some{R}{D} \in \TabLab(s)$, then there is some $t \in \mcS$ such
    that $\tuple{s}{t} \in \Edges(R)$ and $D \in \TabLab(t)$,
  \item [\tab 6] if $\all{R}{D} \in \TabLab(s)$ and $\tuple{s}{t} \in \Edges(R)$, then $D
    \in \TabLab(t)$,
  \item [\tab 7] if $\all{R}{D} \in \TabLab(s)$, $\tuple{s}{t} \in \Edges(R)$ and $\Tr(R)$,
    then $\all{R}{D} \in \TabLab(t)$, and
  \item [\tab 8] $\tuple{s}{t}\in \Edges(R)$ iff $\tuple{t}{s}\in
    \Edges(\Inv(R))$. \eod
  \end{enumerate}

\end{definition}

A tableau $\mcT$ for a concept $C$ is a ``syntactic witness'' for the satisfiability
of $C$:

\begin{lemma} \label{lem:si-tableau}
  An \si-concept $C$ is satisfiable iff there exists a tableau for $C$.
\end{lemma}

\begin{proof}
  For the \emph{if}-direction, if $\mcT = (\mcS,\TabLab,\Edges)$ is a tableau
  for $C$ with $C\in \TabLab(s_0)$, a model $\mathcal{I} = (\domain, \cdot\ifunc)$
  of $C$ can be defined as follows:
  \[
  \begin{array}{rcl} \hspace{2cm}\domain & = & \mcS, \\[0.5ex]
    A\ifunc & = & \{s \mid A \in \TabLab(s)\} \quad\mbox{for all concept names A in
      $\textit{sub}(C)$}, \\[0.5ex] R\ifunc & = & \left\{ \begin{array}{ll} \Edges(R)^+
        & \mbox{if $\Tr(R)$} \\ \Edges(R) & \mbox{otherwise},
      \end{array}
    \right.
  \end{array}
  \]
  where $\Edges(R)^+$ denotes the transitive closure of $\Edges(R)$.  Transitive
  roles are interpreted by transitive relations by definition.  By induction on
  the structure of concepts, we show that, if $D\in \TabLab(s)$, then $s\in
  D\ifunc$. This implies $C\ifunc\neq \emptyset$ because $s_0\in C\ifunc$. Let
  $D\in \TabLab(s)$:
  \begin{enumerate}
  \item If $D = A \in \Names$ is a concept name, then $s\in D^\I$ by definition.
  \item If $D=\neg A$ for $A \in \Names$ then $A \notin \TabLab(s)$ (due to \tab
    2 ), so $s \in \domain \setminus A\ifunc= D\ifunc$.
  \item If $D= (C_1 \sqcap C_2)$, then, due to \tab 3, $C_1 \in \TabLab(s)$ and
    $C_2 \in \TabLab(s)$, and hence, by induction, $s \in C_1\ifunc$ and $s\in
    C_2\ifunc$.  Thus, $s\in (C_1 \sqcap C_2)\ifunc$.
  \item The case $D=(C_1 \sqcup C_2)$ is analogous to the previous one.
  \item If $D= \some{R}{E}$, then, due to \tab 5, there is some $t \in \mcS$
    such that $\tuple{s}{t} \in \Edges(R)$ and $E \in \TabLab(t)$. By definition
    of $\I$, $\tuple{s}{t} \in R\ifunc$ holds as follows. It is immediate, if
    $R \in \Roles$. If $R = S^{-1}$ for $S \in \Roles$, then $\tuple s t \in
    \Edges(R)$ implies $\tuple t s \in \Edges(S)$ by \tab 8. Hence, $\tuple t s
    \in S^\I$ and $\tuple s t \in R^\I$ holds. By induction, $t\in E\ifunc$ and
    hence $s \in (\some{R}{E})\ifunc$.
  \item If $D= (\all{R}{E})$ and $\tuple{s}{t} \in R \ifunc$, then either
    \begin{enumerate}
    \item $\tuple{s}{t} \in\Edges(R)$ and $E\in \TabLab(t)$ (due to \tab 6), or
    \item $\tuple{s}{t} \not\in\Edges(R)$. Due to $\tab 8$, this can only be the
      case if $R$ is transitive and there exists a path of length $n\geq 1$ such
      that $\tuple{s}{s_1},$ $\tuple{s_1}{s_2},\ldots,$ $\tuple{s_n}{t}\in
      \Edges(R)$.  Due to \tab 7, $\all{R}{E}\in \TabLab(s_i)$ for all
      $1\leqslant i \leqslant n$, and we have $E\in \TabLab(t)$, again due to
      \tab 6.
    \end{enumerate}
    In both cases, by induction $t\in E\ifunc$ holds, and hence $s\in
    (\all{R}{E})\ifunc$.
  \end{enumerate}
  
  \bigskip
  
  For the converse direction, if $\mathcal{I} = (\domain, \cdot\ifunc)$ is a
  model of $C$, then a tableau $\mcT = (\mcS,\TabLab,\Edges)$ for $C$ can be
  defined by:
  \begin{eqnarray*}
    \mcS & = & \domain, \\ 
    \Edges(R) & = & R\ifunc, \\ 
    \TabLab(s) & = & \{D \in  \textit{sub}(D) \mid s \in D\ifunc \}.
  \end{eqnarray*}
  
  It  remains to demonstrate that $\mcT$ is a tableau for $C$:
  \begin{enumerate}
  \item $\mcT$ satisfies \tab 1 -- \tab 6 and \tab 8 as a direct consequence of the
    semantics of \si concepts and of inverse roles.
  \item If $s \in (\all{R}{D})\ifunc$, $\tuple{s}{t} \in R\ifunc$ and $\Tr(R)$,
    then $t \in (\all{R}{D})\ifunc$ unless there is some $u$ such that
    $\tuple{t}{u} \in R\ifunc$ and $u \notin D\ifunc$. However, if $\tuple{s}{t}
    \in R\ifunc$, $\tuple{t}{u} \in R\ifunc$ and $\Tr(R)$, then
    $\tuple{s}{u} \in R\ifunc$, which would imply $s \notin (\all{R}{D})\ifunc$.
    $T$ therefore satisfies \tab 7. \qed
  \end{enumerate}
\end{proof}

\subsection{Constructing an \si Tableau}\label{section:si-algo}  

From Lemma~\ref{lem:si-tableau}, it follows that an algorithm that constructs a
tableau for an \si-concept $C$ can be used as a decision procedure for the
satisfiability of $C$. Such an algorithm will now be described.

Like the tableau algorithms that we have studied so far, the algorithm for \si
works by manipulating a constraint system. In the presence of blocking, and
especially in the case of the refined blocking we are using for \si, it is more
convenient to emphasise the graph structure of the constraint system and deal
with an edge- and node-labelled graph instead of an ABox. In case of the
\si-algorithm, the constraint system has the form of a \emph{completion tree}. 

\begin{algorithm}[The \si-algorithm]\label{alg:si}
  Let $C$ be an \si-concept in NNF to be tested for satisfiability and
  $\overline{\Roles}_C$ the set of roles that occur in $C$ together with their
  inverse.  A \emph{completion tree} \index{completion tree!for si@for \si}
  $\Tree = (\bfV, \bfE, \bfL, \BLab)$ is a labelled tree in which each node $x
  \in \bfV$ is labelled with two subsets $\Lab(x)$ and $\BLab(x)$ of
  $\textit{sub}(C)$. Furthermore, each edge $\tuple{x}{y} \in \bfE$ in the tree
  is labelled $\Lab\tuple{x}{y}= R$ for some (possibly inverse) role $R \in
  \overline{\Roles}_C$.  Nodes and edges are added when expanding \some{R}{D}
  and \some{R^{-1}}{D} constraints; they correspond to relationships between
  pairs of individuals and are always directed from the root node to the leaf
  nodes.  The algorithm expands the tree by extending $\Lab(x)$ (and possibly
  $\BLab(x)$) for some node $x$, or by adding new leaf nodes.

  \index{si@\si!tableau algorithm}
  \index{tableau algorithm!for si@for \si}
  
  A completion tree \Tree is said to contain a \emph{clash} if, for a node $x$
  in \Tree, it holds that there is a concept name $A$ such that $\{A, \neg A\}
  \subseteq \Lab(x)$.
  
  If nodes $x$ and $y$ are connected by an edge $\tuple{x}{y} \in \bfE$, then $y$ is
  called a \emph{successor} of $x$ and $x$ is called a \emph{predecessor} of
  $y$. If $\Lab\tuple{x}{y} = R$, then $y$ is called an $R$-\emph{successor} of
  $x$ and $x$ is called an $\Inv(R)$-\emph{predecessor} of $y$.  

  \emph{Ancestor} is the transitive closure of \emph{predecessor} and
  \emph{descendant} is the transitive closure of \emph{successor}.  A node $y$
  is called an $R$\emph{-neighbour} of a node $x$ if either $y$ is an
  $R$-successor of $x$ or $y$ is an $R$-predecessor of $x$.
  
  To define the blocking condition we need the following auxiliary definition.
  For a (possibly inverse) role $S \in \overline{\Roles}_C$, we define the set
  $\Lab(y)/S$ by \index{00lxbyr@$\Lab(y)/R$}
  \[
  \Lab(y)/S = \{ \all{S}{D} \in \Lab(y)\} .
  \]
  A node $y$ is \emph{blocked} if for some ancestor $x$ of $y$, $x$ is blocked or
  \[ 
  \BLab(y) \subseteq \Lab(x) \quad \text{and} \quad \Lab(y)/\Inv(S) =
  \Lab(x)/\Inv(S) 
  \]
  for the unique predecessor $y'$ of $y$ in the completion tree,
   $\Lab\tuple{y'}{y}=S$ holds.
  
  The algorithm initializes a tree \Tree to contain a single node $x_0$, called
  the \emph{root} node, with $\Lab(x_0) = \BLab(x_0) = \{C\}$. \Tree is then
  expanded by repeatedly applying the rules from Figure~\ref{fig:si-rules}.
  
  The \ruleex-rule is called \emph{generating}; all other rules are called
  \emph{non-generating}.

  \begin{figure}[tb!]
    \begin{center}
      \begin{tabular}{@{}l@{$\;$}l@{$\;$}l}
        \ruleand: & \bfif \hfill 1. & $C_1 \sqcap C_2 \in \Lab(x)$ and \\ &
        \hfill 2. & $\{C_1,C_2\} \not\subseteq \Lab(x)$ \\ & \bfthen & $\Lab(x)
        \ruleand \Lab(x) \cup \{C_1,C_2\}$ \\[1ex]
        
        \ruleor: & \bfif \hfill 1. & $C_1 \sqcup C_2 \in \Lab(x)$ and \\ &
        \hfill 2. & $\{C_1,C_2\} \cap \Lab(x) = \emptyset$ \\ & \bfthen & $\Lab(x)
        \ruleor \Lab(x) \cup \{E\}$ for some $E\in \{C_1,C_2 \}$\\[1ex]
        
        \rulefa: & \bfif \hfill 1. & $\all{R}{D} \in \Lab(x)$ and \\ & \hfill
        2. &there is an $R$-successor $y$ of $x$ with $D \notin \BLab(y)$ \\ & \bfthen &
        $\Lab(y) \rulefa \Lab(y) \cup \{D\}$ and $\BLab(y) \rulefa \BLab(y) \cup
        \{D\}$ \bfor\\ & \hfill 2'.&there is an $R$-predecessor $y$ of $x$ 
        with $D \notin \Lab(y)$ \\ & \bfthen & $\Lab(y) \rulefa \Lab(y) \cup
        \{D\}$ and delete \emph{all} descendants of $y$. \\[1ex]
        
        \rulefatr: & \bfif \hfill 1. & $\all{R}{D} \in \Lab(x)$ and $\Tr(R)$
        and \\ & \hfill 2. &there is an $R$-successor $y$ of $x$ with $\all{R}{D}
        \notin \BLab(y)$ \\ & \bfthen & $\Lab(y) \rulefatr \Lab(y) \cup
        \{\all{R}{D}\}$ and $\BLab(y) \rulefatr \BLab(y) \cup
        \{\all{R}{D}\}$ \bfor \\ & \hfill 2'. &there is an $R$-predecessor $y$ of $x$
        with $\all{R}{D} \notin \Lab(y)$ \\ & \bfthen & $\Lab(y) \rulefatr
        \Lab(y) \cup \{\all{R}{D}\}$ and delete \emph{all} descendants of $y$. \\[1ex]
        
        \ruleex: & \bfif \hfill 1. & $\some{R}{D} \in \Lab(x)$, $x$ is not
        blocked and no non-generating rule \\ && is applicable to $x$ and any of
        its ancestors, and 
        \\ & \hfill 2. & $x$ has no $R$-neighbour $y$ with $D\in \BLab(y)$ \\ &
        \bfthen & create a new node $y$ with $\Lab\tuple{x}{y}=R$ and $\Lab(y)=
        \BLab(y) = \{D\}$ \\[1ex]
      \end{tabular}
      \caption{Tableau expansion rules for \si}
      \label{fig:si-rules}
    \end{center}
    \index{00ruleand@\ruleand}
    \index{00ruleor@\ruleor}
    \index{00rulefa@\rulefa}
    \index{00rulefatr@\rulefatr}
    \index{00ruleex@\ruleex}
    \index{completion rules!for si@for \si}
    \index{si@\si!completion rules}
  \end{figure}
  
  The completion tree is \emph{complete} if, for some node $x$, $\Lab(x)$
  contains a clash or if none of the expansion rules is applicable. If the
  expansion rules can be applied in such a way that they yield a complete,
  clash-free completion tree, then the algorithm returns ``$C$ is satisfiable'';
  otherwise, the algorithm returns ``$C$ is not satisfiable''. \eod
\end{algorithm}

Like for all other tableau algorithms studied in this thesis, it turns out (see
the proof of Lemma~\ref{lemma:si-completeness}) that the choice of which rule to
apply where and when is don't-care non-deterministic---no choice can prohibit the
discovery of a complete and clash-free completion tree for a satisfiable
concept. On the other hand, as before, the choice of the \ruleor-rule is
don't-know non-deterministic---only certain choices will lead to the discovery of
a complete and clash-free completion tree for a satisfiable concept. For an
implementation this means that an arbitrary strategy that selects which rule to
apply where will yield a complete implementation but exhaustive search is
required to consider the different choices of the \ruleor-rule. A similar
situation exist in case of the \shiq-algorithm in
Section~\ref{sec:shiq-practical}, where it follows from the proof of
Lemma~\ref{lemma:shiq-completeness} that choice of which rule to apply where is
don't-care non-deterministic.

Note that in the definition of successor and predecessor, the tree structure is
reflected.  If $y$ is an $R$-successor of $x$ than this implies that $y$ is
successor of $x$ in the completion tree and it is \emph{not} the case that $x$
is an $\Inv(R)$-predecessor of $y$. Successor and predecessor always refer to
the relative position of nodes in the completion tree. This is necessary
because, in the construction of a tableau from a complete and clash-free
completion tree, the edges pointing to blocked successors will be redirected to
the respective blocking nodes, which makes the relative position of nodes in the
completion tree significant.

We are aiming for a \pspace-decision procedure, so, like the\alcqir-algorithm
(Algorithm~\ref{alg:alcqir}), the \rulefa- and \rulefatr-rules delete parts of
the completion tree whenever information is propagated upward in the completion
tree to make tracing possible.

\subsubsection{Correctness}\label{sec:si-soundness}

As before, correctness of the algorithm will be demonstrated by proving that,
for an \si-concept $C$, it always terminates and that it returns ``satisfiable''
if and only if $C$ is satisfiable. To prove this, we follow a slightly different
approach than the one that is indicated by
Theorem~\ref{theo:generic-correctness}. The reason for this is that it is
unclear how to deal with blocked nodes when trying to define a suitable notion
of satisfiability for a completion tree. We will come back to this topic.

Before we start proving the properties we need to establish correctness of the
\si-algorithm, let us state an obvious property of the \si-algorithm:

\begin{lemma}\label{lem:label-subsets}
  Let \Tree be a completion tree generated by the \si-algorithm. Then, for every
  node $x$ of \Tree, $\BLab(x) \subseteq \Lab(x)$.
\end{lemma}

\begin{proof}
  Obviously, $\BLab(x_0) \subseteq \Lab(x_0)$ holds for the only node $x_0$ of
  the initial tree. Subsequently, whenever a concept $D$ is added to $\BLab(x)$
  by an application of one of the rules, then it is always also added to
  $\Lab(x)$. \qed
\end{proof}

We first  show termination of the algorithm:

\begin{lemma}[Termination]\label{lemma:si-termination}
  For each \si-concept $C$, the tableau algorithm terminates.
\end{lemma}

\begin{proof}
  Let $m = \Card{\textit{sub}(C)}$. Obviously, $m$ is linear in the length of
  $C$. Termination is a consequence of the following properties of the expansion
  rules:
  \begin{enumerate}
  \item The expansion rules never remove concepts from node labels.
  \item Successors are only generated for concepts of the form \some{R}{D} and,
    for any node, each of these concepts triggers the generation of at most one
    successor.  Since $\textit{sub}(C)$ contains at most $m$ concepts of the
    form \some{R}{D}, the out-degree of the tree is bounded by $m$.
  \item Nodes are labelled with nonempty subsets of $\textit{sub}(C)$.  If a
    path $p$ is of length $>2^{2m}$, then there are 2 nodes $x,y$ on
    $p$, with $\Lab(x)=\Lab(y)$ and $\BLab(x) = \BLab(y)$, and blocking occurs.
    Since a path on which nodes are blocked cannot become longer, paths are of
    length at most $2^{2m}+1$.
  \end{enumerate}
  
  An infinite run of the completion algorithm can thus only occur due
  to an infinite number of deletions of nodes of the tree. That this
  can never happen can be shown in exactly the same way this has been
  done in the proof of Lemma~\ref{lem:alcqri-alg-terminates}. \qed
\end{proof}

\begin{lemma}[Soundness]\label{lemma:si-soundness}
  If the \si-algorithm generates a complete and clash-free completion tree for a
  concept $C$, then $C$ has a tableau.
\end{lemma}

\begin{proof}
  Let $\Tree = (\bfV, \bfE, \Lab, \BLab)$ be the complete and clash-free
  completion tree constructed by the tableau algorithm for $C$. A tableau $\mcT
  = (\mcS,\TabLab,\Edges)$ can be defined by
  \[
  \begin{array}{@{}r@{\;}c@{\;}l}
    \mcS &=& \{x \mid \mbox{$x$ is a node in \Tree, and $x$ is not
      blocked}\},\\[0.5ex]  
    \TabLab&=& \Lab|_{\mcS},\\[0.5ex] 
    \Edges(R) & = & \{\tuple{x}{y} \in
    \mcS \times \mcS \mid \begin{array}[t]{l@{\,}l} 1. & y \mbox{ is
        an $R$-neighbour of }x\quad\emph{or} \\ 2. & \exists z . \Lab\tuple{x}{z} = R \mbox{
        and $y$ blocks $z$} \quad\emph{or} \\ 3. & \exists z . \Lab\tuple{y}{z} = \Inv(R)
      \mbox{ and $x$ blocks $z$}\}.
    \end{array}
  \end{array}
  \]
  We now show that $\mcT$ is a tableau for $C$. By definition of $\mcT$,
  we have $\Lab = \TabLab$ for all $x \in \mcS$, so it is sufficient to establish
  the required properties for the function $\Lab$.

  \begin{itemize}
  \item $C \in \Lab(x_0)$ for the root $x_0$ of \Tree and, since $x_0$ has no
    predecessors, it cannot be blocked. Hence $C \in \Lab(s)$ for some $s \in
    \mcS$ and \textbf{\tab 1} holds.
  \item \textbf{\tab 2} is satisfied because \Tree is clash-free.
  \item \textbf{\tab 3} and \textbf{\tab 4} are satisfied because neither the \ruleand-rule nor
    the \ruleor-rule apply to any $x \in \mcS$.  Hence $\Lab(x)$ satisfies the
    required properties.
  \item \textbf{\tab 5} is satisfied because, for all $x \in \bfS$, if $\some{R}{D}
    \in \Lab(x)$, then the \ruleex-rule ensures that there is either:
    \begin{enumerate}
    \item an $R$-predecessor $y$ with $D \in \BLab(y) \subseteq \Lab(y)$ (see
      Lemma~\ref{lem:label-subsets}). Because $y$ is a predecessor of $x$ (which
      is an unblocked node), it cannot be blocked, so $y \in \mcS$ and $\tuple{x}{y}
      \in \Edges(R)$.
    \item an $R$-successor $y$ with $D \in \BLab(y) \subseteq \Lab(y)$ (again,
      see Lemma~\ref{lem:label-subsets}). If $y$ is not blocked, then $y \in
      \mcS$ and $\tuple{x}{y} \in \Edges(R)$.  Otherwise, $y$ is blocked
      by some $z$ with $\BLab(y) \subseteq \Lab(z)$.  Hence $D \in \Lab(z)$, $z
      \in \mcS$ and $\tuple{x}{z} \in \Edges(R)$.
    \end{enumerate}
  \item To show that \textbf{\tab 6} is satisfied for all $x \in \mcS$, if $\all{R}{D}
    \in \Lab(x)$ and $\tuple{x}{y} \in \Edges(R)$, we have to consider three
    possible cases:
    \begin{enumerate}
    \item $y$ is an $R$-neighbour of $x$. The \rulefa-rule guarantees $D \in
      \Lab(y)$.
    \item $\Lab\tuple{x}{z} = R$, $y$ blocks $z$. Then by the \rulefa-rule
      we have $D \in \BLab(z)$ and, by the definition of blocking, $\BLab(z)
      \subseteq \Lab(y)$. Hence $D \in \Lab(y)$.
    \item $\Lab\tuple{y}{z} = \Inv(R)$, $x$ blocks $z$. From the definition of
      blocking we have that $\Lab(z)/R = \Lab(x)/R$.
      Hence $\all{R}{D} \in \Lab(z)$ and the \rulefa-rule guarantees $D \in
      \Lab(y)$.
    \end{enumerate}
    In all three cases, $D \in \Lab(y)$ holds.
  \item For \textbf{\tab 7}, let $x \in \mcS$ with $\all{R}{D} \in \Lab(x)$,
    $\tuple{x}{y} \in \Edges(R)$, and $\Tr(R)$. There are three possible cases:
    \begin{enumerate}
    \item $y$ is an $R$-neighbour of $x$. The \rulefatr-rule guarantees
      $\all{R}{D} \in \Lab(y)$.
    \item $\Lab\tuple{x}{z}= R$, $y$ blocks $z$. Then, by the \rulefatr-rule,
      we have $\all{R}{D} \in \BLab(z)$ and, by the definition of blocking,
      $\BLab(z) \subseteq \Lab(y)$. Hence $\all{R}{D} \in \Lab(y)$.
    \item $\Lab\tuple{y}{z} = \Inv(R)$, $x$ blocks $z$. From the definition of
      blocking, we have that $\Lab(z)/R = \Lab(x)/R$. Hence $\all{R}{D} \in
      \Lab(z)$ and the \rulefatr-rule guarantees $\all{R}{D} \in \Lab(y)$.
    \end{enumerate}
  \item \textbf{\tab 8} is satisfied because, for each $\tuple{x}{y} \in \Edges(R)$,
    either:
    \begin{enumerate}
    \item $x$ is an $R$-neighbour of $y$, so $y$ is an $\Inv(R)$-neighbour of
      $x$ and $\tuple{y}{x} \in \Edges(\Inv(R))$.
    \item $\Lab\tuple{x}{z} = R$ and $y$ blocks $z$, so $\Lab\tuple{x}{z} =
      \Inv(\Inv(R))$ and $\tuple{y}{x} \in \Edges(\Inv(R))$.
    \item $\Lab\tuple{y}{z} = \Inv(R)$ and $x$ blocks $z$, so $\tuple{y}{x}
      \in \Edges(\Inv(R))$. \qed
    \end{enumerate}
  \end{itemize}
\end{proof}

We have already mentioned that it is problematic to define a suitable notion of
satisfiability for \si-completion trees (as it would be required by
Theorem~\ref{theo:generic-correctness}) due to blocking. As one can see, the
blocked nodes of the completion tree do not play a r{\^o}le when defining the
tableau, so (hidden) inconsistencies in the labels of indirectly blocked nodes
should not prevent a complete and clash-free tree from being satisfiable. On the
other hand, due to dynamic blocking, blocked nodes may become unblocked during a
run of the algorithm, in which case inconsistencies in these nodes may prevent
the discovery of a complete and clash-free tree. Consequently, for the
completeness proof, we require all nodes, also the blocked ones, to be free from
inconsistencies. Since we have not found a way to uniformly combine these two
different approaches into a single notion of satisfiability for completion
trees, we give a proof for the correctness of the \si-algorithm that does not
rely on Theorem~\ref{theo:generic-correctness}.

\begin{lemma}\label{lemma:si-completeness}
  Let $C$ be an \si-concept in NNF. If $C$ has a tableau, then the
  expansion rules can be applied in such a way that the tableau
  algorithm yields a complete and clash-free completion tree for $C$.
\end{lemma}

\begin{proof}
  Let $\mcT = (\mcS,\TabLab,\Edges)$ be a tableau for $C$. Using $\mcT$, we
  guide the application of the non-deterministic \ruleor-rule such that the
  algorithm yields a completion tree $\Tree$ that is both complete and
  clash-free. The algorithm starts with the initial tree $\Tree$ consisting of a
  single node $x_0$, the root, with $\BLab(x_0) = \Lab (x_0)=\{D\}$.
  
  $\mcT$ is a tableau, hence there is some $s_0\in \mcS$ with $D\in \TabLab
  (s_0)$. When applying the expansion rules to $\Tree$, the application of the
  non-deterministic \ruleor-rule is guided by the labelling in the tableau
  $\mcT$.  We will expand $\Tree$ in such a way that the following invariant
  holds: there exists a function $\pi$ that maps the nodes of $\Tree$ to
  elements of $\mcS$ such that

  \begin{equation}
    \tag{$*$}
    \begin{array}{l}
      \Lab(x) \subseteq \TabLab (\pi(x)) \text{ holds for all nodes  $x$ of \Tree,
        and}\\
      \text{if } \Lab \tuple x y = R \text{ then } \tuple {\pi(x)} {\pi(y)} \in
      \Edges(R) \text{ for all nodes $x,y$ in $\Tree$}.
    \end{array}
  \end{equation}

  \begin{claim}
    If $(*)$ holds for a completion tree $\Tree$ and a rule is applicable to
    $\Tree$, then it can be applied in a way that maintains $(*)$.
  \end{claim}
  
  We have to distinguish the different rules:
  \begin{itemize}
  \item If the \textbf{\ruleand-rule} can be applied to $x$ in $\Tree$ with $D =C_1\sqcap
    C_2 \in \Lab(x) \subseteq \TabLab(\pi(x))$, then $C_1, C_2$ are added to
    $\Lab(x)$. Since $\mcT$ is a tableau, $\{C_1, C_2\} \subseteq
    \TabLab(\pi(x))$, and hence the \ruleand-rule preserves $\Lab(x) \subseteq
    \TabLab(\pi(x))$.
  \item If the \textbf{\ruleor-rule} can be applied to $x$ in $\Tree$ with $D =C_1\sqcup
    C_2 \in \Lab(x) \subseteq \TabLab(\pi(x))$, then there is an $E \in
    \{C_1,C_2\}$ such that $E \in \TabLab(\pi(x))$, and the \ruleor-rule can
    add $E$ to $\Lab(x)$.  Hence the \ruleor-rule can be applied in a way that
    preserves $\Lab(x)\subseteq \TabLab(\pi(x))$.
  \item If the \textbf{\ruleex-rule} can be applied to $x$ in $\Tree$ with $D =
    \some{R}{E} \in \Lab(x) \subseteq \TabLab(\pi(x))$, then $D\in
    \TabLab(\pi(x))$ and there is some $t\in \mcS$ with $\tuple{\pi(x)}{t}\in
    \Edges (R)$ and $E\in \Lab(t)$. The \ruleex-rule creates a new successor $y$
    of $x$ and we extend $\pi$ by setting $\pi' := \pi[y \mapsto t]$, i.e.,
    $\pi'$ is the extension of $\pi$ that maps $y$ to $t$.  It is easy to see
    that the extended completion tree together with the function $\pi'$ satisfy
    $(*)$.
  \item If the \textbf{\rulefa-rule} can be applied to $x$ in $\Tree$ with $D = \all
    {R}{E} \in \Lab(x) \subseteq \TabLab(\pi(x))$ and $y$ is an $R$-neighbour of
    $x$, then $\tuple{\pi(x)}{\pi(y)} \in \Edges (R)$, and thus $E\in
    \Lab(\pi(y))$.  The \rulefa-rule adds $E$ to $\Lab(y)$ and thus preserves
    $\Lab(x)\subseteq \TabLab(\pi(x))$. The deletion of nodes can never violate
    $(*)$.
  \item If the \textbf{\rulefatr-rule} can be applied to $x$ in $\Tree$ with $D = \all
    {R}{E} \in \Lab(x) \subseteq \TabLab(\pi(x))$, $\Tr(R)$, and $y$ an
    $R$-neighbour of $x$, then $\tuple{\pi(x)}{\pi(y)} \in \Edges (R)$, and thus
    $\all{R}{E}\in \TabLab(\pi(y))$.  The \rulefatr-rule adds $\all {R}{E}$ to
    $\Lab(y)$ and thus preserves $\Lab(y)\subseteq \TabLab(\pi(y))$. The
    deletion of nodes can never violate $(*)$.
  \end{itemize}
  
  From this claim, the lemma can be derived as follows. It is obvious that the
  initial tree satisfies $(*)$: since $\mcT$ is a tableau for $C$, there exists
  an element $s_0 \in \mcS$ with $C \in \TabLab(s_0)$ and hence the function
  $\pi$ that maps $x_0$ to $s_0$ satisfies the required properties. The claim
  states that, whenever a rule is applicable, it can be applied in a way that
  preserves $(*)$. Obviously, no completion tree that satisfies $(*)$ 
  contains a clash as this would contradict  \tab 2. Moreover, from
  Lemma~\ref{lemma:si-termination}, we have that the expansion process terminates
  and thus must eventually yield a complete and clash-free completion tree. \qed
\end{proof}

\begin{theorem}\label{theorem:si-decision-procedure}
  The \si-algorithm is a non-deterministic decision procedure for satisfiability
  and subsumption of \si-concepts.
\end{theorem}

\begin{proof}
  Theorem~\ref{theorem:si-decision-procedure} is an immediate consequence of
  Lemma~\ref{lem:si-tableau}, \ref{lemma:si-termination},
  \ref{lemma:si-soundness}, and \ref{lemma:si-completeness}. Moreover, since \si
  is closed under negation, subsumption of concepts $C\sqsubseteq D$ can be
  reduced to the \mbox{(un-)}satisfiability of $C\sqcap \neg D$. \qed
\end{proof}

\subsection{Complexity}

In Lemma~\ref{lemma:si-termination} we have seen that the depths of a completion
tree generated by the \si-algorithm is bounded exponentially in the size of the
input concept. To show that the algorithm can indeed be implemented to run in
polynomial space, we need to carry out a closer analysis of the length of paths
in a completion tree.

In Lemma~\ref{lemma:si-blocking} and \ref{lemma:si-poly_paths}, we establish a
polynomial bound on the length of paths in the completion tree in a manner
similar to that used for the modal logic $\mathsf{S4}$ and $\alc_{R^+}$ in
\cite{HalpernMoses92,Sattler-KI-1996}. It then remains to show that such a tree can
be constructed using only polynomial space.

\begin{lemma}\label{lemma:si-blocking}
  Let $C$ be an \si-concept and $m = \sharp \textit{sub}(C)$, $n > m^3$, and
  $R \in \overline \Roles_C$ be a role with $\Tr(R)$. Let $x_1,\dots,x_n$ be successive nodes of a
  completion tree generated for $C$ by the \si-algorithm with
  $\Lab\tuple{x_i}{x_{i+1}} = R$ for $1 \leq i < n$. If the \rulefa- or the
  \rulefatr-rules cannot be applied to these nodes, then there is a blocked node
  $x_i$ among them.
\end{lemma}

\begin{proof}
  For each node $x$ of the completion tree, $\BLab(x)$ only contains two kinds
  of concepts: the concept that triggered the generation of the node $x$,
  denoted by $C_x$, and concepts which were propagated \emph{down} the
  completion tree by the first alternative of the \rulefa- or \rulefatr-rules.
  Moreover, $\BLab(x) \subseteq \Lab(x)$ holds for any node in the completion
  tree.
  
  Firstly, consider the elements of $\BLab(x_i)$ for $i\geq1$. Let $C_{x_i}$ denote
  the concept that caused the generation of the node $x_i$. Then $\BLab(x_i) -
  \{C_{x_i}\}$ contains only concepts which have been inserted using the
  \rulefa- or the \rulefatr-rule. Let $D \in \BLab(x_i) - \{C_{x_i}\}$.  Then
  either $\all{R}{D} \in \Lab(x_{i-1})$ and the \rulefatr-rule makes sure that
  $\all{R}{D} \in \BLab(x_i)$, or $D$ is already of the from $\all{R}{D'}$ and
  has been inserted into $\BLab(x_i)$ by an application of the \rulefatr-rule to
  $x_{i-1}$. In both cases, it follows that the \rulefa- or the \rulefatr-rule
  yield $D \in \BLab(x_{i+1})$.  Hence we have
  \[ 
  \BLab(x_i) - \{C_{x_i}\} \subseteq \BLab(x_{i+1}) \ \text{for all $1 \leq i <
    n$}, 
  \]
  and, since we have $m$ choices for $C_{x_i}$,
  \[ 
  \sharp \{ \BLab(x_i) \mid 1 \leq i \leq n\} \leq m^2.
  \]
  Secondly, consider $\Lab(x_i)/\Inv(R)$. Again, the \rulefa- and the
  \rulefatr-rules yield
  \[
  \Lab(x_i)/\Inv(R) \subseteq \Lab(x_{i-1})/\Inv(R) \ \text{for all $1 < i \leq
    n$}, 
  \]
  which implies
  \[ 
  \sharp \{ \Lab(x_i)/\Inv(R) \mid 1 \leq i \leq n\} \leq m.
  \]
  Summing up, within $m^3+1$ nodes, there must be at least two nodes $x_j, x_k$
  which satisfy
  \[
  \BLab(x_j) = \BLab(x_k) \quad \text{and} \quad \Lab(x_j)/\Inv(R) =
  \Lab(x_k)/\Inv(R). 
  \] 
  This implies that one of these nodes is blocked by the other. \qed
\end{proof}

We will now use this lemma to give a polynomial bound on the length of paths in
a completion tree generated by the completion rules.

\begin{lemma}\label{lemma:si-poly_paths}
  The paths of a completion tree generated by the \si-algorithm for a
  concept $C$ have a length of at most $m^4$ where $m = \sharp \textit{sub}(C)$.
\end{lemma}

\begin{proof}
  Let $\Tree$ be a completion tree generated for $C$ by the \si-algorithm. For
  every node $x$ of \Tree we define $\ell(x) = \max \{ |D| \mid D \in
  \Lab(x)\}$. If $x$ is a predecessor of $y$ in \Tree, then this implies
  $\ell(x) \geq \ell(y)$.  If not $\Tr(R)$ and $\Lab\tuple{x}{y} = R$, then this
  implies $\ell(x) > \ell(y)$. Furthermore, for $R_1 \neq R_2$ (but possibly
  $R_1 = \Inv(R_2)$), $\Lab\tuple{x}{y} = R_1$ and $\Lab\tuple{y}{z} = R_2$
  implies $\ell(x) > \ell(z)$.
  
  The only way that the maximal length of concepts does not decrease
  is along a pure $R$-path with $\Tr(R)$. However, the \rulefa- and
  the \rulefatr-rule must be applied before the \ruleex-rule may
  generate a new successor.  Together with Lemma~\ref{lemma:si-blocking},
  this guarantees that these pure $R$-paths have a length of at most
  $m^3$.
  
  Summing up, we can have a path of length at most $m^3$ before
  decreasing the maximal length of the concept in the node labels (or
  blocking occurs), which can happen at most $m$ times and thus yields
  an upper bound of $m^4$ on the length of paths in a completion tree.
  \qed
\end{proof}

Note that the extra condition for the \ruleex-rule, which delays its
application until no other rules are applicable, is necessary to prevent the
generation of paths of exponential length. Consider the following example for
some $R$ with $\Tr(R)$:
\begin{align*} 
  C &= \some{R}{D} \sqcap \all{R}{(\some{R}{D})} \sqcap \all{R^{-1}}{A_0} \\
  D &= (\all{R^{-1}}{A_1} \sqcup \all{R^{-1}}{B_1}) \sqcap \dots \sqcap
  (\all{R^{-1}}{A_n} \sqcup \all{R^{-1}}{B_n})
\end{align*}

When started with a root node $x_0$ labelled $\BLab(x_0) = \Lab(x_0) = \{C\}$,
the tableau algorithm generates a successor node $x_1$ with
\[\BLab(x_1) = \{ D, \exists R.D, \all{R}{(\some{R}{D})} \} \]
which, in turn, is capable of generating a further successor $x_2$ with
$\BLab(x_2) = \BLab(x_1)$. Without blocking, this would lead to an infinite path
in the completion tree. Obviously, for $x_1$ and $x_2$, the first part of the
blocking condition is satisfied since $\BLab(x_2) \subseteq \BLab(x_1)$.
However, the second condition causes a problem since, in this example, we can
generate $2^n$ different sets of universal restrictions along $R^{-1}$ for each
node. If we can apply the \ruleex-rule freely, then the algorithm might generate
all of these $2^n$ nodes to find out that (after finally applying the
\rulefatr-rule that causes propagation of the concepts of the form $\all
{R^{-1}} E$ upward in the tree) within the first $n+1$ nodes on this path
there is a blocked one.

\begin{lemma}
  The \si-algorithm can be implemented in \pspace.
\end{lemma}

\begin{proof}
  Let $C$ be the \si-concept to be tested for satisfiability. We can
  assume $C$ to be in NNF because every \si-concept can be turned into
  NNF in linear time. 
  
  Let $m = \sharp \textit{sub}(C)$. For each node $x$ of the completion tree,
  the labels $\Lab(x)$ and $\BLab(x)$ can be stored using $m$ bits for each set.
  Starting from the initial tree consisting of only a single node $x_0$ with
  $\Lab(x_0) = \BLab(x_0) = \{ D \}$, the expansion rules, as given in
  Figure~\ref{fig:si-rules}, are applied. If a clash is generated, then the algorithm
  fails and returns ``$C$ is unsatisfiable''.  Otherwise, the completion tree is
  generated in a depth-first way: the algorithm keeps track of exactly one path
  of the completion tree by memorizing, for each node $x$, which of the
  $\some{R}{D}$-concepts in $\Lab(x)$ successors have yet to be generated. This
  can be done using additional $m$ bits for each node. The ``deletion'' of all
  successors in the \rulefa- or the \rulefatr-rule of a node $x$ is then simply
  realized by setting all these additional bits to ``has yet to be generated''.
  There are three possible results of an investigation of a child of $x$:

  \begin{itemize}
  \item A clash is detected. This stops the algorithm with ``$C$ is
    unsatisfiable''.
  \item The \rulefa- or the \rulefatr-rule leads to an increase of
    $\Lab(x)$. This causes reconsideration of all children of $x$,
    re-using the space used for former children of $x$.
  \item Neither of these first two cases happens. We can then forget
    about this subtree and start the investigation of another child
    of $x$. If all children have been investigated, we consider $x$'s
    predecessor.
  \end{itemize}

  
  Proceeding like this, the algorithm can be implemented using $2m +
  m$ bits for each node, where the $2m$ bits are used to store the two
  labels of the node, while $m$ bits are used to keep track of the
  successors already generated. Since we reuse the memory for the
  successors, we only have to store one path of the completion tree at
  a time. From Lemma~\ref{lemma:si-poly_paths}, the length of this path
  is bounded by $m^4$. Summing up, we can test for the existence of a
  completion tree using $\mathcal{O}(m^5)$ bits.
  
  Unfortunately, due to the \ruleor-rule, the \si-algorithm is a
  non-deterministic algorithm. However, Savitch's theorem \cite{Savitch} tells
  us that there is a deterministic implementation of this algorithm using at
  most $\mathcal{O}(m^{10})$ bits, which is still a polynomial bound. \qed
\end{proof}

Since \alc is a fragment of \si, satisfiability of \si-concepts is \pspace-hard,
which yields:

\begin{theorem}
  Satisfiability and subsumption of \si-concepts is \pspace-complete.
\end{theorem}

There is an immediate optimization of the algorithm which has been omitted for
the sake of the clarity of the presentation. We have only disallowed the
application of the \ruleex-rule to a blocked node, which is sufficient to
guarantee the termination of the algorithm. It is also possible to disallow the
application of more rules to a blocked node without violating the soundness or
the completeness of the algorithm, if the notion of blocking is slightly
adapted.  It then becomes necessary to distinguish directly and indirectly
blocked nodes.  More details can be found in~\cite{HorrocksSattler-DL-1998}. The
technique presented there will stop the expansion of a blocked node earlier
during the runtime of the algorithm and hence will save some work.  



\section{Adding Role Hierarchies and Qualifying Number Restrictions: \shiq}

\label{sec:shiq-worst-case}
 
\def\R{\mcR}

In this section, we study aspects of reasoning with the DL \shiq, i.e., \si
extended with qualifying number restrictions and role hierarchies. Qualifying
number restrictions have already been introduced in Chapter~\ref{chap:alcq} and
require no further discussion. Role hierarchies
\citeyear{HorrocksGough-DL-1997}, which have already been present in early DL
systems like \textsc{back} \cite{TUB-FB13//TUB-FB13-KIT-78} , allow to express inclusion
relationships between roles. For example, role hierarchies can be used to state
that the role $\hasChild$ is a sub-role of $\hasOffspring$.  This makes it
possible to infer that the child of someone whose offsprings are all rich must
also be rich.

The combination of role hierarchies with transitive roles is particularly
interesting because it allows to capture various aspects of part-whole relations
\cite{Sattler-ECAI-2000}. It is also interesting because it is sufficiently
expressive to \emph{internalise} general TBoxes \cite{Baader91c,Schi91,BBNNS93},
i.e., it allows for a reduction of concept satisfiability \wrt general TBoxes to
pure concept satisfiability---always an indication for high expressive power of
a Description Logic.

After defining syntax and semantics of \shiq, we show how internalisation of
general axioms can be accomplished. We then determine the worst-case complexity
of satisfiability for \shiq-concepts as \exptime-complete even if numbers in the
input are in binary coding. This is achieved by a reduction from \shiq to
\alcqir, where role conjunction and general TBoxes are used to simulate role
hierarchies and transitive roles.

While this reduction helps to determine the exact worst-case complexity or the
problem, one cannot expect to obtain an efficient algorithm from it. The reason
for this is that it relies on the highly inefficient automata construction used
to prove Theorem~\ref{theo:alcqir-tbox--exptime-complete}. To overcome this
problem, we present a tableau algorithm that decides satisfiability of \shiq
concepts. In the worst case, this algorithm runs in 2-\nexptime.
Yet, it is  amenable to optimizations and is the basis of the highly
optimized DL system \iFaCT \cite{horrocks99:_fact}, a offspring of the \FaCT
system \cite{Horrocks98c}, which exhibits good performance in system
comparisons~\cite{MassacciDonini-Tableaux2000,Horrocks-Tableaux-2000}.

\subsection{Syntax and Semantics}

\begin{definition}[Syntax and Semantics of \shiq]\label{def:shiq-syntax}
  Let \Roles be a set of atomic \emph{role names}, and $\RolesTrans \subseteq
  \Roles$ a set of \emph{transitive role names}. The set of \shiq-roles is
  defined as the set of \si-roles by $\overline \Roles := \Roles \cup \{ R^{-1}
  \mid R \in \Roles \}$. A role $R \in \overline \Roles$ is called
  \emph{transitive} iff $R \in \RolesTrans$ or $\Inv(R) \in \RolesTrans$.
  
  \index{shiq@\shiq}
  \index{00nrplus@$\RolesTrans$}
  \index{00nr@$\overline{\Roles}$}
  \index{transitive role}
  \index{role!transitive}
  
  A \iemph{role inclusion axiom} is of the form $R\sqsubseteq S$,
  \index{00rsqsubseteqs@$R\sqsubseteq S$} for two \shiq-roles $R$ and $S$. A set
  of role inclusion axioms $\R$ is called a \emph{role hierarchy}. We define
  $\sss$ \index{00sss@$\sss$} as the transitive-reflexive of the relation
  $\R\cup\{ \Inv(R)\sqsubseteq \Inv(S)\mid R\sqsubseteq S \in \R\}$. If $R \sss
  S$ then $R$ is called a \emph{sub-role} of $S$ and $S$ is called a
  \emph{super-role} of $R$ (\wrt $\R$).
  
  \index{sub-role} \index{role!sub-role} \index{super-role} \index{role!super-role}

  A role $R$ is called \emph{simple} with respect to $\R$
  iff $R$ does not have a transitive sub-role.
  
  \index{role!simple role}

  Let \Names be a set of \emph{concept names}.  The set of
  \emph{\shiq-concepts} is built inductively from these using the
  following grammar, where $A \in \Names$, $n \in \N$, $R \in
  \overline \Roles$ is an arbitrary role, and $S \in \overline
  \Roles$ is a simple role:
  \[
  C ::= A  \bnfor \neg C \bnfor C_1 \sqcap C_2 \bnfor C_1 \sqcup C_2 \bnfor
  \all R C \bnfor \some R C \bnfor  \qnrgeq n S C \bnfor \qnrleq n S C .
  \]
  
  An interpretation $\I = ( \Delta^\I, \cdot^\I)$ consists of a non-empty set
  $\Delta^\I$ and a valuation $\cdot^\I$ that maps every concept name $A$ to a
  subset $A^\I \subseteq \Delta^\I$ and every role name $R$ to a binary relation
  $R^\I \subseteq \Delta^\I \times \Delta^\I$ with the additional property that
  every transitive role name $R \in \RolesTrans$ is interpreted by a
  transitive relation.
  
  Such an interpretation is inductively extended to arbitrary \shiq-concepts in
  the usual way. (See Definition~\ref{def:alcqir-syntax+semantics}).
  
  An interpretation $\I$ \emph{satisfies a role hierarchy} $\R$
  iff $R^\I \subseteq S^\I$ for each $R \sqsubseteq S \in \R$; we denote this
  fact by $\I \models \R$ and say that $\I$ is a model of $\R$.
  
  \index{00imodelsr@$\I \models \R$}
  
  A concept $C$ is \emph{satisfiable} with respect to a role hierarchy $\R$ iff
  there is some interpretation $\I$ such that $\I \models \R$ and $C^\I \neq
  \emptyset$.  Such an interpretation is called a \emph{model of} $C$ \wrt $\R$.
  A concept $D$ \emph{subsumes} a concept $C$ \wrt $\R$ (written $C
  \sqsubseteq_{\R} D$) iff $C^\I \subseteq D^\I$ holds for each model $\I$ of
  $\R$.  For an interpretation $\I$, an individual $x \in \Delta^\I$ is called
  an {\em instance} of a concept $C$ iff $x\in C^\I$.
  
  \index{satisfiability!of a concept!w.r.t.\ a role hierarchy}
  \index{satisfiability!of a concept!w.r.t.\ a role hierarchy and a TBox}

  Satisfiability of concepts \wrt TBoxes and role hierarchies is defined in
  the usual way. \eod
\end{definition}

As shown in \cite{HorrocksSattlerTobies-LPAR-99}, the restriction of qualifying
number restrictions to simple roles is necessary to maintain decidability of
\shiq.  Without this restriction, it is possible to reduce an undecidable tiling
problem \cite{berger66:_undecidability_dominoe_problem} to concept
satisfiability.  

\subsubsection{Internalization of TBoxes}

An evidence of \shiq's high expressivity is the fact that it allows for the
\emph{internalization} of TBoxes using a ``universal'' role $U$, that is, a
transitive super-role of all relevant roles. \index{internalization of general axioms}

\begin{lemma}\label{lem:shiq-internalisation}
  Let $C$ be a \shiq-concept, $\R$ a role hierarchy, and $\mcT$ a
  \shiq-TBox. Define
  \[
  C_\mcT:= \mybigsqcap_{C_i\sqsubseteq D_i \in\mcT} \neg C_i\sqcup D_i,
  \]
  and let $U \in \RolesTrans$ be a transitive role that does not occur in
  $\mcT,C$, or $\R$. We set
  \[
  \R_U := \R \cup \{ R\sqsubseteq U, \Inv(R)\sqsubseteq U \mid \text{$R$
    occurs in $\mcT,C$, or $\R$} \}.
  \]
  Then $C$ is satisfiable w.r.t. $\mcT$ and $\R$ iff
  $C\sqcap C_\mcT \sqcap \all{U}{C_\mcT}$
  is satisfiable w.r.t. $\R_U$. 
\end{lemma}

Note that augmenting $\R$ to obtain $\R_U$ in this manner does not turn simple
roles into non-simple roles. The proof of Lemma~\ref{lem:shiq-internalisation}
is similar to the ones that can be found in \cite{Schi91,Baader91c}. Most
importantly, it must be shown that, (a) if a \shiq-concept $C$ is satisfiable
with respect to a TBox $\mcT$ and a role hierarchy $\R$, then $C,\mcT$ have a
\emph{connected} model, i.e., a model where all elements are connected by roles
occurring in $C$ or $\mcT$, and (b) if $y$ is reachable from $x$ via a role path
(possibly involving inverse roles), then $\tuple{x}{y} \in U\ifunc$.  These are
easy consequences of the semantics of \shiq and the definition of $\R_U$. As a
corollary, we get:

\begin{theorem}\label{theorem:shiq-internalisation}
  Satisfiability and subsumption of \shiq-concepts \wrt general TBoxes and role
  hierarchies are polynomially reducible to (un)sat\-is\-fi\-abil\-i\-ty of
  \shiq-concepts \wrt role hierarchies.
\end{theorem}




\subsubsection{Cycle-free Role Hierarchies}

In what we have said so far, a role hierarchy $\R$ may contain a \emph{cycle},
i.e., there may be roles $R,S \in \Roles$ with $R \neq S$, $S \sss R$, and $R
\sss S$. Such cycles would add extra difficulties to the following
considerations. The next lemma shows that, w.l.o.g., we only need to
consider role hierarchies that are \emph{cycle-free}.

\index{role hierarchy!cycle in a}
\index{role hierarchy!cycle-free}

\begin{lemma}\label{lem:shiq-rolehier-no-cycle}
  Let $C$ be a \shiq-concept and $\R$ a role hierarchy. There exists a
  \shiq-concept $C'$ and role hierarchy $\R'$ polynomially computable from $C,
  \R$ such that $\R'$ is cycle free and $C$ is satisfiable \wrt $\R$ iff
  $C'$ is satisfiable \wrt $\R'$.
\end{lemma}

\begin{proof}
  The set $\R$ can be viewed as a directed graph $G = (V,E)$ with vertices $V =
  \{ R \mid R \text{ occurs in $\R$ } \}$ and $E = \{ (S,R) \mid S \sqsubseteq R
  \in \R \}$. The strongly connected components of $G$ can be calculated in
  quadratic time. For every non-trivial strongly connected component $\{
  R_1, \dots R_k \}$, select an arbitrary $S \in \{ R_1, \dots, R_k \}$ such
  that $S \in \RolesTrans$ if $\{ R_1, \dots, R_k \} \cap \RolesTrans \neq
  \emptyset$. For every $1 \leq i \leq k$, replace $R_k$ in $C$ and $\R$ by $S$.
  The results of this replacement are called $C'$ and $\R'$, respectively. It is
  obvious that these can be obtained from $C,\R$ in polynomial time and that
  $(\R')^+$ is cycle-free.
  
  For every $\I$ with $\I \models \R$, it easy to see that, for every strongly
  connected component $\{ R_1, \dots, R_k \}$ of $G$, $R_i^\I = R_j^\I$ holds for
  every $1 \leq i,j\leq k$ and, if $\{ R_1, \dots, R_k \} \cap \RolesTrans \neq
  \emptyset$, then $R_i^\I$ is transitive for every $1 \leq i \leq k$. Hence, $C$
  is satisfiable \wrt $\R$ iff $C'$ is satisfiable \wrt $\R'$ \qed
\end{proof}

Thus, from now on, we only consider cycle-free role hierarchies.

\subsection{The Complexity of Reasoning with \shiq}

So far, the exact complexity of reasoning with \shiq has been an open problem.
It was clear that it is \exptime-hard as a corollary of
Theorem~\ref{theo:alc-general-tboxes-exptime-complete} and this also holds for
pure concept satisfiability by Theorem~\ref{theorem:shiq-internalisation}.
Following De Giacomo's \exptime-completeness result for the DL \ciq
\citeyear{DeGiacomo95a}, it has been conjectured that the problem can be solved
in \exptime. Yet, the results from \cite{DeGiacomo95a} are valid for unary
coding of numbers only and do not easily transfer to \shiq because of the
presence of role hierarchies. Here, we verify the conjecture by giving a
polynomial reduction of \shiq-satisfiability to satisfiability of
\alcqir-concepts \wrt general TBoxes.  In
Theorem~\ref{theo:alcqir-tbox--exptime-complete}, we have already shown that the
latter problem can be solved in \exptime, also for the case of binary coding of
numbers in the input. Our reduction combines two techniques:
\begin{enumerate}
\item To deal with a role hierarchy $\R$, we replace every role $R$ by the
  role conjunction
  \[
  \Ru = \mybigsqcap_{R \sss S} S .
  \]
  Note that, since $R \sss R$, $R$ occurs in $\Ru$.  This usage of role
  conjunction to express role hierarchies is common knowledge in the DL
  community but, to the best of our knowledge, there exists no publication that
  explicitly mentions it.
\item For transitive roles, we shift the technique employed in the
  \si-algorithm to deal with transitive roles into a set of TBox axioms. For
  \si, transitive roles were dealt with by explicitly propagating assertions of
  the form $\all R D$ to all $R$-successors of a node $x$ using the
  \rulefatr-rule. This makes it possible to turn an \si-tableau into a model
  (which must interpret transitive roles with transitive relations) by
  transitively closing the role relations explicitly asserted in the tableau.
  Here, we achieve the same effect using a set of TBox axioms. A similar idea
  can be found in \cite{nivelle00:_trans_s4_k5_gf}.
\end{enumerate}

The usage of $\Ru$ to capture the role hierarchies is motivated by the following
observations.

\begin{lemma}\label{lem:shiq-role-hier}
  Let $\R$ be a role hierarchy. If $S \sss R$ then $(\Su)^\I \subseteq (\Ru)^\I$
  for every interpretation $\I$. Also, for every interpretation $\I$ with $\I
  \models \R$, $(\Ru)^\I = R^\I$
\end{lemma}

\begin{proof}
  If $S \sss R$ then $\{ S' \mid S \sss S'\} \supseteq \{ S' \mid R \sss S' \}$
  and hence $(\Su)^\I \subseteq (\Ru)^\I$. If $\I \models \R$ then $R^\I
  \subseteq S^\I$ for every $S$ with $R \sss S$. Hence $(\Ru)^\I = R^\I$. \qed
\end{proof}

The reduction that captures the transitive roles involves concepts from the
following set:

\begin{definition}\label{def:shiq-closure}
  For a \shiq concept $C$ and a role hierarchy $\R$, the set
  $\clos(C,\R)$ is the smallest set $X$ of \shiq-concepts that satisfies the
  following conditions:
  \begin{itemize}
  \item $C \in X$,
  \item $X$ is closed under sub-concepts and $\nneg$,\footnote{Like in
      Chapter~\ref{chap:alcq}, with $\nneg D$ we
      denote $\NNF(\neg D)$.} and
  \item if $\all R D \in X$, $T \sss R$, and $\Tr(T)$, then $\all T D \in X$. \eod
  \end{itemize}
  \index{00closcr@$\clos(C,\R)$}
  \index{00nnegc@$\nneg C$}
\end{definition}

It is easy to see that, for a concept $C$ and a role hierarchy $\R$, the set
$\clos(C,\R)$ is ``small'':

\begin{lemma}
  For a \shiq-concept $C$ in NNF and a role hierarchy $\mcR$,
  $\sharp \clos(C,\R) = \mcO(|C| \times |\mcR|)$.
\end{lemma}

\begin{proof}
  Like in the proof of Lemma~\ref{lem:alcq-closure-linear}, it is easy to see that
  the smallest set $X'$ that contains $C$ and is closed under sub-concepts and
  $\nneg$ contains $\mcO(|C|)$ concepts. For \shiq, we additionally have to add
  concepts $\all T D$ to $X'$ if $\all R D \in X'$ with $T \sss R$ and $\Tr(T)$,
  and again close $X'$ under sub-concepts and $\nneg$ to obtain $\clos(C,\R)$..
  This may yield at most two concepts for every concept in $X'$ and every role
  $T$ that occurs in $\mcR$ because, for every such concept $\all R D \in X'$,
  it suffices to add $\all T D$ and $\some T \nneg D$. Since it is a sub-concept
  of $all R D$, $D$ does not need to be reconsidered in the closure process. \qed
\end{proof}

We now formally introduce the employed reduction from \shiq to \alcqir.

\begin{definition}\label{def:transitive-roles-reduction}
  Let $C$ be a \shiq-concept in NNF and $\R$ a role hierarchy. For every concept
  $\all R D \in \clos(C)$ let $X_{R,D} \in \Names$ a be unique concept name that
  does not occur in $C$. We define the function $\cdot^\trans$ inductively on
  the structure of concepts (in NNF) by setting
  \begin{align*}
    A^\trans & = A \text{ for all $A \in \Names$}\\
    (\neg A)^\trans & = \neg A \text{ for all $A \in \Names$}\\
    (C_1 \sqcap C_2)^\trans & = C_1^\trans \sqcap C_2^\trans\\
    (C_1 \sqcup C_2)^\trans & = C_1^\trans \sqcup C_2^\trans\\
    \qnrgleq n R D^\trans & = \qnrgleq n \Ru {D^\trans}\\
    (\all R D)^\trans & = X_{R,D} \\
    (\some R D)^\trans & = \neg X_{R,\nneg D}
  \end{align*}
  The TBox $\mcT_C$ is defined by
  \begin{align*}
    \mcT_C = & \{ X_{R,D} \doteq \all \Ru {D^\trans} \mid \all R D \in
    \clos(C,\R) \} \cup \\
    & \Bigl \{ X_{R,D} \sqsubseteq 
       \mybigsqcap_{T \sss R, \Tr(T)} \all \Tu {X_{T,D}} \Bigm | \all R D \in
      \clos(C,\R) \Bigr \}
  \end{align*} \eod
\end{definition}
                                
\begin{lemma}\label{lem:shiq-to-alcqir}
  Let $C$ be a \shiq-concept in NNF, $\R$ a role hierarchy, and $\cdot^\trans$
  and $\mcT_C$ as defined in Definition~\ref{def:transitive-roles-reduction}.
  $C$ is satisfiable \wrt $\R$ iff the \alcqir-concept $C^\trans$ is satisfiable
  \wrt $\mcT_C$.
\end{lemma}

\begin{proof}
  For the \emph{only-if}-direction, let $C$ be a \shiq-concept and $\R$ a set of
  role axioms. Assume that $\I$ is a \mbox{(\shiq-)}model of $C$ \wrt $\R$. Let $\mcX
  = \{ X_{R,D} \mid \all R D \in \clos(C,\R) \}$ bet the set of freshly introduced
  concept names from Definition~\ref{def:transitive-roles-reduction}.

  We will construct an \alcqir-model $\I'$ for $C^\trans$ and $\mcT_C$ from $\I$
  by setting
  \[
  (X_{R,D})^{\I'} = (\all R D)^\I 
  \]
  for every $X_{R,D} \in \mcX$, and maintaining the interpretation of all other
  concept and role names.
  
  \begin{claim}\label{claim:shiq-1}
    For all $D \in \clos(C,\R)$, $(D^\trans)^{\I'} = D^\I$.
  \end{claim}
  
  This claim is proved by induction on the structure of concepts. For the base
  case $D = A \in \Names \setminus \mcX$, $A^\trans = A$ holds, and hence $(A^\trans)^{\I'} = A^\I$. For
  all other cases, except for $D = \all R E$ and $D = \some R E$, the claim follows
  immediately by induction because $R^\I = (\Ru)^\I = (\Ru)^{\I'}$ for every $R
  \in \Roles$, since $\I \models \R$ and because of Lemma~\ref{lem:shiq-role-hier}.
   
  For $D = \all R E$, $D^\trans = X_{R,E}$ and by construction of $\I'$,
  $(X_{R,E})^{\I'} = (\all R E)^\I$.  For $D = \some R E$, $D^\trans = \lnot
  X_{R,\nneg E}$ and
  \[
  (D^\trans)^{\I'} = \Delta^{\I'} \setminus (X_{R,\nneg E})^{\I'} = \Delta^{\I}
  \setminus (\all R {\nneg E})^\I = (\some R E)^\I ,
  \]
  which finishes the proof of the claim.
   
  In particular, since $C^\I \neq \emptyset$, also $(C^\trans)^{\I'} \neq
  \emptyset$. It remains to show that $\I' \models \mcT_C$ holds. For the first
  set of axioms, this holds because 
  \[
  (X_{R,D})^{\I'} = (\all R D)^\I = (\all \Ru {D^\trans})^{\I'} ,
  \]
  since $R^\I = (\Ru)^{\I'}$ because of Lemma~\ref{lem:shiq-role-hier}, and $D^\I =
  (D^\trans)^{\I'}$ due to Claim~\ref{claim:shiq-1}.
  
  For the second set of axioms, let $T$ be a role with $T \sss R$ and $\Tr(T)$.
  Then
  \[
  (X_{R,D})^{\I'} \subseteq (\all \Tu {X_{T,D}})^{\I'} ,
  \]
  unless there is an $x \in (X_{R,D})^{\I'}$ and an $y \in \Delta^{\I'}$ with
  $(x,y) \in (\Tu)^{\I'} = T^\I$ and $y \not \in (X_{T,D})^{\I'} = (\all T
  D)^\I$.  This implies the existence of an element $z \in \Delta^\I$ with
  $(y,z) \in T^{\I}$ and $z \not \in D^\I$. Since $(x,y) \in T^\I$ and $(y,z)
  \in T^\I$, transitivity of $T^\I$ implies $(x,z) \in T^\I \subseteq R^\I =
  (\Ru)^{\I'}$. Thus $(x,z) \in (\Ru)^{\I'}$ and $z \not \in (D^\trans)^{\I'}$
  because $D^\I = (D^\trans)^{\I'}$. This implies $x \not \in (\all \Ru
  {D^\trans})^{\I'}$, which is a contradiction because $(\all \Ru
  {D^\trans})^{\I'} = (\all R D)^\I = (X_{R,D})^{\I'}$ and $x \in
  (X_{R,D})^{\I'}$.  Summing up, $(X_{R,D})^{\I'}$ is contained in the
  interpretation of every conjunct that appears on the right-hand side of the
  axioms, and hence
  \[
  (X_{R,D})^{\I'} \subseteq \Bigl ( \mybigsqcap_{T \sss R, \Tr(T)} \all \Tu
    {X_{T,D}} \Bigr )^{\I'} .
  \]
  This holds for every $X_{R,D} \in \mcX$ and hence $\I' \models \mcT_C$.  Thus,
  we have shown that $C^\trans$ is satisfiable \wrt $\mcT_C$.

  For the \emph{if}-direction, let $\I$ be an interpretation with $(C^\trans)^\I
  \neq \emptyset$ and $\I \models \mcT_C$. From $\I$ we construct an
  interpretation $\I'$ such that $C^{\I'} \neq \emptyset$ and $\I' \models
  \R$. To achieve the latter, we define $R^{\I'}$ as follows:
  \[
  R^{\I'} = \begin{cases}
    ((\Ru)^\I)^+ & \text{if $\Tr(R)$}, \\
    (\Ru)^\I \cup \displaystyle \bigcup_{S\sss R, S\neq R} S^{\I'} &
    \text{otherwise} . 
  \end{cases}
  \]
  Since $\R$ is cycle-free, $R^{\I'}$ is well-defined for every $R$ and it is
  obvious that, for every $R$ with $\Tr(R)$, $R^{\I'}$ is transitive.  First, we
  check that $\I'$ indeed satisfies $\R$. 

  \begin{claim}\label{claim:shiq-2}
    $\I' \models S \sqsubseteq  R$ for every $S \sss R$.
  \end{claim}
  
  If $\lnot \Tr(R)$, this is immediate from the construction. If $\Tr(R)$, then the
  proof is more complicated. It is by induction on the number $\| S \| = \sharp
  \{ S' \mid S' \sss S, S' \neq S \}$, where the case for $\Tr(S)$ does not make
  use of the induction hypothesis.
  \begin{itemize}
  \item If $\| S \| = 0$ and $\lnot \Tr(S)$, then $S^{\I'} = (\Su)^\I \subseteq
    (\Ru)^\I \subseteq ((\Ru)^\I)^+$ due to Lemma~\ref{lem:shiq-role-hier}
    because $S \sss R$.
  
  \item If $\| S \| = n \geq 0$ and $\Tr(S)$, then $(\Su)^\I \subseteq (\Ru)^\I$
    since $S \sss R$, and hence $S^{\I'} = ((\Su)^\I)^+ \subseteq ((\Ru)^\I)^+ =
    R^{\I'}$.
    
  \item If $\| S \| = n > 0$ and $\lnot \Tr(S)$, then $S^{\I'} = (\Su)^\I \cup
    \bigcup_{S' \sss S, S' \neq S} S'^{\I'}$. For every $S' \sss S$ with $S'
    \neq S$, $\| S' \| < \| S \|$ because $\R$ is cycle-free. Since $S' \sss R$
    holds by the definition of $\sss$, induction yields $(S')^{\I'}
    \subseteq R^{\I'}$. Also, since $S \sss R$, $(\Su)^\I \subseteq
    (\Ru)^\I$ and hence $S^{\I'} \subseteq R^{\I'}$.
  \end{itemize}

  \begin{claim}\label{claim:shiq-3}
    If $(x,y) \in R^{\I'}$, then $(x,y) \in (\Ru)^\I$ or there exists a role $T
    \sss R$ with $\Tr(T)$ and a path $x_0, \dots, x_k$ such that $k > 1, x =
    x_0, y=x_k$, and $(x_i,x_{i+1}) \in (\Tu)^\I$ for $0 \leq i < k$.
  \end{claim}
  
  Again, then proof is by induction on $\| \cdot \|$ and, if $\Tr(R)$ holds, then we do
  not need to make use of the induction hypothesis. 

  \begin{itemize}
  \item If $\| R \| = 0$ and $\lnot \Tr(R)$, then $R^{\I'} = (\Ru)^\I$ and thus
    $(x,y) \in R^{\I'}$ implies $(x,y) \in (\Ru)^\I$.
  \item If $\| R \| = n \geq 0$ and $\Tr(R)$, then $R^{\I'} = ((\Ru)^\I)^+$. If
    $(x,y) \in (\Ru)^\I$, then we are done. Otherwise, there exists a path $x_0,
    \dots, x_k$ with $k>1, x = x_0, y=y_k$, and $(x_i, x_{i+1}) \in
    (\Ru)^\I$. Also, by definition, $R \sss R$.
  \item If $\| R \| = n > 0$ and $\lnot \Tr(R)$, then either $(x,y) \in (\Ru)^\I$
    or $(x,y) \in S^{\I'}$ for some $S \sss R$ with $S \neq R$. In the latter
    case, $\| S \| < \| R \|$ and, by induction, either $(x,y) \in (\Su)^\I
    \subseteq (\Ru)^\I$, or there exists a role $T \sss S$ with $\Tr(T)$ and a path
    $x_0, \dots, x_k$ with $k>1, x = x_0, y=x_k$, and $(x_i,x_{i+1}) \in (\Tu)^\I$ for $0
    \leq i  < k$. Since $T \sss S \sss R$, also $T \sss R$.
  \end{itemize}

  \begin{claim}\label{claim:shiq-4}
    For a simple role $R$, $R^{\I'} = (\Ru)^\I$.
  \end{claim}
  
  The proof is by induction on the number of sub-roles of $R$. If $R$ is simple
  and has no sub-roles, then $R^{\I'} = (\Ru)^\I$ holds by definition of $\I'$.
  If $R$ is simple, then every role $S$ with $S \sss R$ and $S \neq R$ has less
  sub-roles than $R$ because $\R$ is cycle free, and must be simple because
  otherwise $R$ would not be simple. Hence, the induction hypothesis is
  applicable to each such $S$, which yields $S^{\I'} = (\Su)^\I$. Also, since $S
  \sss R$, $S^{\I'} = (\Su)^\I \subseteq (\Ru)^\I$ holds by
  Lemma~\ref{lem:shiq-role-hier} and hence $R^{\I} = (\Ru)^{\I'}$.

  \begin{claim}\label{claim:shiq-5}
    $D^{\I'} = (D^\trans)^\I$ for every $D \in \clos(C,\R)$.
  \end{claim}
  
  The proof is by induction on the value $\lceil \cdot \rceil$ of concepts in
  $\clos(C,\R)$, where the function $\lceil \cdot \rceil$ 
  \index{00cnorm@$\lceil C \rceil$} is defined by
  \[
  \lceil D \rceil = \begin{cases}
    2 \times \| D \| + 1 & \text{if $D = \some R E$}\\
    2 \times \| D \| & \text{otherwise}
  \end{cases}
  \]
  where the definition of the norm $\| \cdot \|$ \index{00normc@$"\"|C"\"|$} of a
  \shiq-concept is similar to the definition for \alcq extended to universal and
  existential restrictions:
  \[
  \begin{array}{lclcl}
    \|A\| & := & \|\lnot A\| & := & 0 \quad \text{for $A\in \Names$}\\ 
    \|C_1 \sqcap C_2 \| & := &  \| C_1 \sqcup C_2\| & := & 
    1+\|C_1\|+\|C_2\|\\
    \| \forall R.C \| & := & \| \exists R.C \| & := & 1+ \| C \|\\
    \| (\bowtie \; n \; S \; C) \| & & & := & 1 +\| C \|
  \end{array} 
  \] 
  
  The purpose of this (seemingly rather strange) definition of $\lceil \cdot
  \rceil$ is to reduce the case for an existential restriction to
  its dual universal restriction. Except for existential, universal, and number
  restrictions, all cases are straightforward.
  \begin{itemize}
  \item If $D = \all R E$, then $D^\trans = X_{R,E}$. If $x \not \in
    (X_{R,E})^\I$ then, since $\I \models X_{R,E} = \all \Ru {E^\trans}$, also
    $x \not \in \all \Ru {E^\trans}$. By induction, $(E^\trans)^\I = E^{\I'}$ and
    $(\Ru)^\I \subseteq R^{\I'}$, and hence $x \not \in (\all R E)^{\I'}$.
    
    If $x \in (X_{R,E})^\I$ and $(x,y) \in R^{\I'}$, then, by
    Claim~\ref{claim:shiq-2}, there are two possibilities.
    \begin{itemize}
    \item If $(x,y) \in (\Ru)^\I$, then $y \in (E^\trans)^\I$ holds because $\I
      \models X_{R,E} = \all \Ru {E^\trans}$. By induction, $(E^\trans)^\I =
      E^{\I'}$, and hence $y \in E^{\I'}$.
    \item There is a role $T \sss R$ with $\Tr(T)$ and a path $x_0, \dots, x_k$
      with $k>1, x = x_0, y=x_k$, and $(x_i,x_{i+1}) \in (\Tu)^\I$ for $0 \leq i
      < k$. Since $\I \models X_{R,E} \sqsubseteq \all \Tu {X_{T,E}}$ and $\I
      \models X_{T,E} = \all \Tu {X_{T,E}}$, we have $x_i \in (X_{T,E})^\I$ for
      every $1 \leq i \leq k$, and $x_{k-1} \in (X_{T,E})^\I$ in particular.
      Since $\I \models X_{T,E} \doteq \all \Tu {E^\trans}$, it follows that $y = x_k
      \in (E^\trans)^\I$ and, by induction, $y \in E^{\I'}$.
    \end{itemize}
    In any case, we have shown that $y \in E^{\I'}$ and, since $y$ has been chosen
    arbitrarily with $(x,y) \in R^{\I'}$, $x \in (\all R E)^{\I'}$ holds.
  \item If $D = \some R E$, then $D^\trans = \lnot X_{R,\nneg E}$ and 
    \[
    D^{\I'} = \Delta^{\I'} \setminus (\all R {\nneg E})^{\I'} = \Delta^{\I}
    \setminus (X_{R,\nneg E})^\I = (\lnot X_{R,\nneg E})^\I ,
    \]
    where $(\all R {\nneg E})^{\I'} = (X_{R,\nneg E})^\I$ follows by induction
    since $\lceil \all R {\nneg E} \rceil < \lceil \some R E \rceil$.
  \item If $D = \qnrgleq n R E$, then $D^\trans = \qnrgleq n \Ru {E^\trans}$ and
    $R$ is simple. By Claim~\ref{claim:shiq-4}, $R^{\I'} = (\Ru)^\I$ holds.
    Also, by induction, $(E^\trans)^\I = E^{\I'}$ and hence $D^{\I'} =
    (D^\trans)^\I$.
  \end{itemize}
  
  This finishes the proof of Claim~\ref{claim:shiq-5}, which yields $C^{\I'} =
  (C^\trans)^{\I} \neq \emptyset$. Since we have already shown that $\I' \models
  \R$, we have proved satisfiability of $C$ \wrt $\R$. \qed
\end{proof}

Since the reduction from Definition~\ref{def:transitive-roles-reduction} is
obviously polynomial in $|C|$ and $|\R|$, Lemma \ref{lem:shiq-to-alcqir}
together with Theorem~\ref{theo:alcqir-tbox--exptime-complete} and
Theorem~\ref{theorem:shiq-internalisation} yield the following corollary.

\begin{corollary}\label{cor:shiq-exptime}
  The following problems are \exptime-complete even in the case of binary coding
  of numbers in the input:
  \begin{itemize}
  \item Satisfiability and subsumption of \shiq-concepts \wrt role hierarchies.
  \item Satisfiability and subsumption of \shiq-concepts \wrt general TBoxes and
    role hierarchies.
  \end{itemize}
\end{corollary}

Obviously, the reduction from Definition~\ref{def:transitive-roles-reduction}
works also for \shiq-ABoxes and so, from
Theorem~\ref{theo:alcqir-kb-exptime-complete}, we get that also \shiq-knowledge
bases can be handled in \exptime.

\begin{corollary}\label{cor:shiq-kb-exptime}
  Knowledge base satisfiability and instance checking for \shiq are
  \exptime-complete, even in the case of binary coding of numbers in the input.
\end{corollary}

Finally, it is easy to see how to extend the reduction from
Definition~\ref{def:transitive-roles-reduction} to \shiqo, \index{shiqo@\shiqo}
the extension of \shiq with nominals. Simply set $i^\trans = i$ for every
individual $i \in \Individuals$. Since \shiqo strictly contains \alcqio, we get
the following.

\begin{corollary}\label{cor:shiqo-nexptime}
  Concept satisfiability, satisfiability \wrt general TBoxes, and knowledge base
  satisfiability for \shiqo are \nexptime-hard. The problems are
  \nexptime-complete if unary coding of numbers in the input is assumed.
\end{corollary}

\begin{proof}
  The lower bound is immediate from Corollary~\ref{cor:alcqio-concept-nexptime},
  since \alcqio is strictly contained in \shiqo. For the upper bound, the reduction
  from Definition \ref{def:transitive-roles-reduction} extended to \shiqo by
  setting $i^\trans = i$ for every individual $i \in \Individuals$ yields a
  reduction from \shiqo to \alcqio, for which the corresponding problems are
  solvable in \nexptime if unary coding of numbers in the input is assumed. \qed
\end{proof}

\section{Practical Reasoning for \shiq}

\label{sec:shiq-practical}

The previous \exptime-completeness results for \shiq rely on the highly
inefficient automata construction of Definition~\ref{def:alcqir-atomaton} used
to prove Theorem~\ref{theo:alcqir-tbox--exptime-complete} and, in the case of
knowledge base reasoning, also on the wasteful pre-completion technique used to
prove Theorem~\ref{theo:alcqir-kb-exptime-complete}. Thus, we cannot expect to
obtain an implementation from these algorithms that exhibits acceptable runtimes
even on relatively ``easy'' instances. This, of course, is a prerequisite for
using \shiq in real-world applications.

For less expressive DLs, some of the implementations of reasoners that perform
fastest in system comparisons \cite{MassacciDonini-Tableaux2000} are based on
tableau calculi similar to the ones we have already studied in this thesis.
Among them are \FaCT \cite{Horrocks98c} for the DL \shf, \index{shf@\shf} RACE
\cite{Haarslev99a} for \shn, \index{shn@\shn} and DLP
\cite{Patel-Schneider-Tableaux2000} for an extension of $\alc_{\text{reg}}$
\index{alcreg@$\alc_{\text{reg}}$} with number restrictions. The efficiency of
these implementations is due to a number of optimizations
\cite{BaaderFranconi+-OptJournal-94,Horrocks97b,horrocks99:_optim_descr_logic_subsum,HorrocksTobies-KR-2000,HaarslevMoeller-DL-2000-PSEUDO}
for which tableau algorithms proved to be amenable.

To make these optimizations applicable and to allow for an easy extension of
existing implementations to \shiq, we develop a tableau algorithm that decides
concept satisfiability for \shiq. By Theorem~\ref{theorem:shiq-internalisation},
such an algorithm can also be used to decide concept satisfiability \wrt general
TBoxes. This algorithm can be seen as the culmination point of the development of
tableau-based decision procedures for more and more expressive DLs. To mention
only the more recent ones: Sattler \citeyear{Sattler-KI-1996} describes an
algorithm for $\s$ that is subsequently extended to deal with role hierarchies
$(\sh)$ by Horrocks~\citeyear{Horrocks98c}.  Haarslev and M{\"o}ller
\citeyear{HaarslevMoeller-KR-2000} add number restrictions $(\shn)$ while
Horrocks and Sattler \citeyear{HorrSat-JLC-99} add inverse roles and functional
restrictions (\shif). Here, we extend the latter algorithm to deal with
qualifying number restrictions to obtain a tableau based decision procedure for
\shiq.

Many techniques required for this extension are already present in the
\shif-algorithm \cite{HorrSat-JLC-99} and in the \alcq-algorithm presented in
Chapter \ref{chap:alcq}. In addition to these techniques, we develop a novel way
to construct a model from a completion tree to prove soundness of the
\shiq-algorithm. This is necessary because \shiq no longer has the finite model
property.

\subsection{A \shiq-Tableau}

For the tableau algorithm, it will be helpful to have a syntactic satisfiability
criterion for satisfiability that deals with the extra complexity caused by
transitive roles, similar to the tableau for \si defined in
Definition~\ref{def:si-tableau}. The \shiq-algorithm will then search for
\shiq-tableaux rather than for models. Like for \si, elements of a \shiq-tableau
are labelled with sets of ``relevant'' concepts.  Due to the presence of
qualifying number restrictions, not only the sub-concepts of the input concepts
are of relevance but also their negations (in NNF). Also, propagation of
universal restrictions along transitive roles is slightly more complicated in
the presence of a role hierarchy compared to the case of \si, and may involve
universal restrictions that are not present as sub-concepts of the input
concept. Hence, elements of the tableau are labelled not only with sub-concepts
of the input concept but rather from the larger set $\clos(C,\R)$ that is
defined in Definition \ref{def:shiq-closure}.

Based on this set, the definition of a tableau for \shiq is now similar to the
one for \si in Definition~\ref{def:si-tableau}.

\begin{definition}[A Tableau for \shiq] \label{def:shiq-tableau}
  Let $C$ be a \shiq-concept in NNF, $\R$ a role hierarchy, and $\overline{\Roles}_{C,\R}$
  the set of roles occurring in $C,\R$, together with their inverses.  A
  \emph{tableau} $\mcT$ for $C$ \wrt $\R$ is a triple $(\mcS,\TabLab,\Edges)$
  such that $\mcS$ is a non-empty set, $\TabLab:\mcS \to 2^{\clos(C,\R)}$ maps each
  element to a subset of $\clos(C,\R)$, $\Edges:\overline{\Roles}_{C,\R} \to 2^{\mcS \times
    \mcS}$ maps each role in $\overline{\Roles}_{C,\R}$ to a set of pairs of individuals,
  and the following conditions are satisfied:
  \index{tableau!for shiq@for \shiq}
  \begin{itemize}
    \itemindent2ex
  \item [\tab 1] There is an $s \in \mcS$ with $C \in \TabLab(s)$, and
  \end{itemize}
  for all $s,t \in \mcS$, $A,C_1,C_2,D \in \clos(C,\R)$, and $R,S\in \overline{\Roles}_{C,\R}$,
  \begin{itemize}
    \itemindent2ex
  \item [\tab 2] if $A \in \TabLab(s)$, then $\neg A \not\in
    \TabLab(s)$ for $A \in \Names$,
  \item [\tab 3] if $C_1 \sqcap C_2 \in \TabLab(s)$, then $C_1 \in\TabLab(s)$ and
    ${C_2\in\TabLab(s)}$, 
  \item [\tab 4] if $C_1 \sqcup C_2 \in \TabLab(s)$, then $C_1 \in \TabLab(s)$ or $C_2 \in \TabLab(s)$,
  \item [\tab 5] if $\some{R}{D} \in \TabLab(s)$, then there is some $t \in \mcS$ such
    that $\tuple{s}{t} \in \Edges(R)$ and $D \in \TabLab(t)$,
  \item [\tab 6] if $\all{R}{D} \in \TabLab(s)$ and $\tuple{s}{t} \in \Edges(R)$, then $D
    \in \TabLab(t)$,
  \item [\tab 7] if $\all{R}{D} \in \TabLab(s)$, $\tuple{s}{t} \in
    \Edges(T)$ for some $T \sss R$ with $\Tr(T)$, then $\all{T}{D} \in
    \TabLab(t)$, 
  \item [\tab 8] $\tuple{s}{t}\in \Edges(R)$ iff $\tuple{t}{s}\in
    \Edges(\Inv(R))$.
  \item [\tab 9] if $\tuple s t \in \Edges(S)$ and $S \sss R$, then
    $\tuple s t \in \Edges(R)$,
  \item [\tab {10}] if $\qnrgeq n S D \in \TabLab(s)$, then $\sharp S^\mcT(s,D) \geq n$,
  \item [\tab {11}] if $\qnrleq n S D \in \TabLab(s)$, then $\sharp S^\mcT(s,D) \leq n$,
  \item [\tab {12}] if $\qnrgleq n S D \in \TabLab(s)$ and $\tuple s t \in
    \Edges(S)$, then $D \in \TabLab(t)$ or $\nneg D \in \TabLab(t)$,
  \end{itemize}
  where we use $\bowtie$ as a placeholder for both $\leq$ and $\geq$ and we
  define
  \[
  S^\mcT(s,D) := \{ t \in \mcS \mid \tuple s t \in \Edges(S) \
  \text{and} \ D \in \TabLab(t) \} .
  \]
  \index{00rt@$R^\mcT(s,D)$}
\end{definition}

The existence of a tableau is a necessary and sufficient criterion for
satisfiability:

\begin{lemma}\label{lem:shiq-tableau-sat-criterion}
  A \shiq-concept $C$ is satisfiable with respect to a role hierarchy $\R$
  iff there exists a tableau for $C$ with respect to $\R$.
\end{lemma}

\begin{proof}
  For the \emph{if}-direction, the construction of a model of $C$ from a tableau
  for $C$ is similar to the one presented in the proof of
  Lemma~\ref{lem:si-tableau} where the interpretation of the roles is defined
   in the same manner as it was done in the proof of
  Lemma~\ref{lem:shiq-to-alcqir}.  To be more precise, if $\mcT = (\mcS,\TabLab,\Edges)$ is a
  tableau for $C$ \wrt $\R$ and $C\in \TabLab(s_0)$, a model $\I = (\domain,
  \cdot\ifunc)$ of $C$ can be defined as follows:
  \begin{eqnarray*}
    \domain & = & \mcS \\
    A\ifunc & = & \{s \mid A \in \TabLab(s)\} \mbox{ for all
      concept names $A$ in $\clos(C,\R)$} \\
    R\ifunc & = & \left\{ \begin{array}{ll}
        \Edges(R)^+ & \mbox{if $\Tr(R)$} \\
        \Edges(R) \cup \displaystyle \bigcup\limits_{S\sss R, S\neq R}S\ifunc &
        \mbox{otherwise} 
      \end{array}
    \right.
  \end{eqnarray*}
  
  Like in the proof of Lemma~\ref{lem:shiq-to-alcqir}, it is easy to see that $\I
  \models \R$ and that $(s,t) \in R^\I$ iff $(s,t) \in \Edges(R)$ or there exists
  a role $T \sss R$ with $\Tr(T)$ and a path $s_0, \dots, s_k$ such that $k>1$,
  $s = s_0$, $t = s_k$, and $(s_i,s_{i+1}) \in \Edges(T)$ for $0 \leq i <
  k$. Moreover, if $R$ is simple, then $R^\I = \Edges(R)$.

  
  It remains to show that $C^\I \neq \emptyset$. This is done by proving that $D
  \in \TabLab(s)$ implies $s \in D\ifunc$ for each $D \in \clos(C,\R)$ and $s
  \in \mcS$.  Since $C \in \TabLab(s_0)$, we then have $s_0 \in C^\I$
  and hence $\I$ is a model of $C$. The proof is by induction on the \emph{norm}
  $\|\cdot\|$ of concepts as defined in the proof of
  Lemma~\ref{lem:shiq-to-alcqir}.
  The two base cases of the induction are $D = A$ or $D = \neg A$ for $A \in
  \Names$. If $A \in \TabLab(s)$, then, by definition, $s \in A^\I$. If $\neg A
  \in \TabLab(s)$, then, by \tab 2, $A \not \in \TabLab(s)$ and hence $s \not
  \in A^\I$. For the induction step, we have to distinguish several cases:
  \begin{itemize}
  \item The cases $D = C_1 \sqcap C_2$, $D = C_1 \sqcup C_2$, and $D = \some R
    E$ are exactly as for \si in the proof of Lemma~\ref{lem:si-tableau}
  \item $D = \all R E$. Let $s \in \mcS$ with $D \in
    \TabLab(s)$, let $t \in \mcS$ be an arbitrary individual
    such that $\tuple s t \in R^\I$. There are two possibilities:
    \begin{itemize}
    \item $\tuple s t \in \Edges(R)$. Then \tab 6 implies $E \in \TabLab(t)$ and,
      by induction, $t \in E^\I$.
    \item $\tuple s t \not \in \Edges(R)$. Due to \tab 8, this can only be the
      case if there is a role $T\sss R$ with $\Tr(T)$ and a path
      $\tuple{s}{s_1}, \tuple{s_1}{s_2},\ldots,$ $\tuple{s_{k-1}}{t}\in
      \Edges(T)$ with $k>1$. Then \tab 7 implies $\all T E \in \TabLab(s_i)$ for
      all $1 \leq i \leq k-1$ and particularly $\all T E \in \TabLab(s_{k-1})$.
      Due to \tab 6, $E \in \TabLab(t)$ also holds.  Again, by induction, this
      implies $t \in E^\I$.
    \end{itemize}
    In both cases, we have $t \in E^\I$ and, since $t$ has been chosen arbitrarily,
    $s \in D^\I$ holds.
  \item $D = \qnrgeq n S E$. For an $s$ with $D \in \TabLab(s)$, we
    have $\sharp S^\mcT(s,E) \geq n$ by \tab {10}. Hence there are
    $n$ individuals $t_1,\dots,t_n$ such that $t_i \neq t_j$ for $i
    \neq j$, $\tuple s {t_i} \in \Edges(S)$, and $E \in \TabLab(t_i)$
    for all $i$. By induction, we have $t_i \in E^\I$ and, since
    $\Edges(S) \subseteq S^\I$, also $s \in D^\I$.
  \item $D = \qnrleq n S E$. For this case, it is crucial that $S$ is a simple
    role because this implies $S^\I = \Edges(S)$. Let $s$ be an individual with
    $D \in \TabLab(s)$. Due to \tab {12}, we have $E \in \TabLab(t)$ or $\nneg E
    \in \TabLab(t)$ for each $t$ with $\tuple s t \in \Edges(S)$. Moreover,
    $\sharp S^\mcT(s,E) \leq n$ holds due to \tab {11}. We show that $\sharp
    S^\I(s,E) \leq \sharp S^\mcT(s,E)$: assume $\sharp S^\I(s,E) > \sharp
    S^\mcT(s,E)$.  This implies the existence of some $t$ with $\tuple s t \in
    S^\I$ and $t \in E^\I$ but $E \not\in \TabLab(t)$ (because $S^\I =
    \Edges(S)$). By \tab {12}, this implies $\nneg E \in \TabLab(t)$, which, by
    induction, yields $t \in (\nneg E)^\I$, in contradiction to $t \in E^\I$.
  \end{itemize}
  
  For the \emph{only-if}-direction, we have to show that satisfiability
  of $C$ \wrt $\R$ implies the existence of a
  tableau $T$ for $C$ \wrt $\R$.
  
  Let $\mathcal{I} = (\domain, \cdot\ifunc)$ be a model of $C$ with
  $\I \models \R$. A tableau $\mcT =
  (\mcS,\TabLab,\Edges)$ for $C$ can be defined by:
  \begin{align*}
    \mcS & = \domain, \\
    \Edges(R) & = R\ifunc, \\
    \TabLab(s) & = \{D \in \clos(C,\R) \mid s \in D\ifunc \} .
  \end{align*}
  
  It remains to demonstrate that $\mcT$ is a tableau for $D$:
  \begin{itemize}
  \item Except for \tab 7 and \tab 9, all conditions are satisfied
    as a direct consequence of the definition of the semantics of
    \shiq-concepts.
  \item For \tab 7, if $s \in (\all{R}{D})\ifunc$ and $\tuple{s}{t} \in T\ifunc$
    for $T$ with $\Tr(T)$ and $T\sss R$, then $t \in (\all{T}{D})\ifunc$ unless
    there is some $u$ such that $\tuple{t}{u} \in T\ifunc$ and $u \not\in
    D\ifunc$. In this case, since $\tuple{s}{t} \in T\ifunc$, $\tuple{t}{u} \in
    T\ifunc$, and $\Tr(T)$, it holds that $\tuple{s}{u} \in T\ifunc$. Hence
    $\tuple{s}{u} \in R\ifunc$ and $s \not\in (\all{R}{D})\ifunc$---in
    contradiction to the assumption. $\mcT$ therefore satisfies \tab 7.
  \item Condition \tab 9 is satisfied because $\I \models\R$ and set-inclusion
    is a transitive property. \qed
  \end{itemize}
\end{proof}

\subsection{A Tableau Algorithm for \shiq}

In the following, we present an algorithm that, given a \shiq-concept $C$ and a
role hierarchy $\R$, decides the existence of a tableau for $C$ \wrt $\R$. As
before, we assume that $\R$ is cycle-free.  Like the \si-algorithm, the
\shiq-algorithm works on a finite completion tree, and employs a blocking
technique to guarantee termination.

\subsubsection{Pair-wise Blocking}

From the fact that \shiq no longer has the finite model property, it is
immediately clear that the tableau construction we have
employed for \si will not work without modification for \shiq, as this technique
always resulted in finite tableaux and hence in finite models.

Horrocks and Sattler \citeyear{HorrSat-JLC-99} show that for the fragment \shif
of \shiq, dynamic blocking no longer is sufficient and describe the
\iemph{pair-wise blocking} \index{blocking!pair-wise blocking} technique that can
successfully be applied also for \shiq: if a path contains two pairs of
successive nodes that have pair-wise identical labels and whose connecting edges
have identical labels, then the path beyond the second pair is no longer
expanded---it is blocked. Blocked paths are then ``unraveled'' to construct an
infinite tableau from a finite completion tree.  The identical labels make sure
that copies of the blocking node and its descendants can be substituted for the
blocked node and its respective descendants.  Note the similarity between this
pair-wise blocking condition and the condition imposed by the combination of the
blocking condition and the cut rule by De Giacomo and
Massacci~\citeyear{degiacomo00:_combin_deduc_model_check_tableaux} for \CPDL.

Figure~\ref{fig:pairblock} shows that pair-wise blocking is crucial in
order to ensure that the algorithm discovers the unsatisfiability of
the concept
\[
\neg A \sqcap \nrleq 1 F \sqcap \some{F^{-1}}{D} \sqcap \all{R^{-1}}{(\some{F^{-1}}{D})} ,
\]
where $\Tr(R)$, $F \sqsubseteq R$, and $D$ represents the concept
\[
A \sqcap \nrleq 1 F \sqcap \some{F}{\neg A}.
\]
Using dynamic blocking, $z$ would be blocked by $y$. The resulting tree cannot
represent a cyclic model in which $y$ is related to itself by an $F^{-1}$ role
as this would conflict with $\nrleq 1 F \in \Lab(y)$. The tree must therefore
represent the infinite model generated by recursively replacing each occurrence
of $z$ with a copy of the tree rooted at $y$. However, this also does not lead
to a valid model, since, if $z$ is substituted by a copy of $y$, then the
constraint $\some{F}{\neg A} \in \Lab(y)$, which was satisfied because of $\neg
A \in \Lab(x)$, is no longer satisfied in its new location.

\begin{figure}[t!]
  \begin{center}
    \vspace{2ex}
    \parbox[t]{3.75in}{\input{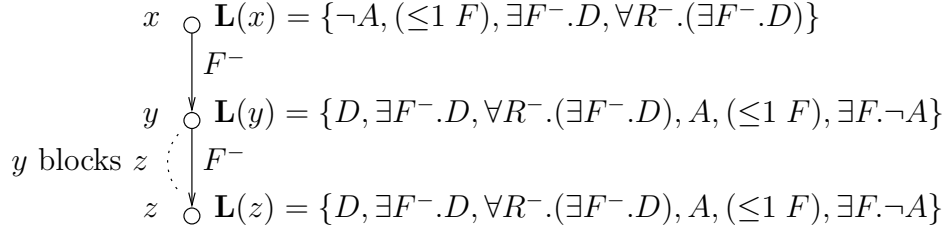}}
    \caption{A tableau where pair-wise blocking is crucial}
    \label{fig:pairblock}
  \end{center}
\end{figure}

When pair-wise blocking is used, $z$ is no longer blocked by $y$ as the labels
of their predecessors ($y$ and $x$ respectively) are not equal, and the
algorithm continues to expand $\Lab(z)$. The expansion of $\some{F}{\neg A} \in
\Lab(z)$ calls for the existence of a node whose label includes $\neg A$ and
that is connected to $z$ by an $F$-labelled edge. Because of $\nrleq 1 F \in
\Lab(z)$, this node must be $y$, and this results in a contradiction as both $A$
and $\neg A$ will be in $\Lab(y)$.

To extend the \shif-algorithm from \cite{HorrSat-JLC-99} to \shiq, we add rules
that deal with  qualifying number restrictions, similar to the ones used by
the standard algorithm for \alcq (Algorithm~\ref{alg:alcq-standard}), namely a
\rulegeq-rule that introduces new successor nodes to satisfy
$\geq$-restrictions, a \ruleleq-rule that identifies nodes as required by
$\leq$-restrictions, and a \rulechoose-rule that makes sure that all
relevant concepts at a node are either positively or negatively asserted
for that node (refer to \tab {12}).

In order to guarantee the termination of the algorithm, we have to make sure
that the \rulegeq- and \ruleleq-rules cannot be applied in a way that would
yield an infinite sequence of rule applications, generating and identifying
successors indefinitely.  This is enforced by recording in a relation ``$\ndoteq$''
which nodes have been introduced by an application of the \rulegeq-rule and by
prohibiting the identification of these nodes by the \ruleleq-rule.

\begin{algorithm}[The \shiq-algorithm]\label{alg:shiq}
  \label{def:shiq-tableau-algorithm}
  Let $C$ be a \shiq-concept in NNF to be tested for satisfiability \wrt a role
  hierarchy $\R$ and $\overline{\Roles}_{C,\R}$
  \index{00nrcr@$\overline{\Roles}_{C,\R}$} the set of roles that occur in $C$
  and $\R$ together with their inverses.  A \emph{completion tree}
  \index{completion tree!for shiq@for \shiq} $\Tree = (\bfV, \bfE, \bfL)$ is a
  labelled tree in which each node $x \in \bfV$ is labelled with a set $\Lab(x)
  \subseteq \clos(C,\R)$ and each edge $(x,y) \in E$ is labelled with a set
  $\Lab(x,y) \subset \overline{\Roles}_{C,\R}$. The algorithm expands the tree
  by extending $\Lab(x)$ for some node $x$, or by adding new leaf nodes.
  Additionally, we keep track of inequalities between nodes of the tree with a
  symmetric binary relation $\ndoteq$ between nodes in $\bfV$.
  
  \index{shiq@\shiq!tableau algorithm}
  \index{tableau algorithm!for shiq@for \shiq}

  Given a completion tree, a node $y$ is called an $R$-\emph{successor} of a
  node $x$ if $y$ is a successor of $x$ and $S\in \Lab\tuple{x}{y}$ for some
  $S$ with $S\sss R$; $y$ is called an $R$\emph{-neighbour} of $x$ if it is an
  $R$-successor of $x$, or if $x$ is an $\Inv(R)$-successor of y.
  
  For a role $R$, a concept $D$, and a node $x \in \bfV$, we define
  $R^\Tree(x,D)$ by
  \[
   R^\Tree(x,D) =  \{ y \mid \text{$y$ is an $R$-neighbour of $x$ and
    $D \in \Lab(y)$} \} .
  \]
  
  \index{00rt@$R^\Tree(x,D)$}

  A node $x$ is \emph{directly blocked} if none of its ancestors is
  blocked, and it has ancestors $x'$, $y$, and $y'$, such that
  \begin{enumerate}
  \item $x$ is a successor of $x'$ and $y$ is a successor of $y'$,
    and
  \item $\Lab(x) = \Lab(y)$ and $\Lab(x') = \Lab(y')$ \emph{and}
  \item $\Lab\tuple{x'}{x} = \Lab\tuple{y'}{y}$.
  \end{enumerate}
  In this case we will say that $y$ blocks $x$.
  
  A node is \emph{indirectly blocked} if its predecessor is directly or
  indirectly blocked, and in order to avoid wasted expansion after an
  application of the \ruleleq-rule, a node $y$ will also be taken to be
  indirectly blocked if $y$ is a successor of a node $x$ and $\Lab\tuple{x}{y} =
  \emptyset$.
    
  For a node $x$, $\Lab(x)$ contains a \emph{clash} if, for some
  concept name $A \in \Names$, $\{A, \neg A\} \subseteq \Lab(x)$, or if,
  for a some concept $D$, some role $S$, and some $n \in \N$: $\qnrleq
  n S D \in \Lab(x)$ and there are $n+1$ nodes $y_0,\dots,y_n$ such
  that $D \in \Lab(y_i)$, $y_i$ is an $S$-neighbour of $x$, and $y_i
  \ndoteq y_j$ for all $0 \leq i < j \leq n$.
  
  The algorithm initializes the tree \Tree to contain a single node
  $x_0$, called the \emph{root} node, with $\Lab(x_0) = \{C\}$. The
  inequality relation $\ndoteq$ is initialized with the empty set.
  \Tree is then expanded by repeatedly applying the rules from
  Figure~\ref{fig:shiq-rules}.
  
  The completion tree is complete if, for some node $x$, $\Lab(x)$
  contains a clash, or if none of the rules is applicable. If, for
  an input concept $C$, the expansion rules can be applied in such a
  way that they yield a complete, clash-free completion tree, then the
  algorithm returns ``$C$ is satisfiable'', and ``$C$ is
  unsatisfiable'' otherwise. \eod
\end{algorithm}

For a discussion of the different kinds on non-determinism present in the
\shiq-algorithm, compare below Algorithm~\ref{alg:si}.

Like for \si, the definition of \emph{successor} and \emph{predecessor} reflects
the relative position of two nodes in the completion tree: if $x$ is an
$R$-successor of $y$ then this implies that $(x,y) \in \bfE$ and it is not the
case that $y$ is an $\Inv(R)$-successor of $x$. It is necessary to make this
pedantic distinction because when we construct a tableau from a complete and
clash-free completion tree in the proof of Lemma~\ref{lemma:shiq-soundness},
a blocked successor is replaced by a copy of the sub-completion tree
consisting of the respective blocking node and its descendants. This makes the
distinction between an $R$-successors and an $\Inv(R)$- predecessors significant
has has to be reflected in the completion rules.

Note that the definition of blocking is recursive because the status of a node
depends, among other things, on the status of its predecessor. Since the
dependency is on the predecessor and the ancestors, one can determine the status
every node starting at the root, which has no predecessor or ancestor and hence
is never blocked. Once the blocking status of a node has been determined, one
can then determine the status of its successors.

Since we only block along a path in the completion tree, for every directly
blocked node there is a uniquely determined blocking node. Assume there would be
a directly blocked node $x$ and two distinct unblocked nodes $y_1,y_2$ blocking
$x$. Since both $y_1$ and $y_2$ must be ancestors of $x$, w.o.l.g., $y_1$ is an
ancestor of $y_2$. Yet, this implies that $y_1$ directly blocks $y_2$ and $x$
cannot be directly blocked because it has the blocked ancestor $y_2$, a
contradiction.

\begin{figure}[tb!]
  \begin{center}
    \begin{tabular}{@{}l@{$\;$}l@{$\;$}l}
      \ruleand: & \bfif \hfill 1. & $C_1 \sqcap C_2 \in \Lab(x)$ and \\ &
      \hfill 2. & $\{C_1,C_2\} \not\subseteq \Lab(x)$ \\ & \bfthen & $\Lab(x)
      \ruleand \Lab(x) \cup \{C_1,C_2\}$ \\[1ex]
      
      \ruleor & \bfif \hfill 1. & $C_1 \sqcup C_2 \in \Lab(x)$ and \\ &
      \hfill 2. & $\{C_1,C_2\} \cap \Lab(x) = \emptyset$ \\ & \bfthen & $\Lab(x)
      \ruleor \Lab(x) \cup \{E\}$ for some $E\in \{C_1,C_2 \}$\\[1ex]
      
      \rulefa: & \bfif \hfill 1. & $\all{R}{D} \in \Lab(x)$, $x$ is not 
      indirectly blocked,  and \\ & \hfill
      2. &there is an $R$-neighbour $y$ of $x$ with $D \not\in \Lab(y)$ \\ &
      \bfthen & 
      $\Lab(y) \rulefa \Lab(y) \cup \{D\}$ \\[1ex]
      
      \rulefatr: & \bfif \hfill 1. & $\all{R}{D} \in \Lab(x)$, $x$ is not
      indirectly blocked, and \\  
      & \hfill 2. & there is some $T$ with $\Tr(T)$ and $T \sss R$ \\
      & \hfill 3. & there is a $T$-neighbour $y$ of $x$ with $\all{T}{D}
      \not\in \Lab(y)$ \\ 
      & \bfthen & $\Lab(y) \rulefatr \Lab(y) \cup  \{\all{T}{D}\}$ \\[1ex]
      
      \ruleex: & \bfif \hfill 1. & $\some{R}{D} \in \Lab(x)$, $x$ is not
      blocked \\
      & \hfill 2. & $x$ has no $R$-neighbour $y$ with $D\in \Lab(y)$ \\ &
      \bfthen & create a new successor $y$ of $x$ with $\Lab\tuple{x}{y}=\{ R \}$ and $\Lab(y)=
      \{D\}$ \\[1ex]
      
      \rulechoose: & \bfif \hfill 1. & $(\bowtie \; n \; S \; D) \in \Lab(x)$,
      $x$ is not indirectly blocked, and\\
      & \hfill 2. & there is an $S$-neighbour $y$ of $x$ with $\{ D,\nneg D\}
      \cap  \Lab(y)= \emptyset$\\
      & \bfthen & $\Lab(y) \rulechoose \Lab(y) \cup \{E\}$  for some $E \in \{D,\nneg D \}$\\[1ex]

      \rulegeq: & \bfif \hfill 1. & $\qnrgeq n S D \in \Lab(x)$, $x$ is not
      blocked, and\\
      & \hfill 2. & there are not $n$ $S$-neighbours  $y_1,\dots,y_n$ of $x$
      with \\
      && $D \in \Lab(y_i)$ and  $y_i \ndoteq y_j$ for $1 \leq i < j \leq n$\\
      
      & \bfthen & create $n$ new successors $y_1,\dots,y_n$ of $x$ with $\Lab\tuple
      x {y_i} =  \{S\}$,\\
      & & $\Lab(y_i) = \{D\}$, and $y_i \ndoteq y_j$ for $1 \leq i < j \leq
      n$.\\[1ex] 
      
      \ruleleq: & \bfif \hfill 1. &  $\qnrleq n S D \in \Lab(x)$ with $n \geq
      1$,  $x$ is not indirectly
      blocked, and\\
      & \hfill 2. & $\sharp S^\Tree(x,D) > n$ and there are two $S$-neighbours
      $y,z$ of $x$ with \\
      & &  $D \in \Lab(y), D \in \Lab(z)$, $y$ is a successor
      of $x$, and not $y \ndoteq z$\\
      & \bfthen & 1.\ $\Lab(z) \ruleleq \Lab(z) \cup \Lab(y)$ and\\
      & & 2.\ \begin{tabular}[t]{rl}
        \multicolumn{2}{l}{if $z$ is a predecessor of $x$} \\
        then &
        $\begin{array}[t]{rcl}
          \Lab \tuple{z}{x}
          &\ruleleq&  \Lab \tuple{z}{x} \cup \Inv(\Lab \tuple{x}{y})
        \end{array}$\\
        else \hfill &
        $\begin{array}[t]{rcl}
          \Lab \tuple{x}{z}
          &\ruleleq&  \Lab \tuple{x}{z} \cup \Lab \tuple{x}{y}
        \end{array}$
      \end{tabular}\\
      & & 3.\ $\Lab\tuple{x}{y} \ruleleq \emptyset$ \\
      & & 4.\ Set $u \ndoteq z$ for all $u$ with $u \ndoteq y$
    \end{tabular}
  \end{center}
  \caption{Tableau expansion rules for \shiq}
  \label{fig:shiq-rules}
  \index{00ruleand@\ruleand}
  \index{00ruleor@\ruleor}
  \index{00rulefa@\rulefa}
  \index{00rulefatr@\rulefatr}
  \index{00ruleex@\ruleex}
  \index{00rulechoose@\rulechoose}
  \index{00rulegeq@\rulegeq}
  \index{00ruleleq@\ruleleq}
  \index{shiq@\shiq!completion rules}
  \index{completion rules!for shiq@for \shiq}
\end{figure}

Before we prove the correctness of the \shiq-algorithm, we discuss the
intuition behind the expansion rules and their correspondence to the
constructors of \shiq. Roughly speaking,\footnote{For the following
  considerations, we employ a simpler view of the correspondence
  between completion trees and models, and do not bother with the
  unraveling construction mentioned above.}
the completion tree is a partial description of a model whose individuals
correspond to nodes and whose interpretation of concept and role names is
determined by the node and edge labels. Since the completion tree is a tree,
this would not yield a correct interpretation of transitive roles, and thus the
interpretation of transitive roles is built via the transitive closure of the
relations induced by the corresponding edge labels.

The \ruleand-, \ruleor-, \ruleex-, and \rulefa-rules are the standard tableau
rules for \alc from Algorithm~\ref{alg:alc}, with the exception that we limit
the applicability of the \rulefa- and \rulefatr-rule to those nodes that are not
blocked or directly blocked. The \rulefatr-rule is similar to the \rulefatr-rule
for \si without refined blocking extended to deal with role-hierarchies as
follows.  Assume a situation that satisfies the precondition of the
\rulefatr-rule, i.e., $\all{R}{D} \in \Lab(x)$, and there is a $T$-neighbour $y$
of $x$ with $\Tr(T)$, $T\sss R$, and $\all{R}{D} \not\in \Lab(y)$.  If $y$ has a
$T$-neighbour $z$, then, due to the transitivity of $T$, $x$ and $z$ are also
related via $T$.  Since $T \sss R$, it is also an $R$-neighbour of $x$ and hence
must satisfy $D$. This is ensured by adding $\all{T}{D}$ to $\Lab(y)$, which, in
turn, causes $D$ to be added to $\Lab(z)$.

The rules dealing with qualifying number restrictions work similarly to the
rules of the standard algorithm for \alcq (Algorithm~\ref{alg:alcq-standard}).
For a concept $\qnrgeq n S D \in \Lab(x)$, the \rulegeq-rule generates $n$
$S$-successors $y_1,\dots,y_n$ of $x$ with $D \in \Lab(y_i)$. To prevent the
\ruleleq-rule from indentifying these new nodes, it also sets $y_i \ndoteq y_j$
for each $1\leq i < j \leq n$ .  Conversely, if $\qnrleq n S D \in \Lab(x)$ and
$x$ has more than $n$ $S$-neighbours that are labelled with $D$, then the
\ruleleq-rule chooses two of them, say $y,z$, that are not required to be
distinct by $\ndoteq$ and merges them, together with the edges connecting them
with $x$. The algorithm constructs a completion \emph{tree} so at least one of
$y,z$ must be a successor of $x$. Let this be $y$. If $z$ is a predecessor of
$x$ in the completion tree, then it is necessary that we join $y$ onto $z$ and
not vice vesus because otherwise $x$ would become detached in the completion tree.

The definition of a clash takes care of the situation where the $\ndoteq$
relation makes it impossible to merge any two $S$-neighbours of $x$, while the
\rulechoose-rule ensures that all $S$-neighbours of $x$ are labelled with either
$D$ or $\nneg D$. The relation $\ndoteq$ is used to prevent infinite sequences
of rule applications for contradicting number restrictions of the form $\qnrgeq
n S D$ and $\qnrleq m S D$, with $n>m$.

Labeling edges with sets of roles allows a single node to be both an $S$- and
$R$-successor of $x$ even if $S$ and $R$ are not comparable with respect to
$\sss$. An example for a concept that enforces such a situation is $\qnrgeq 2
{S_1} A \sqcap \qnrgeq 2 {S_2} A \sqcap \qnrleq 3 R A$ with $S_i \sqsubseteq R$,
which enforces a successor reachable both via $S_1$ and $S_2$.

We will now prove correctness of the tableau algorithm in a manner similar to
the one for the \si-algorithm.

\subsubsection{Termination}

Like for \si, termination of the algorithm is ensured by blocking, which
prevents the creation of unbounded paths in the completion tree.

\begin{lemma}[Termination]\label{lem:shiq-termination}
  For each \shiq-concept $C$ and role hierarchy $\R$, the tableau
  algorithm terminates.
\end{lemma}

\begin{proof} 
  Let $m = \sharp \clos(C,\R)$, $k = \sharp \overline{\Roles}_{C,\R}$, and
  $n_{\textit{max}}$ the maximum $n$ that occurs in a concept of the
  form $(\bowtie \; n \; S \; D) \in \clos(C,\R)$.  Termination is a
  consequence of the following properties of the expansion rules:

  \begin{itemize}
  \item The expansion rules never remove nodes from the tree or
    concepts from node labels. Edge labels can only be changed by the
    \ruleleq-rule which either expands them or sets them to
    $\emptyset$; in the latter case, the node below the
    $\emptyset$-labelled edge is blocked and this block is never
    broken.
  \item Each successor of a node $x$ is the result of the application of the
    \ruleex-rule or the \rulegeq-rule to $x$. (Note that the \ruleleq-rule, does
    not move nodes in the tree.) For a node $x$, each concept in $\Lab(x)$ can
    trigger the generation of successors at most once.
    
    For the \ruleex-rule, if a successor $y$ of $x$ was generated
    for a concept $\some R D \in \Lab(x)$ and later $\Lab\tuple x y$
    is set to $\emptyset$ by the \ruleleq-rule, then there is some
    $R$-neighbour $z$ of $x$ with $D \in \Lab(z)$.
    
    For the \rulegeq-rule, if $y_1, \dots, y_n$ were generated by
    the \rulegeq-rule for $\qnrgeq n S D\in\Lab(x)$, then $y_i
    \ndoteq y_j$ holds for all $1 \leq i < j \leq n$.  This implies
    that there are always $n$ $S$-neighbours $y'_1, \dots, y'_n$ of
    $x$ with $D \in \Lab(y'_i)$ and $y'_i \ndoteq y'_j$ for all $1
    \leq i < j \leq n$, since the \ruleleq-rule never merges two
    nodes $y'_i, y'_j$ with $y'_i \ndoteq y'_j$ and, whenever an
    application of the \ruleleq-rule sets $\Lab\tuple x {y'_i}$
    to $\emptyset$, there is some $S$-neighbour $z$ of $x$ which
    ``inherits'' both $D$ and all inequalities from $y'_i$.
    
    Since $\clos(C,\R)$ contains a total of at most $m$ concept of the form
    \some{R}{D} and $\qnrgeq n S D$, the out-degree of the tree is bounded by
    $m\cdot n_{\textit{max}}$.
    
  \item Nodes are labelled with non-empty subsets of $\clos(C,\R)$ and
    edges with subsets of $\overline{\Roles}_{C,\R}$, so there are at most $2^{2mk}$
    different possible labellings for a pair of nodes and an edge.
    Therefore, if a path is of length  $>2^{2mk}$, then,
    from the pair-wise blocking condition, there must be two nodes
    $x,y$ on this path such that $x$ is directly blocked by $y$.

    Since a path on which nodes are blocked cannot become longer,
    paths are of length at most $2^{2mk}$. \qed
  \end{itemize}
\end{proof}

\subsubsection{Completeness}

To prove completeness of the \shiq-algorithm, we proceed as for the
\si-algorithm and guide the application of the non-deterministic \ruleor-,
\rulechoose-, and \ruleleq-rule using a function that maps nodes of the
completion tree to elements of a tableau.

\begin{lemma}\label{lemma:shiq-completeness}
  Let $C$ be a \shiq-concept in NNF. If $C$ has a tableau \wrt $\R$, then the
  expansion rules can be applied in such a way that the tableau algorithm yields
  a complete and clash-free completion tree for $C$ \wrt $\R$.
\end{lemma}

\begin{proof}
  Let $T = (\mcS,\TabLab,\Edges)$ be a tableau for $C$ w.r.t.  $\R$. We use this
  tableau to guide the application of the non-deterministic rules. To do this,
  we will inductively define a function $\pi$, mapping the nodes $\bfV$ of the
  tree \Tree to $\mcS$ such that, for each $x,y \in \bfV$:
  \[
  \left.
    \begin{array}{l}
      \Lab(x) \subseteq \TabLab(\pi(x))\\
      \text{if $y$ is an $R$-neighbour of $x$, then $\tuple {\pi(x)} {\pi(y)} \in
        \Edges(R)$}\\
      \text{$x \ndoteq y$ implies $\pi(x) \neq \pi(y)$}
    \end{array}
    \qquad \right \} (*)
  \]

  \begin{claim}
    Let \Tree be a completion-tree and $\pi$ a function that satisfies $(*)$. If
    a rule is applicable to $\Tree$, then the rule is applicable to $\Tree$ in a
    way that yields a completion-tree $\Tree'$ and an extension of $\pi$ that
    satisfies $(*)$.  \vspace{1ex}
  \end{claim}
  
  Let \Tree be a completion-tree and $\pi$ a function that satisfies $(*)$.  We
  have to consider the various rules.
  \begin{itemize}
  \item \textbf{For the \ruleand-, \ruleor-, and \ruleex-rule}, this is analogous to the proof
    of Lemma~\ref{lemma:si-completeness} for \si.
  \item \textbf{The \rulefa-rule:} If $\all R D \in \Lab(x)$, then
    $\all R D  \in \TabLab(\pi(x))$, and if $y$ is an $R$-neighbour of
    $x$, then also $\tuple{\pi(x)}{\pi(y)} \in \Edges(R)$ due to
    $(*)$. \tab 6 implies $D \in \Lab(\pi(y))$ and hence
    the \rulefa-rule can be applied without violating $(*)$.
  \item \textbf{The \rulefatr-rule:} If $\all R D \in \Lab(x)$, then
    $\all R D \in \TabLab(\pi(x))$, and if there is some $T \sss R$
    with $\Tr(T)$ and $y$ is an $T$-neighbour of $x$, then also
    $\tuple{\pi(x)}{\pi(y)} \in \Edges(T)$ due to $(*)$. \tab 7
    implies $\all T D \in \TabLab(\pi(y))$ and hence the
    \rulefatr-rule can be applied without violating $(*)$.
  \item \textbf{The \rulechoose-rule:} If $\qnrgleq n S D \in
    \Lab(x)$, then $\qnrgleq n S D \in \TabLab(\pi(x))$, and, if there
    is an $S$-neighbour $y$ of $x$, then $\tuple {\pi(x)} {\pi(y)} \in
    \Edges(S)$ due to $(*)$. \tab {12} implies $\{ D, \nneg D \} \cap
    \TabLab(\pi(y) \neq \emptyset$. Hence the \rulechoose-rule can add
    an appropriate concept $E \in \{ D, \nneg D \}$ to $\Lab(x)$ such
    that $\Lab(y) \subseteq \TabLab(\pi(y))$ holds.
  \item \textbf{The \rulegeq-rule:} If $\qnrgeq n S D \in \Lab(x)$,
    then $\qnrgeq n S D \in \TabLab(\pi(x))$ and \tab {10} implies
    $\sharp S^\mcT(\pi(x),D) \geq n$.  Hence there are elements
    $t_1,\dots,t_n \in \mcS$ such that $\tuple {\pi(x)} {t_i} \in
    \Edges(S)$, $D \in \TabLab(t_i)$, and $t_i \neq t_j$ for $1 \leq i
    < j \leq n$. The \rulegeq-rule generates $n$ new nodes $y_1,
    \dots, y_n$. By extending $\pi := \pi [y_1 \mapsto t_1, \dotsm y_n
    \mapsto t_n]$, one obtains a function $\pi'$ that satisfies $(*)$
    for the extended tree.
  \item \textbf{The \ruleleq-rule:} If $\qnrleq n S D \in \Lab(x)$, then
    $\qnrleq n S D \in \TabLab(\pi(x))$ and \tab {11} implies $\sharp
    S^\mcT(\pi(x),D) \leq n$. If the \ruleleq-rule is applicable, we have
    $\sharp S^\Tree(x,D) > n$, which implies that there are at least $n+1$
    $S$-neighbours $y_0,\dots,y_n$ of $x$ such that $D \in \Lab(y_i)$. Thus,
    there must be two nodes $y,z \in \{ y_0, \dots, y_n \}$ such that $\pi(y) =
    \pi(z)$ (because otherwise $\sharp S^\mcT(\pi(x),D) > n$ would hold). Since
    $\pi(y) = \pi(z)$, we have that $y \ndoteq z$ cannot hold because of $(*)$,
    and $y,z$ can be chosen such that $y$ is a successor of $x$ because $x$ has
    at most one predecessor. Hence the \ruleleq-rule can be applied without
    violating $(*)$.
  \end{itemize}
  
  Why does this claim yield the completeness of the tableau algorithm? For the
  initial completion-tree consisting of a single node $x_0$ with $\Lab(x_0) =
  \{C \}$ and ${\ndoteq} = {\emptyset}$, the function $\pi = [x_0 \mapsto s_0]$
  for some $s_0 \in \mcS$ with $C \in \Lab(s_0)$ satisfies $(*)$. Such an $s_0$
  exists due to \tab 1.  Whenever a rule is applicable to \Tree, it can be
  applied in a way that maintains $(*)$, and, since the algorithm terminates, we
  have that any sequence of rule applications must terminate.  Property $(*)$
  implies that any tree \Tree generated by these rule-applications must be
  clash-free as there are only two possibilities for a clash, and it is easy to
  see that neither of these can hold in \Tree:
  \begin{itemize}
  \item \Tree cannot contain a node $x$ such that $\{ A , \neg A \}
    \in \Lab(x)$ because $\Lab(x) \subseteq \TabLab(\pi(x))$ and hence
    \tab 2 would be violated for $\pi(x)$.
  \item \Tree cannot contain a node $x$ with $\qnrleq n S D \in
    \Lab(x)$ and $n+1$ $S$-neighbours $y_0, \dots y_n$ of $x$ with $D
    \in \Lab(y_i)$ and $y_i \ndoteq y_j$ for $0 \leq i < j \leq n$
    because $\qnrleq n S D \in \TabLab(\pi(x))$, and, since $y_i
    \ndoteq y_j$ implies $\pi(y_i) \neq \pi(y_j)$, $\sharp
    S^\mcT(\pi(x),D) > n$, in contradiction to \tab {11}. \qed
  \end{itemize}
\end{proof}

\subsubsection{Soundness}

Due to the presence of qualifying number restrictions and the lack of the finite
model property, the construction of a tableau from a complete and clash-free
completion tree is much more involved than this has been the case for \si where,
in case of a block, it was possible to generate a cyclic model. Here, the
completion tree is unraveled into an infinite tree by successively substituting
a blocked node by the subtree rooted at the blocking node. The presence of
qualifying number restrictions makes it necessary, in case of a blocking
situation, to record the pair of blocking and blocked node (see the case for
\tab {10} in the proof below).

\begin{lemma}[Soundness]\label{lemma:shiq-soundness}
  If the \shiq-algorithm generates a complete and clash-free completion tree for
  a concept $C$ and a role hierarchy $\R$, then $C$ has a tableau,
  \wrt $\R$.
\end{lemma}

\begin{proof}
  Let $\Tree = (\bfV, \bfE, \Lab)$ be a complete and clash-free completion tree.
  A \emph{path} is a sequence of pairs of nodes of \Tree of the form $p =
  [\frac{x_0}{x'_0}, \dots, \frac{x_n}{x'_n}]$. We define auxiliary functions
  $\Tail, \Tail'$ by setting, for such a path $p$, $\Tail(p) = x_n$ and
  $\Tail'(p) = x'_n$. With $[p | \frac{x_{n+1}}{x'_{n+1}}]$ we denote the path
  $[\frac{x_0}{x'_0}, \dots, \frac{x_n}{x'_n}, \frac{x_{n+1}}{x'_{n+1}}]$. The
  set $\Paths(\Tree)$ is defined inductively as follows:
  \begin{itemize}
  \item For the root node $x_0$ of \Tree, $[\frac{x_0}{x_0}] \in
    \Paths(\Tree)$, and
  \item For a path $p \in \Paths(\Tree)$ and a node $z$ in \Tree:
    \begin{itemize}
    \item if $z$ is a successor of $\Tail(p)$ and $z$ is not blocked, then
      $[p|\frac{z}{z}] \in \Paths(\Tree)$, or
    \item if, for some node $y$ in \Tree, $y$ is a successor of $\Tail(p)$ and
      $z$ blocks $y$, then $[p|\frac{z}{y}] \in \Paths(\Tree)$.
    \end{itemize}
  \end{itemize}
  
  Please note that, due to the construction of \Paths, for $p \in \Paths(\Tree)$
  with $p=[p'|\frac{x}{x'}]$, we have that $x$ is not blocked, $x'$ is blocked
  iff $x \neq x'$, and $x'$ is never indirectly blocked---it is either directly
  blocked or unblocked. Furthermore, the blocking condition implies $\Lab(x) =
  \Lab(x')$.

  Now we can define a tableau $\mcT = (\mcS,\TabLab,\Edges)$ with:
  \[
  \begin{array}{r@{\,}c@{\,}l}
    \mcS & = & \Paths(\Tree),\\[0.5ex]
    \TabLab(p) & = & \Lab(\Tail(p)),\\[0.5ex]
    \Edges(R) & = & \{\tuple{p}{q} \in \mcS \times \mcS \mid
    \begin{array}[t]{l@{\,}l}
      \mbox{Either $q = [p|\frac{x}{x'}]$ and $x'$ is an
          $R$-successor of $\Tail(p)$,}\\ 
      \mbox{or $p = [q|\frac{x}{x'}]$  and $x'$ is an
          $\Inv(R)$-successor of $\Tail(q)$}\}. 
    \end{array}
  \end{array}
  \]
  \begin{claim}
    $\mcT$ is a tableau for $C$ \wrt $\R$.
  \end{claim}
  
  We show that $\mcT$ satisfies all the properties from
  Definition~\ref{def:shiq-tableau}.

  \begin{itemize}
  \item $C \in \TabLab [ \frac{x_0}{x_0} ]$ because $C \in
    \Lab(x_0)$, hence \textbf{\tab 1} holds.
  \item \textbf{\tab 2} holds because \Tree is clash-free.
  \item \textbf{\tab 3, \tab 4} hold because \Tree is complete.
  \item \textbf{\tab 5}: assume $\some R D \in \TabLab(p)$ and let $x =
    \Tail(p)$.  In \Tree, there is an $R$-neighbour $y$ of $x$ with $D \in
    \Lab(y)$ because $x$ is not blocked and the \ruleex-rule is not applicable.
    There are two possibilities:
    \begin{itemize}
    \item $y$ is a successor of $x$ in \Tree. If $y$ is not blocked, then $q :=
      [p|\frac{y}{y}] \in \mcS$ and $\tuple p q \in \Edges(R)$ as well as $D \in
      \TabLab(q)$. If $y$ is blocked by some node $z$ in \Tree, then $q :=
      [p|\frac{z}{y}] \in \mcS$, $(p,q) \in \Edges(R)$ and, since
      $\TabLab(q) = \Lab(z) = \Lab(y)$, $D \in \TabLab(q)$.
      
    \item $y$ is a predecessor of $x$. Again, there are two possibilities:
      \begin{itemize}
      \item $p$ is of the form $p = [q|\frac{x}{x}]$ with $\Tail(q) = y$.
      \item $p$ is of the form $p = [q|\frac{x}{x'}]$ with $\Tail(q) = u \neq
        y$. Since $x$ only has one predecessor in $\Tree$, the node  $u$ cannot
        be the predecessor of $x$. Then it must be the predecessor of $x'$ in
        \Tree, $x' \neq x$, and $x$ blocks $x'$, all due to the construction of
        $\Paths(\Tree)$.  Together with the definition of the blocking
        condition, this implies $\Lab\tuple u {x'} = \Lab\tuple y x$ as well as
        $\Lab(u) = \Lab(y)$, due to the pair-wise blocking condition.
      \end{itemize}
      In both cases, $\tuple p q \in \Edges(R)$ and $D \in \TabLab(q)$.
    \end{itemize}
  \item \textbf{\tab 6}: assume $\all R D \in \TabLab(p)$ and $\tuple p q \in
    \Edges(R)$. If $q = [p | \frac{x}{x'}]$, then $x'$ is an $R$-successor of
    $\Tail(p)$, and thus $D \in \Lab(x')$ because the \rulefa-rule is not
    applicable. Since $\TabLab(q) = \Lab(x) = \Lab(x')$, we have $D \in
    \TabLab(q)$.  If $p = [q | \frac{x}{x'}]$, then $x'$ is an
    $\Inv(R)$-successor of $\Tail(q)$, $\Tail(q)$ an $R$-neighbour of $x'$ and
    thus $D \in \TabLab(q) = \Lab(\Tail(q))$ because $x'$ is not indirectly
    blocked and the \rulefa-rule is not applicable.
  \item \textbf{\tab 7}: assume $\all R D \in \TabLab(p)$ and $\tuple p q \in
    \Edges(T)$ for some $T \sss R$ with $\Tr(T)$. If $q = [p|\frac{x}{x'}]$,
    then $x'$ is a $T$-successor of $\Tail(p)$ and thus $\all T D \in \Lab(x')$
    because otherwise the \rulefatr-rule would be applicable. From $\TabLab(q) =
    \Lab(x) = \Lab(x')$, it follows that $\all T D \in \TabLab(q)$. If $p =
    [q|\frac{x}{x'}]$, then $x'$ is an $\Inv(R)$-successor of $\Tail(q)$, and
    hence $\Tail(q)$ is a $T$-neighbour of $x'$. Because $x'$ is not indirectly
    blocked, this implies $\all T D \in \TabLab(q) = \Lab(\Tail(q))$.
  \item \textbf{\tab {12}}: assume $\qnrgleq n S D \in \TabLab(p)$ and $\tuple p
    q \in \Edges(S)$. If $q = [p | \frac{x}{x'}]$, then $x'$ is an $S$-successor
    of $\Tail(p)$ and thus $\{D,\nneg D\} \cap \Lab(x') \neq \emptyset$ because
    the \rulechoose-rule is not applicable.  Since $\TabLab(q) = \Lab(x) =
    \Lab(x')$, we have $\{D, \nneg D\} \cap \TabLab(q) \neq \emptyset$.  If $p =
    [q | \frac{x}{x'}]$, then $x'$ is an $\Inv(S)$-successor of $\Tail(q)$,
    $\Tail(q)$ is an $S$-neighbour of $x'$, and thus $\{D, \nneg D\} \cap
    \TabLab(q) = \Lab(\Tail(q)) \neq \emptyset$ because $x'$ is not indirectly
    blocked and the \rulechoose-rule is not applicable.
  \item \textbf{\tab 8} is satisfied due to the symmetric definition of
    $\Edges$. 
  \item \textbf{\tab 9} is satisfied due to the definition of $R$-successors
    that takes into account the role hierarchy $\sss$.
  \item \textbf{\tab {10}}: assume $\qnrgeq n S D \in \TabLab(p)$. Completeness
    of \Tree implies that there exist $n$ distinct individuals $y_1, \dots, y_n$ in \Tree
    such that each $y_i$ is an $S$-neighbour of $\Tail(p)$ and $D \in
    \Lab(y_i)$. We claim that, for each of these individuals, there is a path
    $q_i$ such that $\tuple p {q_i} \in \Edges(S)$, $D \in \TabLab(q_i)$, and
    $q_i \neq q_j$ for all $1 \leq i < j \leq n$. Obviously, this implies
    $\sharp S^\mcT(p,D) \geq n$.  For each $y_i$, there are three possibilities:
    \begin{itemize}
    \item $y_i$ is an $S$-successor of $x$ and $y_i$ is not blocked in \Tree.
      Then $q_i = [p|\frac{y_i}{y_i}]$ is a path with the desired properties.
    \item $y_i$ is an $S$-successor of $x$ and $y_i$ is blocked in \Tree by some
      node $z$. Then $q_i = [p|\frac{z}{y_i}]$ is the path with the desired
      properties. Since the same $z$ may block several of the $y_j$s, it is
      indeed necessary to include the blocking nodes explicitly into the path
      construction to make these paths distinguishable.
    \item $x$ is an $\Inv(S)$-successor of $y_i$. Since $\Tree$ is a tree, there
      may be at most one such $y_i$. This implies that $p$ is of the form $p =
      [q | \frac{x}{x'}]$ with $\Tail(q) = y_i$.  The path $q$ has the desired
      properties and, obviously, $q$ is distinct from all other paths $q_j$.
    \end{itemize}
  \item Assume \textbf{\tab {11}} is violated. Hence there is some $p \in \mcS$
    with $\qnrleq n S D \in \TabLab(p)$ and $\sharp S^\mcT(p,D) > n$. We show
    that this implies $\sharp S^\Tree(\Tail(p),D) > n$, in contradiction to
    either  clash-freeness or completeness of \Tree.  Define $x = \Tail(p)$
    and $\mbP = S^\mcT(p,D)$.  Due to the assumption, we have $\sharp \mbP > n$.  We
    distinguish two cases:
    \begin{itemize}
    \item $\mbP$ contains only paths of the form $q = [p|\frac{y}{y'}]$. We claim
      that the function $\Tail'$ is injective on $\mbP$. Assume that there are two
      paths $q_1,q_2 \in \mbP$ with $q_1 \neq q_2$ and $\Tail'(q_1) = \Tail'(q_2) =
      y'$. Then $q_1$ is of the form $q_1 = [p | \frac{y_1}{y'}]$ and $q_2$ is
      of the form $q_2 = [p | \frac{y_2}{y'}]$ with $y_1 \neq y_2$. If $y'$ is
      not blocked in \Tree, then $y_1 = y' = y_2$, contradicting $q_1 \neq q_2$.
      If $y'$ is blocked in \Tree, then both $y_1$ and $y_2$ block $y'$, which
      implies $y_1 = y_2$, again a contradiction.

      Since $\Tail'$ is injective on $\mbP$, it holds that $\sharp \mbP = \sharp
      \Tail'(\mbP)$. Also for each $y' \in \Tail'(\mbP)$, $y'$ is an $S$-successor
      of $x$, and $D \in \Lab(y')$. This implies $\sharp S^\Tree(x,D) > n$.
    \item $\mbP$ contains a path $q$ where $p$ is of the form $p = [q |
      \frac{x}{x'}]$.  Obviously, $\mbP$ may only contain one such path. As in the
      previous case, $\Tail'$ is an injective function on the set $\mbP' := \mbP
      \setminus \{ q \}$, each $y' \in \Tail'(\mbP')$ is an $S$-successor of $x$
      and $D \in \Lab(y')$ for each $y' \in \Tail'(\mbP')$. To show that indeed
      $\sharp S^\Tree(x,D)>n$ holds, we have to prove the existence of a further
      $S$-neighbour $u$ of $x$ with $D \in \Lab(u)$ and $u \not\in \Tail'(\mbP')$.
      We distinguish two cases:
      \begin{itemize}
      \item $x=x'$. Hence $x$ is not blocked. This implies that $x$ is an
        $\Inv(S)$-successor of $z$ in \Tree. Since $\Tail'(\mbP')$ contains only
        successors of $x$, we have that $z \not\in \Tail'(\mbP')$ and, by
        construction, $z$ is an $S$-neighbour of $x$ with $C \in \Lab(z)$.
      \item $x \neq x'$. This implies that $x'$ is blocked in \Tree by $x$ and
        that $x'$ is an $\Inv(S)$-successor of $z$ in \Tree. The definition of
        pair-wise blocking implies that $x$ is an $\Inv(S)$-successor of some
        node $u$ in \Tree with $\Lab(u) = \Lab(z)$. Again, since $\Tail'(\mbP')$
        contains only successors of $x$, we have that $u \not\in \Tail'(\mbP')$ and,
        by construction, $u$ is an $S$-neighbour of $x$ with $D\in \Lab(u)$. \qed
      \end{itemize}
    \end{itemize}
  \end{itemize}
\end{proof}

By showing termination (Lemma~\ref{lem:shiq-termination}), soundness
(Lemma~\ref{lemma:shiq-completeness}), and completeness
(Lemma~\ref{lemma:shiq-soundness}) of the \shiq-algorithm, we have established
its correctness:

\begin{theorem}\label{theorem:shiq-decision-procedure}
  The \shiq-algorithm is a non-deterministic decision procedure for
  satisfiability and subsumption of \shiq-concepts \wrt a role hierarchy.
\end{theorem}

Of course, due to Theorem~\ref{theorem:shiq-internalisation}, the tableau
algorithm can also be used to decide satisfiability and subsumption of
\shiq-concepts \wrt a general TBox.  To apply the algorithm to \shiq-knowledge
bases, one can either use a pre-completion approach similar to the one used to
prove Theorem~\ref{theo:alcqir-kb-exptime-complete} (probably with
catastrophic effects on the runtime of the algorithm), or one can integrate
the ABox directly into the tableau algorithm. Horrocks, Sattler, and Tobies
\citeyear{HorrSattTob-CADE-2000} present an algorithm that follows the latter
approach.

We have already mentioned that we do not expect to obtain a worst-case optimal
solution for \shiq-satisfiability from the tableau approach---such an algorithm
has already been given in order to prove Theorem~\ref{cor:shiq-exptime}.
Instead, Algorithm~\ref{alg:shiq} is intended as a \emph{practical} decision
procedure that can be optimized so that it performs well for reasoning tasks
occurring in applications. Nevertheless, it is interesting to know how far our
tableau approach exceeds the worst-case complexity.

\begin{lemma}
  The \shiq-algorithm runs in 2-\nexptime.
\end{lemma}

\begin{proof}
  Let $C$ be a \shiq-concept and $\R$ a role hierarchy. Let $m = \sharp
  \clos(C,\R)$, $k = \sharp \overline{\Roles}_{C,R}$, and $\nmax$ be the maximum
  $n$ that occurs in a qualifying number restriction in $\clos(C,\R)$. If we set
  $n = |C| + |\R|$, the it holds that $m = \mcO(|C| \cdot |\R|) = \mcO(n^2)$, $k
  = \mcO(|C| + |R|) = \mcO(n)$, and $\nmax = \mcO(2^{|C|}) = \mcO(2^n)$. In the
  proof of Lemma~\ref{lem:shiq-termination}, we have shown that paths in a
  completion tree for $C$ become no longer than $2^{2mk}$ and that the
  out-degree of a completion tree is bounded by $m \cdot \nmax$.  Hence, the
  \shiq-algorithm will construct a tree with no more than
  \[
  (m \cdot \nmax)^{2^{2mk}} = \mcO((n^2 \cdot 2^n)^{2^{2n^3}}) = \mcO(2^{n \cdot
    2^{2 n^3}}) = \mcO(2^{2^{n^4}})
  \]
  nodes. Each node of this tree is labelled with a subset of $\clos(C,\R)$ and
  each edge is labelled with a subset of $\overline{\Roles}_{C,\R}$. Since
  every application of a rule either adds a node to the tree,  a concept or
  role to one of the labels, or sets the label of an edge to $\emptyset$ (in
  which case the corresponding successor is blocked forever), the
  \shiq-algorithm runs in 2-\nexptime. \qed
\end{proof}

This seems to be a discouraging result: the tableau algorithm runs in
2-\nexptime while the worst-case complexity is only \exptime. On the other hand,
there exist DL systems like \iFaCT \cite{horrocks99:_fact}, which is based on
Algorithm~\ref{alg:shiq}, or RACE \cite{Haarslev99a}, which is based on a
similar algorithm \cite{HaarslevMoeller-KR-2000}. These systems show good
performance in system comparisons \cite{MassacciDonini-Tableaux2000} and are
successfully utilized in a number of applications
\cite<e.g.,>{HaarslevMoeller-DL-2000-HPR,FranconiNg-KRDB-2000}.  This can be
explained by the fact that tableau algorithms seem to be particularly amenable
to optimizations
\cite{BaaderFranconi+-OptJournal-94,Horrocks97b,horrocks99:_optim_descr_logic_subsum,HorrocksTobies-KR-2000,HaarslevMoeller-DL-2000-PSEUDO}.
It is these optimizations that cause the good behaviour of implementations based
on tableau algorithms like Algorithm~\ref{alg:shiq}.



\cleardoublepage


\chapter{Guarded Fragments}

\label{chap:gf}

\newcommand{\gfruleand}{\genericrule{\wedge}}
\newcommand{\gfruleor}{\genericrule{\vee}}
\newcommand{\ruleeq}{\genericrule{{=}}}
\newcommand{\rulepropa}{\genericrule{\updownarrow}}
\newcommand{\rulepropfa}{\genericrule{{\updownarrow \forall}}}

The Guarded Fragment \index{guarded fragment} of first-order logic, introduced
by Andr\'eka, van Benthem, and N\'emeti \citeyear{AndrekaBenNem98}, is a
successful attempt to transfer many good properties of modal, temporal, and
description logics to a large, naturally defined fragment of predicate logic.
Among these are decidability, the finite model property, invariance under an
appropriate variant of bisimulation, and other nice model theoretic
properties~\cite{AndrekaBenNem98,Graedel99a}.

The Guarded Fragment (GF) \index{guarded fragment} is obtained from full first-order logic through
relativization of quantifiers by so-called guard formulas.
Every appearance of a quantifier in \GF\ must be of the form
\[
\ex \bfy (\alpha(\bfx,\bfy) \wedge \phi(\bfx,\bfy)) \ \text{or} \ \fa
\bfy (\alpha(\bfx,\bfy) \rightarrow \phi(\bfx,\bfy)) ,
\]
where $\alpha$ is a positive atomic formula, the \emph{guard}, that contains all
free variables of $\phi$. This generalises quantification in description, modal,
and temporal logics, where quantification is restricted to those elements
reachable via some accessibility relation. For example, in DLs, quantification
occurs in the form of existential and universal restrictions like $\all
\hasChild \rich$, which expresses that those individuals \emph{reachable via the
  role} (guarded by) \hasChild must be rich.

By allowing for more general formulas as guards while preserving the idea of
quantification only over elements that are \emph{close together} in the model,
one obtains generalisations of \GF\ which are still well-behaved in the above
sense.  Most importantly, one can obtain the \iemph{loosely guarded fragment}
(\LGF)~\cite{Benthem97} and the \iemph{clique guarded fragment}
(\CGF)~\cite{Graedel99b}, for which decidability, invariance under clique
guarded bisimulation, and some other properties have been shown in
\cite{Graedel99b}. 

Guarded fragments have spawned considerable interest in the DL community, mainly
for two reasons. On the one hand, many DLs can be embedded into suitable guarded
fragments, which allows the transfer, e.g., of decidability results from guarded
logics to DLs.  Goncalves and Gr{\"a}del \citeyear{GoncalvesGra00} prove
decidability of the guarded fragment $\mu$ACGFI, which, among other DLs, allows
a simple embedding of \alcqi and \alcqir, proving the decidability of these
logics. On the other hand, guarded fragments generalise DLs and add expressive
power that is not present in classical DLs, but interesting for knowledge
representation. For example, Lutz, Sattler, and Tobies
\citeyear{LutzSattlerTobies-DL-99} present a restriction of \GF\ that strictly
contains \alci and allows for $n$-ary relations instead of the binary roles of
most DLs.

\GF, \LGF, and \CGF\ are decidable and known to be 2-\EXPTIME\ complete, which
is shown by Gr{\"a}del~\citeyear{Graedel99b,Graedel99a} using game and
automata-based approaches. For these guarded fragments, the automata approach
has the same problems as it has for modal and description logics: it is unclear how to
turn the automata decision procedures into efficient implementations and a naive
implementation has every-case exponential complexity, which makes it unusable
for applications.  So, while these approaches yield (worst-case) optimal
complexity results for many logics, they appear to be unsuitable as a starting
point for an efficient implementation. As we have seen, many decidability
results for modal or description logics are based on tableau algorithms and some
of the fastest implementations of modal satisfiability procedures are based on
tableau calculi~\cite{Horrocks-Tableaux-2000,Patel-Schneider-Tableaux2000}.
Unlike automata algorithms, the average-case behaviour in practice is so good
that finding \emph{really} hard problems to test these implementations has
become a problem in itself \cite{HorrocksPatelSSebastiani-IGPL-2000}.
In this chapter, we generalise the principles of the tableau algorithms
encountered in this thesis to develop a tableau algorithm for \CGF.  


Recall the conjecture by Vardi that the tree model property is the main reason
for the decidability of many modal style logics \cite{Vardi97}. As pointed out
in \cite{Graedel99a}, the generalised tree model property explains the similarly
robust decidability of guarded logics, and can be seen as a strong indication
that guarded logics are a generalisation of modal logics that retain the essence
of modal logics. This becomes even more evident when regarding the respective
fixed-point extensions \cite{GraedelWal99} and is the foundation of general
decidability results for guarded logics via reduction to the modal
$\mu$-calculus and the monadic theory of countable trees ($S\omega S$)
\cite{graedel-tcs-2001}.  The generalised tree model property of \CGF\ is also
essential for our tableau algorithm. Indeed, as a corollary of the constructions
used to show the soundness of our algorithm, we obtain an alternative proof for
the fact that \CGF\ has the generalised tree model property.


\section{Syntax and Semantics}

For the definitions of \GF\ and \LGF\ we refer the reader to
\cite{Graedel99b}. The \iemph{clique guarded fragment} \CGF\ of first-order logic can
be obtained in two equivalent ways, by either semantically or syntactically
restricting the range of the first-order quantifiers. In the following we will use bold
letters to refer to tuples of elements of the universe ($\bfa,\bfb,\dots$)
resp.\ tuples of variables ($\bfx,\bfy,\dots$).

\begin{definition}[Semantic CGF]
  Let $\tau$ be a relational vocabulary. For a $\tau$-structure $\mfA$ with universe $A$, the
  \iemph{Gaifman graph} of $\mfA$ is defined as the undirected graph $G(\mfA) =
  (A, E^\mfA)$ with
  \begin{align*}
    E^\mfA = \{ & (a,a') \;:\;   a \neq a', \text{there exists } R
    \in \tau \text{ and} \\  &  \bfa \in R^\mfA  \text{ which contains both } a  \text{ and } a' \} .
  \end{align*}
  Under \iemph{clique guarded semantics} we
  understand the modification of standard first-order semantics, where, instead
  of ranging over all elements of the universe, a quantifier is restricted to
  elements that form a clique in the Gaifman graph, including the binding for
  the free variables of the matrix formula.
  More precisely, let $\mfA$ be a $\tau$-structure
  and $\rho$ an environment mapping variables to elements of $A$.  We define
  the model relation inductively over the structure of formulas as the usual
  \FO\ semantics with the exception
  \[
  \begin{array}{@{}p{\textwidth}@{}}
    \multicolumn{1}{l}{\mfA,\rho \models \fa y . \phi(\bfx,y)  \mbox{ iff,
        for all } a \in A, \mbox{ such that }}\\
    \multicolumn{1}{c}{\rho(\bfx) \cup \{ a \} \mbox{ forms a clique in } G(\mfA),}\\ 
    \multicolumn{1}{r}{\mbox{it is the case that } \mfA,\rho[x \mapsto a]\models \phi},
  \end{array}
  \]
  and a similar definition for the existential case.  With \CGF\ we denote
  first-order logic restricted to clique guarded semantics. \index{clique
    guarded fragment!semantic} \eod
\end{definition}

\begin{definition}[Syntactic CGF]
  Let $\tau$ be a relational vocabulary.  A formula $\alpha$ is a
  \iemph{clique-formula} for a set $\bfx \subseteq \free(\alpha)$ if $\alpha$
  is a (possibly empty if $\bfx$ contains only one variable) conjunction of atoms (excluding equality statements)
  such that each two distinct elements from $\bfx$ coexist in
  at least one atom, each atom contains at least an element from $\bfx$, and
  each element from $\free(\alpha) \setminus \bfx$ occurs exactly once in $\alpha$.
  In the following, we will identify a clique-formula $\alpha$ with the set of its conjuncts.
  
  The \emph{syntactic \CGF} \index{clique guarded fragment!syntactic} is inductively defined as follows.
  \begin{enumerate}
  \item Every relational atomic formula $Rx_{i_1}\dots x_{i_m}$ or $x_i = x_j$ 
    belongs to \CGF.
  \item \CGF\ is closed under Boolean operations.
  \item If $\bfx,\bfy,\bfz$ are tuples of variables, $\alpha(\bfx,\bfy,\bfz)$
    is a \emph{clique-formula} for $\bfx \cup \bfy$ and $\phi(\bfx,\bfy)$ is a
    formula in $\CGF$ such that $\free(\phi) \subseteq \bfx \cup \bfy$,
    \[
    \begin{array}{ll}
      \mbox{then} & \quad \ex \bfy \bfz . (\alpha(\bfx,\bfy,\bfz) \wedge \phi(\bfx,\bfy)) \\
      \mbox{and} & \quad \fa \bfy \bfz . (\alpha(\bfx,\bfy,\bfz) \rightarrow \phi(\bfx,\bfy))
    \end{array}
    \]
    belong to \CGF. 
  \end{enumerate}
  
  We will use $(\ex \bfy \bfz . \alpha(\bfx,\bfy,\bfz)) \phi(\bfx,\bfy)$ and
  $(\fa \bfy \bfz . \alpha(\bfx,\bfy,\bfz)) \phi(\bfx,\bfy)$ as alternative
  notations for $\ex \bfy \bfz . (\alpha(\bfx,\bfy,\bfz) \wedge
  \phi(\bfx,\bfy))$ and $\fa \bfy \bfz .  (\alpha(\bfx,\bfy,\bfz) \rightarrow
  \phi(\bfx,\bfy))$, respectively. A formula of the form $\fa \bfy \bfz .
  (\alpha(\bfx,\bfy,\bfz) \rightarrow \phi(\bfx,\bfy))$ is called
  \emph{universally quantified}. \eod
  
\end{definition}

The following Lemma can be shown by elementary formula manipulations that
exploit that every $z \in \bfz$ occurs exactly once in $\alpha$.

\begin{lemma}\label{lem:guard-ex-fa}
  Let $\alpha(\bfx,\bfy,\bfz)$ be a clique-formula for $\bfx,\bfy$. Then
  \[
  \fa \bfy \bfz . (\alpha(\bfx,\bfy,\bfz) \rightarrow \phi(\bfx,\bfy)) 
  \equiv  \; \fa \bfy .  (\exists \bfz . \alpha(\bfx,\bfy, \bfz) \rightarrow
  \phi(\bfx,\bfy)) .
  \]
\end{lemma}

The use of the name \CGF\ for both the semantic and the syntactic clique guarded 
fragment is justified by the following Lemma.

\begin{lemma}
  Over any finite relational vocabulary
  the syntactic and semantic versions of the \CGF\ are equally expressive.
\end{lemma}

\noindent \textbf{Proof sketch:} By some elementary equivalence transformations, every
syntactically clique guarded formula can be brought into a form where
switching from standard semantics to clique guarded semantics does not change
its meaning.  Conversely, for any finite signature there is a finite disjunction
$clique(\bfx,y,\bfz)$ of clique-formulas for $\bfx,y$ such that $\bfa,b$ form
a clique in $G(\mfA)$ iff $\mfA \models \ex \bfz . clique(\bfa,b,\bfz)$. By
guarding every quantifier with such a formula and applying some elementary
formula transformations and Lemma~\ref{lem:guard-ex-fa}, we get, for every FO
formula $\psi$, a syntactically clique guarded formula that is equivalent to
$\psi$ under clique guarded semantics. If we fix a finite relational vocabulary,
this transformation is polynomial in the number of variables of the formula, or,
more precisely, the maximal number of free variables of all sub-formulas. \qed

\vspace{1ex}

In the following we will only consider the syntactic variant of the clique
guarded fragment.

\begin{definition}[NNF, Closure, Width]
  In the following,  all
  formulas are assumed to be in negation normal form (NNF), \index{negation
    normal form} where negation occurs only
  in front of atomic formulas. Every formula in \CGF\ can be transformed into
  NNF in linear time by pushing negation inwards using DeMorgan's law and the
  duality of the quantifiers.
  
  For a sentence $\psi \in \CGF$ in \NNF, let $\clos(\psi)$
  \index{00clospsi@$\clos{\psi}$} be the smallest set
  that contains $\psi$ and is closed under sub-formulas. Let $C$ be a set of
  constants. With $\clos(\psi,C)$ \index{00clospsic@$\clos(\psi,C)$} we denote the set
  \[
  \clos(\psi,C) = \{ \phi(\bfa) \;:\; \bfa \subseteq C, \phi(\bfx) \in \clos(\psi)
  \} .
  \]
  The \emph{width} \index{width of a formula} of a formula $\psi \in \CGF$ is defined by
  \[ 
  \width(\psi) := \max\{|\free(\phi)| \;:\; \phi \in \clos(\psi) \} .
  \]
   \eod
\end{definition}


\section{Reasoning with Guarded Fragments}

Many of the approaches for decision procedures for description and modal logics
described in Section~\ref{sec:reasoning-methodologies} have been successfully
applied for \CGF: Gr{\"a}del \citeyear{Graedel99b} shows decidability of (a
fixed point extension of) \CGF\ using translation to the monadic second-order theory
of countable trees $S\omega S$
\cite{rabin69:_decid_secon_order_theor_autom_infin_trees} and to the modal
$\mu$-calculus with backwards modalities
\cite{vardi98:_reason_about_past_two_way_autom}. Also in \cite{Graedel99b}, it
is shown that \CGF\ is 2-\exptime-complete and \exptime-complete for sentences of
bounded width, where the upper bound is based on a reduction to emptiness of
alternating two-way automata\cite{vardi98:_reason_about_past_two_way_autom}.
Resolution based decision procedures for guarded fragments are described in
\cite{GanzNiv99,NivelleRij00} where the approach in \cite{GanzNiv99} can be
extended to \CGF. Finally, a tableau decision procedure for
a fragment of \GF\ is given in \cite{marx00:_label_deduc_guard_fragm}. Yet, to
the best of our knowledge, there does not exist a tableau decision procedure
that is capable of deciding the full \GF, let alone \CGF. In the following, we
will supply such an algorithm. As it turns out, the algorithm as well as the
proof of its correctness mainly employ ideas we have already encountered in the
algorithms and proofs for DLs in this thesis---another indication for the modal
nature of \CGF, since it is amenable to the same techniques successfully used for
description and modal logics.

Let us briefly recall the main ``ingredients'' of tableau algorithms for modal
or description logics like the ones encountered in this thesis. Satisfiability
of a concept $C$ is decided by a syntactically guided search for a model for
$C$.  Models are usually represented by a graph in which the nodes correspond to
elements and the edges correspond to the role relations in the model.  Each node
is labelled with a set of concepts that this node must satisfy, and new edges and
nodes are created as required by existential restrictions. Since many modal and
description logics have the tree model property, the graphs generated by these
algorithms are trees, which allows for simpler algorithms and easier
implementation and optimization of these algorithms.  Indeed, some of the
fastest implementations of modal or description logics satisfiability algorithms
use tableau calculi~\cite{Horrocks-Tableaux-2000,Patel-Schneider-Tableaux2000}.

For many modal or description logics, e.g. \textsf{K} or \alc, termination of
these algorithms is due to the fact that the nesting of universal or existential
restrictions of the concepts appearing at a node strictly decreases with every
step from the root of the tree (e.g., compare Lemma~\ref{lem:alc-tree-bounds}).
For other logics, e.g., \textsf{K4}, \textsf{K} with the universal modality, or
the DLs \si and \shiq, this is no longer true and termination has to be enforced
by other means. One possibility for this is \emph{blocking}, i.e., stopping the
creation of new successor nodes below a node $v$ if there already is an ancestor
node $w$ that is labelled with similar concepts as $v$ (e.g., compare
Lemma~\ref{lemma:si-termination}).  Intuitively, in this case the model can fold
back from the predecessor of $v$ to $w$, creating a cycle.  Unraveling of these
cycles recovers an (infinite) tree model. Since the algorithms guarantee that
the concepts occurring in the label of the nodes stem from a finite set (usually
the sub-concepts of the input concept), every growing path will eventually
contain a blocked node, preventing further growth of this path and (together
with a bound on the degree of the tree) ensuring termination of the algorithm.

\subsection{Tableau Reasoning for \CGF}

Our investigation of a tableau algorithm for \CGF\ starts with the
observation that \CGF\ also has some kind of tree model property.

\begin{definition}
  Let $\tau$ be a relational vocabulary. A $\tau$-structure $\mfA$ has
  \iemph{tree width} $k$ if $k \in \mathbb{N}$ is minimal with the following property.
  
  There exists a directed tree $T=(V,E)$ and a function $f : V \rightarrow
  2^A$ such that
  \begin{itemize}
  \item for every $v \in V$, $|f(v)| \leq k+1$,
  \item for every $R \in \tau$ and $\bfa \in R^\mfA$, there exists $v \in V$ 
    with $\bfa \subseteq f(v)$, and
  \item for every $a \in A$, the set $V_a = \{ v \in V \;:\; a \in f(v) \}$
    induces a subtree of $T$.
  \end{itemize}
  Every node $v$ of $T$ induces a substructure $\mfF(v) \subseteq \mfA$ of
  cardinality at most $k+1$.  The tuple $\langle T, (\mfF(v))_{v \in T} \rangle$
  is called a \iemph{tree decomposition} of $\mfA$.
  
  A logic $\mcL$ has the \iemph{generalised tree model property} if there
  exists a computable function $t$, assigning to every sentence $\psi \in
  \mcL$ a natural number $t(\psi)$ such that, if $\psi$ is satisfiable, then
  $\psi$ has a model of tree width at most $t(\psi)$. \eod
\end{definition}

\begin{fact}[Tree Model Property for \CGF]\label{fact:tree-model-property}
  \index{guarded fragment!clique guarded fragment!tree model property of the}
  Every satisfiable sentence $\psi$ $\in \CGF$ of width $k$ has a countable
  model of tree width at most $k-1$.
\end{fact}

This is a simple corollary of \cite[, Theorem 4]{Graedel99b}, where the same
result is given for $\mu$\CGF, that is \CGF\ extended by a least fixed point
operator.

Fact~\ref{fact:tree-model-property} is the starting point for our definition
of a \emph{completion tree} for a formula $\psi \in \CGF$. A node $v$ of such
a tree no longer stands for a single element of the model (as in the modal case),
but rather for a substructure $\mfF(v)$ of a tree decomposition of the model. To
this purpose, we label every node $v$ with a set $\bfC(v)$ of constants (the
elements of the substructure) and a subset of $cl(\psi,\bfC(v))$, reflecting
the formulas that must hold true for these elements.

To deal with \emph{auxiliary elements}---elements helping to form a clique in
$G(\mfA)$ that are not part of this clique themselves---we will use the auxiliary constant symbol
$*$ as a placeholder for unspecified elements in atoms.  The intention is to
keep the number of constants at each node as small as possible.  The $*$
will be used for the extra elements occurring in clique formulas that are not
part of the clique itself.

The following definitions are useful when dealing with these generalised atoms.

\begin{definition}\label{def:any-element}
  Let $K$ denote an infinite set of constants and $* \not \in K$.  For any set
  of constants $C \subseteq K$ we set $C^* = C \cup \{ * \}$. We use
  $t_1, t_2, \dots$ to range over elements of $K^*$. The relation $\gtast$ is
  defined by \index{00geqstar@$\gtast$} \index{00inast@$\inast$}
  \[
  R t_1 \dots t_n \gtast R t'_1 \dots t'_n \text{ iff for  all $i \in \{ 1
    \dots n \}$} \text{ either } t_i = {*} \text{ or } t_i = t'_i .
  \]
  For an atom $\beta$ and a set of formulas $\Phi$ we define $\beta \inast \Phi$ iff
  there is a $\beta' \in \Phi$ with $\beta \gtast \beta'$.
 
  For a set of constants $C \subseteq K$ and an atom $\beta = R t_1 \dots t_n$, we define
  \[
  \beta \capast C = R t'_1 \dots t'_n \; \mbox{ where } \; t'_i = \begin{cases}
    t_i & \mbox{if } t_i \in C,\\
    * & \mbox{otherwise.}
  \end{cases}
  \] \index{00capast@$\capast$} \eod
\end{definition}

We use the notation $\bfa^*$ to indicate that the tuple $\bfa^*$ may contain
$*$'s. Obviously, $\gtast$ is transitive and reflexive, and $\beta \capast C \gtast
\beta$ for all atoms $\beta$ and sets of constants $C$.

While these are all syntactic notions, they have a semantic counterpart that
clarifies the intuition of $*$ standing for an unspecified element. Let
$\bfa'$ denote the tuple obtained from a tuple $\bfa^*$ by replacing every
occurrence of $*$ in $\bfa^*$ with a distinct fresh variable, and let $\bfz$
be precisely the variables used in this replacement. For an atom $\beta$, we define
\[
\mfA \models \beta(\bfa^*) \; \mbox{ iff } \; \mfA \models \exists \bfz . \beta(\bfa') .
\]
It is easy to see that
\[
\begin{array}{rcl}
  \beta(\bfa) \gtast \beta(\bfb) & \text{implies} &  \beta(\bfb) \models
  \beta(\bfa), \text{ and}\\
  \beta(\bfa) \inast \Phi & \text{implies} & \Phi \models \beta(\bfa) 
\end{array}
\]
because, if $\bfa \gtast \bfb$, then $\bfb$ is obtained from $\bfa$ by
replacing some $*$ with constants, which provide witnesses for the existential
quantifier.

We further write $\Phi|_C$ \index{00Phirestrict@$\Phi"|_C$} to denote the subset
of $\Phi$ containing all formulas that only use constants in $C$.

\begin{algorithm}[The \CGF-algorithm]\label{alg:cgf}
  Let $\psi \in \CGF$ be a closed formula in \NNF. A \emph{completion tree}
  \index{completion tree!for the clique guarded fragment}
  $\bfT = (\bfV,\bfE,\bfC,\Delta,\bfN)$ for $\psi$ is a node labelled tree
  $(\bfV,\bfE)$ with the labelling function $\bfC$ labelling each node $v \in
  \bfV$ with a subset of $K$, $\Delta$ labelling each node $v \in \bfV$ with a
  subset of $cl(\psi,\bfC(v)^*)$ where all formulas $\beta(\bfx, \ast,\dots,\ast)
  \in \Delta(v)$ using $\ast$ are atoms (excluding equality statements),
  and the function $\bfN : \bfV \to \setN$ mapping each node to a distinct
  natural number, with the additional property that, if $v$ is an ancestor of
  $w$, then $\bfN(v) < \bfN(w)$.
  
  \index{tableau algorithm!for the clique guarded fragment}
  \index{clique guarded fragment!tableau algorithm}
  
  A constant $c \in K$ is called \emph{shared} \index{shared constant} between
  two nodes $v_1, v_2 \in \bfV$, if $c \in \bfC(v_1) \cap \bfC(v_2)$, and $c \in
  \bfC(w)$ for all nodes $w$ on the (unique, undirected, possibly empty)
  shortest path connecting $v_1$ to $v_2$.
  
  A node $v \in \bfV$ is called \emph{directly blocked}\footnote{The definition
    of blocking is recursive. Like for the \si- and the \shiq-algorithm, this
    does not cause any problems because the status of a node $v$ only depends on
    its label and the status of nodes $w$ with $\bfN(w) < \bfN(v)$. The
    recursion terminates at the root node, where the $\bfN$-value is minimal.}
  by a node $w \in \bfV$, if $w$ is not blocked, $\bfN(w) < \bfN(v)$, and there
  is an injective mapping $\pi$ from $\bfC(v)$ into $\bfC(w)$ such that, for all
  constants $c \in \bfC(v)$ that are shared between $v$ and $w$, $\pi(c) = c$,
  and $\pi(\Delta(v)) = \Delta(w)|_{\pi(\bfC(v)^*)}$. Here and throughout this
  thesis we use the convention $\pi(*) = *$ for every function $\pi$ that
  verifies a blocking.
  
  A node is called \emph{blocked} if it is directly blocked or if its
  predecessor is blocked.
  
  A completion tree $\bfT$ \emph{contains a clash} if there is a node $v \in
  \bfV$ such that
  \begin{itemize}
  \item for a constant $c \in \bfC(v)$, $c \neq c \in \Delta(v)$,
    or 
  \item there is an atomic formula $\beta$ and a tuple $\bfa
    \subseteq \bfC(v)$ such that $\{ \beta(\bfa), \neg \beta(\bfa) \} \subseteq
    \Delta(v)$.
  \end{itemize}
  Otherwise, $\bfT$ is called \emph{clash-free}.  A completion tree $\bfT$ is
  called \emph{complete} if none of the \emph{completion rules} given in
  Figure~\ref{fig:completion-rules} can be applied to $\bfT$. A complete and
  clash-free completion tree for $\psi$ is called a \emph{tableau} for $\psi$.
  
  To test $\psi$ for satisfiability, the tableau algorithm creates an initial
  tree with only a single node $v_0$, $\Delta(v_0) = \{ \psi \}$ and $\bfC(v_0)
  = \{ a_0 \}$ for an arbitrary constant $a_0$. The rules from
  Figure~\ref{fig:completion-rules} are applied until either a clash occurs,
  producing output ``$\psi$ is not satisfiable'', or the tree is complete, in
  which case ``$\psi$ is satisfiable'' is output. \eod
\end{algorithm}

The set $\bfC(v_0)$ is initialized with a non-empty set of constants to make
sure that empty structures are excluded. For a discussion of the different kinds
of non-determinism that occur in the \CGF-algorithm, see below
Lemma~\ref{lem:completeness-alt}.

\begin{figure}[b]
  \footnotesize
  \begin{tabular}{lll}
    \gfruleand: & \bfif \hfill 1. & $\phi \wedge \theta \in \Delta(v)$\\
    & \hfill 2.& $\{\phi,\theta \} \not    \subseteq \Delta(v)$\\
    & \bfthen & $\Delta(v) \gfruleand \Delta(v) \cup \{ \phi,\theta \}$\\[1ex]

    \gfruleor: & \bfif \hfill 1. & $\phi \vee \theta \in \Delta(v)$ and\\
    & \hfill 2. & $\{\phi,\theta \} \cap \Delta(v) = \emptyset$\\
    & \bfthen & $\Delta(v) \gfruleor \Delta(v) \cup \{ \chi \}$ for some $\chi \in 
    \{ \phi,\theta \}$\\[1ex]

    \ruleeq: & \bfif \hfill & $a=b \in \Delta(v)$ and  $a \neq b$ \\
    & \bfthen & for all $w$ that share $a$  with  $v$, $\bfC(w) \ruleeq (\bfC(w)
    \setminus \{ a \}) \cup \{ b \}$\\
    && and $\Delta(w) \ruleeq \Delta(w)[a \mapsto  b]$\\[1ex]

    \rulefa : & \bfif \hfill 1. & $(\fa
    \bfy\bfz.\alpha(\bfa,\bfy,\bfz))\phi(\bfa,\bfy) \in 
     \Delta(v)$, and \\
     & \hfill 2. & there exists  $\bfb \subseteq \bfC(v)$ such that for all
     $\beta(\bfx,\bfy,\bfz) \in \alpha$,  $\beta(\bfa,\bfb,* \dots *) \inast  
     \Delta(v)$, and \\ 
     & \hfill 3. & $\phi(\bfa,\bfb) \not \in \Delta(v)$\\
     & \bfthen & $\Delta(v) \rulefa \Delta(v) \cup \{ \phi(\bfa,\bfb) \}$\\[1ex]

    \ruleex: & \bfif \hfill 1. & $(\ex
    \bfy\bfz.\alpha(\bfa,\bfy,\bfz))\phi(\bfa,\bfy) \in 
    \Delta(v)$, and\\
    & \hfill 2. &  for every  $\bfb, \bfc \subseteq \bfC(v),\{
    \alpha(\bfa,\bfb,\bfc), \phi(\bfa,\bfb) \} \not\subseteq \Delta(v)$, and\\
    & \hfill 3. & there is no child $w$ of $v$ with $\{ \alpha(\bfa,\bfb,\bfc),
    \phi(\bfa,\bfb) \} \subseteq \Delta(w)$ for some  $\bfb,\bfc \subseteq
    \bfC(w)$, and\\
    & \hfill 4. & $v$ is not blocked\\
    & \bfthen & $\bfV \ruleex \bfV \cup \{w \}$, $\bfE \ruleex \bfE \cup \{
    (v,w)\}$ for a fresh node $w$\\
    && let  $\bfb,\bfc$ be sequences of distinct and fresh constants that match\\
    &&the lengths of $\bfy, \bfz$ and set\\ 
    && $\bfC(w) = \bfa \cup \bfb \cup
    \bfc$,  and \\ 
    && $\Delta(w) = \{\alpha(\bfa,\bfb,\bfc), \phi(\bfa,\bfb)  \}$, and\\
    && $\bfN(w) =  1 + \max \{ \bfN(v) \;:\; v \in \bfV \setminus\{w\} \}$\\[1ex]

    \rulepropa: & \bfif \hfill 1. & $\beta(\bfa^*)  \in \Delta(v)$, $\beta$
    atomic, not an equality, and\\
    & \hfill 2. & $w$ is a neighbour of  $v$  with $\bfa^* \cap \bfC(w) \neq
    \emptyset$, and\\
    & \hfill 3. & $\beta(\bfa^*)\capast{\bfC(w)} \not \in \Delta(w)$\\
    & \bfthen & $\Delta(w) \rulepropa \Delta(w) \cup \{
    \beta(\bfa^*)\capast{\bfC(w)}\}$\\[1ex]

     \rulepropfa: & \bfif \hfill 1.& $\phi(\bfa) \in  \Delta(v)$, $\phi(\bfa)$
     is universally. quantified, and\\
     & \hfill 2. & $w$ is  a neighbour of $v$ with  $\bfa \subseteq \bfC(w)$,
     and\\
     & \hfill 3. & $\phi(\bfa) \not \in \Delta(w)$\\
     & \bfthen & $\Delta(w) \rulepropfa \Delta(w) \cup \{ \phi(\bfa) \}$
  \end{tabular}
  \caption{The completion rules for \CGF}
  \label{fig:completion-rules}
  \index{clique guarded fragment!tableau algorithm!completion rules}
  \index{completion rules!for the clique guarded fragment}
  \index{00ruleand@\gfruleand}
  \index{00ruleor@\gfruleor}
  \index{00ruleeq@\ruleeq}
  \index{00rulefa@\rulefa}
  \index{00ruleex@\ruleex}
  \index{00rulepropa@\rulepropa}
  \index{00rulepropfa@\rulepropfa}
\end{figure}

While our notion of tableaux has many similarities to the tableaux appearing in
\cite{GraedelWal99}, there are two important differences that make the version
used here more suitable as basis for a tableau algorithm.  We will see that
every completion tree generated by the tableau algorithm is finite. Conversely,
tableaux in~\cite{GraedelWal99}, in general, can be infinite.  Also,
in~\cite{GraedelWal99} every node is labelled with a complete
$(\psi,\bfC(v))$-type, i.e., every formula $\phi \in \clos(\psi,\bfC(v))$ is
explicitly asserted true of false at $v$. Conversely, a completion tree contains
only assertions about relevant formulas. This implies a lower degree of
non-determinism in the algorithm, which is important for an efficient
implementation.

\subsection{Correctness}

The techniques used to establish correctness of the \CGF-algorithm
bear a strong resemblance to the techniques we have employed for the
tableau algorithms for description logics in the previous chapters.
Extra complexity is added by the fact that completion trees for \CGF
are more complex objects than the completion trees for Description
Logics, mainly because each node now stands for a substructure rather
than for a single element of the model.




\subsubsection{Termination}

The following technical lemma is a consequence of the completion rules
and the blocking condition.

\begin{lemma}\label{lem:tree-bounds}
  Let $\psi \in \CGF$ be a sentence in NNF with $|\psi| = n$,
  $\width(\psi)$ $= m$, and $\bfT$ a completion tree generated for $\psi$ by
  application of the rules in Figure~\ref{fig:completion-rules}. For every
  node $v$ in $\bfT$,
  \begin{enumerate}
  \item $|\bfC(v)| \leq m$,
  \item 
    $|\Delta(v)| \leq n \times (m+1)^m$, and
  \item 
    any $\ell > 2^{n \times (m+1)^m}$ distinct nodes in $\bfT$ contain a
    blocked node.
  \end{enumerate}
\end{lemma}

\begin{proof}
  Nodes are only generated when initializing the tree (with  a single
  constant) and by the \ruleex-rule and no constants are added to a $\bfC(v)$
  once $v$ has been generated (but some may be removed by application of the
  \ruleeq-rule).
    
  When triggered by the formula $(\ex \bfy \bfz . \alpha(\bfa,\bfy,
  \bfz))\phi(\bfa, \bfy)$, the \ruleex-rule initializes $\bfC(w)$ such that it
  contains $\bfa$ and another constant for every variable in $\bfx$ and $\bfy$.
  Hence,
  \[
  |\bfC(w)| \leq |\bfa \cup \bfy \cup \bfz| \leq |\free(\alpha)| \leq \width(\psi) . 
  \]
  
  The set $\Delta(v)$ is a subset of $cl(\psi,\bfC(v)^*)$, for which
  $|cl(\psi,\bfC(v))| \leq n \times (m+1)^m$ holds because there are at most $n$
  formulas in $cl(\psi)$, each of which has at most $m$ free variables.  There
  are at most $(|\bfC(v)|+1)^m$ distinct sequences of length $m$ with constants
  from $\bfC(v)^*$.
    
  Let $v_1, \dots, v_\ell$ be $\ell > 2^{n \times (m+1)^m}$ distinct nodes.  For
  every $v_i$, we will construct an injective mapping $\pi_i : \bfC(v_i)
  \rightarrow \{1, \dots m \}$ such that, if a constant $a$ is shared between
  two nodes $v_i,v_j$, then $\pi_i(a) = \pi_j(a)$.
    
  Let $u_1, \dotsm , u_k$ denote the nodes of a subtree of $\bfT$ that contains
  every node $v_i$ and that is rooted at $u_1$.  By induction over the distance
  to $u_1$, we define an injective mapping $\nu_i : \bfC(u_i) \rightarrow \{ 1,
  \dots, m \}$ for every $i \in \{1, \dots, k \}$ as follows.  For $\nu_1$ we
  pick an arbitrary injective function from $\bfC(u_1)$ to $\{1, \dots, m\}$.
  For a node $u_i$ let $u_j$ be the predecessor of $u_i$ in $\bfT$ and $\nu_j$
  the corresponding function, which has already been defined because $u_j$ has a
  smaller distance to $u_1$ than $u_i$. For $\nu_i$ we choose an arbitrary
  injective function such that $\nu_i(a) = \nu_j(a)$ for all $a \in \bfC(u_i)
  \cap \bfC(u_j)$.
    
  All mappings $\nu_i$ are injective. For any constant $a$ the set $\bfV_a := \{
  v \in \bfV \mid a \in \bfC(v) \}$ induces a subtree of $\bfT$.  If $u_i, u_j
  \in \bfV_a$ are neighbours, the definition above ensures $\nu_i(a) =
  \nu_j(a)$. By induction over the length of the shortest connecting path we
  obtain the same for arbitrary $u_i, u_j \in \bfV_a$.
    
  For every node $v_i$ there is a $j_i$ such that $v_i = u_{j_i}$ and we set
  $\pi_i = \nu_{j_i}$.  There are at most $2^{n \times (m+1)^m}$ distinct
  subsets of $cl(\psi,\{ 1, \dots, m, * \})$.  Hence, there must be two nodes
  $v_i, v_j$ such that $\pi_i(\Delta(v_i)) = \pi_j(\Delta(v_j))$ and, w.l.o.g.,
  $\bfN(v_i) < \bfN(v_j)$.  This implies that $v_j$ is blocked by $v_i$ via $\pi
  := \pi_i^{-1} \circ \pi_j$. Note that for $\pi$ to be well-defined, $\pi_i$
  must be injective.  By construction, $\pi$ preserves shared constants.  Since
  $\pi_i(\Delta(v_i)) = \pi_j(\Delta(v_j))$, $\pi(\Delta(v_j)) =
  \Delta(v_i)|_{\pi(\bfC(v_j))}$ holds. \qed
\end{proof}

\begin{lemma}[Termination]\label{theo:termination}
  Let $\psi \in \CGF$ be a sentence in NNF. Any sequence of rule
  applications of the tableau algorithm starting from the initial tree
  terminates.
\end{lemma}

\begin{proof}
  For any completion tree $\bfT$ generated by the tableau algorithm, we define
  $\|\cdot\| : \bfV \mapsto \mathbb{N}^3$ by 
  \begin{align*}
    \|v\| := (& |\bfC(v)|, \ \ n\times (m+1)^m - |\Delta(v)|, \\
    & |\{ \phi \in \Delta(v)  :  \phi \mbox{
      triggers the  \ruleex-rule for } v \}|) .
  \end{align*}
  The lexicographic order $\prec$ on $\mathbb{N}^3$ is well-founded, i.e.\ it
  has no infinite decreasing chains. Any rule application decreases $\|v\|$
  w.r.t.\ $\prec$ for at least one node $v$, and never increases $\|v\|$ w.r.t.\ 
  $\prec$ for an existing node $v$. However it may create new successors, one at
  a time. Since $\prec$ is well-founded, there can only be a finite number of
  applications of rules to every node in $\bfT$ and hence a finite number of
  successors and an infinite sequence of rule applications would generate a tree
  of infinite depth.
  
  Yet, as a corollary of Lemma~\ref{lem:tree-bounds}, we have that the depth of
  $\bfT$ is bounded by $2^{n \times (m+1)^m}$. For assume that there 
  is a path of length $> 2^{n \times (m+1)^m}$ in $\bfT$ with deepest node
  $v$. By the time $v$ has been created (by an application of the $\ruleex$-rule
  to its predecessor $u$), the path from the root of $\bfT$ to $u$ contains at
  least $2^{n \times (m+1)^m}$ nodes, and hence a blocked node.  This implies
  that $u$ is blocked too, and the $\ruleex$-rule cannot be applied to create
  $v$. \qed
\end{proof}



\subsubsection{Completeness}

\begin{lemma}\label{lem:completeness-alt}
  Let $\psi \in \CGF$ be a closed formula in \NNF. If $\psi$ is satisfiable,
  then there is a sequence of rule applications starting from the initial
  tree that yields a tableau.
\end{lemma}

\begin{proof}
  Since $\psi$ is satisfiable, there is a model $\mfA$ of $\psi$. We will use
  $\mfA$ to guide the application of the non-deterministic \gfruleor-rule. For
  this we incremently define a function $g : \bigcup \{ \bfC(v) \mid v \in
  \bfV \} \rightarrow A$ such that $\mbox{for all } v \in \bfV \;:\; \mfA
  \models g(\Delta(v))$. We refer to this property by $(\S)$.
  
  The set $\Delta(v)$ can contain atomic formulas $\alpha(\bfa^*)$, where $*$
  occurs at some positions of $\bfa^*$. The constant $*$ is not mapped to an
  element of $A$ by $g$. We deal with this as described just after
  Definition~\ref{def:any-element} by setting 
  \[
  \mfA \models g(\alpha(\bfa^*)) \; \mbox{ iff } \; \mfA \models \exists \bfz . g(\alpha(\bfa')) .
  \]

\begin{claim}\label{claim:comp1}
  If, for a completion tree $\bfT$, there exists a function
  $g$, such that $(\S)$ holds and a rule is applicable to $\bfT$, then it can be 
  applied in a way that maintains $(\S)$.
\end{claim}



  \begin{itemize}
  \item For the $\gfruleand$- and the $\gfruleor$-rule this is obvious.
  \item If the \ruleeq-rule is applicable to $v \in \bfV$ with $a = b \in
    \Delta(v)$, then, since $\mfA \models g(a) = g(b)$, $g(a)=g(b)$ must hold.
    Hence, for every node $w$ that shares $a$ with $v$, $g(\Delta(w)) =
    g(\Delta(w)[a \mapsto b])$ and the rule can be applied without violating
    $(\S)$.
  \item If the \rulefa-rule is applicable to $v \in \bfV$ with $(\fa \bfy \bfz .
    \alpha(\bfa,\bfy, \bfz))\phi(\bfa,\bfy) \in \Delta(v)$ and $\bfb \subseteq
    \bfC(v)$ with $\beta(\bfa,\bfb,* \dots *) \inast \Delta(v)$ for all atoms
    $\beta(\bfx,\bfy,\bfz) \in \alpha$, then, from the definition of $\inast$,
    there is a tuple $\bfc^* \subseteq \bfC(v)^*$, such that $\beta(\bfa,\bfb,*
    \dots *) \gtast \beta(\bfa,\bfb,\bfc^*)$ and $\beta(\bfa,\bfb,\bfc^*) \in
    \Delta(v)$.  From $(\S)$ we get that $\mfA \models \exists \bfz .
    \beta(g(\bfa),g(\bfb),\bfz)$ and since every $z \in \bfz$ appears exactly once
    in $\alpha$, also $\mfA \models \exists \bfz .  \alpha(g(\bfa),g(\bfb),
    \bfz)$. Hence, we have
    \[
    \{ \mfA \models \{ \fa \bfy  \bfz . \alpha(g(\bfa),\bfy,\bfz))
    \rightarrow \phi(g(\bfa),\bfy), \exists \bfz .  \alpha(g(\bfa),g(\bfb),
    \bfz) \},
    \]
    which, by Lemma~\ref{lem:guard-ex-fa}, implies $\mfA \models \phi(g(\bfa),
    g(\bfb))$ and hence $\phi(\bfa,\bfb)$ can be added to $\Delta(v)$ without
    violating $(\S)$.

  \item If the \ruleex-rule is applicable to  $v \in \bfV$ with $(\ex
    \bfy \bfz . \alpha(\bfa,\bfy, \bfz))\phi(\bfa,\bfy)$, then this implies
    \eqa \mfA \models g((\ex \bfy \bfz . \alpha(\bfa,\bfy,
    \bfz))\phi(\bfa,\bfy)).\neqa Hence, there are sequences $\bfb',
    \bfc' \subseteq A$  such that $\mfA \models \{ \alpha(g(\bfa),
    \bfb', \bfc'), \phi(g(\bfa),\bfb') \}$.  If we define $g$ such that
    $g(\bfb) = \bfb'$ and $g(\bfc) = \bfc'$, then  $\mfA
    \models \{ g(\alpha(\bfa,\bfb,\bfc), g(\phi(\bfa,\bfb)) \}$. Note, that this
    might involve setting $g(b_1) = g(b_2)$ for some $b_1,b_2 \in \bfb$.  With
    this construction the resulting extended completion-tree $\bfT$ and
    extended function $g$ again satisfy $(\S)$.
  \item If the \rulepropa-rule is applicable to $v \in \bfV$ with $\beta(\bfa^*)
    \in \Delta(v)$ and a neighbour $w$ with $\bfa^* \cap \bfC(w) \neq
    \emptyset$, then it adds $\beta(\bfa^*)\capast{\bfC(w)}$ to $\Delta(w)$.
    From $(\S)$ we get that $\mfA \models \beta(g(\bfa^*))$, and since
    $\beta(\bfb^*) := \beta(\bfa^*)\capast{\bfC(w)} \gtast \beta(\bfa^*)$, this
    implies $\mfA \models \beta(g(\bfb^*))$. Hence, adding $ \beta(\bfa^*)
    \capast \bfC(w) = \beta(\bfb^*)$ to $\Delta(w)$ does not violate $(\S)$.
  \item If the \rulepropfa-rule is applicable to a node $v \in \bfV$ with a
    universally quantified formula $\phi(\bfa) \in \Delta(v)$ and a neighbour $w$ which
    shares $\bfa$ with $v$, $(\S)$ yields $\mfA \models \phi(g(\bfa))$. Hence,
    adding $\phi(\bfa)$ to $\Delta(w)$ does not violate $(\S)$.
  \end{itemize}

\begin{claim}\label{claim:comp2} A completion-tree $\bfT$ for which a function
  $g$ exists such that $(\S)$ holds is clash free.
\end{claim}

  
  Assume that $\bfT$ contains a clash, namely, there is a node $v \in \bfV$
  such that either $a \neq a \in \bfV(v)$---implying $\mfA \models g(a)
  \neq g(a)$---, or that there is a sequence $\bfa \subseteq \bfC(v)$, and
  an atomic formula $\beta$ such that $\{ \beta(\bfa), \neg \beta(\bfa) \}
  \subseteq \Delta(v)$.  From $(\S)$, $\mfA \models \{
  \beta(g(\bfa)), \neg \beta(g(\bfa)) \}$ would follow, also a contradiction.
  
  These claims yield Lemma~\ref{lem:completeness-alt} as follows. Let $\bfT$ be
  a tableau for $\psi$. Since $\mfA \models \psi$, $(\S)$ is satisfied for the
  initial tree together with the function $g$ mapping $a_0$ to an arbitrary
  element of the universe of $\mfA$. By Lemma~\ref{theo:termination}, any
  sequence of applications is finite, and from Claim~\ref{claim:comp1} we get
  that there is a sequence of rule-applications that maintains $(\S)$. By
  Claim~\ref{claim:comp2}, this sequence results in a tableau. This completes
  the proof of Lemma~\ref{lem:completeness-alt}. \qed
\end{proof}

Lemma~\ref{lem:completeness-alt} involves two different kinds of
non-determinism, namely, the choice which rule to apply to which constraint
(as several rules might be applicable simultaneously), and which
disjunct to choose in an application of the $\gfruleor$-rule. While the latter
choice is \emph{don't-know} non-de\-ter\-min\-istic, i.e., for a satisfiable formula
only certain choices will lead to the discovery of a tableau, the former
choice is \emph{don't-care} non-deterministic.  This means that arbitrary
choices of which rule to apply next will lead to the discovery of a tableau
for a satisfiable formula. For an implementation of the tableau algorithm this
has the following consequences. Exhaustive search is necessary to deal with
all possible expansions of the $\gfruleor$-rule, but arbitrary strategies of
choosing which rule to apply next, and where to apply it, will lead to a
correct implementation, although the efficiency of the implementation will
strongly depend on a sophisticated strategy.




\subsubsection{Soundness}

In order to prove the correctness of the tableau algorithm we have to show that
the existence of a tableau for $\psi$ implies satisfiability of $\psi$. To this
purpose, we will construct a model from a tableau.  From the construction
employed in the proof we obtain an alternative proof of
Fact~\ref{fact:tree-model-property}.

\begin{lemma}\label{lem:treewidth}
  Let $\psi \in \CGF[\tau]$ with $k = \width(\psi)$ and let $\bfT$ be a
  tableau for $\psi$ generated by the tableau algorithm. Then $\psi$ is
  satisfiable and has a model of tree width at most $k-1$.
\end{lemma}

\begin{proof}
  Let $\bfT = (\bfV,\bfE,\bfC,\Delta,\bfN)$ a tableau for $\psi$.  For every
  direct blocking situation we fix a mapping $\pi$ verifying this blocking.
  Using an unraveling construction, we will construct a model $\mfA$ for $\psi$
  of width at most $k-1$ from $\bfT$. First, we ``unravel'' blocking situations
  in $\bfT$ by successively replacing every blocked node with a copy of the
  subtree of $\bfT$ rooted at the blocking node. Formally, this is achieved by
  the following path construction.  We define
  \[
  \bfV_u = \{ v \in \bfV \;:\; v \mbox{ is not blocked or directly blocked } \} .
  \]
  Since from now on we only deal with nodes from $\bfV_u$, every blocking is
  direct and we will no longer explicitly mention this fact.
  
  The set $\Paths(\bfT)$ is inductively defined by\footnote{This complicated
    form of unraveling, where we record both blocked and blocking node is
    necessary because there might be a situation where two successors $v_1,v_2$
    of a node are directly blocked by the same node $w$.}

  \begin{itemize}
  \item $[\ubpair{v_0}] \in \Paths(\bfT)$ for the root $v_0$ of $\bfT$,
  \item if $[\bpair{v_1}\dots \bpair{v_n}] \in \Paths(\bfT)$, the node $w$ is a
    successor of $v_n$ and $w$ is not blocked, then $[\bpair{v_1}\dots
    \bpair{v_n} \ubpair{w}] \in \Paths(\bfT)$,
  \item if $[\bpair{v_1}\dots \bpair{v_n}] \in \Paths(\bfT)$, $w$ is a successor
    of $v_n$  blocked by the node $u \in \bfV$, then $[ \bpair{v_1}
    \dots \bpair{v_n} \pair{u}{w}] \in \Paths(\bfT)$.
  \end{itemize}

  The set $\Paths(\bfT)$ forms a tree, with $p'$ being a successor of $p$ if
  $p'$ is obtained from $p$ by concatenating one element $\pair{u}{w}$ at the
  end.  We define the auxiliary functions $\Tail, \Tail'$ by setting
  $\Tail(p) = v_n$ and $\Tail'(p) = v_n'$
  for every path $p = [\bpair{v_1} \dots \bpair{v_n}]$. 
  
  Intuitively, for every node $v$ of $\bfT$, the paths $p \in \Paths(\bfT)$ with
  $v = \Tail(p)$ stand for distinct copies of $v$ created by the unraveling.  The
  universe of $\mfA$ consists of (classes of) constants labelling nodes in $\bfT$
  paired with the paths at whose $\Tail$ they appear to distinguish constants
  occurring at different copies of a node of $\bfT$. Formally, we define
  \[
  \bfC(\bfT) = \{ (a,p) \;:\; p \in \Paths(\bfT)  \wedge a \in \bfC(\Tail(p)) \}
  .
  \]
  Constants appearing at consecutive nodes of $\bfT$ stand for the same element
  and the same holds for constants related by a mapping $\pi$ verifying a block.
  Hence, to obtain the universe of $\mfA$, we factorize $\bfC(\bfT)$ as follows.
  Let $\sim$ be the smallest symmetric relation on $\bfC(\bfT)$ satisfying
  \begin{itemize}
  \item $(a,p) \sim (a,q)$ if $q$ is a successor of $p$ in $\Paths(\bfT)$,
    $\Tail'(q)$ is an unblocked successor of $\Tail(p)$, and $a \in
    \bfC(\Tail(p)) \cap \bfC(\Tail'(q))$,
  \item $(a,p) \sim (b,q)$ if $q$ is a successor of $p$ in $\Paths(\bfT)$,
    $\Tail'(q)$ is a blocked successor of $\Tail(p)$, $a \in \bfC(\Tail(p)) \cap
    \bfC(\Tail'(q))$, and $\pi(a) = b$ for the function $\pi$ that verifies that
    $\Tail'(q)$ is blocked by $\Tail(q)$.
  \end{itemize}
  With $\approx$ we denote the reflexive, transitive closure of $\sim$ and with
  $\apclass a p$ the  class of $(a,p)$, i.e., the set $\{ (b,q) \in
  \bfC(\bfT) \mid (b,q) \approx (a,p) \}$. Since
  $(a,p) \sim (b,q)$ iff $p,q$ are neighbours in $\Paths(\bfT)$, for every
  $(a,p)$, the set 
  \[
  \Paths(\apclass{a}{p}) := \{ q \mid \exists b. (b,q) \in \apclass a p \}
  \]
  is a subtree of  $\Paths(\bfT)$.
  
  The classes of $\bfC(\bfT)/\approx$ will be the elements of the universe of
  $\mfA$.  First we need to prove some technicalities for this construction.

  \begin{claim}\label{claim:tree1} Let $p \in \Paths(\bfT)$ and $a,b \in
    \bfC(\Tail(p))$. Then 
    $(a,p) \approx (b,p)$ iff $a = b$.
  \end{claim}
  

    Assume the claim does not hold and let $a \neq b $ with $(a,p) \approx
    (b,p)$. By definition of $\sim$, $(a,p) \not \sim (b,p)$ must hold.  Hence,
    there must be a path $(c_1,p_1) \sim \dots \sim (c_k, p_k)$ such that
    $a=c_1$, $b = c_k$, and $p= p_1 = p_k$. W.l.o.g., assume we have picked
    $a,b,p$ such that this path has minimal length $k$. Such a minimal path must
    be of length $k=3$, for if we assume a path of length $k > 3$, there must be
    $2 \leq i < j \leq k-1$ such that $p_i = p_j$, because the relation $\sim$
    is defined along paths in the tree $\Paths(\bfT)$. If $c_i = c_j$ then we
    can shorten the path between position $i$ and $j$ and obtain a shorter path.
    If $c_i \neq c_j$, then the path $(c_i,p_i) \sim \dots \sim (c_j,p_j)$ is
    also a shorter path with the same properties. Hence, a minimal path must be
    of the form $(a,p) \sim (c,q) \sim (b,p)$. If $\Tail'(q)$ is not blocked,
    by the definition of $\sim$, $a = c = b$ must hold. Hence, since $a \neq b$,
    $\Tail'(q)$ must be blocked by $\Tail(q)$. From the definition of $\sim$ we
    have $a,b \in \bfC(\Tail'(q))$ and $\pi(a) = c = \pi(b)$ for the function
    $\pi$ verifying that $\Tail'(q)$ is blocked by $\Tail(q)$.  Since $\pi$ must
    be injective, this is a contradiction.
  
  Since the set $\Paths(\bfT)$ is a tree, and as a consequence of
  Claim~\ref{claim:tree1}, we get the following.

  \begin{claim}\label{claim:tree2}
    Let $p,q \in \Paths(\bfT)$ with $p = [ \bpair{v_1} \dots \bpair{v_n}]$, $q
    = [ \bpair{v_1} \dots \bpair{v_n} \bpair{w}]$. If, for $a \in \bfC(v_n), b
    \in \bfC(w)$, $(a,p) \approx (b,q)$ then $(a,p) \sim (b,q)$.
  \end{claim}
  
    If $(a,p) \approx (b,q)$ then there must be a path $(c_1, p_1) \sim \dots
    \sim (c_k,p_k)$ such that $a = c_1$, $b = c_k$, $p = p_1$, and $q = p_k$.
    Since $\sim$ is only defined along paths in the tree $\Paths(\bfT)$, there
    must be a step from $p$ to $q$ (or, dually, from $q$ to $p$) in this path,
    more precisely, there must be an $i \in \{1, \dots k-1\}$ such that $p_i =
    p$ and $p_{i+1} = q$ holds. Hence, we have the situation
    \[
    (a,p) \approx (c_i,p) \sim (c_{i+1},q) \approx (b,q) .
    \]
    Claim~\ref{claim:tree1} implies $a = c_i$ and $b = c_{i+1}$ and hence $(a,p)
    \sim (b,q)$.
  
  Using Claim~\ref{claim:tree2}, we can show that the blocking condition and the
  $\rulepropa$- and $\rulepropfa$-rule work as desired.
  
  \begin{claim}\label{claim:tree3} Let $p, q \in \Paths(\bfT)$, $\bfa \subseteq
    \bfC(\Tail(p)), \bfb \subseteq \bfC(\Tail(q))$, $\bfa,\bfb$ non-empty
    tuples, and $(\bfa,p) \approx (\bfb, q)$.
    \begin{itemize}
    \item For every atom $\beta$, $\beta(\bfa,* \dots *) \inast
      \Delta(\Tail(p))$  iff $\beta(\bfb,* \dots *) \inast \Delta(\Tail(q))$.
    \item For every
      universally quantified $\phi$, $\phi(\bfa) \in \Delta(\Tail(p))$  iff $\phi(\bfb)
      \in \Delta(\Tail(q))$. 
    \end{itemize}
  \end{claim}
  
    Since both propositions are symmetric, we only need to prove one direction.
    If  $(\bfa,p) \approx (\bfb,q)$ with $\bfa = a_1 a_2 \dots a_m$ and $\bfb =
    b_1 b_2 \dots b_m$, then  
    \[
    \{p,q\} \subseteq \bigcap_{i=1}^m \Paths(\apclass{a_i} p) 
    \]
    and, as an intersection of subtrees of $\Paths(\bfT)$, $\bigcap_{i=1}^m
    \Paths(\apclass{a_i} p)$ is itself a subtree of $\Paths(\bfT)$. Hence, in
    $\Paths(\bfT)$ there is a path $p_1, \dots, p_k$ for which there exist
    tuples of constants $\bfc_1, \dots, \bfc_k$ with $(\bfc_1,p_1) \approx \dots
    \approx (\bfc_k,p_k)$, $p = p_1$, $q = p_k$, $\bfa = \bfc_1$, and $\bfb =
    \bfc_k$.  Since $\bfa,\bfb$ are non-empty, so are the $\bfc_i$. From
    Claim~\ref{claim:tree2}, we get that for any two neighbours $p_i,p_{i+1}$ in
    $\Paths(\bfT)$, $(\bfc_i,p_i) \approx (\bfc_{i+1},p_{i+1})$ implies
    $(\bfc_i,p_i) \sim (\bfc_{i+1},p_{i+1})$.
    
    By two similar inductions on  $i$ with $1 \leq i \leq k$ we show that if
    $\beta(\bfa, * \dots * ) \inast \Delta(\Tail(p))$ then $\beta(\bfc_i, *
    \dots * ) \inast \Delta(\Tail(p_i))$ and if $\phi(\bfa) \in
    \Delta(\Tail(p))$ then $\phi(\bfc_i) \in \Delta(\Tail(\bfc_i))$.
    
    For $i=1$ in both cases nothing has to be shown. Now assume that the we have
    shown these properties up to $i$. W.l.o.g., assume $p_{i+1}$ is a successor
    of $p_i$ in the tree $\Paths(\bfT)$. The other case is handled dually. There
    are two possibilities:
    \begin{itemize}
    \item $\Tail'(p_{i+1})$ is not blocked. Then $\Tail(p_{i+1}) =
      \Tail'(p_{i+1})$ and by the definition of $\sim$, $\Tail(p_{i+1})$ is a
      successor of $\Tail(p_i)$ in $\bfT$ and $\bfc_i = \bfc_{i+1}$ holds.
      
      If $\beta(\bfa, * \dots * ) \inast \Delta(\Tail(p))$ then $\beta(\bfc_i, *
      \dots * ) \inast \Delta(\Tail(p_i))$ holds by induction and due to the
      $\rulepropa$-rule, this implies $\beta(\bfc_{i+1}, * \dots *) \inast
      \Delta(\Tail(p_{i+1}))$. The $\rulepropa$-rule is applicable because, for
      the the non-empty tuple $\bfc_i$, $\bfc_i = \bfc_{i+1} \subseteq
      \bfC(\Tail(p_{i+1}))$ holds.
      
      If $\phi(\bfa) \in \Delta(\Tail(p))$ then by induction $\phi(\bfc_i) \in
      \Delta(\Tail(\bfp_i))$ and due to the $\rulepropfa$-rule this implies
      $\phi(\bfc_{i+1}) \in \Delta(\Tail(p_{i+1}))$.
    \item $\Tail'(p_{i+1})$ is blocked by $\Tail(p_{i+1})$ (with function
      $\pi$) and $\Tail'(p_{i+1})$ is a successor of $\Tail(p_i)$ in $\bfT$.
      Then, by definition of $\sim$, we have $\bfc_{i+1} = \pi(\bfc_i)$ and
      $\bfc_i \subseteq \bfC(\Tail(p_i)) \cap \bfC(\Tail'(p_{i+1}))$. 
      
      If $\beta(\bfa, * \dots * ) \inast \Delta(\Tail(p))$ then $\beta(\bfc_i, *
      \dots * ) \inast \Delta(\Tail(p_i))$ holds by induction and due to the
      $\rulepropa$-rule this implies $\beta(\bfc_i, * \dots *) \inast
      \Delta(\Tail'(p_{i+1}))$. The $\rulepropa$-rule is applicable because, for
      the non-empty tuple $\bfc_i$, $\bfc_i \subseteq \bfC(\Tail'(p_{i+1}))$
      holds.  The node $\Tail(p_{i+1})$ blocks $\Tail'(p_{i+1})$, which implies
      \[
      \pi(\beta(\bfc_i, * \dots *)) = \beta(\bfc_{i+1},* \dots *) \inast
      \Delta(\Tail(p_{i+1})) .
      \]

      If $\phi(\bfa) \in \Delta(\Tail(p))$ then by induction $\phi(\bfc_i) \in
      \Delta(\Tail(p_i))$ and due to the $\rulepropfa$-rule this implies
      $\phi(\bfc_i) \in \Delta(\Tail'(p_{i+1}))$. Since $\Tail(p_{i+1})$ blocks
      $\Tail'(p_{i+1})$, $\pi(\phi(\bfc_i)) = \phi(\bfc_{i+1}) \in
      \Delta(\Tail(p_{i+1}))$ holds. 
    \end{itemize}
  
    We now define the structure $\mfA$ over the universe $A =
    \bfC(\bfT)/{\approx}$. For a relation $R \in \tau$ of arity $m$, $R^\mfA$ is
    defined to be the set of tuples $(\apclass{a_1}{p_1}, \dots,
    \apclass{a_m}{p_m})$ for which there exists a path $p \in \Paths(\bfT)$ and
    constants $c_1, \dots c_m$ such that $(c_i,p) \approx ({a_i},{p_i})$ for all
    $1 \leq i \leq m$, and $Rc_1\dots c_m \in \Delta(\Tail(p))$.
  
  
  It remains to show that this construction yields $\mfA \models \psi$. This
  is a consequence of the following claim.
  
  \begin{claim}\label{claim:tree4} For every path $p \in \Paths(\bfT)$ and $\bfa
    \subseteq \bfC(\Tail(p))$, if $\phi(\bfa) \in \Delta(\Tail(p))$, then $\mfA
    \models \phi(\apclass{\bfa}{p})$.
  \end{claim}

  We show this claim by induction on the structure of $\phi$.  If $\phi(\bfa) =
  Ra_1 \dots a_m \in \Delta(\Tail(p))$, then the claim holds immediately by
  construction of $\mfA$.
    
  Assume $\phi(\bfa) = \neg R\bfa \in \Delta(\Tail(p))$, but $\apclass{\bfa}{p}
  \in R^\mfA$. Then, by the definition of $\mfA$, there must be a path $p'$ and
  constants $\bfc$ such that $(\bfa,p) \approx (\bfc,p')$ and $R \bfc \in
  \Delta(\Tail(p'))$. From Claim~\ref{claim:tree3} we have that $(\bfa,p)
  \approx (\bfc,p')$ implies $R \bfa \inast \Delta(\Tail(p))$ and, since $\bfa$
  contains no occurrence of $*$, $R \bfa \in \Delta(\Tail(p))$. Hence $\bfT$
  contains the clash $\{ R\bfa, \neg R\bfa \} \subseteq \Delta(\Tail(p))$, a
  contradiction to the fact that $\bfT$ is clash-free. Thus, $\apclass{\bfa}{p}
  \not \in R^\mfA$.
 
  Assume $\phi(\bfa) = a \neq b \in \Delta(\Tail(p))$ but $\apclass{a}{p} =
  \apclass{b}{p}$. From Claim~\ref{claim:tree1} we get that this implies $a=b$
  and hence $\bfT$ contains the clash $a \neq a \in \Delta(\Tail(p))$. Again,
  this is a contradiction to the fact that $\bfT$ is clash-free and
  $\apclass{a}{p} \neq \apclass{b}{p}$ must hold.
 
  For positive Boolean combinations the claim is immediate due to the $\gfruleand$- and
  $\gfruleor$-rule.
  
  Let $\phi(\bfa) = (\fa \bfy \bfz .  \alpha(\bfa,\bfy,\bfz))\chi(\bfa,\bfy) \in
  \Delta(\Tail(p))$ and $\bfb, \bfp, \bfc, \bfq$ arbitrarily chosen with
  \begin{equation}
    \mfA \models
    \alpha(\apclass{\bfa}{p},\apclass{\bfb}{\bfp}, \apclass{\bfc}{\bfq}) .
    \label{eqn:guard-satisfied}
  \end{equation}
  We need to show that also $\mfA \models \chi(\apclass{\bfa}{p},
  \apclass{\bfb}{\bfp})$ holds. In order to bring completeness of $\bfT$ and the
  $\rulefa$-rule into play, we show that information about the fact that
  (\ref{eqn:guard-satisfied}) holds is present at a \emph{single} node in $\bfT$
  where it triggers the \rulefa-rule. We rely on the fact that universal
  quantifiers must be guarded.

  Every $y_i \in \bfy$ coexists with every other variable $y_j \in \bfy$ in at
  least one atom $\beta^{(i,j)} \in \alpha(\bfa,\bfy,\bfz)$ and with every
  element $a_\ell \in \bfa$ in at least one atom $\gamma^{(i,\ell)} \in
  \alpha(\bfa,\bfy,\bfz)$.  For any two distinct variables $y_i,y_j$, $\mfA
  \models \beta^{(i,j)}(\apclass{\bfa}{p},\apclass{\bfb}{\bfp},
  \apclass{\bfc}{\bfq})$ holds and this can only be the case if there is a path
  $q^{(i,j)}$ and constants $d^{(i,j)}, e^{(i,j)}$ such that $(b_i,p_i) \approx
  (c^{(i,j)},q^{(i,j)})$ and $(b_j,p_j) \approx (d^{(i,j)},q^{(i,j)})$.
    
  Similarly, for every element $\apclass{b_i}{p_i} \in \apclass{\bfb}{\bfp}$ and
  every element $(a_\ell,p)$ there exists a path $r^{(i,\ell)}$ and constants
  $f^{(i,\ell)}, g^{(i,\ell)}$ such that $(b_i,p_i) \approx
  (f^{(i,\ell)},r^{(i,\ell)})$ and $(a_\ell,p) \approx
  (g^{(i,\ell)},r^{(i,\ell)})$. For every $i$ and $\ell$,
  $\Paths(\apclass{b_i}{p_i})$ and $\Paths(\apclass{a_\ell}{p})$ are subtrees of
  $\Paths(\bfT)$.
  
  The tree $\Paths(\apclass{b_i}{p_i})$ overlaps with the tree
  $\Paths(\apclass{b_j}{p_j})$ at $q^{(i,j)}$ and with the tree
  $\Paths(\apclass{a_\ell}{p})$ at $r^{(i,\ell)}$.  From this it follows
  \cite[Proposition 4.7]{Go80}  that there exists a common path 
  \[
  s \in \bigcap_{i} \Paths(\apclass{b_i}{p_i}) \cap \bigcap_\ell
  \Paths(\apclass{a_\ell}{p}) .
  \]
  Thus, there are tuples $\bfa'$, $\bfb'$ such that
  \begin{equation}
    \label{eqn:approx_lgf}
    (\bfa',s) \approx (\bfa,p) \mbox{ and } (\bfb',s) \approx (\bfb,\bfp)   .
  \end{equation}
  
  We now show that the preconditions of the $\rulefa$-rule are satisfied at
  $\Tail(s)$ for the formula $(\fa \bfy \bfz .
  \alpha(\bfa',\bfy,\bfz))\chi(\bfa',\bfy)$ and the tuple $\bfb'$. First, due to
  Claim~\ref{claim:tree3}, it holds that $(\forall \bfy \bfz.  \alpha(\bfa',\bfy,
  \bfz))\chi(\bfa',\bfy) \in \Delta(\Tail(s))$ because $(\bfa,p) \approx
  (\bfa',s)$ and $(\forall \bfy \bfz.  \alpha(\bfa,\bfy, \bfz))\chi(\bfa,\bfy)
  \in \Delta(\Tail(p))$.
  
  For every $\beta(\bfx,\bfy,\bfz) \in \alpha(\bfx,\bfy,\bfz)$,
  $\beta(\bfa',\bfb',* \dots *) \inast \Delta(\Tail(s))$ holds as follows:
  from (\ref{eqn:guard-satisfied}, \ref{eqn:approx_lgf}) we get
  \[
  \mfA \models
  \beta(\apclass{\bfa'}{s},\apclass{\bfb'}{s}, \apclass{\bfc}{\bfq}) .
  \]
  Since $\beta$ is an atom, this implies the existence of a path $t$ and
  tuples $\bfa'', \bfb'', \bfc'$ with
  \begin{equation}
    \begin{array}{cc}
      & (\bfa',s) \approx (\bfa'',t), \ (\bfb',s) \approx
      (\bfb'',t), \  (\bfc,\bfq) \approx (\bfc',t) \\
      \mbox{and} &
      \beta(\bfa'',\bfb'',\bfc') \in \Delta(\Tail(t))
      \label{eqn:approc_cgf2}
    \end{array}
  \end{equation}
  Clearly, $\beta(\bfa'',\bfb'',* \dots *) \gtast \beta(\bfa'', \bfb'', \bfc')$
  and, since $\beta(\bfa'', \bfb'', \bfc') \in \Delta(\Tail(t))$, it holds that
  $\beta(\bfa'',\bfb'',* \dots *) \inast \Delta(\Tail(t))$. Thus, by
  Claim~\ref{claim:tree3} it holds that $\beta(\bfa', \bfb', * \dots *) \inast
  \Delta(\Tail(s))$.
  
  Since this is true for every atom $\beta$, the preconditions of the
  $\rulefa$-rule are satisfied and the completeness of $\bfT$ yields
  $\chi(\bfa',\bfb') \in \Delta(\Tail(s))$. By induction, $\mfA \models
  \chi(\apclass{\bfa'}{s},\apclass{\bfb'}{s})$ holds and together with
  (\ref{eqn:approx_lgf}) this implies $\mfA \models
  \chi(\apclass{\bfa}{p},\apclass{\bfb}{\bfp})$. Since $\bfa,\bfp,\bfc,\bfq$
  have been chose arbitrarily, $\mfA \models \phi(\apclass{\bfa}{p})$ holds.
  
  If $\phi(\bfa) = (\ex \bfy \bfz . \alpha(\bfa,\bfy, \bfz)) \chi(\bfa,\bfy) \in
  \Delta(\Tail(p))$, there are two possibilities:
  \begin{itemize}
  \item there are $\bfb,\bfc \subseteq \bfC(\Tail(p))$ with
    $\{\alpha(\bfa,\bfb,\bfc),\chi(\bfa,\bfb) \} \subseteq \Delta(\Tail(p))$.  Then, by induction, we have
    \[
    \mfA \models \{ \alpha(\apclass{\bfa}{p},\apclass{\bfb}{p},
    \apclass{\bfc}{p}), \chi(\apclass{\bfa}{p},\apclass{\bfb}{p}) \}
    \]
    and hence $\mfA \models \phi(\apclass{\bfa}{p})$. 
  \item there are no such $\bfb,\bfc \subseteq \bfC(\Tail(p))$, then there is a
    successor $w$ of $\Tail(p)$ and  $\bfb,\bfc \subseteq
    \bfC(w)$ with $\{ \alpha(\bfa,\bfb, \bfc), \chi(\bfa,\bfb) \} \subseteq
    \Delta(w)$. The node $w$ can be blocked or not.
    
    If $w$ is not blocked, then $p' = [p,\pair{w}{w}] \in \Paths(\bfT)$ and by
    induction
    \[
    \mfA \models 
    \{ \alpha(\apclass{\bfa}{p'},\apclass{\bfb}{p'}, \apclass{\bfc}{p'}),
    \chi(\apclass{\bfa}{p'},\apclass{\bfb}{p'}) \} .
    \]
    From the definition of $\approx$ we have, $(\bfa,p') \approx (\bfa,p)$ and
    hence $\mfA \models \phi(\apclass{\bfa}{p})$.
   
    If $w$ is blocked by a node $u$ (with function $\pi$) then $p' = [p,
    \pair{u}{w}] \in \Paths(\bfT)$. From the blocking condition, we have that
    $u$ is unblocked and $\pi \{ \alpha(\bfa,\bfb, \bfc), \chi(\bfa,\bfb) \})
    \subseteq \Delta(u)$. Hence, by induction
    \[
    \begin{array}{rll}
      \mfA \models \{ &
      \alpha(\apclass{\pi(\bfa)}{p'},\apclass{\pi(\bfb)}{p'}, \apclass{\pi(\bfc)}{p'}),\\
      & \chi(\apclass{\pi(\bfa)}{p'},\apclass{\pi(\bfb)}{p'}) & \} ,
    \end{array}
    \]
    and, by definition of $\approx$, we have that $(\bfa,p) \approx (\pi(\bfa),
    p')$ and hence, $\mfA \models \phi(\apclass{\bfa}{p})$. 
  \end{itemize}

As a special instance of Claim~\ref{claim:tree4} we get that $\mfA \models
\psi$.  From Lemma~\ref{lem:tree-bounds}, we get that, for every node $v \in
\bfV$, $|\bfC(v)| \leq \width(\psi)$ and hence the tree $\Paths(\bfT)$ together
with the function $f : \Paths(\bfT) \rightarrow \bfC(\bfT)/{\approx}$ with $f(p) =
\bfC(\Tail(p))/{\approx}$ provides a tree decomposition of $\mfA$ of width $\leq
\width(\psi)-1$.  This completes the proof of Lemma~\ref{lem:treewidth}. \qed
\end{proof}

As an aside, together with Lemma~\ref{lem:completeness-alt}, the construction
used to prove Lemma~\ref{lem:treewidth} yields an alternative proof of
Fact~\ref{fact:tree-model-property}:

\begin{corollary}\label{cor:tmp-cgf}
\CGF, and hence also \LGF\ and \GF\ have the generalised tree model property.
\index{guarded fragment!clique guarded fragment!tree model property of the|textbf}
\index{guarded fragment!tree model property of the}
\end{corollary}

\begin{proof}
  Let $\psi \in \CGF[\tau]$ be satisfiable. Then, from
  Lemma~\ref{lem:completeness-alt} we get that there is a tableau $\bfT$ for
  $\psi$. By Lemma~\ref{lem:treewidth}, $\bfT$ induces a model for $\psi$ of
  tree width at most $\width(\psi) - 1$. Note that we have never relied on
  Fact~\ref{fact:tree-model-property} to obtain any of the results in this thesis
  and hence have indeed given an alternative proof for the generalised tree
  model property of \CGF. For \LGF\ and \GF, observe that the embedding of these
  logics into \CGF\ may increase the width of the sentence but not by more than
  a recursive amount. \qed
\end{proof}

Lemma~\ref{theo:termination}, \ref{lem:completeness-alt}, and
Lemma~\ref{lem:treewidth} yield correctness of the tableau algorithm for \CGF.

\begin{theorem}\label{theo:cgf-correct}
  The tableau algorithm is a decision procedure for \CGF-satisfiability.
\end{theorem}

An optimized implementation of this tableau algorithm is part of ongoing work.
It will be interesting to see if the tableau algorithm is amenable to the same
optimizations developed for modal or description logic tableau algorithms and how
it performs in comparison with the resolution based approach from
\cite{GanzNiv99}.



\cleardoublepage


\chapter{Summary}

The two major subjects of this thesis were (i) the worst-case complexity of
reasoning with expressive description logics, particularly in the presence of
counting operators; and (ii) the development of practical algorithms for
description and guarded logics. This chapter summarizes and comments on the main
results obtained on these topics.

\paragraph{Local Counting} Qualifying number restrictions introduce a form of
counting into DLs, which is \emph{local} because only statements about the
number of role successors of an individual are expressible.  Until now, the
impact of qualifying number restrictions on the complexity of reasoning was
unknown, if binary coding of numbers in the input is assumed. In this thesis, we
have shown that---in terms of worst-case complexity---qualifying number
restrictions do not lead to a rise in complexity of the reasoning problems.

Like for \alc, concept satisfiability for \alcq is \pspace-complete, even for
the case of binary coding of numbers in the input (Theorem
\ref{theo:completeness-binary-coding}).  The same applies to\alcqir
(Theorem~\ref{theo:alcqir-pspace-complete}), which extends \alcq with inverse
roles and safe role expressions, and is one of the most expressive DLs for which
qualifying number restrictions have been studied.

For the case of the other inference problems, we have shown
(Theorem~\ref{theo:alcqir-kb-exptime-complete}) that knowledge base
satisfiability for \alcqir is \exptime-complete also in the case of binary
coding of numbers in the input (Theorem~\ref{theo:alcqir-kb-exptime-complete}),
and hence again has the same complexity as the same problem for \alc. 

It is also possible to add qualifying number restrictions to DLs that allow for
transitive roles and role hierarchies without an increase in worst-case
complexity. Concept satisfiability (with or without general TBoxes) for \shiq is
\exptime-hard (Theorem~\ref{cor:shiq-exptime}), also if binary coding of numbers
in the input is assumed. This matches the complexity of the same problems for
the DL $\sh$, i.e., the fragment of \shiq that does not allow for inverse roles
or number restrictions.  To maintain satisfiability of the decision problems for
\shiq, it was necessary to allow qualifying number restrictions only over roles
that are neither transitive nor have transitive sub-roles. In the absence of
role hierarchies, the effect of number restrictions over transitive roles on
complexity and decidability is open.

\paragraph{Global Counting} If we additionally consider
constructors that allow to express \emph{global} counting statements like
cardinality restrictions or nominals,\footnote{At first sight, it might look odd
  that we subsume nominals under global counting. Yet, the requirement that
  nominals must be interpreted by singletons can be seen as a form of global
  counting and, indeed, Lemma~\ref{lem:cardres-vs-nominals} exhibits a close
  connection between reasoning with nominals and with cardinality restrictions.}
then this increases the complexity of the inference problems, independent on the
coding of numbers in the input.

We have shown that knowledge base satisfiability becomes \nexptime-hard if
cardinality restrictions are added to \alcq
(Theorem~\ref{theo:alcq-sat-is-nexptime-hard}) or \alcqi
(Theorem~\ref{theo:alcqi-sat-is-nexptime-hard}), while knowledge base
satisfiability for \alcq and \alcqi without cardinality restrictions is
\exptime-complete (as a Corollary of
Theorem~\ref{theo:alcqir-kb-exptime-complete}). The ``gap'' in the complexity is
even wider if we consider nominals: the complexity of concept satisfiability
rises from \pspace-complete to \nexptime-hard if nominals are added to \alcqi
(Corollary~\ref{cor:alcqio-concept-nexptime}).

A special case is the DL \alcqib, for which nominals or cardinality restrictions
can be added without a change in the complexity of the inference problems. Yet,
a closer look shows that cardinality restrictions (and hence nominals) can
already be expressed by \alcqib-concepts
(Lemma~\ref{lem:cadres-to-boolean-roles}), which explains that they do not have
an impact on the complexity.

\paragraph{Coding of Numbers} One of the recurring themes of the thesis has been
the impact that coding of numbers in the input has on the complexity of the
reasoning problems. With respect to this topic, we have obtained only an
incomplete picture.

All our results for local counting are independent on the coding of numbers and
one of the main contributions of this thesis is the development of algorithms
that deal with binary coding of numbers in the input without an additional
exponential overhead.

For logics that allow for global counting, we obtain tight complexity result
only if unary coding of numbers in the input is assumed. This is because the
upper \mbox{(\nexptime-)} bounds rely on a reduction to \ctwo, the two-variable
fragment of FOL with counting quantifiers, for which the exact complexity is
also known only for unary case. Of course, reasoning does not become easier in
the binary case, and so the lower bounds hold independently of the coding. For the
upper bounds, we only know that all problems can be solved in 2-\nexptime but we
do not have matching hardness-results. It is an interesting open question
whether exponential blow-up is necessary or whether are algorithms that can
deal with the binary case without an increase in complexity.

Until such algorithms are developed (or 2-\nexptime-hardness is proved), it is
open if the complexity of these reasoning problems rises when switching from
unary to binary coding of numbers in the input. The only case for which an
increase in complexity is certain is \alcq with cardinality restrictions, for
which satisfiability is \exptime-complete in the unary case
(Corollary~\ref{cor:alcq-exptime-complete}) and \nexptime-hard in the binary
case (Theorem~\ref{theo:alcq-sat-is-nexptime-hard}).

\paragraph{Practical Algorithms}

The practicality of inference algorithms, i.e., how easily they can be
implemented and optimized and how they behave on ``real world'' instances, is
important for their application in DL systems. In general, tableau algorithms
have proven to be amenable to a number of powerful optimization techniques and
are successfully employed in many DL systems. However, a tableau algorithm is
not practical \emph{just because} it is a tableau algorithm. One important
criterion for the practicality of an algorithm seems to be the degree to which
is depends on non-deterministic choices because, in a (necessarily
deterministic) implementation, the different possibilities have to be searched
exhaustively in order to obtain a complete algorithm.  This search is what
is responsible for most of the runtime of tableau algorithms and most of the
aforementioned optimizations aim to reduce the size of the search space.

So, one of the major design principles of the \si-, \shiq-, and \CGF-algorithm
in this thesis was to avoid non-deterministic choices as much as possible.  The
\shiq-algorithm developed in this thesis (Algorithm~\ref{alg:shiq}), forms the
basis of the highly optimized DL system \iFaCT \cite{horrocks99:_fact} that
shows good performance in system comparisons \cite{Horrocks-Tableaux-2000} and
is successfully applied in applications \cite<see, e.g.,>{FranconiNg-KRDB-2000}.
One problem of the \shiq-algorithm lies in the non-deterministic identification
of nodes due to its \ruleleq-rule. The development of optimization techniques
that specifically deal with this problem is part of ongoing work. Moreover, it
will be interesting to extend the refined blocking strategy developed for the
\si-algorithm to \shiq and implement it in the \iFaCT system. We have claimed
that the tableau algorithm for \CGF\ (Algorithm~\ref{alg:cgf}) is useful as the
basis of an efficient reasoner and the implementation of such a system is in
progress. It will be interesting to see how this implementation performs in
comparison with the existing decision procedures for guarded fragments based on
refinements of general FOL theorem proving techniques.

In the \alcq-algorithm (Algorithm~\ref{alg:alc}) and the \alcqir-algorithm
(Algorithm~\ref{alg:alcqir}), we have freely used non-determinism in order to
obtain a space-efficient algorithm. This implies that they seem to be less
suited for an implementation because of their highly non-deterministic
\rulegeq-rule.  

For the remaining algorithms developed in this thesis, it is at
least questionable if they can serve as the basis of an efficient implementation.
This is especially the case for the decision procedure used to prove
\exptime-completeness of concept satisfiability of \alcqir \wrt general TBoxes
(see Theorem~\ref{theo:alcqir-tbox--exptime-complete}), which is based on a
highly inefficient automata-construction. It is even less likely that an
efficient decision procedure can be obtained from our decision procedures for
knowledge base satisfiability for \alcqir (see
Theorem~\ref{theo:alcqir-kb-exptime-complete}) or from the worst-case optimal
decision procedures for \shiq (see Corollary~\ref{cor:shiq-exptime} and
Corollary~\ref{cor:shiq-kb-exptime}) because these add a wasteful pre-completion
technique and various translations on top of the already inefficient algorithm
for \alcqir with general TBoxes.

All decision procedures for the \nexptime-hard DLs presented in this thesis
employ a reduction to \ctwo, for which the only known decision procedures work
by model enumeration and so there exists no decision procedure for these logics
that could be of practical use. This situation is particularly unsatisfactory
for the DL \shiqo, for which such a decision procedure would be of high interest
due to \shiqo's r\^ole for inferences for the semantic web
\cite{fensel00:_oil_nutsh,horrocks01:_ontol_reason_seman_web}. Maybe the most
intriguing question left open by this thesis is how practical decision
procedures for \nexptime-complete modal and description logics can be developed.


\cleardoublepage



\NoCommaBetweenTwoAuthors
\bibliographystyle{theapa}
\bibliography{strings,mybib,proceedings}

\cleardoublepage


\clearpage\addcontentsline{toc}{chapter}{\protect List of Figures}
\listoffigures

\cleardoublepage

\end{document}
